\RequirePackage{pdf14}

\documentclass[letterpaper]{article}
\usepackage[usenames,dvipsnames,table]{xcolor}
\usepackage{graphicx}
\usepackage{caption}
\usepackage{subcaption}
\usepackage{array,setspace,mathrsfs,amsfonts,amsmath,yfonts,dsfont,bbm,colonequals,mathtools}
\usepackage{./aux/tikz-cd}
\usepackage{./aux/jheppub}
\usepackage{afterpage}

\newcommand\qq{\mathbbmtt{Q}}

\newcommand\eea{\end{eqnarray}}
\newcommand\ee{\end{equation}}

\newcommand{\goodchi}{\protect\raisebox{1pt}{$\chi$}}

\DeclarePairedDelimiter\floor{\lfloor}{\rfloor}

\newcommand\mf{\mathfrak}

\newcommand\nn{\nonumber}
\newcommand\ph{\phantom}
\newcommand\wt{\widetilde}
\newcommand\eg{{\it e.g.}}
\newcommand\ie{{\it i.e.}}
\newcommand\cf{{\it cf.}}

\newtheorem{conj}{Conjecture}

\newcommand\tr{\text{Tr}}

\newcommand\D{\Delta}
\renewcommand\aa{\alpha}
\newcommand\bb{\beta}

\newcommand\aad{\dot{\aa}}
\newcommand\bbd{\dot{\bb}}

\newcommand\Cb{\mathbb{C}}
\newcommand\Rb{\mathbb{R}}
\newcommand\Zb{\mathbb{Z}}
\newcommand\Nb{\mathbb{N}}
\newcommand\Pb{\mathbb{P}}

\renewcommand\AA{\mathcal{A}}
\renewcommand\SS{\mathcal{S}}

\newcommand\BB{\mathcal{B}}
\newcommand\CC{\mathcal{C}}
\newcommand\DD{\mathcal{D}}
\newcommand\EE{\mathcal{E}}

\newcommand\GG{\mathcal{G}}
\newcommand\HH{\mathcal{H}}
\newcommand\II{\mathcal{I}}
\newcommand\JJ{\mathcal{J}}
\newcommand\KK{\mathcal{K}}

\newcommand\MM{\mathcal{M}}
\newcommand\NN{\mathcal{N}}
\newcommand\OO{\mathcal{O}}

\newcommand\PP{\mathcal{P}}
\newcommand\QQ{\mathcal{Q}}
\newcommand\RR{\mathcal{R}}

\newcommand\XX{\mathcal{X}}

\newcommand\zb{{\bar z}}

\newcommand\dce{h^{\vee}}

\def\QQ{\mathcal{Q}}

\def\ad{\dot{\alpha}}
\def\bd{\dot{\beta}}

\def\Tr{\mbox{Tr}}

\def\a{\alpha}
\def\b{\beta}

\def\ph{\phantom}

\def\QQ{\mathcal{Q}}

\def \sof{{\mf{so}}}
\def \suf{{\mf{su}}}
\def \slf{{\mf{sl}}}
\def \ef{{\mf{e}}}
\def \gf{{\mf{g}}}
\def \hhf{{\mf{h}}}
\def \Rf{{\mf{R}}}

\def\SU{\mathrm{SU}}

\def\hf{\frac{1}{2}}

\def\del{\partial}

\newcommand{\Gt}{{\wt G}}

\newcommand\restr[2]{{% we make the whole thing an ordinary symbol
  \left.\kern-\nulldelimiterspace % automatically resize the bar with \right
  #1 % the function
  \vphantom{\big|} % pretend it's a little taller at normal size
  \right|_{#2} % this is the delimiter
  }}
\newcommand\fverb{\setbox\fverbbox=\hbox\bgroup\verb}
\newcommand\fverbdo{\egroup\medskip\noindent%
			\fbox{\unhbox\fverbbox}\ }
\newcommand\fverbit{\egroup\item[\fbox{\unhbox\fverbbox}]}
\newbox\fverbbox

\title{The \texorpdfstring{$\NN=2$}{N=2} superconformal bootstrap}
\author[1]{Christopher Beem,}
\author[2]{Madalena Lemos,}
\author[3]{Pedro Liendo,}
\author[2]{Leonardo Rastelli,}
\author[4]{Balt C. van Rees}

\affiliation[1]{Institute for Advanced Study, Einstein Drive, Princeton, NJ 08540, USA}
\affiliation[2]{C.~N.~Yang Institute for Theoretical Physics, Stony Brook University, Stony Brook, NY 11794-3840, USA}
\affiliation[3]{IMIP, Humboldt-Universit{\"a}t zu Berlin, IRIS Adlershof, Zum Gro{\ss}en Windkanal 6, 12489 Berlin, Germany}
\affiliation[4]{Theory Group, Physics Department, CERN, CH-1211 Geneva 23, Switzerland}

\preprint{HU-EP-14/61, CERN-PH-TH-2014-269, YITP-SB-14-54}

\bigskip
\abstract{
In this work we initiate the conformal bootstrap program for $\NN=2$ superconformal field theories in four dimensions. We promote an abstract operator-algebraic viewpoint in order to unify the description of Lagrangian and non-Lagrangian theories, and formulate various conjectures concerning the landscape of theories. We analyze in detail the four-point functions of flavor symmetry current multiplets and of $\NN=2$ chiral operators. For both correlation functions we review the solution of the superconformal Ward identities and describe their superconformal block decompositions. This provides the foundation for an extensive numerical analysis discussed in the second half of the paper. We find a large number of constraints for operator dimensions, OPE coefficients, and central charges that must hold for any $\NN=2$ superconformal field theory.
}


\begin{document}
\setcounter{tocdepth}{2}
\maketitle
\setcounter{page}{0}

%!TEX root = ../draft_maxi_Neq2.tex

\section{Introduction}
\label{sec:intro}

In this work we initiate the conformal bootstrap program for four-dimensional conformal field theories with $\NN=2$ supersymmetry. These theories are extraordinarily rich, both physically and mathematically, and have been studied intensively from many viewpoints. Nevertheless, we feel that a coherent picture is still missing. We hope that the generality of the conformal bootstrap framework will allow such a picture to be developed. We also feel the time is ripe for such an investigation -- the recent explosion of results for $\NN=2$ superconformal field theories (SCFTs) calls out for a more systematic approach, while the methods first introduced in \cite{Rattazzi:2008pe} have reinvigorated the conformal bootstrap \cite{Ferrara:1972cq,Ferrara:1973vz,Ferrara:1973yt,Ferrara:1974pt,Ferrara:1974nf,Ferrara:1974ny,Polyakov:1974gs} with a powerful and flexible toolkit for studying conformal field theories.

The first examples of $\NN=2$ superconformal field theories (SCFTs) were relatively simple gauge theories with matter representations chosen so that the beta functions for all gauge couplings would vanish. Since then, the library of known theories has grown in size, with the new additions including many Lagrangian models \cite{yuji}, but remarkably also many theories that appear to admit no such description. In particular, the class $\SS$ construction of \cite{Gaiotto:2009we,Gaiotto:2009hg} gives rise to an enormous landscape of theories, most of which resist description by conventional Lagrangian field theoretic techniques. Despite this abundance, the current catalogue seems fairly structured, and one may reasonably suspect that a complete classification of $\NN=2$ superconformal field theories (SCFTs) will ultimately be possible. The development of the $\NN=2$ superconformal bootstrap seems an indispensable step towards this ambitious goal.

Our first task is to introduce an abstract operator-algebraic language for $\NN=2$ SCFTs. In this reformulation, we retain only the vector space of local operators (organized into representations of the superconformal algebra), and the algebraic structure on this vector space defined by the operator product expansion. From this viewpoint, we can see that a theory is free (or contains a free factor) if its operator spectrum includes higher spin currents; we can see that a theory has a Higgs branch of vacua if its operator algebra includes an appropriate chiral ring that is the coordinate ring of an affine algebraic variety; and so on and so forth. Representation theory of the $\NN=2$ superconformal algebra proves an invaluable tool, as its shortened representations neatly encode different facets of the physics. This algebraic viewpoint is remarkably rich, and we have have dedicated the next section to its extensive presentation. 

Once equipped with the proper language, we can make an informed decision on where and how to employ numerical bootstrap methods. We explain that there are three classes of four-point functions that should be the starting point for any systematic exploration of this type: the stress-tensor four-point function; the moment map four-point function; and the four-point function of $\NN=2$ chiral operators. In the present work, we report on numerical investigations into specific examples of the latter two classes. The requisite superconformal block expansion for the first correlator, which is the most universal, is not yet available, so this case is left for future work. The moment map four-point function is related to the flavor symmetry of the theory, and we focus on the cases of $\suf(2)$ and $\ef_6$. The $\suf(2)$ case is clearly the simplest and is a natural starting point, while $\ef_6$ case is interesting because exceptional flavor symmetries cannot appear in any Lagrangian field theory, and $\ef_6$ is (among others) the simplest case to bootstrap after $\suf(2)$. On the other hand, the four point function of $\NN=2$ chiral operators gives us access to a very different aspect of the physics, namely the Coulomb branch chiral ring.

There are two broad types of questions that we can hope to address by bootstrap methods. First of all, we can \emph{constrain the space of consistent $\NN=2$ SCFTs}. There are a number of universal structures that appear throughout the $\NN=2$ catalogue that cannot be satisfactorily explained in the abstract bootstrap language. Are Coulomb branch chiral rings always freely generated? Are central charges bounded from below by those of free theories, or are there exotic theories with even lower central charges? Is every $\NN=2$ conformal manifold parametrized by gauge couplings? As we will see, these questions can sometimes be connected with the constraints of crossing symmetry, and then numerical analysis can provide (partial) answers.

Our second motivation is to \emph{learn more about specific $\NN=2$ SCFTs}. There are many cases where supersymmetry can tell us a lot about an $\NN=2$ SCFT even when we have no Lagrangian description. In many examples we know, \eg, the central charges (including flavor central charges), the spectrum of protected operators, and some OPE coefficients associated with protected operators. This partial knowledge can be used as \emph{input} for a numerical bootstrap analysis. Optimistically, we may hope that this protected data and the constraints of crossing symmetry are enough to determine the theory uniquely. The bootstrap may then allow us to effectively solve the theory along the lines of what has been done for the three dimensional Ising CFT \cite{ElShowk:2012ht,El-Showk:2014dwa,Kos:2014bka}. Because the bootstrap is completely nonperturbative in nature, it is a natural tool for studying intrinsically strongly coupled (non-Lagrangian) theories. In fact, when it comes to studying unprotected operators in a non-Lagrangian theory, the bootstrap is really the only game in town.

The detailed organization of the paper can be found in the table of contents. In the first part (sections 2-4) we develop the algebraic viewpoint and the details of the superconformal block expansion for the two classes of correlators that we consider, while in the second part (sections 5-8) we present our numerical investigations. Several appendices complement the main text with technical and reference material.
%!TEX root = ../draft_maxi_Neq2.tex

\section{The \texorpdfstring{$\NN=2$}{N=2} superconformal bootstrap program}
\label{sec:philosophy}

In the bootstrap approach to conformal field theories, one adopts an abstract viewpoint that takes the algebra of local operators as the primary object. On the other hand, the majority of conventional wisdom and communal intuition about $\NN=2$ field theories arises from a Lagrangian -- or at least quasi-Lagrangian -- perspective. This leads to something of a disconnect. The bootstrap perspective is likely to be unfamiliar to many experts in supersymmetric field theory, while amongst readers with a background in the conformal bootstrap the additional structure that follows from $\NN=2$ supersymmetry may not be well known. In this section we will try to bridge this divide. 

%%%%%%%%%%%%%%%%%%%%%%%%%%%%%%%%%%%%%%%%%%%%%%%%%%%%%%%%%%%%%%%%%%%%%%%%%%%%%%%%%%%%%%%%%%%%%%%%%%%%%%%%%%%%
\subsection{The insufficiency of Lagrangians}
\label{subsec:N2_lagrangians}
%%%%%%%%%%%%%%%%%%%%%%%%%%%%%%%%%%%%%%%%%%%%%%%%%%%%%%%%%%%%%%%%%%%%%%%%%%%%%%%%%%%%%%%%%%%%%%%%%%%%%%%%%%%%

Let us recall some aspects of Lagrangian $\NN = 2$ field theories, which provide a historical foundation of the subject and help to guide our thinking even for the non-Lagrangian theories discussed below. The building blocks of an $\NN=2$ four-dimensional Lagrangian are vector multiplets, transforming in the adjoint representation of a gauge group $G$, and hypermultiplets (the \emph{matter content}), transforming in some representation $R$ of $G$.\footnote{More generally, for appropriate choices of gauge group one can allow for ``half-hypermultiplets'', \ie, $\NN=1$ chiral multiplets, transforming in pseudo-real representations of $G$. See, \eg, \cite{yuji} for a recent discussion.} For the theory to be microscopically well-defined, the gauge group should contain no abelian factors,\footnote{An exception is when no hypermultiplet is charged under the abelian factors, in which case there are decoupled copies of the free vector multiplet SCFT in the theory.} so we can take $G$ to be semi-simple,
%%%%%%
\begin{equation}
G= G_1 \times G_2 \times \cdots G_n~.
\end{equation} 
%%%%%%
To each simple factor $G_i$ is associated a complexified gauge coupling $\tau_i \in \Cb$, ${\rm Im}\; \tau_i > 0$, and for each choice of $(G, R, \{ \tau_i\})$  there is a unique, classically conformally invariant $\NN=2$ Lagrangian. For the \emph{quantum} theory to be conformally invariant, the matter content must be chosen so that the one loop beta functions for the gauge couplings vanish. Thanks to $\NN=2$ supersymmetry, this is also a sufficient condition at the full quantum level. 

The classification of the pairs $(G,R)$ that lead to $\NN=2$ SCFTs can therefore be reduced to a purely combinatorial problem, whose complete solution has been described recently in \cite{yuji}. The simplest examples are $\NN=2$ superconformal QCD, which has gauge group $G = SU(N_c)$ and $N_f = 2 N_c$ hypermultiplets in the fundamental representation, and $\NN=4$ super Yang-Mills theory (which can be regarded as an $\NN=2$ SCFT), for which $G$ is any simple group and the hypermultiplets transform in the adjoint representation.
 
The \emph{conformal manifold} of a CFT is the space of theories that can be realized by deforming a given CFT by exactly marginal operators. In a slight abuse of terminology we often refer to the conformal manifold of an $\NN=2$ SCFT as the (not necessarily proper) submanifold of the full conformal manifold where in addition the full $\NN=2$ supersymmetry is preserved. For a Lagrangian theory this submanifold coincides with the space of gauge couplings $\{\tau_i\}$, up to the discrete identifications induced by generalized $S$-dualities.\footnote{Because the action of $S$-duality can have fixed points in the space of gauge couplings, the conformal manifold may have orbifold points, so it may not really be a \emph{manifold}.} The conformal manifold comes endowed with a metric -- the Zamolodchikov metric -- which is K\"ahler and with respect to which the weak coupling points (where some $\tau_i\to\infty$ in some $S$-duality frame) are at infinite distance as measured from the interior. Thus the conformal manifold of any $\NN=2$ Lagrangian SCFT is non-compact with boundaries where gauge couplings are turned off.
 
Lagrangian theories also always possess nontrivial moduli spaces of supersymmetric vacua. The simplest parts of the moduli space are the \emph{Coulomb branch} and the \emph{Higgs branch}. The Coulomb branch consists of vacua where the complex scalar fields $\varphi_i$ in the vector multiplets acquire nonzero vacuum expectation values (vevs), while the complex scalars $(q,\tilde q)$ in the hypermultiplets are set to zero -- this branch is characterized by the fact that $SU(2)_R$ is unbroken, while $U(1)_r$ is broken. Alternatively, on the Higgs branch only the hypermultiplet scalars get nonzero vevs, and this branch is characterized by $SU(2)_R$ breaking with $U(1)_r$ preserved. There can also be \emph{mixed branches} where the entire $R$-symmetry is broken, though we will not have much to say about mixed branches in this paper.

The best way to parametrize these moduli spaces is by the vevs of gauge-invariant combinations of the elementary fields. The Coulomb branch is parametrized by the vevs of operators of the form $\{\Tr\,\varphi^k \}$. These operators form a freely generated ring, called the \emph{Coulomb branch chiral ring}, with generators in one-to-one correspondence with the Casimir invariants of the gauge group. Similarly, the Higgs branch can be parametrized by the vevs of gauge invariant composites of the hypermultiplet scalars. These operators also form a finitely generated ring, the \emph{Higgs branch chiral ring}. The Higgs branch chiral ring is generally \emph{not} freely related, but rather has relations so that the Higgs branch acquires a description as an affine complex algebraic variety. Alternatively, the Higgs branch can be expressed as a Hyperk\"ahler quotient \cite{Hitchin:1986ea}.

%%%%%%%%%%%%%%%%%%%%%%%%%%%%%%%%%%%
\subsubsection*{Isolated SCFTs and quasi-Lagrangian theories}
%%%%%%%%%%%%%%%%%%%%%%%%%%%%%%%%%%%

Lagrangian SCFTs make up only small subset of all SCFTs. A wealth of strongly coupled $\NN=2$ SCFTs with no marginal deformations are known to exist -- by virtue of being isolated, they cannot have a conventional Lagrangian description. One particularly elegant way to find such isolated theories is through generalized $S$-dualities of the kind discussed in \cite{Argyres:2007cn}. By taking a Lagrangian theory and dialing a marginal coupling all the way to infinite strength, one may recover a weakly gauged \emph{dual} description which involves one or more isolated SCFTs and a set of vector multiplets to accomplish the gauging. In this dual description the gauging procedure is described in what we may call a \emph{quasi-Lagrangian} fashion: the isolated SCFT is treated as a non-Lagrangian black box with a certain flavor symmetry, which is allowed to talk to the vector multiplets through minimal coupling of the conserved flavor current of the isolated SCFT to the gauge field. The one-loop beta function for each simple gauge group factor is given by
%%%%%%
\begin{equation}
\beta = -h^\vee + 4 k~,
\end{equation}
%%%%%%
where $h^\vee$ is the dual Coxeter number of the group and $k$ the flavor central charge, defined from the two-point function of the conserved flavor current. This determines a simple condition for when non-Lagrangian theories can be.  (Of course, this expression for $\beta$ applies also to the Lagrangian case, where the flavor current is a composite operator made of the hypermultiplet fields.) 

The web of generalized $S$-dualities for large classes of theories can be elegantly described through the class $\SS$ constructions of \cite{Gaiotto:2009we,Gaiotto:2009hg}. These theories arise from twisted compactifications of the six-dimensional $(2,0)$ theories on a punctured Riemann surface, with additional discrete data specified at each puncture. The marginal deformations of the four-dimensional theory correspond to the moduli of the Riemann surface, and weakly gauged theories arise if the Riemann surface degenerates. In this picture the isolated theories correspond to three-punctured spheres which have no continuous moduli. They do, however, depend on the discrete data at the three punctures as well as on a choice of $\gf \in \{ A_n, D_n, E_n\}$ for the six-dimensional ancestor theory. In this way several infinite classes of isolated theories can be constructed. A few of these theories turn out to be equal to theories of free hypermultiplets, but most cases do not admit a Lagrangian description.

Another large class of isolated theories are the Argyres-Douglas fixed points \cite{Argyres:1995jj} which describe the infrared physics at special points on the Coulomb branch of another $\NN = 2$ theory. At these distinguished points several BPS particles with mutually non-local charges become simultaneously massless, which precludes any Lagrangian description of the infrared theory. Alternatively, many Argyres-Douglas fixed points can be constructed in class $\SS$ by allowing for irregular singularities on the UV curve \cite{Xie:2012hs}. Argyres-Douglas theories have also recently been used as building blocks in a quasi-Lagrangian set-up \cite{Buican:2014hfa}.

In order to describe the currently known landscape of $\NN = 2$ SCFTs, then, it is clearly not sufficient to only consider Lagrangians with hypermultiplets and vector multiplets. We can certainly accommodate any theory in a framework which takes as fundamental the spectrum and algebra of local operators. This is the basic starting point for the bootstrap approach that we take in this paper. The remainder of this section is dedicated to the development of such a framework.

%%%%%%%%%%%%%%%%%%%%%%%%%%%%%%%%%%%%%%%%%%%%%%%%%%%%%%%%%%%%%%%%%%%%%%%%%%%%%%
\subsection{The bootstrap philosophy}
\label{subsec:abstract_CFT_viewpoint}
%%%%%%%%%%%%%%%%%%%%%%%%%%%%%%%%%%%%%%%%%%%%%%%%%%%%%%%%%%%%%%%%%%%%%%%%%%%%%%

In the bootstrap approach, we take a (super)conformal field theory to be characterized by its local operator algebra.\footnote{In adopting this perspective, we are therefore willfully ignoring the complications associated with including non-local observables -- such as Wilson line operators in conformal gauge theory -- and non-trivial spacetime geometries.} The aim is then to understand the constraints imposed upon such algebras by (super)conformal invariance, associativity, and unitarity. This approach dates back to the foundational papers of \cite{Ferrara:1972cq,Ferrara:1973vz,Ferrara:1973yt,Ferrara:1974pt,Ferrara:1974nf,Ferrara:1974ny,Polyakov:1974gs}. See, \eg, \cite{Rattazzi:2008pe, slavalectures} for modern expositions. We will briefly recall the general logic, while placing particular emphasis on the role played by short representations of the conformal algebra. In the next subsection we describe the special features that arise in the $\NN=2$ superconformal case.

The local operators $\{\OO_i(x)\}$ of a CFT form a vector space that is endowed with a product that gives it something like an associative algebra structure. The product for local operators is known as the Operator Product Expansion (OPE), and takes the schematic form
%%%%%%
\begin{equation}\label{eq:OPE}
\OO_1 (x) \OO_2 (y) = \sum_k c_{12 k} (x-y) \OO_k (y)~.
\end{equation}
%%%%%%
Any correlation function of separated local operators in flat spacetime $\mathbb{R}^d$ can be evaluated by successive applications of the OPE, which is an absolutely convergent expansion. The OPE follows as a straightforward consequence of the \emph{state/operator correspondence}.\footnote{See \cite{Pappadopulo:2012jk} for a recent discussion.} To each local operator is associated a state, obtained by acting on the vacuum with the operator inserted at the origin,
%%%%%%
\begin{equation}
\OO (x) \to | \OO \rangle \colonequals \OO(0) | 0 \rangle~,
\end{equation}
%%%%%%
and conversely each state defines a unique local operator,
%%%%%% 
\begin{equation}
|\psi \rangle \to \OO_\psi (x)~.
\end{equation}
%%%%%%
As customary, we will use the language of operators or states interchangeably. 
  
To completely specify a CFT at the level of correlators of local operators, it is therefore sufficient to list the set of local operators (that is, the set of their quantum numbers) and the structure constants appearing in their OPEs. Conformal invariance streamlines the presentation of this information. First, it allows the local operators to be assembled into conformal families, each of which transforms as a highest weight representation of the conformal algebra $\sof(d,2)$. The highest weight state, known as the conformal primary, is annihilated by all raising operators in the conformal algebra, notably the special conformal generators $K_\mu$. Specializing to the four-dimensional case, a representation $\RR [\Delta,j_1,j_2]$ of $\sof(4,2) \cong \suf(2,2)$ is labelled by the quantum numbers of the primary, namely its conformal dimension $\Delta$ and its Lorentz spins $(j_1,j_2)$. If the theory enjoys an additional global symmetry $G_F$, then the local operators can be further organized into $G_F$ representations, labelled by some flavor symmetry quantum numbers ${f}$, and the full representations are then denoted as $\RR [\Delta,j_1,j_2;f]$. Conformal symmetry also restricts the spacetime dependence of the functions $c_{ijk} (x)$ appearing in the OPE \eqref{eq:OPE}. In particular, the functions $c_{ijk}(x)$ are uniquely determined in terms of the quantum numbers of the representations $\RR_i$, $\RR_j$, and $\RR_k$ and the coefficients $\lambda_{ijk}^s$ that parametrize their three-point functions.%
\footnote{In the simplest case of three spacetime scalars (with no additional flavor charges), the three-point function is completely fixed up to a single overall coefficient $\lambda_{ijk}$. In general there are multiple parameters $\lambda_{ijk}^s$, $s =1, \dots {\rm mult}(ijk)$, where the (finite) multiplicity ${\rm mult}(ijk)$ is given by the number of independent conformally covariant tensor structures that can be built from the three reps $\RR_{i,j,k}$.} %
All told, the data that fully specify the local theory amount to a countably infinite list 
%%%%%%
\begin{equation}
\{ a_i, \lambda^s_{ijk} \}~, \quad a_i \colonequals (\Delta, j_1, j_2, {f})_i ~.
\end{equation}
%%%%%%
These data are constrained by the requirements that the theory be unitary and that the OPE be associative. The hypothesis underlying the conformal bootstrap is that these constraints are so powerful that they can completely determine the local data given some minimal physical input. In practice, one expects that the input will include the global symmetry of the theory and some simple spectral assumptions such as the number of relevant operators.
 
%%%%%%%%%%%%%%%%%%%%%%%%%%%%%%%%%%%
\subsubsection*{Unitarity and shortening}
%%%%%%%%%%%%%%%%%%%%%%%%%%%%%%%%%%%
 
We first recall the constraints imposed by unitarity. Non-trivial\footnote{We use the qualification ``non-trivial'' to exclude the vacuum representation, which consists of a single state with $\Delta = j_1 = j_2 = 0$.} unitary representations of $\sof(4,2)$ are required to satisfy the following \emph{unitarity bounds},
%%%%%%
\begin{eqnarray}
\makebox[1in][l]{$\Delta \geqslant j_1 + j_2 +2$}	\, \quad &{\rm for}& \quad\, j_1 j_2 \neq 0~,\nn\\
\makebox[1in][l]{$\Delta \geqslant j_2 + 1$} 		\, \quad &{\rm for}& \quad\, j_1 = 0~,\\
\makebox[1in][l]{$\Delta \geqslant j_1 + 1$} 		\, \quad &{\rm for}& \quad\, j_2 = 0~.\nn
\end{eqnarray}
%%%%%%
Generic representations are denoted as $\AA_{\Delta,j_1,j_2}$. Non-generic, or \emph{short}, representations occur when the norm of a conformal descendant state in the Verma module built over some conformal primary is rendered null by a conspiracy of quantum numbers. This happens precisely when the unitarity bounds are saturated, leading the following list of short representations:
%%%%%%
\begin{equation}
\begin{split}
\makebox[.35in][l]{$\CC_{j_1,j_2}~$}&: \quad~ \Delta = j_1+j_2+2~,			\\
\makebox[.35in][l]{$\BB^L_{j_1}~$} 	&: \quad~ \Delta = j_1+1~,\quad j_2=0~,	\\
\makebox[.35in][l]{$\BB^R_{j_2}~$} 	&: \quad~ \Delta = j_2+1~,\quad j_1=0~,	\\
\makebox[.35in][l]{$\BB~$}			&: \quad~ \Delta = 1~,\quad j_1=j_2=0~.
\end{split}
\end{equation}
%%%%%%
All of these representations have null states at level one with the exception of $\BB$, which has a null state at level two.

The presence of short representations in the spectrum of a CFT is connected to the existence of free fields and symmetries in the theory. In particular, the primaries of $\BB$-type representations are decoupled free fields, and as such are not of much interest when studying interacting CFTs. For example, the primary of a $\BB$ representation is a free scalar field $\phi(x)$. Modding out by the null state at level two imposes the operator constraint 
%%%%%%
\begin{equation}
P^\mu P_\mu \phi = \Box \phi (x)= 0~,
\end{equation} 
%%%%%%
which is nothing but the free scalar equation of motion. Similarly, $\BB^{\star}_{\frac{1}{2}}$ multiplets have as their primaries free Weyl fermions; the null state level one imposes the free equation of motion
%%%%%%
\begin{equation}
\begin{split}
\BB^L_{\frac{1}{2}}~:\qquad &\partial^{\alpha \dot \alpha}\psi_{\alpha}(x)=0~,\\
\BB^R_{\frac{1}{2}}~:\qquad &\partial^{\alpha \dot \alpha}\tilde\psi_{\dot\alpha}(x)=0~.
\end{split}
\end{equation}
%%%%%%

On the other hand, $\CC$-type representations have various \emph{conserved currents} as their primaries; their level-one null state is the consequence of a conservation equation,
%%%%%%
\begin{equation}\label{eq:C_multiplet_conservation}
\partial^{\alpha_1 \dot \alpha_1} J_{\alpha_1 \cdots \alpha_{2 j_1} \, \dot \alpha_1 \cdots \dot \alpha_{2 j_2} } (x) = 0~.
\end{equation}
%%%%%%
Conserved currents with spin $j_1 + j_2 >2$ are \emph{higher-spin currents}, which are a hallmark of free CFTs \cite{Maldacena:2011jn,Alba:2013yda}. For the purposes of the bootstrap, we will usually impose by hand that no such multiplets appear. Conserved currents with $(j_1, j_2) = (1, \frac{1}{2})$ and $(j_1, j_2) = (\frac{1}{2}, 1)$ give rise to an enhancement of the conformal algebra to a \emph{super}conformal algebra -- when these operators are present one should therefore be taking full advantage of the power of superconformal symmetry. 

Thus, amongst the short representations of $\sof(4,2)$, those which may be present in an interacting non-supersymmetric CFT are $\CC_{1,1}$ and $\CC_{\frac12,\frac12}$. In the former case, the conformal primary is the stress tensor $T_{\mu\nu}$. In the latter case, the conformal primary is a conserved current $J_\mu$, so the presence of such multiplets portend the existence of continuous global symmetries.

%%%%%%%%%%%%%%%%%%%%%%%%%%%%%%%%%%% 
\subsubsection*{Locality in the operator algebra}
%%%%%%%%%%%%%%%%%%%%%%%%%%%%%%%%%%%
 
An important remark is in order. When characterizing CFTs by their local operator algebra, certain ingredients which are usually automatically present in a Lagrangian context are no longer necessarily compulsory. For example, one need not assume that the local algebra includes a stress tensor at all. Indeed, there are interesting local algebras, such as the algebra of local operators supported on conformal defects in a higher-dimensional CFT, in which the stress tensor is \emph{not} present. The presence of a stress tensor is clearly connected with the notion of \emph{locality} in the CFT, and we will take the existence of a unique stress tensor (that is, the existence of a unique conformal representation of type $\CC_{1, 1}$) as part of the \emph{definition} of a local CFT. 

Similarly, in the Lagrangian context a continuous global symmetry implies the existence of a conserved current in the operator spectrum. We will assume the validity of this claim even in the non-Lagrangian context:
%%%
\begin{conj}[CFT Noether ``theorem'']\label{conj:Noether}
In a local CFT, to any continuous global symmetry is associated a conserved current in the operator algebra that generates the symmetry.
\end{conj}
%%%
Clarifying the conceptual status of this ``theorem'' is an important open problem. On one hand, one may take it as part of the definition of what it means for a CFT to be local, in which case this is a tautology. Alternatively, it is possible that the theorem may be derived from general principles in a suitable axiomatic framework.\footnote{It is unclear whether the axioms for the algebra of local operators should be sufficient for this purpose. It is possible that the existence of a conserved current could follow from the assumptions that the operator algebra is invariant under a continuous symmetry \emph{and} that there is a stress tensor.  Alternatively, the framework may need to be enlarged, perhaps allowing for correlation functions in non-trivial geometries, subject to suitable locality assumptions.} Whatever the case may be, the proof of such a statement is of interest in part due to its reinterpretation via AdS/CFT, which is the statement that there are no continuous global symmetries in AdS quantum gravity.

%%%%%%%%%%%%%%%%%%%%%%%%%%%%%%%%%%%
\subsubsection*{Canonical data}
%%%%%%%%%%%%%%%%%%%%%%%%%%%%%%%%%%%

The data associated to short representations of the conformal algebra carries particular physical significance. The three-point function of the stress tensor depends on three parameters, two of which can be identified with the two coefficients appearing in the conformal anomaly, conventionally denoted by $a$ and $c$. The $a$ coefficient gives a measure of the degrees of freedom of the theory and serves as a height function in theory space: for two CFTs connected by RG flow, $a_{\rm UV} > a_{\rm IR}$ \cite{Cardy:1988cwa,Komargodski:2011vj}. However, since $a$ can only be extracted from the stress tensor three-point function, it is rather difficult to access by bootstrap methods -- one would generally need to consider correlation functions involving external stress tensors, which are very complicated \cite{Dymarsky:2013wla}. By contrast, if one uses the canonical normalization for the stress tensor, its \emph{two-point} function is proportional to $c$. The $c$ coefficient will then appear in any four-point function containing an intermediate stress tensor, making its presence ubiquitous in the bootstrap literature. Using ``conformal collider'' observables, it was argued in \cite{Hofman:2008ar} that in a general unitarity CFT the ratio of conformal anomaly coefficients must obey the bounds\footnote{The argument uses positivity of energy correlators in a unitarity theory, which is a reasonable physical assumption (see also \cite{Kulaxizi:2010jt}). It would be interested to recover the HM bounds by conformal bootstrap methods. This will likely have to await for the complete conformal block analysis of the stress tensor four-point function, a challenging technical problem.}
%%%%%%
\begin{equation}\label{eq:HM_bounds}
\frac{1}{3} \leqslant \frac{a}{c} \leqslant \frac{31}{18}~.
\end{equation}
%%%%%%
The lower bound is saturated by the free scalar CFT, the upper bound by the free vector CFT. There is strong evidence that these free CFTs are the \emph{only} theories saturating the bounds \cite{Zhiboedov:2013opa}.

Similarly, the two-point function of canonically normalized currents depends on a parameter $k$ often called the \emph{flavor central charge} that can be identified with an 't Hooft anomaly for the corresponding global symmetry \cite{Erdmenger:1996yc,Anselmi:1997rd}. This parameter appears in the OPE of conserved currents as follows,
%%%%%%
\begin{equation}\label{eq:flavor_charge}
J_\mu^A(x)J_\nu^B(0)\sim\frac{3k}{4\pi^4}\delta^{AB}\frac{x^2g_{\mu\nu}-2x_\mu x_\nu}{x^8}+\frac{2}{\pi^2}\frac{x_\mu x_\nu f^{AB}_{\ph{AB}C}x\cdot J^C(0)}{x^6}+\ldots~.
\end{equation}
%%%%%%
Like the $c$ central charge, the flavor central charge makes frequent appearances in the bootstrap because it controls the contribution of the conserved current in a correlation function of charged operators.

In a sense, the data associated to the spectrum of conserved currents and stress tensors and their associated anomaly coefficients is the most basic data associated to a conformal field theory. We designate this data as the \emph{canonical data} for the CFT. It is natural to organize an exploration of the space of conformal field theories in terms of these parameters, and if one wants to study a particular theory in detail this data is an obvious starting point. This has not always been the approach in the existing bootstrap literature thus far, but that is at least in part because the natural observables through which to pursue such a strategy would be the four point functions of conserved currents and stress tensors. At a technical level, these are much more complex observables than the correlators of spacetime scalars.

\subsubsection*{The numerical bootstrap approach}

Intuitively, associativity of the operator algebra is a tremendous constraint. However, aside from the case of two-dimensional CFTs where the global conformal symmetry algebra enhances to two copies of the infinite-dimensional Virasoro algebra, it seems very difficult to extract useful information from these conditions. The way forward was shown in \cite{Rattazzi:2008pe}, where the focus was shifted away from trying to \emph{solve} the associativity problem and towards obtaining \emph{constraints} for, \eg, the spectrum of local operators or their OPE coefficients in a unitary CFT. The prototypical bounds that can be obtained in this way are  uppers bound for the dimension of lowest-lying operator of a given spin, or a lower bound on the $c$ central charge of a theory, all given some input about the spectrum of scalar operators.

In order to test associativity it suffices to investigate four-point functions in a given CFT, where the OPE can be taken in three essentially inequivalent ways by fusing different pairs of operators together. For each choice one finds a representation of the four-point function as a sum over conformal blocks \cite{slavalectures}, with one block for each conformal multiplet that appears in both OPEs. The statement that these three decompositions have to sum to exactly the same result is known as \emph{crossing symmetry}. It was shown in \cite{Rattazzi:2008pe} that useful bounds can be extracted already from the requirement of crossing symmetry for a single four-point function involving four identical scalar operators. Such an analysis is conspicuously tractable -- as opposed to trying to solve all of the infinitely many crossing symmetry constraints simultaneously, we simply find the conditions that follow from a finite subset of those constraints. The structure of four-point functions and their OPE decompositions are severely constrained by conformal symmetry -- see, \eg, \cite{slavalectures} for an introductory exposition.

The work of \cite{Rattazzi:2008pe} has been extended in numerous directions, and bounds have been obtained in theories with and without supersymmetry and in various spacetime dimensions. Further numerical bootstrap results can be found for example in \cite{Rychkov:2009ij,Vichi:2009zz,Caracciolo:2009bx,Poland:2010wg,Rattazzi:2010gj,Rattazzi:2010yc,Vichi:2011ux,Poland:2011ey,ElShowk:2012ht,Liendo:2012hy,ElShowk:2012hu,Beem:2013qxa,Gliozzi:2013ysa,Kos:2013tga,El-Showk:2013nia,Alday:2013opa,Gaiotto:2013nva,Berkooz:2014yda,El-Showk:2014dwa,Gliozzi:2014jsa,Nakayama:2014lva,Nakayama:2014yia,Alday:2014qfa,Chester:2014fya,Kos:2014bka,Caracciolo:2014cxa,Paulos:2014vya,Bae:2014hia}. An essential ingredient in the numerical analysis is the (super)conformal block decomposition of a four-point functions. These structure have been investigated in various cases in, \emph{e.g.}, \cite{Costa:2011mg,Dolan:2011dv,Costa:2011dw,SimmonsDuffin:2012uy,Siegel:2012di,Osborn:2012vt,Pappadopulo:2012jk,Hogervorst:2013sma,Fitzpatrick:2013sya,Hogervorst:2013kva,Fitzpatrick:2014oza,
Khandker:2014mpa,Elkhidir:2014woa,Costa:2014rya,Dymarsky:2013wla}. In related work, \cite{Fitzpatrick:2012yx,Komargodski:2012ek,Alday:2013cwa,Fitzpatrick:2014vua,Vos:2014pqa} obtained nontrivial constraints for the operator spectrum by considering in particular the OPE in the limit where operators become lightlike separated.

%%%%%%%%%%%%%%%%%%%%%%%%%%%%%%%%%%%
\subsection{Operator algebras of \texorpdfstring{$\NN=2$}{N=2} SCFTs}
\label{subsec:N2_algebras}
%%%%%%%%%%%%%%%%%%%%%%%%%%%%%%%%%%%
 
The superconformal case follows largely the same conceptual blueprint as the non-supersymmetric case, where we replace the conformal algebra $\sof(4,2)$ with the superconformal algebra is $\suf(2,2|2)$. The maximal bosonic subalgebra is just the conformal algebra $\sof(4,2)\equiv\suf(2,2)$ times the R-symmetry algebra $SU(2)_R \times U(1)_r$. Additionally there are sixteen fermionic generators -- eight Poincar\'e supercharges and eight conformal supercharges -- denoted as $\{\QQ^{\II}_{\aa},\,\wt\QQ_{\II\aad},\,\SS_{\JJ}^{\aa},\,\wt\SS^{\JJ\aad}\}$ where $\II=1, 2$, $\alpha = \pm$, and $\dot\alpha = \dot\pm$ are $SU(2)_R$, $\suf(2)_1$, and $\suf(2)_2$ indices, respectively.

The spectrum of local operators can be organized in highest weight representations of $\suf(2,2|2)$ whose highest weight states, known as superconformal primaries, are annihilated by all lowering operators of the superconformal algebra -- in particular, by all the conformal supercharges $\SS$. These representations are labelled by the quantum numbers $[\Delta,j_1,j_2,R,r]$ of the superconformal primary; the additional labels $R$ and $r$ that extend the ordinary conformal case are the eigenvalues of the Cartan generators of $SU(2)_R$ and $U(1)_r$. We will also consider theories that are invariant under additional flavor symmetry ${\mf g}_F$ (a semi-simple Lie algebra commuting with $\suf(2,2|2)$), which introduces additional flavor quantum numbers $f$. In summary, the local data for an $\NN=2$ SCFT are
%%%%%%
\begin{equation}
\{ a_i, \lambda^s_{ijk} \}~, \quad a_i \colonequals [\Delta, j_1, j_2, R, r; f]_i~.
\end{equation}
%%%%%%
In analogy with the conformal case, the coefficients $\lambda_{ijk}^s$ encode the information needed to completely reconstruct the \emph{superspace} three-point functions\footnote{In the conformal case, the $\lambda^s_{ijk}$ can be extracted from the three-point function of the conformal primaries, because descendant operators are simply derivatives of the primaries and their three-point functions contain no extra information. In general this is no longer the case with superconformal symmetry: knowledge of the three-point functions of the superconformal primaries does not always suffice. But at an abstract level there is no difference: what matters are superconformally covariant structures that can be built from the three representations.} $\langle \RR_i (x_1, \theta_1) \, \RR_j(x_2, \theta_2) \, \RR_k (x_3, \theta_3) \rangle$.

%%%%%%%%%%%%%%%%%%%%%%%%%%%%%%%%%%%
\subsubsection*{Unitarity and shortening}
%%%%%%%%%%%%%%%%%%%%%%%%%%%%%%%%%%%

The unitary representation theory of the $\NN=2$ superconformal algebra is more elaborate than that of the ordinary conformal algebra. The unitarity bounds are now given by
%%%%%%
\begin{equation}\label{eq:unit_bounds}
\begin{alignedat}{3}
\Delta	&\geqslant \Delta_i~,&\qquad				&&				 j_i&\neq0~,\\
\Delta	&=	\Delta_i-&2~~\mbox{~or~}~~\Delta&\geqslant& \Delta_i~,	\qquad j_i&=0~,\\
\end{alignedat}
\end{equation}
%%%%%%
where we have defined
%%%%%%
\begin{equation}
\Delta_1 \colonequals 2+2j_1+2R+r~, \qquad \Delta_2 \colonequals 2+2j_2+2R-r~.
\end{equation}
%%%%%%
The unitary representations of $\suf(2,2|2)$ have been classified in \cite{Dobrev:1985qv,Dolan:2002zh,Kinney:2005ej}. Short representations occur when one or more of these bounds are saturated, and the different ways in which this can happen correspond to different combinations of Poincar\'e supercharges that can annihilate the highest weight state of the representation. There are again two types of shortening conditions, the $\BB$ type and the $\CC$ type. Each type now has four incarnations corresponding to the choice of chirality (left or right-moving) and the choice of $SU(2)_R$ component:
%%%%%%
\begin{eqnarray}\label{eq:constitutent_shortening_conditions}
\BB^\II&:&\qquad \QQ^\II_{\alpha}|\psi\rangle=0~,\quad\alpha=1,2~,\\
{\bar \BB}_\II&:&\qquad \wt\QQ_{\II\dot\alpha}|\psi\rangle=0~,\quad\dot\alpha=1,2~,\\
\CC^\II&:&\qquad
\begin{cases}
\epsilon^{\a\b}\QQ^\II_{\alpha}					|\psi\rangle_\beta=0~,\quad &j_1\neq0~,\\
\epsilon^{\a\b}\QQ^\II_{\alpha}\QQ^\II_{\beta}	|\psi\rangle=0~,\quad &j_1=0~,
\end{cases}\\
{\bar\CC}_\II&:&\qquad
\begin{cases} 
\epsilon^{\ad\bd}\wt\QQ_{\II\ad}				|\psi\rangle_\beta=0~,\quad &j_2\neq0~,\\
\epsilon^{\ad\bd}\wt\QQ_{\II\ad}\wt\QQ_{\II\bd}	|\psi\rangle=0~,\quad &j_2=0~.
\end{cases}
\end{eqnarray}
%%%%%%
Some authors refer to $\BB$-type conditions as shortening conditions, and to $\CC$-type conditions as \emph{semi}-shortening conditions, to highlight the fact that a $\BB$-type condition is twice as strong. We refer to Appendix \ref{App:representations} for a tabulation of all allowed combinations of (semi-)shortening conditions and for naming conventions for the resulting representations. 

Because of the proliferation of short representations in the $\NN=2$ context, there is potentially much more ``canonical data'' than in the non-supersymmetric case. Indeed, these many short representations are closely related to various nice features theories with $\NN=2$ supersymmetry. Here we focus primarily on three classes of short representations that have particularly straightforward connections to familiar physical characteristics of $\NN=2$ theories. These representations have the distinction of obeying the maximum number of shortening or semi-shortening conditions that can simultaneously be imposed (two and four, respectively). In the notations of \cite{Dolan:2002zh}, they are: 
%%%%%%
\begin{itemize}
\item 
$\EE_r$: Half-BPS multiplets ``of Coulomb type''. These obey two $\BB$-type shortening conditions of the same chirality: $\BB^1\cap\BB^2$. In other terms,
they are $\NN=2$ chiral multiplets, annihilated by the action of \emph{all} left-handed supercharges.\footnote{We are focusing on the \emph{scalar} $\EE_r$ multiplets -- $\EE_r \colonequals \EE_{r(0,0)}$ in the notations of Table \ref{Tab:shortening}. Representation theory allows for $\NN=2$ chiral multiplets $\EE_{r(0,j_2)}$ with $j_2 \neq 0$, but such exotic multiplets do not occur in any known $\NN=2$ SCFT. See \cite{Buican:2014qla} for a recent discussion.}
%%%
\item
$\hat{\BB}_R$: Half-BPS multiplets ``of Higgs type''. These obey two $\BB$-type shortening conditions of opposite chirality: $\BB^1\cap\bar\BB_{2}$. These types of operators are sometimes called ``Grassmann-analytic''.
%%%
\item 
$\hat{\CC}_{0(j_1,j_2)}$: The stress tensor multiplet (the special case $j_1 = j_2= 0$) and its higher spin generalizations. These obey the maximal set of semi-shortening conditions: $\CC^1\cap \CC^2 \cap \bar\CC_1 \cap \bar\CC_2$.
\end{itemize}
%%%%%%
The CFT data associated to these representations encodes some of the most basic physical information about an $\NN=2$ SCFT. We now look at each in more detail, starting from the third and most universal class, which contains the stress tensor multiplet.

%%%%%%%%%%%%%%%%%%%%%%%%%%%%%%%%%%%
\subsubsection*{Stress tensor data}
%%%%%%%%%%%%%%%%%%%%%%%%%%%%%%%%%%%

The maximally semi-short multiplets $\hat{\CC}_{0(j_1,j_2)}$ contain conserved tensors of spin $2 + j_1 + j_2$. For $j_1 + j_2 > 0$, such multiplets are not allowed in an interacting CFT, and we will always impose their absence from the double OPE of the four-point functions under consideration.

The $\hat{\CC}_{0(0,0)}$ representation includes a conserved tensor of spin two, which we identify as the stress tensor of the theory. By definition, a \emph{local} $\NN=2$ SCFT will contain exactly one $\hat{\CC}_{0(0,0)}$ multiplet.\footnote{A caveat to this definition of locality is that in the tensor product of two local theories there will be two stress tensor multiplets. For the purposes of the conceptual discussion here we restrict our attention to theories that are not factorizable in this manner -- we might call such theories \emph{simple}.} We will usually assume that the theories that we study are local, but we'll also briefly explore non-local theories, which have no stress tensor and thus no $\hat{\CC}_{0(0,0)}$ multiplet. 

The superconformal primary of $\hat{\CC}_{0(0,0)}$ is a scalar operator of dimension two that is invariant under all $R$-symmetry transformations. The other bosonic primaries in the multiplet are the conserved currents for $SU(2)_R \times U(1)_r$ and the stress tensor itself. An analysis in $\NN=2$ superspace \cite{Kuzenko:1999pi} reveals that three-point function of $\hat{\CC}_{0(0,0)}$ multiplets involves two independent structures, whose coefficients can be parametrized in terms of the $a$ and $c$ anomalies. The $\NN=2$ version of the Hofman-Maldacena bounds reads
%%%%%%
\begin{equation}\label{eq:Neq2HM_bounds}
\frac{1}{2} \leqslant \frac{a}{c} \leqslant \frac{5}{4}~.
\end{equation}
%%%%%%
The lower bound is saturated by the free hypermultiplet theory, and the upper bound by the free vector multiplet theory. By a generalization of the analysis of \cite{Zhiboedov:2013opa}, one should be able to argue that these are the only $\NN=2$ SCFTs saturating the bounds.

In this paper we will not study the four-point function of the stress tensor multiplet, because the requisite superconformal block expansion has not yet been worked out. We will, however, have indirect access to the $c$ anomaly coefficient. As in the non-supersymmetric case, if one chooses the canonical normalization for the stress tensor then two-point function of $\hat{\CC}_{0(0,0)}$ multiplets will depend on $c$ only. The $c$ coefficient will make an appearance in all four point functions that we study, since $\hat{\CC}_{0(0,0)}$ appears in their double OPE.

%%%%%%%%%%%%%%%%%%%%%%%%%%%%%%%%%%%
\subsubsection*{Coulomb and Higgs branches}
%%%%%%%%%%%%%%%%%%%%%%%%%%%%%%%%%%%

As indicated by our choice of terminology, the two types of half-BPS multiplets -- $\EE_r$ and $\hat{\BB}_R$ -- are closely related to the Coulomb and Higgs branches of the moduli space of vacua, respectively. In Lagrangian theories, the superconformal primaries in the $\EE_r$ multiplets are the gauge-invariant composites of vector multiplet scalars that parameterize the Coulomb branch, and the superconformal primaries in the $\hat{\BB}_R$ multiplets are the gauge-invariant composites of hypermultiplet scalars that parameterize the Higgs branch.

We should call attention to the fact that a satisfactory understanding of the phenomenon of spontaneous conformal symmetry breaking has not yet been developed in the language of CFT operator algebras. In principle, the local data should contain all necessary information to describe the phases of the theory where conformal symmetry is spontaneously broken. A method to extract this information is, however, presently not known. Even the basic question of \emph{whether} a given CFT possesses nontrivial vacua remains out of reach. Since all known examples of vacuum manifolds in CFTs occur in supersymmetric theories, one might speculate that supersymmetry is a necessary condition for spontaneous conformal symmetry breaking.

We are now ready to look in more detail at the CFT data encoded in the two classes of BPS multiplets.

%%%%%%%%%%%%%%%%%%%%%%%%%%%%%%%%%%%
\subsubsection*{Coulomb branch data}
%%%%%%%%%%%%%%%%%%%%%%%%%%%%%%%%%%%

We will refer to the data associated to $\EE_{r}$ multiplets as \emph{Coulomb branch data}. By passing to the cohomology of the left-handed Poincar\'e supercharges, one finds a commutative ring of operators known as the \emph{Coulomb branch chiral ring}, the elements of which can be identified with the superconformal primaries of $\EE_{r}$ multiplets. In all known examples, this ring is exceedingly simple, and it is natural to formulate a conjecture that the ring is always as simple as it is in the examples:\footnote{To the best of our knowledge, this conjecture was first explicitly stated in the literature by Yuji Tachikawa in \cite{Tachikawa:2013kta}.}
\begin{conj}[Free generation of the Coulomb chiral ring]\label{conj:free_coulomb_ring}
In any $\NN=2$ SCFT, the Coulomb branch chiral ring is freely generated.
\end{conj}
This conjecture can in principle be translated into a statement about the OPE coefficients of the $\EE_r$ multiplets. For instance, a simple consequence is that no $\EE_r$ superconformal primary can square to zero in the chiral ring, so an $\EE_{2r}$ operator must appear with nonzero coefficient in the OPE of the $\EE_r$ with itself. Precisely this kind of statement can be tested by numerical bootstrap methods, as we will describe in Section \ref{sec:eps_results}.

The number of generators of the Coulomb branch chiral ring is usually referred to as the \emph{rank} of the theory. The set $\{ r_1, \dots r_{\rm rank} \}$ of $U(1)_r$ charges of these chiral ring generators is one of the most basic invariants of an $\NN=2$ SCFT. Unitarity implies $r \geqslant 1$, with $r=1$ only in the case of the free vector multiplet, so we will always assume $r>1$. In Lagrangian SCFTs, the $r_i$ are all integers, but there are several non-Lagrangian models that possess $\EE_r$ multiplets with interesting fractional values of $r$. We are not aware of any examples where $U(1)_r$ charges take irrational values.

It is widely believed that the Coulomb branch of the moduli space of any $\NN=2$ SCFT is parameterized by assigning independent vevs to each of the Coulomb branch chiral ring generators. We will generally operate under the assumption that this statement is true, which amounts to assuming the validity of the following conjecture.
%%%
\begin{conj}[Geometrization of the Coulomb chiral ring]\label{conj:coulomb_geometrization}
The Coulomb chiral ring is isomorphic to the holomorphic coordinate ring on the Coulomb branch.
\end{conj}
%%%
We note that the union of Conjecture \ref{conj:free_coulomb_ring} and Conjecture \ref{conj:coulomb_geometrization} implies that the Coulomb branch of any $\NN=2$ SCFT just $\Cb^r$, with $r$ the rank of the theory.

At present we are not sure how one might establish Conjecture \ref{conj:coulomb_geometrization} using bootstrap methods due to the obstacle of spontaneous conformal symmetry breaking discussed above. However, once one has found their way onto the Coulomb branch, the powerful technology of Seiberg-Witten (SW) theory becomes applicable. The effective action for the low-energy $U(1)^{\rm rank}$ gauge theory on the Coulomb branch is characterized by geometric data (in the simplest cases, this is the SW curve, more generally it is some abelian variety). There are well-developed techniques to determine the SW geometry, which apply to most Lagrangian examples and to several non-Lagrangian cases as well. In turn, the SW geometry determines a wealth of physical information, such as the spectrum of massive BPS states. Unfortunately, how to translate this information into CFT data remains an unsolved problem.\footnote{See however \cite{Cecotti:2010fi} for a relation between the spectrum of BPS states on the Coulomb branch and a certain partition function (evaluated at the conformal point), which appears to be closely related to the superconformal index.}

In \cite{Shapere:2008zf}, Shapere and Tachikawa (ST) proved a remarkable formula that relates the $a$ and $c$ central charges to the generating $r$-charges $\{ r_1, \dots r_{\rm rank}\}$,
%%%%%%
\begin{equation} \label{eq:ST_sum_rule}
2a - c = \frac{1}{4} \sum_{i=1}^{\rm rank} (2 r_i - 1)~.
\end{equation}
%%%%%%
The ST sum rule holds in all known examples, and it is tempting to conjecture that it is a general property of all $\NN=2$ SCFTs. The derivation of \cite{Shapere:2008zf} requires that the SCFTs in question be realized at a point on the moduli space a Lagrangian theory. The result can then be extended to all SCFTs connected to that class of theories by generalized $S$-dualities. In particular, this includes a large subset of theories of class $\SS$.
 
According to the ST sum rule, a theory with zero rank necessarily has $a/c = 1/2$, which is the value saturating the lower HM bound. As remarked above, there are strong reasons to believe that the only SCFT saturating this bound is the free hypermultiplet theory. However, since the whole logic of \cite{Shapere:2008zf} relies on the existence of a Coulomb branch, this reasoning is circular. An interacting SCFT of zero rank would be rather exotic, but we do not know how to rule it out with present methods. 

The special case of the $\EE_{2}$ multiplet is particularly significant. The top component of the multiplet, obtained by acting with four right-moving supercharges on the superconformal primary,\footnote{In an abuse of notation, we are denoting the superconformal primary with the same symbol $\EE_2$ that represents the whole multiplet.} $\OO_4 \sim \tilde Q^4 \EE_2$ is a scalar operator of dimension four. This operator provides an exactly marginal deformation of the SCFT that preserves the full $\NN=2$ supersymmetry. (By CPT symmetry, there is also a complex conjugate operator $\overline{\OO}_4 \sim Q^4 \bar{\EE}_{-2}$). The converse is also true: any $\NN=2$ supersymmetric exactly marginal operator $\OO_4$ must be the top component of an $\EE_2$ multiplet. It follows that the number of $\EE_2$ multiplets is equal to the (complex) dimension of the conformal manifold of the theory. In a Lagrangian theory, there is an $\EE_2$ multiplet for each simple factor of the gauge group, and the exactly marginal operator $\OO_4 \sim {\rm Tr} (F^2 + i \tilde F^2)$ (where $F$ is the Yang-Mills field strength) is dual to the complexified gauge coupling.

Another true feature of all Lagrangian SCFTs (and many non-Lagrangian ones in class $\SS$) is that they can be constructed by taking isolated building blocks with no marginal deformations (such as hypermultiplets in the Lagrangian case, or $T_N$ theories in the class $\SS$ case) and gauging global symmetry groups for which the beta function will vanish. A natural conjecture is that this feature is indeed universal:
%%%
\begin{conj}[Decomposability]\label{conj:decomposability}
Any $\NN=2$ SCFT with an $n$-dimensional conformal manifold can be constructed by gauging $n$ simple factors in the global symmetry group of a collection of isolated $\NN=2$ SCFTs.
\end{conj}
%%%
Of course such a decomposition need not be unique -- the existence of inequivalent decompositions of the same theory is what is often called ``generalized $S$-duality''. Note that the validity of this conjecture would imply the absence of compact conformal manifolds for $\NN=2$ SCFTs.\footnote{In the $\NN=1$ case the existence of compact conformal manifolds has recently been established in \cite{Buican:2014sfa}. The methods used there cannot easily be generalized to the $\NN=2$ case.}

%%%%%%%%%%%%%%%%%%%%%%%%%%%%%%%%%%%
\subsubsection*{Higgs branch data}
%%%%%%%%%%%%%%%%%%%%%%%%%%%%%%%%%%%

In a similar vein, the $\hat{\BB}_R$ multiplets are expected to encode the information about the Higgs branch of the theory. The $\hat{\BB}_R$ superconformal primaries, which are also $SU(2)_R$ highest weights, form the \emph{Higgs branch chiral ring}. In all known examples this ring describe by a finite set of generators obeying polynomial relations. The algebraic variety defined by this ring is then expected to coincide with the Higgs branch of vacua. This expectation can be formalized as follows:
%%%
\begin{conj}[Geometrization of the Higgs chiral ring]\label{conj:higgs_geometrization}
In any $\NN=2$ SCFT, the Higgs branch chiral ring is isomorphic to the holomorphic coordinate ring on the Higgs branch of vacua.
\end{conj}
%%%
The Higgs branch of vacua is hyperk\"ahler, so there are actually a $\Cb\Pb^1$ worth of holomorphic coordinate rings on it depending on the choice of complex structure. The choice of complex structure corresponds to a choice of Cartan element in $SU(2)_R$, so we have implicitly made the choice already.

In this paper we will focus on the simplest non-trivial\footnote{$\hat{\BB}_{\frac{1}{2}}$ describes a free hypermultiplet.} case of these multiplets, the $\hat{\BB}_1$ multiplet. This multiplet plays a distinguished role, because it encodes the information about the continuous global symmetries of the theory. Indeed, the multiplet contains a conserved current,
%%%%%%
\begin{equation}
J_{\alpha \dot \alpha} =\epsilon^{\JJ \KK } \QQ^{\II}_\alpha \wt{\QQ}_{\JJ \dot \alpha} \phi_{\II \KK}~,
\end{equation}
%%%%%%
where $\phi_{\II\JJ}$ is the operator of lowest dimension in the $\hat{\BB}_1$ multiplet. It is an $SU(2)_R$ triplet and is often referred to as the \emph{moment map} operator. (The superconformal primary is the highest $SU(2)_R$ weight $\phi_{11}$.) The current $J_{\alpha \dot \alpha}$ generates a continuous symmetry global symmetry, and is thus necessarily in the adjoint representation of some Lie group $G_F$. Vice versa, if the theory enjoys a continuous global symmetry, it follows from Conjecture \ref{conj:Noether} that the CFT contains an associated conserved current $J_{\alpha \dot \alpha}$, and one can show that in an interacting $\NN =2$ SCFT such a current must necessarily belong to a $\hat{\BB}_1$ multiplet. Indeed, one can survey the list of superconformal representations and identify all the ones that contain conserved spin one currents that are also $SU(2)_R \times U(1)_r$ singlets. The list is very short: $\hat{\BB}_1$ and $\hat{\CC}_{0(\frac{1}{2}, \frac{1}{2})}$. The latter multiplet has a conserved current as its superconformal primary, but also contains conserved a spin three conserved current among its descendants, so by our usual criterion it is not allowed in an interacting SCFT. What's more, $\hat{\BB}_1$ representations cannot combine with other short representations to form long representations, so the $\hat{\BB}_1$ content of a theory is an invariant on the conformal manifold. To reiterate,  a SCFT may have a flavor
symmetry  enhancement only in a singular limit where some free subsector decouples (such as the zero coupling limit of a gauge theory) and   $\hat{\CC}_{0(\frac{1}{2},\frac{1}{2})}$ multiplets split off from long multiplets hitting the unitarity bound. In the ``bulk'' of the conformal manifold, flavor symmetries are always associated to $\hat {\BB}_1$ multiplets.

As we have already mentioned in the context of exactly marginal gauging of SCFTs, to each simple non-abelian factor of the global symmetry group is associated a \emph{flavor central charge} $k$, defined from the OPE coefficient of the conserved current with itself \eqref{eq:flavor_charge}. Thus the most basic data associated to the $\hat{\BB}_1$ representations in an SCFT are the global symmetry group $G_F = G_1 \times \dots G_k$ and the corresponding flavor central charges.

%%%%%%%%%%%%%%%%%%%%%%%%%%%%%%%%%%%
\subsubsection*{Chiral algebra data}
%%%%%%%%%%%%%%%%%%%%%%%%%%%%%%%%%%%

It was recognized in \cite{Beem:2013sza} (see also \cite{Beem:2014rza,Lemos:2014lua}) that the local operator algebra of any $\NN=2$ SCFT admits a closed subsector isomorphic to a two-dimensional chiral algebra. The operators that play a role in the chiral algebra are the so-called \emph{Schur operators}, which (by definition) obey the conditions\footnote{In fact one can show that the first condition implies the second in a unitary theory.}
%%%%%%
\begin{equation}\label{eq:schur_conditions}
\Delta - (j_1 + j_2) - 2R = 0~, \qquad j_2 - j_1 - r = 0~.
\end{equation}
%%%%%%
Schur operators are found in the following short representations,
%%%%%%
\begin{equation}\label{eq:schur_multiplets}
\hat{\BB}_R~,\quad\DD_{R(0,j_2)}~,\quad\bar{\DD}_{R(j_1,0)}~,\quad\hat{\CC}_{R(j_1,j_2) }~.
\end{equation}
%%%%%%
One should in particular note the absence of the $\EE_r$ multiplets from this structure. Each supermultiplet in this list contains precisely one Schur operator: for the $\hat{\BB}_R$ multiplets, the Schur operator is the superconformal primary itself, while for the other multiplets in \eqref{eq:schur_multiplets} it is a superconformal descendant.\footnote{For example, the Schur operator in a $\hat{\CC}_{0(0,0)}$ multiplet is a single component of the $SU(2)_R$ conserved current.} When inserted on a fixed plane $\Rb^2 \subset \Rb^4$, parametrized by the complex coordinate $z$ and its conjugate $\zb$, and appropriately \emph{twisted} (the twist identifies the right-moving global conformal algebra $\overline{\slf(2)}$ acting on $\bar z$ with the complexification of $\suf(2)_R$ algebra), Schur operators have \emph{meromorphic} correlation functions. The rationale behind this construction is that twisted Schur operators are closed under the action of a certain nilpotent supercharge, $\qq \colonequals \QQ_-^1 + \wt{\SS}^1_{\dot -}$, and they have well-defined meromorphic OPEs at the level of $\qq$ cohomology. This is precisely the structure that defines a two-dimensional chiral algebra. 

We refer the reader to \cite{Beem:2013sza} for a comprehensive explanation of this construction. Here we mainly wish to emphasize that the \emph{chiral algebra data} (\ie, the Schur operators and their three-point functions) are a very natural generalization of the Higgs data. Since they are subject to associativity conditions expressed by meromorphic equations, the chiral algebra data can be often determined exactly given some minimum physical input.
 
The simplest example, and the one that will play a role in this paper, is the case of moment maps. Moment maps transform in the adjoint representation of the flavor symmetry group, and in the associated chiral algebra they correspond to affine Kac-Moody currents, where the level $k_{2d}$ of the affine current algebra is related to the four-dimensional flavor central charge $k$ by the universal relation 
%%%%%%
\begin{equation}
k_{2d} = - \frac{k}{2}~.
\end{equation}
%%%%%%
The four-point function of affine currents completely determined by meromorphy and crossing symmetry. In the present context, it admits a reinterpretation as a certain meromorphic piece of the full moment map four-point function. Crucially, this meromorphic piece contains the complete information about the contribution of short representations to the double OPE of the four-point function.\footnote{To be able to uniquely reconstruct the contribution of the short representations from the meromorphic function, one must make the now-familiar assumption that the theory does not contain higher-spin conserved currents.} All in all, combining the constraints of four-dimensional unitarity with the ability to solve exactly for the contributions of short representations leads to novel unitarity bounds for the level $k$ and the trace anomaly coefficient $c$ that are valid in any interacting $\NN=2$ SCFT. These bounds will play a significant role in the analysis of Section \ref{Sec:Bhatresults}. 
%!TEX root = ../draft_maxi_Neq2.tex

\subsection{A first look at the landscape: theories of low rank}
\label{subsec:landscape_of_theories}

The ultimate triumph of the $\NN=2$ bootstrap program would be the classification of $\NN=2$ SCFTs. If the decomposability conjecture of Section \ref{subsec:N2_algebras} holds true, then this problem is reduced to the enumeration of elementary building block theories with no conformal manifold. Still, this is completely out of reach at present, and any attempt at a direct attack on the classification problem would be premature. We are still very much in an exploratory phase. 

To organize our explorations we may characterize theories by their \emph{rank} -- \ie, the dimension of their Coulomb branch or the number of generators in the Coulomb branch chiral ring. Theories with low rank by and large have smaller values for their central charges than their higher-rank counterparts, so this may be a reasonable measure of the complexity of a theory. From the bootstrap point of view, theories with small central charges are attractive as targets for numerical study.

The rank zero case is probably trivial. The simplest conjecture is that the only $\NN=2$ SCFT with no Coulomb branch is the free hypermultiplet theory. This would be compatible with the universal validity of the Shapere-Tachikawa bound. 

For rank one, we can start by reviewing the list of established theories. This survey will prove useful in our efforts to interpret the numerical bootstrap results reported in later sections. The classic rank one theories are the SCFTs that arise on a single D3 brane probing an $F$-theory singularity with constant dilaton \cite{Sen:1996vd,Banks:1996nj,Dasgupta:1996ij,Minahan:1996fg,Minahan:1996cj,Aharony:1998xz}. There are seven such singularities, denoted by $H_0$, $H_1$, $H_2$, $D_4$, $E_6$, $E_7$, $E_8$. With the exception of the theory associated to the $D4$ singularity, which is an $SU(2)$ gauge theory with $N_f=4$ fundamental flavors, these theories are all isolated non-Lagrangian SCFTs. They have an alternative realization is in class $\SS$, where they are associated to punctured spheres with certain special punctures -- see, \eg, \cite{Gaiotto:2009we, Bonelli:2011aa,Gaiotto:2012sf,Xie:2012hs,Benini:2009gi}.
 
Basic properties of these rank one SCFTs are summarized in Table \ref{Tab:rank1theories}. Their flavor symmetry algebra $\hhf$ is given by the Lie algebra of the same name (with $H_i \to A_i$; the $H_0$ theory has no flavor symmetry). From the $F$-theory realization it is manifest that the Higgs branch of each theory is the one-instanton moduli space for the corresponding flavor symmetry group. As algebraic varieties, these Higgs branches are generated by the $\hhf$ moment maps subject to a set of quadratic relations known as the Joseph relations. Relatedly, the flavor central charge $k$ and the $c$ anomaly saturate the unitarity bounds derived in \cite{Beem:2013sza}. It was argued in Section 4 of \cite{Beem:2013sza} that this is strong evidence that the protected chiral algebra is the affine Lie algebra $\hat \hhf_{k_{2d}}$ at level $k_{2d} = - \frac{k}{2}$.\footnote{We mention in passing, as this will play a role later, that each of these theories admits a rank $N$ generalization, physically realized on the worldvolume of $N$ parallel D3 branes probing the same $F$-theory singularity. The Higgs branches of the higher rank theories are the moduli spaces of rank-$N$ $\hhf$-instantons, with global symmetry $\hhf \otimes \suf(2)$ for $N\geqslant2$.}

Another well-known rank one $\NN=2$ SCFT is $\NN=4$ super Yang-Mills theory with gauge group $SU(2)$. Regarded as an $\NN=2$ theory, it has flavor symmetry $\hhf = \suf(2)$, the commutant of $SU(2)_R \times U(1)_r$ in the full $SU(4)$ R-symmetry. There are three more recent additions to the list of rank one theories. They were initially discovered in \cite{Argyres:2007tq} by considering the strong coupling limit of Lagrangian theories and then given a class $\SS$ re-interpretation in \cite{Chacaltana:2012ch,Chacaltana:2011ze}. In these theories the Coulomb branch is generated by an $\EE_{r}$ multiplet with $r = 3\,,4\,,6$. These are the same values as in the $E_6$, $E_7$ and $E_8$ theories in Table \ref{Tab:rank1theories}, but the flavor symmetries for these new theories are smaller. Given the serendipitous discovery of these ``new'' rank one theories, one may rightly view with suspicion the claim that the list of known rank one theories is complete. How could we find out?

A systematic study of rank one $\NN=2$ SCFTs has been undertaken by Argyres and collaborators \cite{Argyres:2010py, Argyres} using Seiberg-Witten technology.\footnote{The rank two case is considerably more involved \cite{Argyres:2005pp,Argyres:2005wx}.} Let us give a quick informal summary of this approach. The Coulomb branch chiral ring of a rank one theory is by definition generated by a single operator $\EE_{r_0}$. Assuming the validity of Conjecture \ref{conj:free_coulomb_ring}, this operator should not be nilpotent, and further assuming Conjecture \ref{conj:coulomb_geometrization}, its vacuum expectation value
%%%%%%
\begin{equation}
u \colonequals \langle \EE_{r_0} \rangle~,
\end{equation}
%%%%%%
parametrizes the Coulomb branch of vacua. For $u \neq 0$, the theory admits a low-energy description in terms of an effective $U(1)$ gauge theory, whose data are encoded in a family of elliptic curves \cite{Seiberg:1994rs,Seiberg:1994aj},
%%%%%%
\begin{equation}
y^2 = x^3 + f(u, m_i) x + g(u, m_i)~,
\end{equation}
%%%%%%
and in a meromorphic one form $\lambda_{\rm SW} (u, m_i)$, subject to certain consistency conditions. The complex parameters $\{ m_i \}$ are mass parameters, dual to the Cartan generators of the flavor symmetry algebra $\hhf$ of the theory. For zero masses, the curve must take a scale invariant form, \ie, it must transform homogeneously if one rescales $x$, $y$ and $u$ with the appropriate weights. The scaling weight of $u$ is nothing but the conformal dimension $\Delta = r_0$ of $\EE_{r_0}$. The possible scale-invariant curves are then given by a subset of Kodaira's classification of stable degenerations of elliptic curves depending holomorphically on a single complex parameter. There turn out to be seven cases, and they are the same as the $F$-theory singularities with constant dilaton. Starting from the scale-invariant curve, one can construct its mass deformations (which must be compatible with the existence of the meromorphic one-form $\lambda_{\rm SW}$), and infer the flavor symmetry algebra $\hhf$. It turns out that for a given scale invariant curve there can be numerous inequivalent mass deformations \cite{Argyres:2010py, Argyres}. The ``canonical'' rank one theories of Table \ref{Tab:rank1theories} correspond to the \emph{maximal} mass deformation, but submaximal deformations with smaller flavor symmetry are also possible. An example of this phenomenon that we have already implicitly encountered is the submaximal deformation of the $D4$ singularity, with $\hhf = \suf(2) \subset \sof(8)$, which corresponds to $\NN=4$ SYM with gauge group $SU(2)$. The ``new'' rank one theories of \cite{Argyres:2007tq, Chacaltana:2012ch} are recognized as submaximal deformations of the $E_6$, $E_7$ and $E_8$ Kodaira singularities, but several other possibilities also appear to be consistent\footnote{We are grateful to P. Argyres for sharing some of the results of \cite{Argyres} with us prior to publication.} \cite{Argyres}. In the absence of an independent physical construction (in class $\SS$ or otherwise), it is \emph{a priori} unclear whether the mere existence of a Seiberg-Witten geometry guarantees the existence of a full fledged SCFT. The bootstrap approach should be able to shed light on this question, at the very least by providing some consistency checks of the candidate models.

In summary, even for rank one the situation is not completely settled. There are several established theories and a growing list of possible additional models. A complete elucidation of the rank one case should be a benchmark for our understanding of the $\NN=2$ landscape.

\begin{table}
\centering
\renewcommand{\arraystretch}{1.3}
\begin{tabular}{|c||c|c|c|c|c|c|c|}
\hline
$G$		&	$H_0$	      & $H_1$ 	       & $H_2$ 	        & $D_4$ 	      & $E_6$ 		    & $E_7$ 		  & $E_8$ 			\\ \hline\hline 
$\hhf$	&	--		      & $\suf(2)$      & $\suf(3)$ 	    & $\sof(8)$       & $\ef_6$ 		& $\ef_7$ 		  & $\ef_8$ 		\\ \hline
$\dce$	& 	--		      &	$2$ 		   & $3$ 		    & $6$ 		      & $12$			& $18$			  & $30$			\\ \hline
$k$ 	& $\frac{12}{5}$  & $\frac{8}{3}$  & $3$ 		    & $4$ 		      & $6$ 			& $8$ 			  & $12$ 			\\ \hline
$c$ 	& $\frac{11}{30}$ & $\frac{1}{2}$  & $\frac{2}{3}$  & $\frac{7}{6}$   & $\frac{13}{6}$  & $\frac{19}{6}$  & $\frac{31}{6}$  \\ \hline
$a$   	& $\frac{43}{120}$& $\frac{11}{24}$& $\frac{7}{12}$ & $\frac{23}{24}$ & $\frac{41}{24}$ & $\frac{59}{24}$ & $\frac{95}{24}$ \\ \hline
$r_0$ 	& $\frac{6}{5}$   & $\frac{4}{3}$  & $\frac{3}{2}$  & $2$             & $3$             & $4$             & $6$             \\ \hline
\end{tabular}
\caption{Properties of rank one SCFTs associated to maximal mass deformations of the Kodaira singularities \cite{Argyres:2007cn,Cheung:1997id,Aharony:2007dj}. We list the name of the singularity, the flavor symmetry algebra $\hhf$ and its dual Coxeter number $h^\vee$, the flavor central charge $k$, the $c$ and $a$ anomaly coefficients, and the $U(1)_r$ charge $r_0$ of the Coulomb branch chiral ring generator.\label{Tab:rank1theories}}
\end{table}
%!TEX root = ../draft_maxi_Neq2.tex

\section{The moment map four-point function}
\label{sec:moment_map_correlator}

As our first observable of interest we take the four-point function of moment map operators. 
As explained in the previous section, these are the superconformal primaries for representations containing conserved currents for global symmetries (the $\hat{\BB}_1$ multiplets). This is in some sense the paradigmatic observable by means of which we can investigate SCFTs with flavor symmetries. The moment map operators are spacetime scalars of conformal dimension two, and they transform in the adjoint representation of $SU(2)_R$ while being neutral with respect to $U(1)_r$. Like the conserved currents in the same multiplet, they transform in the adjoint representation of the flavor symmetry group $G_F$. We denote these operators $\phi_{(\II\JJ)}^{A}(x)$, where $\II,\JJ=1,2$ are fundamental indices for $SU(2)_R$ and $A=1,\ldots,\dim G_F$ is an adjoint index for $G_F$.

The purpose of the present section is to describe the structure of this correlation function and to formulate its (super)conformal block decomposition. Let us briefly outline the general trajectory of this analysis. The four-point function of moment map operators can initially be organized to reflect the constraints of conformal symmetry, $SU(2)_R$ symmetry, and $G_F$ flavor symmetry. In practice this means decomposing the general correlator into a number of functions of conformal cross ratios that encode the contributions of operators with fixed transformation properties under $SU(2)_R$ and $G_F$ in the conformal block expansion. These functions are further constrained by superconformal Ward identities \cite{Dolan:2001tt} (see also \cite{Nirschl:2004pa,Dolan:2004mu}). The ultimate result of these Ward identities is that the functions corresponding to different $SU(2)_R$ channels are not independent, but rather the full four-point function is algebraically determined in terms of a set of meromorphic functions $f_i(z)$ and unconstrained functions $\GG_i(z,\bar z)$, where the index $i$ runs over the irreps that appear in the tensor product of two copies of the adjoint representation of $G_F$,
%%%%%%
\begin{equation}\label{eq:adjoint_squared_decomp}
\mathrm{Adj}(G_F)\otimes\mathrm{Adj}(G_F)\equalscolon\bigoplus_{i=1}^n \Rf_i(G_F)~.
\end{equation} 
%%%%%%
The meromorphic functions are identical to the four-point functions of affine currents in two dimensions \cite{Beem:2013sza}, and are completely determined by the flavor central charge. The unconstrained functions $\GG_i(z,\bar z)$ are best considered in a superconformal partial wave expansion. They can be split into two parts which we call $\GG^{\rm short}_i(z,\bar z)$ and $\GG^{\rm long}_i(z,\bar z)$. The former functions encode the contributions of protected operators appearing in the OPE of two moment maps, and under mild assumptions%
%%%
\footnote{The assumption in question is that there are no higher spin conserved currents appearing in the conformal block decomposition. This is expected to hold true for any interacting theory.} %
%%%
they can be completely determined in terms of the central charges $k$ and $c$ by reading off the relevant CFT data from the (now fixed) meromorphic functions. The latter functions encode the spectrum and OPE coefficients of unprotected operators, about which we generally have scant knowledge. The point of the numerical analysis of Section \ref{Sec:Bhatresults} will be to constrain the CFT data encoded in the functions $\GG^{\rm long}_i(z,\zb)$ using crossing symmetry.

\subsection{Structure of the four-point function}
\label{Sec:Bhat4ptfunc}

The appearance of the four-point function in question can be cleaned up a bit by introducing some auxiliary structure. Following \cite{Nirschl:2004pa}, we eliminate the explicit $SU(2)_R$ indices on $\phi_{(\II\JJ)}^{A}(x_i)$ in favor of complex polarization vectors $t^{\II}$ in terms of which we define 
%%%%%%
\begin{equation}
\varphi^A(t,x)\colonequals\phi_{(\II\JJ)}^{A}(x)t^{\II}t^{\JJ}~.
\end{equation} 
%%%%%%
With these conventions, conformal symmetry and $R$-symmetry demand that the four-point function of moment map operators be of the form
%%%%%%
\begin{equation}\label{eq:four_point_form}
\langle \varphi^A(t_1,x_1)  \varphi^B(t_2,x_2) \varphi^C(t_3,x_3) \varphi^D(t_4,x_4)\rangle = \frac{(t_1\cdot t_2)^2(t_3 \cdot t_4)^2}{x_{12}^4 x_{34}^4}  G^{ABCD}(u,v;w)~,
\end{equation}
%%%%%%
where $u$ and $v$ are (standard) conformally invariant cross-ratios,
%%%%%%
\begin{equation}\label{eq:cross_ratio_def}
u\colonequals\frac{x_{12}^2 x_{34}^2}{x_{24}^2x_{13}^2}\equalscolon z \zb~,\qquad v\colonequals\frac{x_{14}^2 x_{23}^2}{x_{24}^2x_{13}^2}\equalscolon(1-z)(1- \zb)~,
\end{equation}
%%%%%%
$w$ is the unique $SU(2)_R-$invariant ``cross-ratio'' of the auxiliary variables,
%%%%%%
\begin{equation}\label{eq:aux_cross_ratio}
w\colonequals\frac{(t_1 \cdot t_2)\, (t_3 \cdot t_4)}{(t_1 \cdot t_3)\, (t_2 \cdot t_4)}~, %= \frac{w_{12} w_{34}}{w_{13} w_{24}}~.
\end{equation}
%%%%%%
and we have defined the contraction $t_i\cdot t_j\colonequals \epsilon_{\II\JJ}t_{i}^{\II}t_{j}^{\JJ}$.

The flavor symmetry of the correlator can be captured by introducing a complete basis $P^{ABCD}_i$ of invariant tensors. We can always choose this basis such that the label $i$ runs over the various irreducible representations $\Rf_i$ of $G_F$ that appear in the tensor product of two copies of the adjoint representation of $G_F$, with the $P_{i}^{ABCD}$ projectors onto this representation. We may then write
%%%%%%
\begin{equation}\label{eq:flavor_channel_decomp}
G^{ABCD}(u,v;w) = \sum_{i \in \mathrm{Adj}\otimes \mathrm{Adj}} G_i(u,v;w)P_{i}^{ABCD}~,
\end{equation}
%%%%%%
and the projectors themselves satisfy
%%%%%%
\begin{equation}\label{eq:projector_condition}
P_{i}^{ABCD} P_{j}^{DCEF}= \delta_{ij} P_{i}^{ABEF}~, \qquad P_{i}^{ABBA}=\mathrm{dim}(R_i)~.
\end{equation}
%%%%%%

For each representation $\Rf_i$ one can decompose the corresponding $G_i(u,v;w)$ into three terms corresponding to the three $SU(2)_R$ channels. In terms of the auxiliary variable $w$ we find
%%%%%%
\begin{equation}\label{eq:R_symmetry_decomp}
G_i(u,v;w)=\sum_{R=0}^{2} a_{i,R} (u,v)P_R(y)~,
\end{equation}
%%%%%%
where we have defined $y=\frac{2}{w}-1$, and the $P_R(y)$ are Legendre polynomials
%%%%%%
\begin{equation}\label{eq:legendre_polys}
P_0(y)=1~,\qquad P_1(y)=y~, \qquad P_2(y)= \frac{1}{2}\left(3 y^2-1\right)~.
\end{equation}
%%%%%%
Each of the $a_{i,R}(u,v)$ has a conventional conformal block decomposition that encodes the exchanged conformal families in the appropriate flavor and $R$-symmetry representations. These conformal blocks are actually grouped together in \emph{super}conformal blocks, as we will explain further below.

The consequences of superconformal covariance for this four-point function have been analyzed in detail in \cite{Dolan:2001tt,Nirschl:2004pa,Dolan:2004mu}. Because supersymmetry transformations commute with flavor symmetries, the superconformal Ward identities apply to each $G_i(u,v;w)$ independently. The end result of the analysis in those papers is neatly encapsulated in the following specialization condition,
%%%%%%
\begin{equation}\label{WardIdentityTwist}
\restr{G_i(u,v;w)}{w=\zb} = f_i( z)~, \qquad \restr{G_i(u,v;w)}{w=z} = f_i(\zb)~,
\end{equation}
%%%%%%
where it is the same meromorphic function $f_i$ appearing in both expressions. We note here that this condition can also be seen to follow from the existence of the superconformal twist introduced in \cite{Beem:2013sza}. In terms of these meromorphic functions, one then finds that the the $G_i(u,v;w)$ take the following form \citep{Nirschl:2004pa},
%%%%%%
\begin{equation}\label{WardIentitySol}
G_i(u,v;w)=\frac{z (w-\bar z)f_i(\bar z)-\bar z (w-z)f_i(z)}{w(z-\bar z)} + \left(1-\frac{z}{w}\right)\left(1- \frac{\bar z}{w}\right)\GG_i(u,v)~.
\end{equation}
%%%%%%
Upon decomposing this expression in the basis of Legendre polynomials of $y$, one recovers expressions for the various $R$-symmetry channels in terms of $f_i$ and $\GG_i$,
%%%%%%
\begin{eqnarray}\label{aR_amplitudes}
a_{i,2}(u,v)&=& \frac{u\,\GG_i(u,v)}{6}~,\\
a_{i,1}(u,v)&=&\frac{u(f_i(z)-f_i(\zb))}{2(z-\zb)}-\frac{(1-v)\GG_i(u,v)}{2}~,\nn\\
a_{i,0}(u,v)&=&\GG_i(u,v) \left(\frac{v+1}{2}-\frac{u}{6}\right)-\frac{u}{2 (z-\zb)} \left(\frac{(2-z) f_i(z)}{z}-\frac{(2-\zb) f_i(\zb)}{\zb}\right)~.\nn
\end{eqnarray}
%%%%%%
We see that (for a given flavor symmetry channel) the functions $a_{i,R}(u,v)$ are not independent; instead they are all determined in terms of the meromorphic function $f_i(z)$ and a single unconstrained function $\GG_i(u,v)$.

%%%%%%%%%%%%%%%%%%%%%%%%%%%%%%%%%%%%%%%%%%%%%%%%%%%%%%%%%%%%%%%%%%%%%%%%%%%%%%%%%%%%%%%%%%%%%%%%%%%%%%%%%%%%
\subsubsection{Constraints of crossing symmetry}\label{subsec:su2_crossing}
%%%%%%%%%%%%%%%%%%%%%%%%%%%%%%%%%%%%%%%%%%%%%%%%%%%%%%%%%%%%%%%%%%%%%%%%%%%%%%%%%%%%%%%%%%%%%%%%%%%%%%%%%%%%

As a consequence of Bose symmetry, the four-point function must be invariant under arbitrary permutations of the four inserted operators. For the functions $G_i(u,v;y)$, these permutations lead to the following relations,
%%%%%%
\begin{alignat}{3}
\label{braiding}
&(x_1,t_1)\longleftrightarrow(x_2,t_2)&& \qquad\implies\qquad &&G_i \left(\frac{u}{v},\frac{1}{v};-y\right)=(-1)^{\mathrm{symm}(i)} G_i (u,v;y)~,\\
\label{crossing}
&(x_1,t_1)\longleftrightarrow(x_3,t_3)&& \qquad\implies\qquad &&\frac{1-2y+y^2}{4} G_i \left(v,u;\frac{y+3}{y-1}\right)=\frac{v^2}{u^2}G_j(u,v;y) F^{\ph{i}j}_i~.
\end{alignat}
%%%%%%
The first of these is called the \emph{braiding relation}, while we refer to the second as the \emph{crossing symmetry equation}. We have introduced the notation $\mathrm{symm}(i)$ which is equal to zero or one if representation $i$ appears in the symmetric or antisymmetric tensor product of two copies of the adjoint, respectively. The matrix $F^{\ph{i}j}_i$ relates the projectors in one channel with the projectors in the crossed channel:
%%%%%%
\begin{equation}
P_i^{ABCD}=P_j^{CBAD}  F^{\ph{i}j}_i~,
\label{project_diff_chan}
\end{equation}
%%%%%%
and is related to ``Wigner's 6j-coefficients'' (see, \eg, \cite{Cvitanovic:2008zz}). In the cases considered in the present work this matrix satisfies $F^{\ph{i}j}_iF^{\ph{i}k}_j=\delta_i^{\ph{i}k}$.

The corresponding constraints for the functions $f_i(z)$ and $\GG_i(u,v)$ are obtained from \eqref{braiding} and \eqref{crossing} by using the solution for $G_i(u,v;y)$ from \eqref{WardIentitySol} and reading off the constraints term by term in a $y$-expansion. This exercise was already worked out without the flavor symmetry structure in \citep{Nirschl:2004pa}. Upon including flavor symmetry indices we find two sets of relations involving only the meromorphic functions,
%%%%%%
\begin{equation}\label{eq:crossingf}
f_i(z)=(-1)^{\mathrm{symm}(i)}f_i\left(\frac{z}{z-1}\right)~,\qquad z^2 f_i (1-z)=(1-z)^2 f_j (z) F^{\ph{i}j}_i~,
\end{equation}
%%%%%%
one braiding equation involving only the two-variable functions,
%%%%%%
\begin{equation}\label{braidG}
\GG_i(u,v)=(-1)^{\mathrm{symm}(i)}\frac{1}{v}\GG_i\left(\frac{u}{v},\frac{1}{v}\right)~.
\end{equation}
%%%%%%
There is one additional non-trivial crossing symmetry relation for the unconstrained function,
%%%%%%
\begin{eqnarray}\label{bootstrapeqn}
&&(z-\zb)(1-z)^2(1-\zb)^2 F^{\ph{j}i}_j \GG_j(u,v)+z^2 \zb^2 (\zb-z) \GG_i(v,u)\nn\\
&&+z \zb \Big(z(\zb-1)f_i(1-z)-\zb(z-1)f_i(1-\zb)\Big)=0~.
\end{eqnarray}
%%%%%%
This is the equation that we will investigate numerically. Before doing so we have to first compute its superconformal block decomposition and solve the other crossing symmetry equations, in particular the last equation in \eqref{eq:crossingf}. We will discuss these two topics in the next two subsections.

%%%%%%%%%%%%%%%%%%%%%%%%%%%%%%%%%%%%%%%%%%%%%%%%%%%%%%%%%%%%%%%%%%%%%%%%%%%%%%%%%%%%%%%%%%%%%%%%%%%%%%%%%%%%
\subsubsection{Fixing the meromorphic functions}\label{subsec:su2_meromorphic}
%%%%%%%%%%%%%%%%%%%%%%%%%%%%%%%%%%%%%%%%%%%%%%%%%%%%%%%%%%%%%%%%%%%%%%%%%%%%%%%%%%%%%%%%%%%%%%%%%%%%%%%%%%%%

By meromorphicity, the single-variable functions $f_i(z)$ are fixed completely by the structure of their singularities. The only physically allowable singularities occur when two of the inserted operators collide, \ie, at $z=0$, $z=1$, and $z\to\infty$. The equations in \eqref{eq:crossingf} relate the singularities at these three points, so it suffices to specify the singular behavior of $f_i(z)$ near, say, $z = 0$. This simple crossing symmetry problem is reminiscent of what arises in the study of two-dimensional meromorphic conformal field theories. Indeed, a compelling physical picture that explains the relationship between this crossing symmetry problem and the two dimensional case has been presented in \cite{Beem:2013sza}. There it was shown that the functions $f_i(z)$ are precisely equal to the four-point functions of an extended chiral algebra that can be isolated by working at the level of cohomology relative to a particular nilpotent supercharge. Indeed, the equations \eqref{eq:crossingf} are exactly the crossing equations one encounters in studying chiral algebra four-point functions.

In \cite{Beem:2013sza} it was found that the moment maps $\phi_{(\II\JJ)}^{A}(x)$ are related to dimension one affine currents in the corresponding chiral algebra. These affine currents generate an affine Kac-Moody (AKM) algebra $\widehat{G_F}$. The level $k_{2d}$ of this AKM algebra is related to the four-dimensional flavor central charge $k$ as
%%%%%%
\begin{equation}
k_{2d} = - \frac{k}{2}~.
\end{equation}
%%%%%%
Many more details about these chiral algebras can be found in \cite{Beem:2013sza} (see also \cite{Beem:2014kka,Beem:2014rza}). For our purposes here we need only know that the chiral algebra completely determines the one-variable part of the four-point function $f^{ABCD}(z)$ to be the four-point function of affine currents, which for any group $G_F$ takes the form:
%
%%%%%%
\begin{equation}
f^{ABCD}(z)=\delta^{AB}\delta^{CD}+z^2\delta^{AC}\delta^{BD}+\frac{z^2\delta^{AD}\delta^{CB}}{(1-z)^2}+\frac{2z}{k}f^{ACE}f^{BDE}+\frac{2z}{k(z-1)}f^{ADE}f^{BCD}~.
\label{fABCD}
\end{equation}
%%%%%%
Note that the normalization here is such that the current has a canonical two-point function, so the level $k$ appears in the denominator in this expression.

%%%%%%%%%%%%%%%%%%%%%%%%%%%%%%%%%%%%%%%%%%%%%%%%%%%%%%%%%%%%%%%%%%%%%%%%%%%%%%%%%%%%%%%%%%%%%%%%%%%%%%%%%%%%
\subsection{Superconformal partial wave expansion}
\label{Sec:BhatWardID}
%%%%%%%%%%%%%%%%%%%%%%%%%%%%%%%%%%%%%%%%%%%%%%%%%%%%%%%%%%%%%%%%%%%%%%%%%%%%%%%%%%%%%%%%%%%%%%%%%%%%%%%%%%%%

So far we have understood the functional form of the four-point function as follows from $\suf(2,2|2)$ symmetry and an analysis of the associated chiral algebra. The next step is to consider the superconformal partial wave expansion of the correlator.

The supersymmetric OPE of a $\hat{\BB}_1$ representation with itself has been studied in \cite{Arutyunov:2001qw}. The approach taken in that paper was to analyze all possible three-point functions of two $\hat{\BB}_1$ representations with a third \emph{a priori} generic representation in harmonic superspace. The result can be summarized in the following ``fusion rule'',
%%%%%%
\begin{equation}
\label{B1OPE}
\hat{\BB}_1 \times \hat{\BB}_1 \sim \mathbf{1}+\hat{\BB}_1+\hat{\BB}_2+\hat{\CC}_{0(j,j)}+\hat{\CC}_{1(j,j)}+\AA^{\Delta}_{0,0(j,j)}~.
\end{equation}
%%%%%%
This fusion rule can be further refined by taking into account flavor symmetry representations, which lead to some additional constraints. For example, long multiplets can appear in all possible flavor symmetry representations but the stress tensor multiplet $\hat{\CC}_{0(0,0)}$ can only appear as a flavor singlet. The precise selection rules are summarized in Table \ref{tab:flavor_selection}.

Each superconformal multiplet $\XX$ in flavor representation $i$ that appears on the right-hand side of \eqref{B1OPE} must contribute a finite number of conventional conformal blocks to each of the three functions $a_{i,R}(u,v)$ with $0 \leqslant R \leqslant 2$. We denote these contributions as $a^{\XX}_{i\,R}(u,v)$. For this particular four-point function the coefficients of the conventional conformal blocks are all related by the superconformal Ward identities \cite{Dolan:2001tt}, and we end up with just a single undetermined OPE coefficient (squared) for each superconformal block. This leads to the decomposition:
%%%%%%
\begin{equation}
G^{ABCD}(u,v;y)=
\sum_{i} P_i^{ABCD}\sum_{\XX \text{ in rep }i} (-1)^{\text{symm}(i)} \lambda^2_{i\,\XX}  \left( \sum_{R=0}^2 P_R(y) a^{\XX}_{i\,R}(u,v)\right)~,
\label{OPEcoeffsw}
\end{equation}
%%%%%%
where the term in parentheses is the superconformal block. The factor of $(-1)^{\text{symm}(i)}$ follows from reflection positivity. In a unitary theory the $\lambda_{i\,\XX}$ are real and their square is therefore always positive.

The complete set of superconformal blocks for this four-point function was obtained in \cite{Dolan:2001tt}. It is most naturally given in terms of the functions $\GG_i(u,v)$ and $f_i(z)$ introduced above, which is presented in Table \ref{tab:allsuperblocks}, where $\ell=2j_1=2j_2=2j$ since all the multiplets appearing in this OPE have $j_1=j_2$.
%%%%%%
\begin{table}[h!t]
\begin{center}
\renewcommand{\arraystretch}{1.75}
\begin{tabular}{|l|l|l|l|}
\hline
~Multiplet~		&	Contribution to $\GG_i^{\XX}(u,v)$	&	Contribution to $f^{\XX}_i(z)$	&	Restrictions\\
\hline
\hline
${\mathrm{Id.}}$					&	$0$														&$1$		&  \\
\hline
${\AA^{\Delta}_{0,0 j,j)}}$	&	$6 u^{\frac{\Delta-\ell}{2}} G_{\Delta+2}^{(\ell)}(u,v)$	& $0$		& $\Delta \geqslant \ell + 2$ \\
\hline
${\hat \CC_{0(j,j)}}$				&	$0$ 									&	$2g_{2j+2}(z)$				& $j \geqslant 0	$ \\
\hline
${\hat{\BB}_1}$						&	$0$					 				&	$2g_{1}(z)$				&				 \\
\hline
${\hat \CC_{1(j,j)}}$				&	$6 u G_{\ell+5}^{(\ell+1)}(u,v)$		&	$-12 g_{2j+3}(z)$			& $j \geqslant 0$ \\
\hline
${\hat{\BB}_2}$						&	$6 u G_{4}^{(0)}(u,v)$					&	$-12 g_{2}(z)$				&				 \\
\hline
\end{tabular}
\caption{Superconformal blocks for the $\EE_{r_0}$ four point function in the $\hat{2}$ channel.\label{tab:allsuperblocks}}
\end{center}
\end{table}
%%%%%%
In the table $G_{\Delta}^{(\ell)}(u,v)$ denotes the four-dimensional conformal block which is given by \eqref{eq:bos_block} in our conventions, and 
%%%%%%
\begin{equation}
\label{2dblock}
g_{\ell}=\left(-\frac{z}{2}\right)^ {\ell-1} z {}_2 F_1(\ell,\ell;2\ell;z)\,
\end{equation}
%%%%%%
is a two-dimensional conformal block in a chiral algebra, as we discuss in more detail below. Through \eqref{aR_amplitudes}, the contribution of each superconformal multiplet to the $a_{i,R}(u,v)$ is obtained from the contribution of said multiplet to $\GG_i(u,v)$ and $f_i(z)$. This is worked out in detail in Appendix~\ref{App:blocksBhat}.

From inspection of Table \ref{tab:allsuperblocks} it follows that the decomposition into superconformal blocks of a given four-point function can be ambiguous. For example, a long multiplet at the unitarity bound $\Delta = \ell + 2$ contributes in exactly the same manner as a combination of two short multiplets. These ambiguities can be understood from the fact that these two multiplets can recombine to form a long multiplet according to\footnote{Appendix \ref{App:representations} provides an overview of all the recombination rules for the unitary irreps of $\suf(2,2|2)$.}
%%%%%%
\begin{equation}
\AA^{\Delta=2j+2}_{0,0(j,j)} \simeq \hat{\CC}_{0(j,j)} \oplus \hat{\CC}_{\frac{1}{2}(j-\frac{1}{2},j)} \oplus \hat{\CC}_{\frac{1}{2}(j,j-\frac{1}{2})}\oplus \hat{\CC}_{1(j-\frac{1}{2},j-\frac{1}{2})}~.
\label{longdecomp}
\end{equation}
%%%%%%
Only the first and last multiplet are allowed in the OPE of two scalars. For the case $j=0$ we can use $ \hat{\CC}_{1 (- \frac12 ,- \frac12)} =  \hat{\BB}_2$ and we get $\simeq \hat{\CC}_{0 (0,0)} + \hat{\BB}_2$ on the right-hand side. These ambiguities will be fixed below.

The braiding relations \eqref{braidG} together with Table \ref{tab:allsuperblocks} correlate the allowed spins of multiplet $\XX_i$ to $\mathrm{symm}(i)$: only even/odd spins appear in $\GG_i(u,v)$ for a representation appearing in the symmetric/antisymmetric tensor product, respectively. This follows from the braiding relations from the individual conformal blocks, $G_\Delta^{(\ell)}(u,v)=(-1)^{\ell} v^{-\frac{\Delta-\ell}{2}} G_\Delta^{(\ell)}(\frac{u}{v},\frac{1}{v})$. As an example, for flavor singlets the spin of these operators is always even and for the flavor adjoint multiplets these spins are always odd.

While the meromorphic functions $f_i(z)$ receive contributions only from short multiplets, the two-variable functions $\GG_i(u,v)$ include contributions from both long and short multiplets. It is then useful to split the two-variable functions into the long and short contributions appearing in a given channel,
%%%%%%
\begin{equation}
\GG_i(u,v) = \GG_i^{\mathrm{short}}(u,v)+\GG_i^{\mathrm{long}}(u,v)~,
\end{equation}
%%%%%%
where we have
%%%%%%
\begin{equation}
\begin{split}
\makebox[.7in][l]{$\GG_i^{\mathrm{short}}(u,v)$}&=~ 6 \lambda^2_{i\,\hat \BB_2}u G_{4}^{(0)}(u,v) + \sum_{\ell=0,1,\ldots} 6 \lambda^2_{i\,\hat \CC_{1(j,j)}} (-1)^\ell u\,G_{\ell+5}^{(\ell+1)}(u,v)~,\\
\makebox[.7in][l]{$\GG_i^{\mathrm{long}}(u,v)$}	&=~ \sum_{\Delta \geqslant \ell + 2,\ell} 6 \lambda^2_{i\,\AA^{\D}_{\ell}} (-1)^{\ell} u^{\frac{\Delta-\ell}{2}} G_{\Delta+2}^{(\ell)}(u,v)~.
\end{split}
\label{Gshort_Glong}
\end{equation}
%%%%%%
In the next subsection we will show that the coefficients of the short superconformal blocks -- and therefore the complete functional form of $\GG_i^{\mathrm{short}}(u,v)$ -- are completely fixed from the chiral algebra correlator \eqref{fABCD}. All the undetermined information in the four-point function is then contained in $\GG_i^{\mathrm{long}}(u,v)$. These are the functions that will be analyzed numerically in Section \ref{Sec:Bhatresults}.

%%%%%%%%%%%%%%%%%%%%%%%%%%%%%%%%%%%%%%%%%%%%%%%%%%%%%%%%%%%%%%%%%%%%%%%%%%%%%%%%%%%%%%%%%%%%%%%%%%%%%%%%%%%%
\subsubsection{Fixing the short multiplets}
\label{Sec:Bhat4crossing_mini}
%%%%%%%%%%%%%%%%%%%%%%%%%%%%%%%%%%%%%%%%%%%%%%%%%%%%%%%%%%%%%%%%%%%%%%%%%%%%%%%%%%%%%%%%%%%%%%%%%%%%%%%%%%%%

Because the meromorphic functions $f_i(z)$ are completely determined by crossing symmetry (or alternatively by analyzing the associated chiral algebra), their decomposition in chiral blocks of the form \eqref{2dblock} is determined. We can thus write
%%%%%%
\begin{equation}
\label{finblocks}
f_i(z)= \sum_{\ell\geqslant-2} b_{i,\,\ell} (-1)^{\ell} g_{\ell+2}(z)~,
\end{equation}
%%%%%%
with known coefficients $b_{i,\,\ell}$. Upon examining the contributions of general supermultiplets to $f_i(z)$ in Table \ref{tab:allsuperblocks}, we see that the chiral OPE coefficients are related to four-dimensional OPE coefficients of the short multiplets as follows,
%%%%%%
\begin{eqnarray}
b_{\mathbf 1,-2} &=& \lambda^2_{\mathbf{1},\,\text{Id}}~,\nn\\
b_{i,-1} &=& 2 \lambda^2_{i,\,\hat{\BB}_1}~, \label{btolambda}\\
b_{i,\,0} &=& 2\lambda^2_{i,\,\hat{\CC}_{0(0,0)}} - 12 \lambda^2_{i,\,\hat{\BB}_2}~,\nn\\
b_{i,\,\ell} &=& 2\lambda^2_{i,\,\hat{\CC}_{0(j,j)}} -12 \lambda^2_{i,\,\hat{\CC}_{1(j-\frac{1}{2},j-\frac{1}{2})}}~, \qquad \ell=2j \geqslant 1~.\nn
\end{eqnarray}
%%%%%%
Note that in the first line, the identity operator can only appear in the flavor singlet channel $i=\mathbf{1}$. If we now further assume that the theory has (a) no higher spin currents, and (b) a unique stress tensor, then one can actually fix the OPE coefficients of all short multiplets. Indeed, the first assumption implies the absence of any $\hat\CC_{0(j,j)}$ multiplets with $j \geqslant 1$, so in particular
%%%%%%
\begin{equation}
\lambda^2_{i,\,\hat{\CC}_{0(j,j)}}=0 \quad {\rm for} \quad \ell = 2j \geqslant 1~.
\end{equation}
%%%%%%
Our second assumption implies that there is a unique multiplet of type $\hat{\CC}_{0(0,0)}$, which is a flavor singlet, and whose OPE coefficient is fixed in terms of the $c$ central charge according to
%%%%%%
\begin{equation}
\label{chat0coeff}
\lambda^2_{i,\,\hat{\CC}_{0(0,0)}} = \frac{\dim{G_F}}{6c} \delta_{i,\mathbf{1}}~,
\end{equation}
%%%%%%
where $\dim{G_F}$ is the dimension of $G_F$. This numerical value follows from conformal Ward identities upon imposing appropriate normalization conventions which we spell out below.

With these additional conditions, we see that \eqref{btolambda} completely determines the OPE coefficients $\lambda^2_{i,\,\hat{\BB}_1}$, $\lambda^2_{i,\,\hat{\BB}_2}$, and $\lambda^2_{i,\,\hat{\CC}_{1(j,j)}}$, in addition to the coefficient of the identity $\lambda^2_{\mathbf{1},\,\text{Id}}$ which is merely an overall normalization. The remaining undetermined variables in the four-point function are the spectrum of long multiplets $\AA^{\Delta}_{\ell}$ and the corresponding OPE coefficients $\lambda^2_{i,\,\AA^{\Delta}_\ell}$. This demonstrates how the chiral algebra leads to a clear distinction between the contributions of the short multiplets, which we can solve analytically, and the contribution of the long multiplets, which we can determine only numerically.

The precise values of the coefficient $b_{i,\,\ell}$ can be read off from \eqref{fABCD} after decomposing it in the different flavor symmetry channels, using the projectors $P^{ABCD}_i$. The form of these projectors generally depends on $G_F$, see for example \cite{Cvitanovic:2008zz} for many examples. For the singlet and adjoint representation the projectors always have the same universal form, so we can discuss the corresponding decomposition in full generality.

The projector onto the singlet channel is always given by $P_{\bf{1}}^{ABCD} = \frac{1}{\text{dim}G_F} \delta^{AB} \delta^{CD}$, where the normalization is chosen such that the trace of the projector corresponds to the dimension of the representation. The projection of \eqref{fABCD} in the singlet channel then yields:
%%%%%%
\begin{equation}\label{eq:singlet_channel_expansion}
\begin{split}
f_{\mathbf{1}}(z)&=\dim{G_F} + z^2 \left(1+\frac{1}{(1-z)^2}\right) + \frac{2 \psi^2 z^2 h^{\vee}}{k(z-1)} \\
&= \dim{G_F}-\sum_{\ell=0,2,\cdots} \frac{2^{\ell } (\ell +1) (\ell !)^2 \left(2 (\ell +1) (\ell +2) k - 4 \psi^2 \, h^{\vee} \right)}{k(2 \ell +1)!} g_{\ell+2}(z)~,
\end{split}
\end{equation}
%%%%%%
where $h^{\vee}$ is the dual Coxeter number of $G_F$, and $\psi^2$ the length squared of the longest root. In a similar vein, the adjoint projector is fixed to be $P_{\mathrm{Adj.}}^{ABCD}= \frac{1}{\psi^2 h^{\vee}} f^{ABE}f^{EDC}$, which traces to $\dim{G_F}$, and so we find that for any flavor group:
%%%%%%
\begin{equation}\label{eq:adjoint_channel_expansion}
\begin{split}
f_{\mathrm{Adj.}}(z)&=-\frac{(z-2) z \left(\frac{h^{\vee} \psi^2 z}{k}-\frac{h^{\vee} \psi^2}{k}+z^2\right)}{(z-1)^2} \\
&=-\frac{2 \psi^2 h^{\vee}}{k}g_{1}(z)+\sum_{\ell=1,3,\cdots}\frac{2^{\ell } (\ell +1) (\ell !)^2 \left(2 (\ell +1) (\ell +2)-\frac{2 h^{\vee} \psi^2}{k}\right)}{(2 \ell +1)!} g_{\ell+2}(z)~.
\end{split}
\end{equation}
%%%%%%
Equations \eqref{eq:singlet_channel_expansion} and \eqref{eq:adjoint_channel_expansion} determine an infinite number of coefficients $b_{i\,\ell}$. It is worthwhile to analyze the coefficients of the first few low-lying operators in more detail.

Let us begin with the identity operator, which only appears in the singlet channel. From equations \eqref{btolambda} and \eqref{eq:singlet_channel_expansion} we find
%%%%%%
\begin{equation}
\lambda^2_{\mathbf{1},\,\text{Id}} = \dim{G_F}~.
\end{equation}
%%%%%%
The explicit factor $\dim{G_F}$ cancels against the same factor in the projector and therefore the operator is unit normalized, so
%%%%%%
\begin{equation}
\langle\phi^{A}(t_1,x_1)\phi^{B}(t_2,x_2) \rangle = \frac{(t_1 \cdot t_2)\delta^{AB}}{|x_1 - x_2|^4}\,
\end{equation}
%%%%%%
in our conventions.

Next we consider the $\ell=-1$ term in \eqref{finblocks}. This block corresponds to the $\hat{\BB}_1$ multiplet and therefore appears only in the adjoint flavor channel. From \eqref{btolambda} and \eqref{eq:adjoint_channel_expansion} we obtain that
%%%%%%
\begin{equation}
\label{lambda2hatbb1}
\lambda^2_{\hat{\BB}_1} =  \frac{\psi^2 h^{\vee}}{k}~.
\end{equation}
%%%%%%
In Table \ref{Tab:moment_map_blocks} in Appendix \ref{App:blocks} we expanded the superconformal block into a sum of conventional conformal blocks, and with the given coefficient we find the correct contribution of the flavor current conformal block for a four-point function of adjoint fields, see, \eg, \cite{Poland:2010wg,Poland:2011ey}.

At the next order in \eqref{finblocks} we find the coefficients $b_{i,0}$ which according to \eqref{btolambda} get contributions from $\hat{\CC}_{0(0,0)}$ and $\hat \BB_2$ multiplets. 
As we mentioned above, the former multiplet contains the stress tensor and we can fix its coefficient in terms of the central charge. In a general CFT, the contribution of the stress tensor conformal block to the four-point function of a scalar of dimension $2$ is, \eg, \cite{Poland:2010wg,Poland:2011ey}
%%%%%%
\begin{equation}
x_{12}^{4} x_{34}^{4} \langle \phi(x_1)\phi(x_2)\phi(x_3) \phi(x_4) \rangle \sim \frac{4}{90c} u G_{4}^{(2)} \, .
\end{equation}
%%%%%%
According to the third entry in Table \ref{Tab:moment_map_blocks} this conformal block appears in the superconformal block with a factor of $\frac{4}{15}$. After adding an additional factor $\dim{G_F}$ in order to cancel the corresponding factor in the singlet projector we recover \eqref{chat0coeff}.
Using this equation in conjunction with \eqref{btolambda} and the expression of $b_{\mathbf{1},\,0}$ that can be read off from \eqref{eq:singlet_channel_expansion}, we find that for any flavor group
%%%%%%
\begin{equation}
\lambda^2_{\mathbf{1},\,\hat{\BB}_2}= \frac{1}{12}\left(\frac{\dim{G_F}}{3c} -\frac{4\psi^2 h^{\vee}}{k}+4 \right)~.
\end{equation}
%%%%%%
This coefficient must be positive for unitarity theories, and so we obtain a constraint on the allowed values of $k$ and $c$ for a given flavor group $G_F$:
%%%%%%
\begin{equation}
\label{centralchargebound}
\frac{\dim{G_F} }{ c }\geqslant \frac{12 \psi^2 h^{\vee} }{ k }-12~.
\end{equation}
%%%%%%
This is one of the unitary bounds obtained in \cite{Beem:2013sza}. Its saturation corresponds to the absence of the $\hat{\BB}_2$ multiplet in the singlet representation, which implies a relation in the Higgs branch chiral ring of these theories.

Finally, from the last line of \eqref{btolambda} we see that for $j>1$ the two multiplets contributing to the meromorphic function are $\hat{\CC}_{0 (j,j)}$ and $\hat{\CC}_{1(j-\frac{1}{2},j-\frac{1}{2})}$. As we already mentioned before, the $\hat{\CC}_{0 (j,j)}$ multiplets contain conserved currents of spin higher than two and are not expected to be present in an interacting theory and we can set the corresponding OPE coefficients to zero. In the singlet channel this for example directly leads to
%%%%%%
\begin{equation}
\lambda^2_{\mathbf{1},\,\hat{\CC}_{1(0,0)}}=\frac{2}{5} \left(2-\frac{\psi^2 h^{\vee}}{3k}\right)~,
\end{equation}
%%%%%%
whose positivity implies another bound of \cite{Beem:2013sza},
%%%%%%
\begin{equation}
\label{singletkbound}
k \geqslant \frac{\psi^ 2\dce}{6}~.
\end{equation}
%%%%%%
This bound can also be found by using similar arguments in the adjoint channel.

In what follows we will fix the normalization of the longest root to be $\psi^2=2$.

%%%%%%%%%%%%%%%%%%%%%%%%%%%%%%%%%%%%%%%%%%%%%%%%%%%%%%%%%%%%%%%%%%%%%%%%%%%%%%%%%%%%%%%%%%%%%%%%%%%%%%%%%%%%
\subsection{\texorpdfstring{$\suf(2)$ global symmetry}{su(2) global symmetry}}
\label{subsec:su2_case}
%%%%%%%%%%%%%%%%%%%%%%%%%%%%%%%%%%%%%%%%%%%%%%%%%%%%%%%%%%%%%%%%%%%%%%%%%%%%%%%%%%%%%%%%%%%%%%%%%%%%%%%%%%%%

The first special case of the above structure that we will investigate is the case of global symmetry algebra $\suf(2)$. This is quantitatively the simplest case because it is the unique simple algebra for which only three irreducible representations appear in the tensor product of two copies of the adjoint representation -- in particular, we have
%%%%%%
\begin{equation}\label{eq:su2_adjoint_decomp}
\textbf{3} \otimes \textbf{3}= \textbf{1} \oplus \textbf{3} \oplus \textbf{5}~.
\end{equation}
%%%%%%
Interesting examples of $\NN=2$ superconformal theories with $\suf(2)$ flavor symmetry are the theory of a single $D3$ brane probing an $H_1$ singularity in F-theory as well as the theories of any number $n>1$ of $D3$ branes probing any of the F-theory singularities listed in Table \ref{Tab:rank1theories}. (Recall also that the theory of a single free hypermultiplet is invariant under an $\suf(2)_F$ flavor symmetry.)

The projectors onto each of the representations in \eqref{eq:su2_adjoint_decomp} are easy to compute, see, \eg,~\cite{Cvitanovic:2008zz},
\begin{align}
\nn
P_{\text{\bf{1}}}^{A B C D} & = \frac{1}{3}\delta^{A B}\delta^{C D}~,
\\
\nn
P_{\text{\bf{3}}}^{A B C D} & = \frac{1}{2}\left(\delta^{A D}\delta^{C B}-\delta^{A C}\delta^{B D}\right)~,
\\
\nn
P_{\text{\bf{5}}}^{A B C D} & = \frac{1}{2}\left(\delta^{A D}\delta^{C B}+\delta^{A C}\delta^{B D}\right)-P_{\text{\bf{1}}}^{A B C D}~,
\end{align}
where $A=1 \ldots 3$ is an adjoint index. From \cite{Cvitanovic:2008zz} the $F$ matrix can be computed as
%%%%%%
\begin{equation}
F^{\ph{i}j}_i = \frac{1}{\text{dim}(j)}P_i^{B D C A} P_j^{A B D C}~,
\label{Fmat}
\end{equation}
%%%%%%
where $\text{dim}(j) = P_j^{A B B A}$. We will arrange the rows and columns of $F$ such that $i,j=1,2,3$ correspond to the \textbf{1}, \textbf{3}, \textbf{5} channels respectively. The $F$ matrix for $\suf(2)$ is then,
%%%%%%
\begin{equation}
F = \left(
\begin{array}{ccc}
 \frac13 & \ph{-}\frac13 & \ph{-}\frac13 \\
 1 		 & \ph{-}\frac12 & -\frac12 	 \\
 \frac53 & -\frac56 	 & \ph{-}\frac16 \\
\end{array}
\right)~.
\end{equation}
%%%%%%
We can now use equation \eqref{fABCD} and compute the $f_i(z)$ functions,
%%%%%%
\begin{align}
\label{f_su2}
\nn
f_{\text{\bf{1}}}(z) & = \frac{3 - 6 z + (5 - \frac{8}{k}) z^2 - (2 - \frac{8}{k}) z^3 + z^4}{(1 - z)^2}~,
\\
%f_{\text{\bf{3}}}(z) & = -\frac{(-2 + z) z (\tilde k - \tilde k z + z^2)}{(1 - z)^2}
f_{\text{\bf{3}}}(z) & = \frac{-\frac8k z+\frac{12}{k}z^2+(2-\frac4k)z^3-z^4}{(1 - z)^2}~,
\\
\nn
%f_{\text{\bf{5}}}(z) & = \frac{z^2 (2 + \tilde k (-1 + z) - 2 z + z^2)}{(1 - z)^2}
f_{\text{\bf{5}}}(z) & = \frac{(2+\frac4k)(z^2-z^3)+z^4}{(1 - z)^2}~.
\end{align}
%%%%%%
We have chosen conventions in which the flavor central charge of the free hypermultiplet is $k=1$.

As described in the previous subsection, we can use this expression to solve for the $b_{i,\,\ell}$ coefficients in the expansion \eqref{finblocks}. By demanding that the stress tensor is unique and that the theory does not contain higher spin currents we find the OPE coefficients of all the semishort multiplets. After performing the infinite sums in the $\GG^{\mathrm{short}}_i(u,v)$ we will be left with a crossing equation involving only long operators. The final expressions for $\GG^{\mathrm{short}}_i(u,v)$ are given in \eqref{G_short_su2}. The singlet and quintuplet channels are symmetric and so they involve only even spins in the expansion of $f_i(z)$ and $\GG_i(u,v)$, while the triplet channel is antisymmetric and contains only odd spins. 

%%%%%%%%%%%%%%%%%%%%%%%%%%%%%%%%%%%%%%%%%%%%%%%%%%%%%%%%%%%%%%%%%%%%%%%%%%%%%%%%%%%%%%%%%%%%%%%%%%%%%%%%%%%%
\subsection{\texorpdfstring{$\ef_6$ global symmetry}{e6 global symmetry}}
\label{Sec:E6flavor}
%%%%%%%%%%%%%%%%%%%%%%%%%%%%%%%%%%%%%%%%%%%%%%%%%%%%%%%%%%%%%%%%%%%%%%%%%%%%%%%%%%%%%%%%%%%%%%%%%%%%%%%%%%%%

As a second case we consider theories with global symmetry $\ef_6$. This flavor symmetry group also arises in the $F$-theory singularities described above. From the point of view of the crossing symmetry relations, this is actually the second simplest case because five irreducible representations appear in the square of the adjoint representation, 
%%%%%%
\begin{equation}\label{eq:e6_adjoint_decomp}
\textbf{78} \otimes \textbf{78} = \textbf{1} \oplus \textbf{650} \oplus \textbf{2430} \oplus \textbf{78} \oplus \textbf{2925}~,
\end{equation}
%%%%%%
whereas for all other simple groups (aside from $\suf(2)$) there are five or more representations. The projection tensors for $\ef_6$ can be found in \cite{Cvitanovic:2008zz}, in terms of which the $F$ matrix can be computed using \eqref{Fmat}:
%%%%%%
\begin{equation}
F = \left(  \begin{array}{ccccc}
\frac{1}{78} 	& \frac{1}{78} 	& \frac{1}{78} 	& \frac{1}{78} 	  & \frac{1}{78}
\\
\frac{25}{3} 	& -\frac{7}{24} & \frac{5}{24} 	& \frac{25}{12}   & -\frac{1}{6}
\\
\frac{405}{13} 	& \frac{81}{104}& \frac{29}{104}& -\frac{135}{52} & -\frac{9}{26}
\\
1 				& \frac{1}{4} 	& -\frac{1}{12}	& \frac{1}{2} 	  & 0
\\
\frac{75}{2} 	& -\frac{3}{4} 	&  -\frac{5}{12}& 0 			  & \frac{1}{2}
\end{array}\right)~.
\end{equation}
%%%%%%
The indices of the above matrix $F^{\ph{i}j}_i$ run over the ordered set of irreps $i,j \in \{\textbf{1},\textbf{650},\textbf{2430},\textbf{78},\textbf{2925}\}$. 

Positivity of the coefficient of the $\hat{\BB}_2$ multiplet in the $\mathbf{650}$ representation requires
%%%%%%
\begin{equation}\label{E6kbound}
k\geqslant 6~,
\end{equation}
%%%%%%
with saturation of the bound occurring when the coefficient of $\hat{\BB}_2$ goes to zero. The absence of this multiplet corresponds to a relation in the Higgs branch chiral ring \cite{Beem:2013sza}. The only known theory with $k=6$ is the rank one $E_6$ theory.

As before we now compute the $f_i(z)$ functions, which are given in \eqref{f_e6}.
Once again we can use these expressions to solve for the coefficients of the short multiplets and to perform the infinite sums in $\GG^{\mathrm{short}}_i(u,v)$, the final results are given in \eqref{G_short_e6}. We note that the channels $\textbf{1}$, $\textbf{650}$ and $\textbf{2430}$ appear in the symmetric tensor product, while channels $\textbf{78}$ and $\textbf{2925}$ appear in the antisymmetric tensor product. As such the former will only include even spins and the latter only odd spins.
%!TEX root = ../draft_maxi_Neq2.tex

\section{The \texorpdfstring{$\EE_r$}{Er} four-point function}
\label{sec:epsilon_correlator}

Our second observable of interest is the four-point function of $\NN=2$ chiral operators, \ie, the superconformal primaries of $\EE_{r_0}$ multiplets. These multiplets were introduced in Section \ref{sec:philosophy} as being connected to the \emph{Coulomb data} of a theory. We recall that these superconformal primaries are spacetime scalars with non-zero $U(1)_r$ charge $r_0$ that are neutral with respect to $SU(2)_R$ and that have conformal dimension $\Delta=r_0$. We will denote the operator of interest as $\phi_{r_0}$, with the conjugate anti-chiral operator being $\bar \phi_{-r_0}$. Unitarity requires $\Delta \geqslant 1$. In principle, one would like to focus on \emph{generators} of the Coulomb branch chiral ring. Our methods are such that it is not easy to distinguish between generators and composites. However, if we take $r_0 \leqslant 2$, then unitarity dictates that $\phi_{r_0}$ must be a chiral ring generator.

We will be investigating the four-point function of a single chiral operator and its conjugate,
%%%%%%
\begin{equation}
\label{eq:eps_four_point}
\big\langle \phi_{r_0}(x_1) \bar{\phi}_{-r_0}(x_2) \phi_{r_0}(x_3) \bar{\phi}_{-r_0}(x_4)\big\rangle~.
\end{equation}
%%%%%%
The general procedure is now analogous to that of the previous section. We should determine what operators can be exchanged in each channel and find the corresponding superconformal blocks. In contrast to the previous section, here we are dealing with operators that are invariant under any flavor symmetries in the theory but that are nontrivially charged under $U(1)_r$. Although this is an $R$-symmetry, the role it plays in this correlator will be largely that of a $SO(2)$ flavor symmetry, with some minor differences that we discuss below. After obtaining the superconformal blocks in all channels we have to work out the constraints imposed by crossing symmetry. The $\EE_{r}$ multiplets are not involved with the chiral algebra data of a theory. This means that unlike the previous section, we are not able to fix the coefficients of all the short and semi-short multiplets being exchanged. Of all the short and semi-short multiplets appearing in the partial wave expansion, the only coefficient we are able to fix is that of the stress tensor, which must appear in the $\phi_{r_0} \times \bar{\phi}_{-r_0}$ OPE. This gives us a handle on the central charge $c$ of the theory, which together with the dimension of the external operators $\Delta=r_0$ are the two parameters we can tune. We can therefore in principle derive bounds or other constraints as a function of the pair $(r_0,c)$, but we will sometimes leave the central charge arbitrary in order to obtain bounds that are universally valid for all central charges, or alternatively in order to bound $c$ itself.

Another short operator of special interest is the superconformal primary $\phi_{2r_0}$ of the $\EE_{2r_0}$ multiplet, which appears in the $\phi_{r_0} \times \phi_{r_0}$ OPE as part of the Coulomb branch chiral ring. The corresponding conformal block appears with a nontrivial coefficient that is not protected by supersymmetry. As we will see this multiplet is isolated and thus we will be able to bound this coefficient both from below and from above. In this way we will also be able to probe relations on the Coulomb branch chiral ring of the type $\phi_{r_0} \phi_{r_0} \sim 0$.

Finally, let us note that exactly marginal deformations of an $\NN=2$ SCFT that preserve the full $\NN=2$ superconformal invariance lie in $\EE_{2}$ multiplets. More specifically, the deforming operators are obtained by acting with all four anti-chiral supercharges $\wt{Q}_{\II\dot\alpha}$ on the superconformal primary of those multiplets. Theories with $\EE_{2}$ multiplets in their spectrum therefore necessarily lie on a conformal manifold of positive dimension. As we will review below, the coefficient of the $\EE_{4}$ multiplet in the OPE of two $\EE_{2}$ multiplets is related to the curvature of the Zamolodchikov metric on this conformal manifold. This curvature is thus a natural target for numerical investigation.

%%%%%%%%%%%%%%%%%%%%%%%%%%%%%%%%%%%%%%%%%%%%%%%%%%%%%%%%%%%%%%%%%%%%%%%%%%%%%%%%%%%%%%%%%%%%%%%%%%%%%%%%%%%%
\subsection{Structure of the four-point function}
\label{subsec:epsilon_OPE}
%%%%%%%%%%%%%%%%%%%%%%%%%%%%%%%%%%%%%%%%%%%%%%%%%%%%%%%%%%%%%%%%%%%%%%%%%%%%%%%%%%%%%%%%%%%%%%%%%%%%%%%%%%%%

%%%%%%
\begin{figure*}[t!]
    \centering
    \begin{subfigure}[t]{2in}
        \centering
        \includegraphics[height=1.2in]{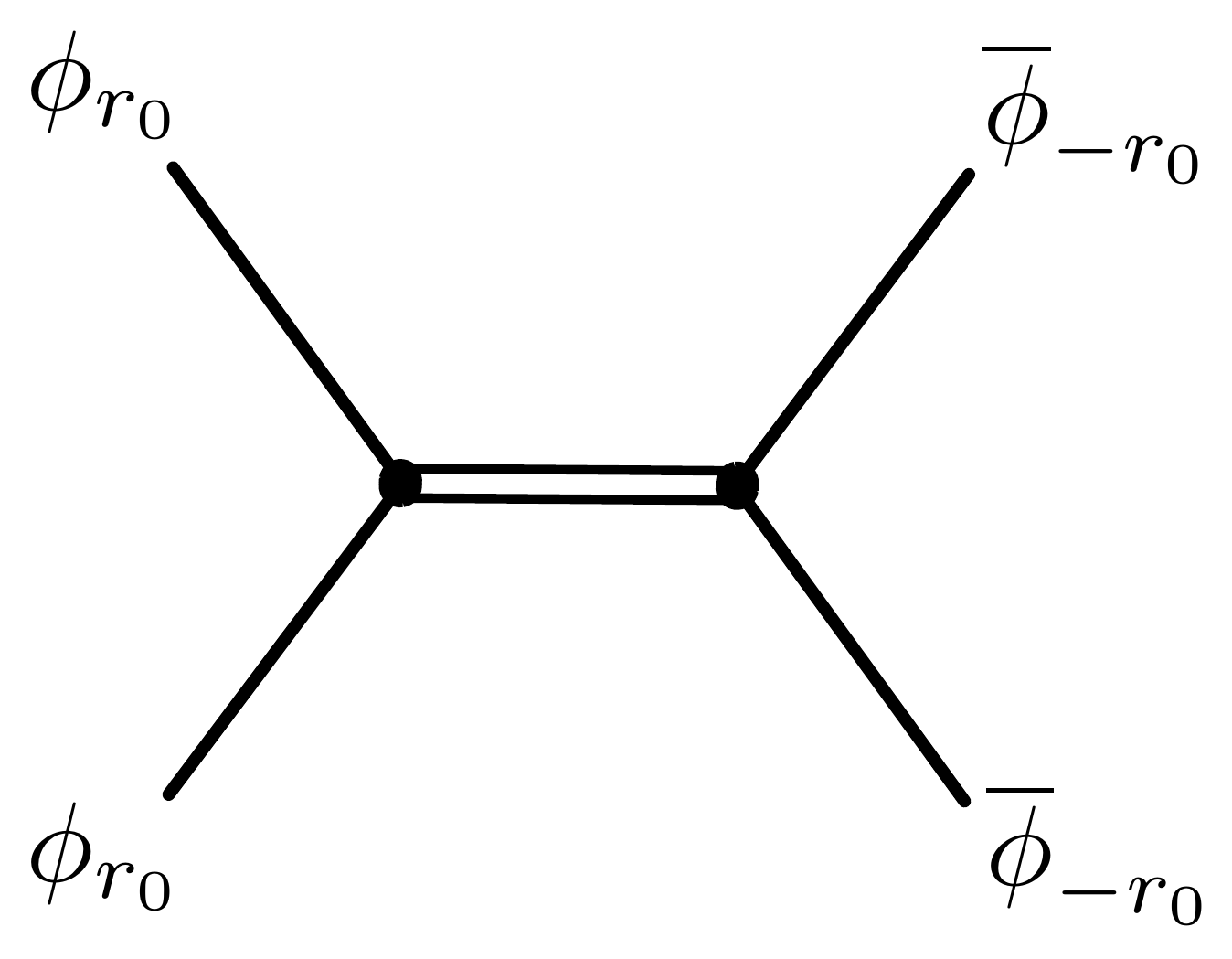}
        \caption{Chiral OPE channel.}
    \end{subfigure}%
    ~ 
    \begin{subfigure}[t]{2in}
        \centering
        \includegraphics[height=1.3in]{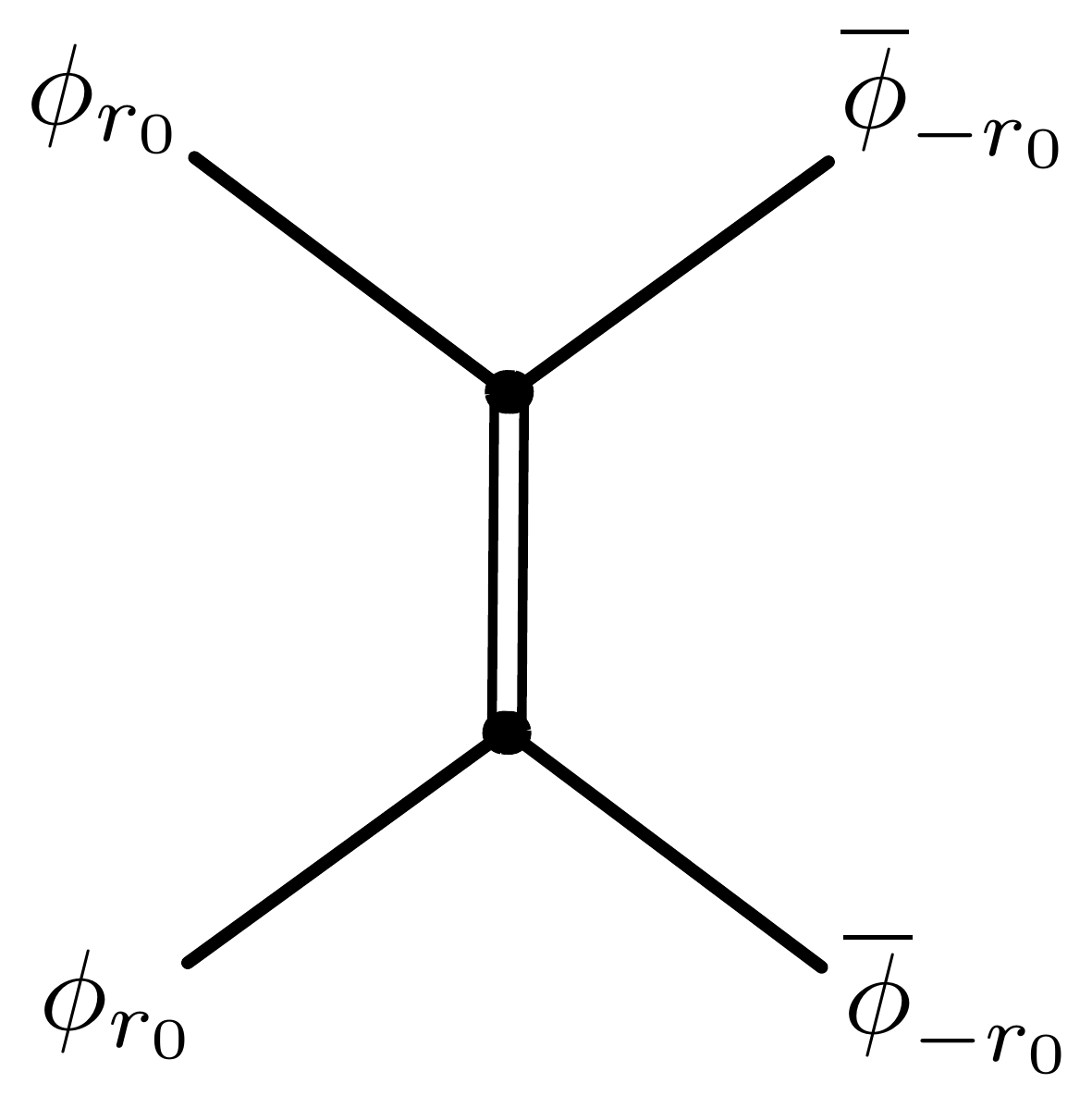}
        \caption{Nonchiral OPE channel.}
    \end{subfigure}
    \caption{The two inequivalent OPE channels for the $\EE_{r}$ four-point function.\label{fig:Eps_OPE_channels}}
\end{figure*}
%%%%%%

In contrast to the case of moment maps, there are now two qualitatively different OPE channels to consider depending on whether we take the non-chiral OPE $\phi_{r_0}(x_1)\times \bar{\phi}_{-r_0}(x_2)$ or the chiral OPE $\phi_{r_0}(x_1)\times \phi_{r_0}(x_2)$ (see Fig. \ref{fig:Eps_OPE_channels}). We now describe the various selection rules for superconformal representations appearing in these two channels, as well as the corresponding superconformal blocks.

%%%%%%%%%%%%%%%%%%%%%%%%%%%%%%%%%%%%%%%%%%%%%%%%%%%%%%%%%%%%%%%%%%%%%%%%%%%%%%%%%%%%%%%%%%%%%%%%%%%%%%%%%%%%
\subsubsection{The $\phi_{r_0}(x_1)\times \bar{\phi}_{-r_0}(x_2)$ channel}
\label{subsubsec:eps_ope_nonchiral}
%%%%%%%%%%%%%%%%%%%%%%%%%%%%%%%%%%%%%%%%%%%%%%%%%%%%%%%%%%%%%%%%%%%%%%%%%%%%%%%%%%%%%%%%%%%%%%%%%%%%%%%%%%%%

We begin with the selection rules for the non-chiral OPE. The problem simplifies due to the fact that an operator $\OO(x_3)$ can participate in a non-zero three-point function $\langle \phi_{r_0}(x_1) \bar{\phi}_{-r_0}(x_2) \mathcal{O}(x_3)\rangle$ only if the superconformal primary of the multiplet to which it belongs also participates in such a non-vanishing three-point function. A sketch of the derivation of this result can be found in Appendix~\ref{subapp:epps_nonchiral_selection}. 

Selection rules for the $U(1)_r$ and $SU(2)_R$ require that any such operator $\OO(x_3)$ be an $SU(2)_R$ singlet and have $r_{\OO}=0$. To appear in the OPE of two scalars they must also have $j_1=j_2 \equalscolon j$. Taken together, these conditions imply the following selection rule:
%%%%%%
\begin{equation}\label{EEbarOPE}
\EE_{r_0} \times \bar{\EE}_{-r_0} \sim \mathbf{1} + \hat{\CC}_{0,(j,j)} + \AA^{\Delta}_{0,0(j,j)}~.
\end{equation}
%%%%%%
Note that the structure of the OPE we present here is \emph{only} for the superconformal primaries of the $\EE_{r_0}$ and $\bar{\EE}_{-r_0}$ multiplets, despite our abuse of notation in using the name of the full multiplet on the left-hand side of the above equation. The superconformal blocks for these multiplets have been computed in \cite{Fitzpatrick:2014oza}. They are given by the general formula
%%%%%%
\begin{eqnarray}\label{ESuperblock}
\GG^{\mathrm{sc}}_{\Delta,\ell}(z,\bar{z}) \colonequals \frac{(z \bar{z})^{\frac{1}{2} \left(\Delta-\ell\right)}}{z-\bar{z}}&&\left(\left(-\frac{z}{2}\right)^{\ell}z\, {}_2F_1\left(\frac{1}{2}\left(\Delta+\ell\right),\frac{1}{2}\left(\Delta+\ell+4\right);\Delta+\ell+2;z)\right) \right.\\
&& \qquad\qquad\quad\left. {}_2F_1\left(\frac{1}{2}\left(\Delta-\ell-2\right),\frac{1}{2}\left(\Delta-\ell+2\right);\Delta-\ell;\bar{z})\right)-z\leftrightarrow\bar{z}\right)~,\nn
\end{eqnarray}
%%%%%%
with $\Delta$ and $\ell=2j$ denoting the dimension and spin of the superconformal primary of each multiplet. The spin can be either even or odd. Note that the superconformal blocks for the $\hat\CC_{0(j,j)}$ representations are simply the specialization of \eqref{ESuperblock} to the case $\Delta=\ell+2$, while the block for the identity operator is just a constant as usual. These superconformal blocks can of course be written as a finite sum of conventional conformal blocks --  we provide such a decomposition in Appendix~\ref{subapp:eps_blocks_nonchiral}.

%%%%%%%%%%%%%%%%%%%%%%%%%%%%%%%%%%%%%%%%%%%%%%%%%%%%%%%%%%%%%%%%%%%%%%%%%%%%%%%%%%%%%%%%%%%%%%%%%%%%%%%%%%%%
\subsubsection{The $\phi_{r_0}(x_1)\times \phi_{r_0}(x_2)$ OPE}
\label{subsubsec:eps_ope_chiral}
%%%%%%%%%%%%%%%%%%%%%%%%%%%%%%%%%%%%%%%%%%%%%%%%%%%%%%%%%%%%%%%%%%%%%%%%%%%%%%%%%%%%%%%%%%%%%%%%%%%%%%%%%%%%

We now turn to the chiral OPE. In this case only $SU(2)_R$ singlets with $r_{\mathcal{O}}=2r_0$ and $j_1=j_2 \equalscolon j$ are allowed, and the spin $\ell \colonequals 2j$ is required to be even because we are considering the OPE of two identical scalars. The complete selection rules for this channel are worked out in Appendix~\ref{subapp:eps_chiral_selection}, where it is shown that only one conformal family per superconformal multiplet can contribute, implying the superconformal blocks are then equal to the standard conformal blocks corresponding to that family. The complete list of superconformal multiplets that can appear in this OPE is derived in the aforementioned appendix.
All told we find the following selection rules, where for simplicity we momentarily assume that $r_0>1$,
%%%%%%
\begin{equation}
\EE_{r_0} \times \EE_{r_0} \sim \AA_{0,2r_0-2(j,j)} + \EE_{2r_0}  + \CC_{0,2r_0-1(j-1,j) }+ \BB_{1,2r_0-1(0,0)} + \CC_{\frac12,2r_0-\frac32(j-\frac12,j)}~.
\end{equation}
%%%%%%
Once again we note that these selection rules only necessarily hold true for the superconformal primaries of the $\EE_{r_0}$ multiplets. The corresponding superconformal blocks for these multiplets are given in Table \ref{tab:EE2blocks}. The blocks for certain additional short multiplets that are allowed when $r_0=1$, are presented in the second part of the table.  
%%%%%%
\begin{table}[h!t]
\begin{center}
\renewcommand{\arraystretch}{1.75}
\begin{tabular}{|l|l|l|}
\hline
~Multiplet~		&	Contribution to $\GG_{\hat{i}=\hat{2}}(u,v)$	&	Restrictions\\
\hline
\hline
$\AA_{0,2r_0-2(j,j)}$			&	$u^{\frac{\Delta-\ell}{2}}\,G_{\Delta}^{(\ell)}(u,v)$ & $\Delta \geqslant 2+2r_0+\ell$ \\
\hline
$\EE_{2r_0}$					&	$u^{r_0}\,G_{2r_0}^{(0)}(u,v)$ &\\
\hline
$\CC_{0,2r_0-1(j-1,j)}$			&   $u^{r_0}\,G_{2r_0+\ell}^{(\ell)}(u,v)$ & $\ell \geqslant 2$\\
\hline
$\BB_{1,2r_0-1(0,0)}$			&	$u^{r_0+1}\,G_{2r_0+2}^ {(0)}(u,v)$ & \\
\hline
$\CC_{\frac12,2r_0-\frac32(j-\frac12,j)}$	&	$u^{r_0+1}\,G_{2r_0+\ell+2}^{(\ell)}(u,v)$ & $\ell \geqslant 2$\\
\hline\hline
$\DD_{1 (0,0)}$				&  $u^2 \,G_{\Delta=4}^ {(0)}$ & $r_0=1$\\
\hline
$\hat{\CC}_{\frac12(j-\frac12,j)}$	&  $u^{2}\,G_{\Delta=\ell+4}^{(\ell)}$ & $\ell\geqslant 2;~r_0=1$\\
\hline
$\hat{\CC}_{0(j-1,j)}$				&  $u\,G_{\Delta=\ell+2}^ {(\ell)}$ & $\ell\geqslant 2;~r_0=1$\\
\hline
\end{tabular}
\caption{Superconformal blocks for the $\EE_{r_0}$ four point function in the chiral channel.\label{tab:EE2blocks}}
\end{center}
\end{table}
%%%%%%
Note that the $\CC_{\frac12,2r_0-\frac32(j-\frac12,j)}$ and $\BB_{1,2r_0-1(0,0)}$ classes of short representations lie at the unitarity bound for long multiplets, and their superconformal blocks are simply the specializations of the long multiplet block to appropriate values of $\Delta$ and $\ell$. The $\EE_{2r_0}$ and $\CC_{0,2r_0-1(j-1,j)}$ representations, on the other hand, are separated from the continuous spectrum of long multiplets by a gap. The three short multiplets that are available only when $r_0=1$ contribute with the same blocks as some of the other blocks appearing in Table \ref{tab:EE2blocks}.

%%%%%%%%%%%%%%%%%%%%%%%%%%%%%%%%%%%%%%%%%%%%%%%%%%%%%%%%%%%%%%%%%%%%%%%%%%%%%%%%%%%%%%%%%%%%%%%%%%%%%%%%%%%%
\subsection{Crossing symmetry}
\label{subsec:Epscrossing}
%%%%%%%%%%%%%%%%%%%%%%%%%%%%%%%%%%%%%%%%%%%%%%%%%%%%%%%%%%%%%%%%%%%%%%%%%%%%%%%%%%%%%%%%%%%%%%%%%%%%%%%%%%%%

To formulate the crossing symmetry condition for this correlator we will treat $U(1)_r$ as an $SO(2)$ global symmetry -- this is similar to the approach used in \cite{Poland:2011ey} to study the four-point function of chiral operators in $\NN=1$ SCFTs. In this approach, the fields $\phi_{r_0}$ and $\bar \phi_{-r_0}$ are combined in the fundamental representation of $SO(2)$ with charge $|r_0|$, which we denote as $\textbf{2}_{|r_0|}$. This representation has the following tensor product with itself,
%%%%%%
\begin{equation}\label{so2tensorprod}
\textbf{2}_{|r_0|} \otimes \textbf{2}_{|r_0|} = \left( \textbf{1} \oplus \textbf{2}_{|2 r_0|} \right)_{\mathrm{symm.}} \oplus \textbf{1}_{\mathrm{antisymm.}}~,
\end{equation}
%%%%%%
where the subscripts denote which representations appear in the symmetrized tensor product and which appear in the antisymmetrized tensor product. The crossing symmetry discussion of Section~\ref{Sec:Bhat4ptfunc} is now directly applicable, with the crossing matrix $F_i^{\ph{i}j}$ given by
%%%%%%
\begin{eqnarray}
\left(
\begin{array}{ccc}
 \frac{1}{2} & \frac{1}{2} & \frac{1}{2} \\
 1 & 0 & -1 \\
 \frac{1}{2} & -\frac{1}{2} & \frac{1}{2} \\
\end{array}
\right)~,
\end{eqnarray}
%%%%%%%
where the ordering of the rows and columns is the same as in \eqref{so2tensorprod}. The crossing equation then takes the form
%%%%%%
\begin{equation}\label{Ecrossing}
(z-\bar{z})((1-z)(1- \bar{z}))^{r_0} F_j^{\ph{j} i} \GG_j(z,\bar{z}) + (\bar{z}-z)(z \bar{z})^{r_0} \GG_i(1-z,1-\bar{z})=0~,
\end{equation}
%%%%%%
where each $ \GG_i(z,\bar{z})$ can be expanded in the blocks relevant for the $SO(2)$ channel $i$. As usual, the braiding relation requires that only operators of even spin appear in the symmetric channels while only operators of odd spin appear in the antisymmetric channel.

In more conventional terms, the symmetric traceless channel $\textbf{2}_{|2 r_0|}$ encodes the operators appearing in the $\phi_{r_0}\times \phi_{r_0}$ OPE. This channel can therefore be expanded entirely in terms of the conformal blocks $G_\Delta^{(\ell)}$ given in Table \ref{tab:EE2blocks} with even spins. The singlet channels, on the other hand, describe the $\phi_{r_0}\times \bar{\phi}_{-r_0}$ OPE, with the symmetric channel getting all the even spin conformal block contributions and the antisymmetric channel getting the odd spin ones. The blocks in these latter two channels are related by supersymmetry because the $U(1)_r$ symmetry is part of the superconformal algebra, and conformal families from the same superconformal multiplet appear in both channels. To wit, in the symmetric singlet channel we have contributions from the superconformal primaries appearing in \eqref{EEbarOPE} with even spin, together with their even spin superconformal descendants, and the even spin superconformal descendants of odd spin superconformal primaries. For the antisymmetric channel the opposite takes place. We have therefore broken the superconformal blocks \eqref{ESuperblock} apart, splitting them by the parity of the spin, with each channel enjoying a ``partial'' superconformal block.

This splitting of superconformal blocks can be ameliorated by a change of basis. Let us define
%%%%%%
\begin{equation}\label{GtoGhat}
\GG_{\hat{1},\hat{3}} \colonequals \GG_{1} \pm \GG_{3}\,,\qquad
\GG_{\hat{2}} \colonequals  \GG_{2}~.
\end{equation}
%%%%%%
All conformal blocks arising from the same superconformal multiplet are now grouped together, with each superconformal multiplet from the singlet channels appearing twice: once each in $\GG_{\hat 1}$ and $\GG_{\hat 3}$. The channels $\hat 1$ and $\hat 3$ are almost identical -- the only difference is an extra minus sign for all the odd spin conformal blocks in $\GG_{\hat{3}}$ due to the extra minus sign in \eqref{GtoGhat}. There are two ways to insert this minus sign. The first option is to decompose the superconformal block \eqref{ESuperblock} into ordinary conformal blocks and insert extra factors of $(-1)^{\ell}$ in front of every block. However the second option is more efficient. We recall that ordinary conformal blocks satisfy the following braiding relation:
%%%%%%
\begin{equation}
\left(\frac{z}{z-1} \frac{\bar{z}}{\bar{z}-1}\right)^{\frac{\Delta-\ell}{2}} G_\Delta^{(\ell)}\left(\frac{z}{z-1},\frac{\bar{z}}{\bar{z}-1}\right)=(-1)^\ell (z \bar{z})^{\frac{\Delta-\ell}{2}} G_\Delta^ {(\ell)}(z,\bar{z})~.
\end{equation}
%%%%%%
We can thus insert the necessary factors of $(-1)^\ell$ by substituting $z \to \frac{z}{z-1}$ and $\bar z \to \frac{\bar z}{\bar z - 1}$ in the superconformal block \eqref{ESuperblock}. We thus find that a supermultiplet in the singlet channel contributes to the four-point function as follows,
%%%%%%
\begin{equation}
\GG_{\hat i = 1}(z,\bar z) \sim \GG_{\Delta,\ell}^{sc}(z,\bar z)\,, \qquad \GG_{\hat i = 3}(z,\bar z) \sim \GG_{\Delta,\ell}^{sc}\left(\frac{z}{z-1},\frac{\bar z}{\bar z -1}\right)~,
\end{equation}
%%%%%%
with the same OPE coefficient appearing in both channels. The operators contributing to the doublet channel still contribute to $\GG_{\hat 2}(z,\bar z)$ as before.

The relevant crossing equation is now the same as in \eqref{Ecrossing} but with $\hat{i}$ and $\hat{j}$ replacing $i$ and $j$, and with the flavor matrix $F_i^{\ph{i}j}$ replaced by
%%%%%%
\begin{eqnarray}
F_{\hat{i}}^{\ph{i}\hat{j}}=\left(
\begin{array}{ccc}
 1 & 0 & 0 \\
 0 & 0 & 2 \\
 0 & \frac{1}{2} & 0 \\
\end{array}
\right)~.
\end{eqnarray}
%%%%%%
This is the same as the crossing equation that was previously derived for chiral operators in $\NN=1$ SCFTs \cite{Poland:2011ey}. This matrix squares to one, which is relevant to the numerical implementation described in the next section.

When $\Delta=\ell=0$ in the $\hat{i}=\hat{1},\hat{3}$ channels we get the contribution of the identity operator, which we set equal to two. This fixes the normalization of the external operators to be one as is conventional (the factor of two arises from the projector onto the singlet, similarly to the discussion in the previous section). The contribution of the stress tensor is contained in the superconformal block \eqref{ESuperblock} with $\Delta=2$ and $\ell=0$. When expanded in ordinary conformal blocks, the contribution is given by
%%%%%%
\begin{equation}
\GG_{i=1}^{\hat{\CC}_{0,(0,0)}}(z,\bar{z})= u G_{2}^{(0)}(u,v) - \frac{u}{2} G_{3}^{(1)}(u,v)+\frac{2u}{30} G_4^{(2)}(u,v)~.
\end{equation}
%%%%%%
The coefficient of this block can be fixed in terms of the central charge $c$ as was done in the previous section. Namely, fixing the coefficient of $uG_4^{(2)}(u,v)$ requires that this superconformal block should appear with coefficient $\frac{r_0^2}{3c}$ (again, a factor of two comes from the projector onto the singlet). The ``braided'' superconformal block appears with the same coefficient.

%%%%%%%%%%%%%%%%%%%%%%%%%%%%%%%%%%%%%%%%%%%%%%%%%%%%%%%%%%%%%%%%%%%%%%%%%%%%%%%%%%%%%%%%%%
\subsubsection{Free theory expansion}
%%%%%%%%%%%%%%%%%%%%%%%%%%%%%%%%%%%%%%%%%%%%%%%%%%%%%%%%%%%%%%%%%%%%%%%%%%%%%%%%%%%%%%%%%%

A simple illustration of the superconformal block decomposition procedure is the explicit analysis of free field theory. Namely we consider the theory of a free vector multiplet, and we study the four-point function of the chiral scalar which has $r_0=1$. Decomposing the free field correlator in terms of the superconformal blocks described above we find the following channel expansions,
%%%%%%
\begin{eqnarray}
\GG_{i=\hat{1}}(z,\bar{z}) &=&\sum_{\ell=0}^{\infty} \frac{(\ell +2) (\ell !)^2 (-2)^\ell}{(2 \ell +1) (2 \ell )!} \GG^{\mathrm{sc}}_{\ell+2,\ell}(z,\bar{z})\,,\nn\\
\GG_{i=\hat{2}}(z,\bar{z}) &=& \sum_{\ell=0}^{\infty} \frac{\left((-1)^{\ell }+1\right) (\ell !)^2 (-2)^\ell}{(2 \ell )!} u G_{\ell+2}^{(\ell)} (u,v)\,,\nn\\
\GG_{i=\hat{3}}(z,\bar{z}) &=&\sum_{\ell=0}^{\infty} \frac{(\ell +2) (\ell !)^2 (-2)^\ell}{(2 \ell +1) (2 \ell )!} \GG^{\mathrm{sc}}_{\ell+2,\ell}\left(\frac{z}{z-1},\frac{\bar{z}}{\bar{z}-1}\right)~.
\label{freevectorexpansion}
\end{eqnarray}
%%%%%%
We can immediately verify that the only difference between channel $\hat 1$ and $\hat 3$ is the replacement $z \to \frac{z}{z-1}$ and $\bar z \to \frac{\bar z}{\bar z - 1}$. We can also ferret out the stress tensor block $ \GG^{\mathrm{sc}}_{2,0}(z,\bar{z})$ and see that it appears with coefficient two. This suggests a central charge of $\frac{1}{6}$, which is correct for the theory of a free vector multiplet. For future reference, we also note that the $\EE_{2}$ block, which is the $u G_{2}^{(0)}$ term in the $\hat 2$ channel, comes with coefficient two.
%!TEX root = ../draft_maxi_Neq2.tex

\section{Operator bounds from crossing symmetry}
\label{Sec:Numerics}

The output from the previous two sections was a collection of crossing symmetry equations and their (super)conformal block decompositions. In this section we describe the numerical methods by which these equations can be used to extract useful information about $\NN=2$ SCFTs. We follow the approach of \cite{Poland:2011ey}, where the original numerical analysis of \cite{Rattazzi:2008pe} was recast as a semidefinite programming problem.

Each of the nontrivial crossing symmetry equations can be put into the general form
%%%%%%
\begin{equation} \label{eq:general_cross_eq}
\HH_i(z,\bar z) + F_i^{\phantom{i}j} \HH_j (1-z,1-\bar z) = 0~.
\end{equation}
%%%%%%
Here and below, summation over repeated indices is always implied. The functions $\HH_i(z,\bar z)$ can always be written as
%%%%%%
\begin{equation}\label{eq:short_long_decomp}
\HH_i(z,\bar z) = \GG_i(z,\bar z) - a_i(z,\bar z)~,
\end{equation}
%%%%%%
where the $a_i(z,\bar z)$ are some known functions that have been fixed analytically, and the $\GG_i(z,\bar z)$ have a decomposition of the form
%%%%%%
\begin{equation}
\GG_i (z,\bar z) = \sum_{k_i} \lambda_{k_i}^2  \Gt_{\Delta_{k_i}}^{(\ell_{k_i})} (z,\bar z)~.
\end{equation}
%%%%%%
The coefficients $\lambda_{k_i}^2$ are real, positive numbers, and the $\Gt_{\Delta}^{(\ell)}(z,\bar z)$ are roughly the superconformal blocks, the precise form of which depends on the crossing symmetry equation in question.\footnote{These $\Gt$ functions are not exactly the superconformal blocks of the previous sections, but rather they include simple prefactors that have been absorbed in their definition. This is not particularly important for the discussion here.} The sum is over all operators that appear in the $i$'th channel, and the matrix $F_i^{\phantom{i}j}$ is related to Wigner's $6j$ symbol for the relevant global symmetry group and in particular is involutory: $F_i^{\ph{i}j} F_j^{\ph{j}k} = \delta_i^{\ph{i}k}$, for the cases considered here.

As in \cite{Rattazzi:2008pe}, we will analyze these equations by considering the action of certain linear functionals upon them. We may introduce one linear functional $\phi^i$ for each channel $i$. The functionals that we consider are defined by taking linear combinations of various numbers of derivatives of the function and evaluating at the symmetric point $z = \zb = 1/2$, \ie,
%%%%%%
\begin{equation}
\phi^i[f_i(z,\bar z)] = \sum_{m,n} \a^i_{mn} \del_z^m \del_{\bar z}^n f_i(z,\bar z) \big|_{z = \bar z = \hf}~.
\end{equation}
%%%%%%

The matrices $\hf (\delta_i^{\ph{i}j} \pm F_i^{\phantom{i}j})$ are projectors onto the positive and negative eigenspaces of $F_i^{\ph{i}j}$, so we can split the coefficients into even and odd parts, $\a^i_{mn} = \a^i_{mn,+} + \a^i_{mn,-}$, satisfying
%%%%%%
\begin{equation}
\label{eigenvecsfunctional}
\hf \a^j_{\pm} (\delta_j^{\ph{j}i} \pm F_j^{\phantom{j}i}) = \a^i_{\pm}\,, \qquad \qquad \hf \a^j_{\pm} (\delta_j^{\ph{j}i} \mp F_j^{\phantom{j}i})  = 0~.
\end{equation}
%%%%%%
Upon acting with our functionals on both sides of \eqref{eq:general_cross_eq}, we find the following equation,
%%%%%%
\begin{equation}
\sum_{m,n} (\a^i_{mn,+} + \a^i_{mn,-})\left(\delta_i^{\ph{i}j} + (-1)^{m + n} F_i^{\ph{i}j}\right) \restr{\del_z^m \del_{\bar z}^n \HH_j(z,\bar z)}{z = \bar z = \hf} = 0~.
\end{equation}
%%%%%%
Only those terms with $m+n$ even in $\phi^i_+$ and those with $m+n$ odd in $\phi^i_-$ have a nontrivial action on the crossing symmetry equation \eqref{eq:general_cross_eq}. Without loss of generality, we can therefore set the other terms to zero. With this choice now implicit, the action of the functional on the crossing symmetry equation can be succinctly written as
%%%%%%
\begin{equation}
\begin{split}
0 &= \phi^i \left[ \HH_i(z,\bar z) + F_i^{\phantom{i}j}\HH_j (1-z,1-\bar z) \right]  \\
&= 2 \sum_{m,n} \left(\a^i_{mn,+} + \a^i_{mn,-}\right) \del_z^m \del_{\bar z}^n \restr{\HH_i(z,\bar z)}{z = \bar z = \hf} \\
&= 2 \phi^i \left[ \HH_i(z,\bar z) \right]\,.
\end{split}
\end{equation}
%%%%%%
The nontrivial relations between the different global symmetry channels have been completely accounted for by the eigenvector constraints \eqref{eigenvecsfunctional}, which are simple algebraic constraints that are easily solved in any given case.

By construction, all the functions appearing in this relation are symmetric under the exchange of $z$ and $\zb$, so we lose nothing by restricting the coefficients of the functionals to obey $m \leqslant n$. We obtain a finite-dimensional functional space by introducing a cutoff $\Lambda\in\Nb$, and demanding that $m + n \leqslant \Lambda$. For each $i$, we then find $(1 + \lfloor \frac{\Lambda}{2} \rfloor)(2 + \lfloor \frac{\Lambda-1}{2} \rfloor)$ independent derivative combinations. The total number of independent coefficients $\a^i_{mn,\pm}$ is then determined by multiplying the number of derivative combinations with $m+n$ even by the number of positive eigenvalues of $F_i^{\ph{i}j}$ and the number of derivative combinations with $m+n$ odd by the number of negative eigenvalues and then taking the sum. Quantitatively, if the total number of channels is $c$ and $F_i^{\phantom{i}j}$ has $b$ positive eigenvalues, then the dimension of the space of functionals is given by
%%%%%%
\begin{equation}
\dim_{\Lambda,c,b}=\frac{c}{2} \left( 1+ \floor*{\frac{\Lambda-1}{2}}\right) \left(2 + \floor*{\frac{\Lambda-1}{2}}\right)  + \frac{b}{2} \left(1 + (-1)^\Lambda\right) \left( 1 + \floor*{\frac{\Lambda}{2}}\right)~.
\end{equation}
%%%%%%
This is the dimension of the space in which we will be performing a numerical search, and is therefore an important measure of the complexity of the numerical problem. For large $\Lambda$ the dimension behaves approximately like $c/2$ times the total number of derivative combinations.

The numerical results presented in subsequent sections are the results of two different strategies. The aim of the first strategy is to provide an upper bound for the lowest dimension operator in a given channel and with a given spin that may appear in a solution of crossing symmetry. For instance, we may want to find an upper bound $\Delta_0^\star$ for the dimension of the first scalar operator in channel $\hat i$. Such a bound follows immediately from the existence of a functional possessing the following properties:
%%%%%%
\begin{equation}
\label{eq:dim_bound_functional}
\begin{split}
\phi^{j} \left[\Gt_{\Delta}^{(\ell)}(z,\bar z)\right] 		&\geqslant 0~, \qquad\qquad\forall\;(\Delta,\ell)\, \text{ in channel }j \neq \hat i~,\\
\phi^{\hat i}\left[\Gt_{\Delta}^{(\ell)}(z,\bar z)\right]	&\geqslant 0~, \qquad\qquad\forall\;(\Delta,\ell)\, \text{ in channel $\hat i$ with $\ell > 0$}~,\\
\phi^{\hat i} \left[\Gt_{\Delta}^{(0)}(z,\bar z)\right] 	&\geqslant 0~, \qquad\qquad\forall\;\Delta \geqslant \Delta_0^\star\, \text{ in channel $\hat i$}~,\\
\sum_i \phi^i \left[a_i(z,\bar z) \right] &\leqslant 0~.
\end{split}
\end{equation}
%%%%%%
It is implicit in this description that the functional need not be positive for scaling dimensions below the unitarity bound, since such operators cannot be present in the type of solution of crossing symmetry that we are aiming to constrain. The optimal bound that can be obtained by this method at a given cutoff will then be the minimal value of $\Delta_0^\star$ for which such a functional exists.

The aim of the second strategy is to provide an upper bound for value of a particular OPE coefficient, say $\lambda_{k_{\hat i}}^2$ which multiplies the block corresponding to an operator with dimension $\Delta_{k_{\hat i}}$ and spin $\ell_{k_{\hat i}}$ in channel $\hat i$. Such a bound follows from performing the following optimization over the space of functionals:
%%%%%%
\begin{equation}
\label{eq:ope_functional}
\begin{split}
\phi^{j} \left[\Gt_{\Delta_{\ph{k_{\hat i}}}}^{(\ell)_{\ph{k_{\hat i}}}}(z,\bar z)\right] &\geqslant 0, \qquad \qquad \forall (\Delta,\ell) \text{ in all channels }j~,\\
\phi^{\hat i} \left[\Gt_{\Delta_{k_{\hat i}}}^{(\ell_{k_{\hat i}})}(z,\bar z)\right] &=1~,\\
\text{minimize }\sum_i \phi^i \left[a_i(z,\bar z) \right]&~.
\end{split}
\end{equation}
%%%%%%
If the minimum is positive and equal to, say, $M$, then we obtain an upper bound
%%%%%%
\begin{equation}
\lambda_{k_{\hat i}}^2 \leqslant M~.
\end{equation}
%%%%%%
If, on the other hand, the minimum turns out to be negative then we are effectively back to the previous case and there can be no solution to crossing symmetry. In our analysis, we often apply this minimization procedure to the block corresponding to the stress tensor multiplet. 
The OPE coefficient for that block is inversely proportional to the $c$ central charge, so an upper bound on the OPE coefficient translates to a lower bound for $c$.

Finally, there are some cases where the quantum numbers $(\Delta_{k_{\hat i}},\ell_{k_{\hat i}})$ of an operator of interest are isolated, in the sense that the corresponding conformal block is not continuously connected to the set of blocks for which the functional is required to be positive in \eqref{eq:ope_functional}. This is common in supersymmetric CFTs because of the various distinguished short multiplets whose scaling dimensions lie strictly below the unitarity bound for generic representations with the same Lorentz and R-symmetry quantum numbers. In such cases we can flip the sign on the second line of \eqref{eq:ope_functional} and instead require 
%%%%%%
\begin{equation}
\phi^{\hat i} [\Gt_{\Delta_{k_{\hat i}}}^{(\ell_{k_{\hat i}})}(z,\bar z)] = - 1~.
\end{equation} 
%%%%%%
In such a case, a negative value for $M$ provides a \emph{lower bound} on the corresponding OPE coefficient, 
%%%%%%
\begin{equation}
\lambda_{k_{\hat i}}^2\geqslant -M~.
\end{equation} 
%%%%%%
Here the result of the optimization is only meaningful if $M\leqslant0$. because unitarity constrains the coefficient is nonnegative.

In each of the cases just described, the search for functionals of appropriate type can be reduced to a semidefinite programming problem \cite{Poland:2011ey}. We review this story in Appendices \ref{App:semidefinite} and \ref{App:polyapprox}, where we also offer additional details about our particular numerical implementation.
%!TEX root = ../draft_maxi_Neq2.tex

\section{Results for the moment map four-point function}
\label{Sec:Bhatresults}

The four-point function of moment map operators depends on a choice of global symmetry $G_F$, the associated flavor central charge $k$, and the trace anomaly coefficient $c$. Under mild assumptions, the contributions of all short multiplets that appear in the conformal block decomposition are completely determined by those parameters. For each such triple $(G_F,k,c)$ there is then a corresponding crossing symmetry relation for the CFT data associated to long multiplets that can be subjected to numerical analysis. 

We have restricted our attention to flavor symmetries $\suf(2)$ and $\ef_6$. From the perspective of the bootstrap equations, these are the least complicated of all simple algebras because the number of irreps appearing the tensor product of two copies of the adjoint is the lowest (three for $\suf(2)$ and five for $\ef_6$). Moreover, since every non-abelian semi-simple Lie algebra has $\suf(2)$ as a subalgebra, the $\suf(2)$ bounds are in a sense universal and must hold for any $\NN=2$ superconformal field theory with a non-abelian flavor symmetry.

Below we will first discuss how in certain regions of the $(c,k)$ plane the crossing symmetry equations can never be satisfied by a unitary theory, irrespective of the precise spectrum of long multiplets. Recall that certain combinations of $c$ and $k$ are already excluded by the unitarity bounds that follow from the chiral algebra \cite{Beem:2013sza}. We will show that the numerical analysis carves out an even smaller region. Within the allowed region in the $(c,k)$ plane we then obtain bounds on operators in various Lorentz and flavor symmetry representations. We finally focus on values of $c$ and $k$ that correspond to known theories and compute more detailed bounds for the scaling dimensions of unprotected operators.

%%%%%%%%%%%%%%%%%%%%%%%%%%%%%%%%%%%%%%%%%%%%%%%%%%%%%%%%%%%%%%%%%%%%%%%%%%%%%%%%%%%%%%%%%%%%%%%%%%%%%%%%%%%%
\subsection{\texorpdfstring{$\suf(2)$}{su(2)} global symmetry}\label{subsec:bhat_results_su2}
%%%%%%%%%%%%%%%%%%%%%%%%%%%%%%%%%%%%%%%%%%%%%%%%%%%%%%%%%%%%%%%%%%%%%%%%%%%%%%%%%%%%%%%%%%%%%%%%%%%%%%%%%%%%

\begin{figure}
\centering
  \includegraphics[scale=0.43]{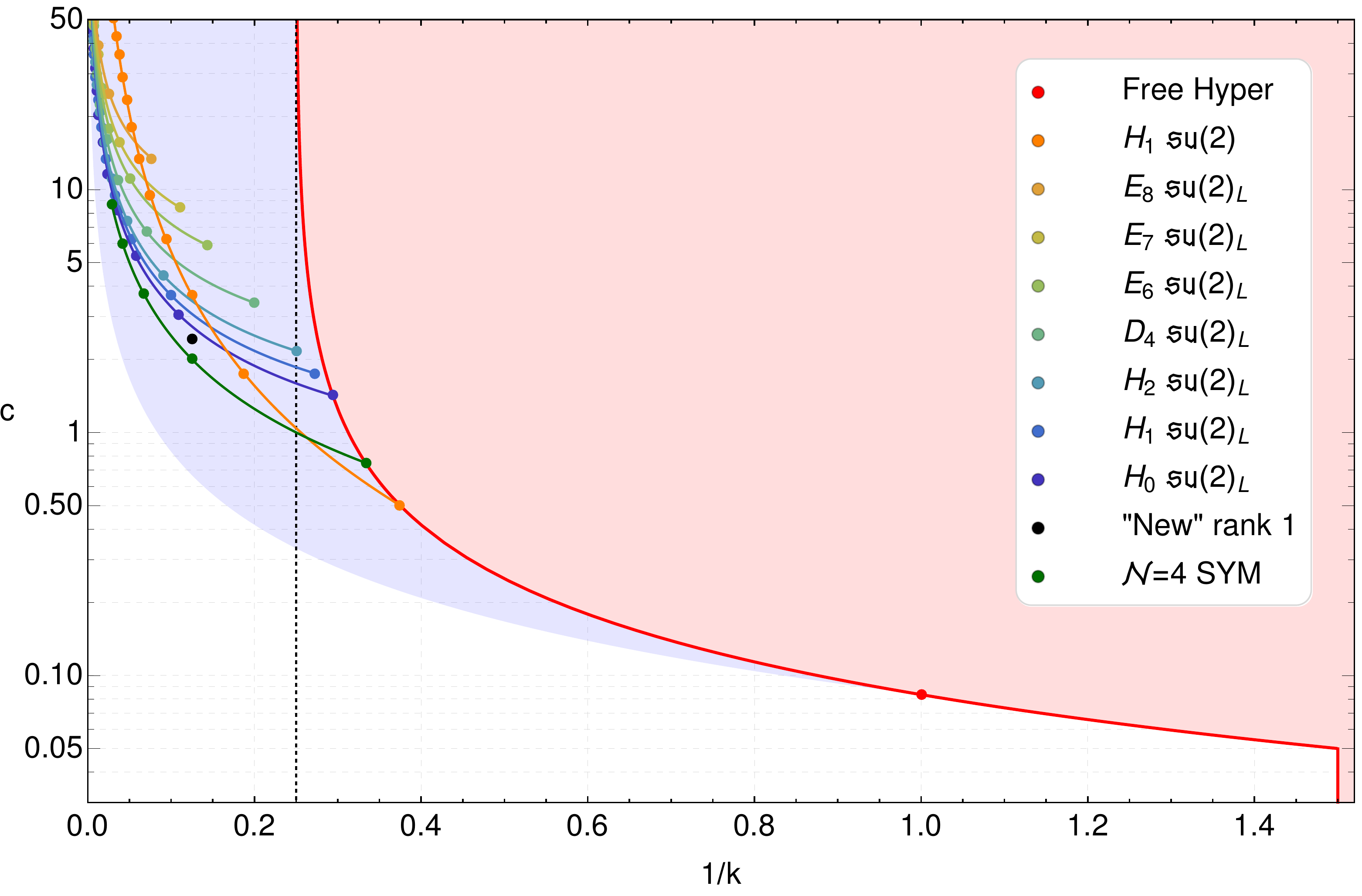}
  \caption{\label{fig:su2landscape}The $(c,k)$ plane for theories with an $\suf(2)$ flavor symmetry. The red region on the right is excluded by analytic unitarity bounds, whereas we are guaranteed to have valid solutions to the crossing symmetry constraints in the blue region. The curves connect points corresponding to theories related to F-theory singularities of different rank, which increases with $c$. We show the $(c,k)$ values corresponding the $\suf(2)_L$ flavor symmetry of all F-theory singularities with rank $N \geqslant 2$, and also to $\suf(2)$ flavor symmetry of the rank $N \geqslant 1$ $H_1$ theory. We also show a curve connecting points corresponding to the $(c,k)$ values of $\NN=4$ SYM with gauge group $SU(N)$. The ``new'' rank one theory is one of the theories obtained in \cite{Argyres:2007tq}. It has a product flavor symmetry with one factor being $\suf(2)$, which is the one whose value of $k$ is shown in the plot. The vertical dotted line corresponds to the value of $k$ for the codimension two defect of the six-dimensional $(2,0)$ theory of type $A_1$, which effectively has $c\to\infty$.}
\end{figure}

Before presenting the results of the numerical analysis, it is useful to review our expectations based on our present knowledge of $\NN = 2$ theories with $\suf(2)$ flavor symmetry. Let us consider the projection of the landscape of such SCFTs to the plane spanned by the two central charges $c$ and $k$. Every point on this plane then falls into one of three categories. First, there are points where a solution to crossing symmetry cannot exist because it would violate a known unitarity bound. Second, there are points where a solution to crossing symmetry is guaranteed to exist because it can in principle be constructed from known theories. All the other points then fall in the third category where we do not \emph{a priori} know if a solution exists. These three regions are charted in Fig.~\ref{fig:su2landscape} and we discuss each of them below.

%%%
%%%

Besides positivity of $c$ and $k$, there are additional unitarity bounds that originate from the chiral algebra construction of \cite{Beem:2013sza}. For $G_F=\suf(2)$ these bounds are given by
%%%%%%
\begin{equation}
\label{analytic_bound}
k \geqslant \frac{2}{3}~, \qquad k \geqslant \frac{16 c}{1+4c}~.
\end{equation}
%%%%%%
We refer to these bounds as the \emph{analytic bounds}, and the regions that they exclude in the $(c,k)$ plane are shaded in red in our plots.

Theories that saturate the analytic bounds have some special properties. For example, if the second of the analytic bounds is saturated then there can be no $\hat{\BB}_2$ multiplet contributing to the moment map four point function in the singlet channel, which implies a relation in the Higgs branch chiral ring. Examples of theories with this feature are the theory of a free hypermultiplet with $(c,k)=(\frac{1}{12},1)$ and the rank one Argyres-Douglas theory with  $(c,k)=(\frac{1}{2},\frac{8}{3})$, which is the rightmost point of type $H_1$ in Fig.~\ref{fig:su2landscape}. Notice that the two bounds in \eqref{analytic_bound} intersect at $(c,k) = (\frac{1}{20},\frac{2}{3})$. The equivalent intersection point for $\ef_6$ flavor symmetry corresponds precisely to the Minahan-Nemeschansky theory \cite{Minahan:1996cj}. It is natural to ask if a theory might exist at the intersection point for $\suf(2)$ flavor symmetry.

The second region contains all pairs $(c,k)$ that correspond to a known $\NN = 2$ SCFT. The region is however not limited to just those points, because we can take \emph{linear combinations} of known solutions as well: the sum of two solutions to crossing symmetry, with relative weights that sum to one, is again a good solution to crossing symmetry (at the level of a single four-point function). Since the central charges appear in four-point functions only through OPE coefficients that are proportional to $1/c$ or $1/k$, a solution constructed in this way has effective central charges
%%%%%%
\begin{equation}
\frac{1}{c_{\text{eff}}} = \frac{\alpha}{c_1} + \frac{1 - \alpha}{c_2}~, \qquad \qquad \frac{1}{k_{\text{eff}}} = \frac{\alpha}{k_1} + \frac{1- \alpha}{k_2}\,,
\end{equation}
%%%%%%
in terms of central charges $(c_i,k_i)$ of the two original solutions and a weight factor $0 \leqslant \alpha \leqslant 1$. In the $(\frac{1}{c},\frac{1}{k})$ plane, the values of $c$ and $k$ that can be realized as linear combinations in this way span the convex hull of all points corresponding to known theories. This region is shaded in blue in Fig.~\ref{fig:su2landscape}. It is effectively spanned by three points: the free hypermultiplet at $(c,k) = (\frac{1}{12},1)$, the generalized free field theory solution where $c$ and $k$ are both infinity, and the four-point function on a codimension two defect in the six-dimensional $(2,0)$ theory of type $A_1$ where $c$ is infinite and $k = 4$. We will discuss these three points in more detail below. We have computed the values of $c$ and $k$ for many other known theories but were not able to find any instance corresponding to a point outside the blue region in Fig.~\ref{fig:su2landscape}.

We should emphasize that taking linear combinations of solutions to crossing symmetry is not the same thing as taking correlation functions of operators in the tensor product of two theories. In particular, there is no guarantee that a linear combination of solutions can be realized in a complete $\NN = 2$ SCFT. We can however be sure that our kind of numerical analysis will not rule out any points corresponding to linear combinations of solutions. A more sophisticated bootstrap analysis might exclude them, but we leave this direction for future work.

Plotting the entire set of known $\NN = 2$ superconformal theories with at least $\suf(2)$ flavor symmetry is a daunting task, so we have opted to show only a subset. In Fig.~\ref{fig:su2landscape} we show in particular the location of the theories that describe the low-energy behavior of $N$ D3 branes probing F-theory singularities. As we discussed in Section \ref{subsec:landscape_of_theories}, there are seven types of these singularities and they are denoted by the corresponding global symmetry group of the SCFT: $H_0$, $H_1$, $H_2$, $D_4$, $E_6$, $E_7$, $E_8$ (with $H_i \rightarrow A_i$). The theories with $N > 1$ have an additional $\suf(2)_L$ flavor symmetry.

The full set of central charges of these theories was calculated in \cite{Aharony:2007dj} as a function of $N$ using holography. Let us denote by $k$ the flavor central charge of the global symmetry group indicated by the name of the theory, and by $k_L$ the level of the additional $\suf(2)_L$ for the theories with rank greater than one. Then the relevant central charges are given by
%%%%%%
\begin{equation}
\label{eq:ckkLFtheories}
\begin{split}
  \makebox[.14in][l]{$c$}    & = \frac{1}{2} N^2 r_0 + \frac{3}{4} N (r_0 - 1) - \frac{1}{12}~,\\
  \makebox[.14in][l]{$k$}    & = 2 N r_0~,\\
  \makebox[.14in][l]{$k_L$}  & = N^2 r_0 - N (r_0 - 1) -1~,\qquad\qquad N \geqslant 2~,
\end{split}
\end{equation}
%%%%%%
where the vale of $r_0$ for each of the seven types is given in Table \ref{Tab:rank1theories}. The resulting values of $c$ and $k$ for these theories are plotted in Fig.~\ref{fig:su2landscape}.

Our plan for the remainder of this subsection can now be formulated as follows. We will first focus on the unshaded, third region in Fig.~\ref{fig:su2landscape}. Could there be theories hidden in this region, or some of these points be excluded? We will see that the latter is true, and we can numerically obtain a universal lower bound on $c$ for each value of $k$. For the remaining allowed region, which includes the entire blue region in Fig.~\ref{fig:su2landscape}, we find upper bounds for the dimension of several unprotected operators as a function of $c$ and $k$. The numerical analysis necessary to generate these bounds was computationally rather demanding because of the two-dimensional parameter space, which limited the value of $\Lambda$ for which the computation was feasible to a maximum of $\Lambda=18$. For restricted values of $c$ and $k$ that are of particular interest, we generated superior bounds by going as high as $\Lambda=22$. In particular, we chose to study the $H_0$ and $H_1$ curves shown in Fig.~\ref{fig:su2landscape}. We also studied the point at $k=4$ and $c= \infty$, which corresponds to an interesting defect SCFT. Bounds for the $\ef_6$ curve are postponed until the next subsection for the purposes of comparison to bounds extracted from the $\ef_6$ moment map four-point function.

%%%%%%%%%%%%%%%%%%%%%%%%%%%%%%%%%%%%%%%%%%%%%%%%%%%%%%%%%%%%%%%%%%%%%%%%%%%%%%%%%%%%%%%%%%%%%%%%%%%%%%%%%%%%
\subsubsection{Constraints on \texorpdfstring{$c$}{c} and \texorpdfstring{$k$}{k}}\label{subsubsec:bhat_su2_ck_bounds}
%%%%%%%%%%%%%%%%%%%%%%%%%%%%%%%%%%%%%%%%%%%%%%%%%%%%%%%%%%%%%%%%%%%%%%%%%%%%%%%%%%%%%%%%%%%%%%%%%%%%%%%%%%%%

%%%
\begin{figure}[h!tbp]
             \begin{center}           
              \includegraphics[scale=0.43]{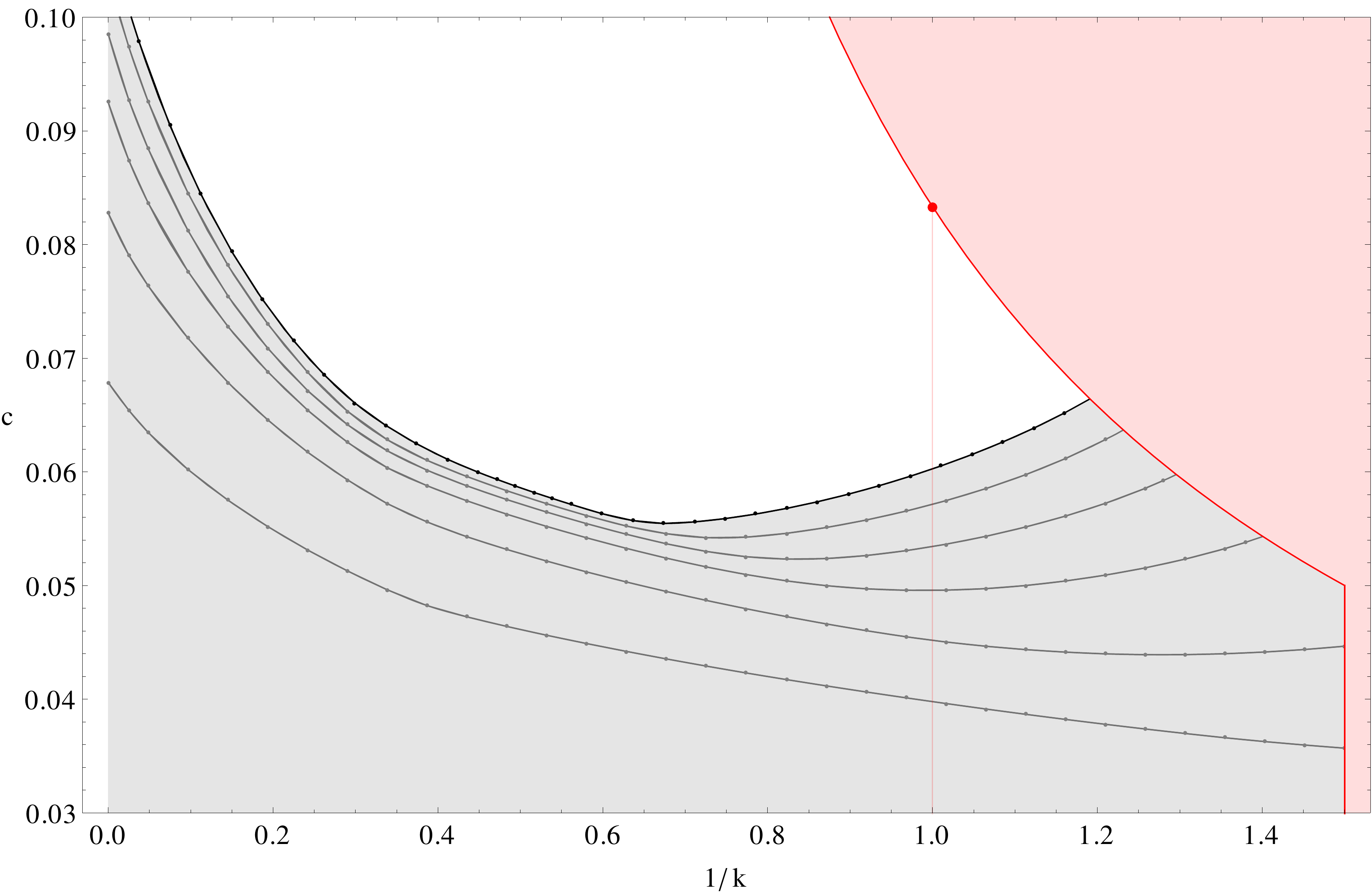}
              \caption{Bounds for the central charge $c$ of a theory with $\suf(2)$ flavor symmetry as a function of the flavor central charge $k$. These bounds are a consequence of crossing symmetry for the $\hat{\BB}_1$ four-point function. The red regions on the right are excluded by the analytic bounds \eqref{analytic_bound}, and the gray region at the bottom is the numerically excluded region. The gray and black lines correspond to the numerical bounds, shown for $\Lambda=10,14,\ldots,30$, with the strongest bound (black line) corresponding to $\Lambda = 30$. The curves are interpolations through the data points shown in the figure. The red dot denotes the free hypermultiplet theory.}
                \label{Fig:SU2-cbound}
            \end{center}
\end{figure}
%%%

%%%
\begin{figure}[h!tbp]
             \begin{center}           
              \includegraphics[scale=0.5]{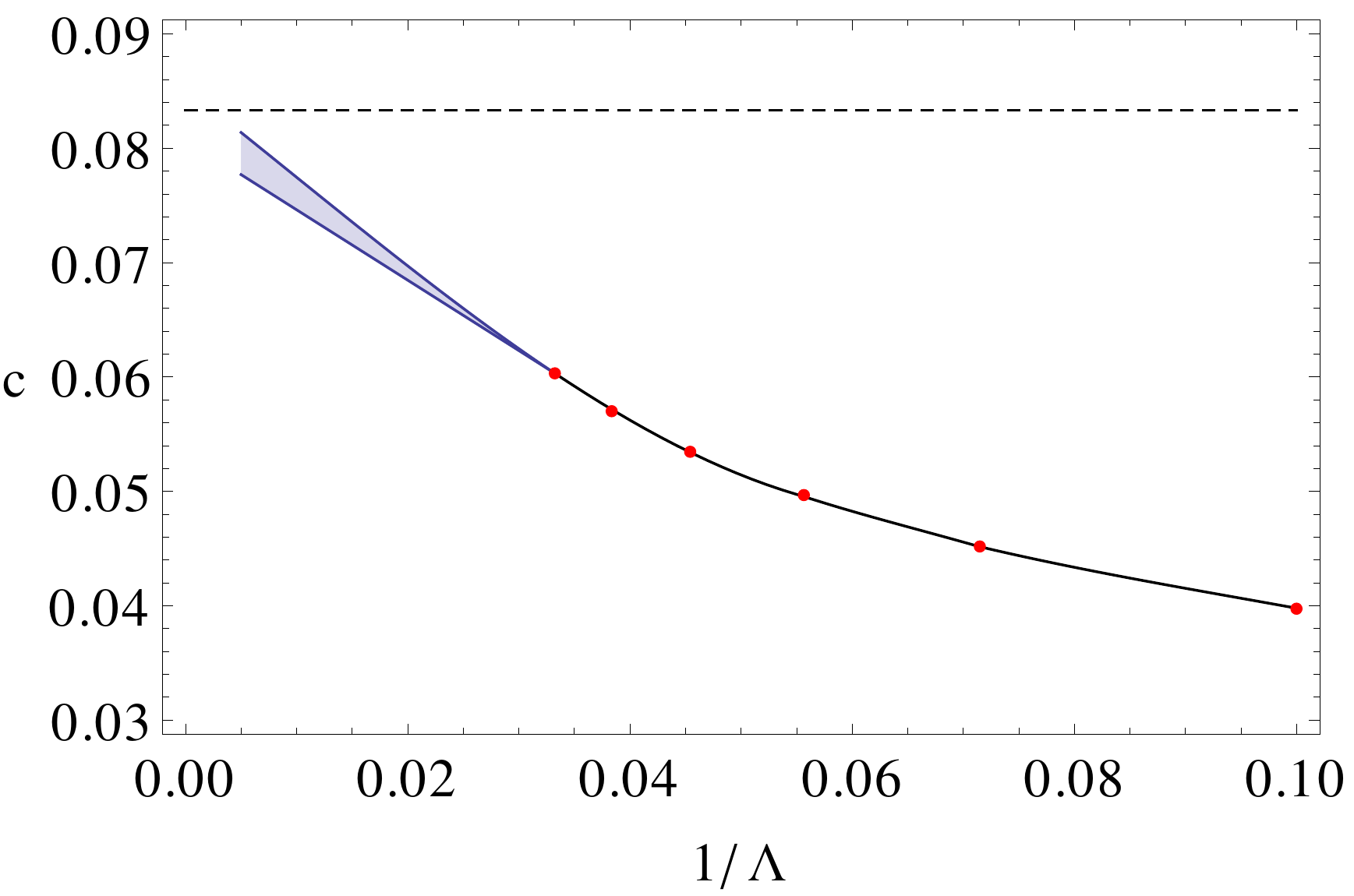}
              \caption{Minimum allowed value of $c$ for a theory with $\suf(2)$ flavor symmetry and $k=1$ as a function of the (inverse of) the maximum number of derivatives. The red dots are our data points, and the blue curves are possible extrapolations to infinite $\Lambda$ intended to guide the eye. The dashed line corresponds to the central charge of the free hypermultiplet $c=\frac{1}{12}$.}
              \label{Fig:su2_cboundkeq1}
            \end{center}
\end{figure}
%%%

To constrain the $(c,k)$ plane we employed the second numerical method described in the previous section. For a given value of $k$ we normalize the functional by demanding that it evaluate to one on the contribution of protected operators to \eqref{bootstrapeqn} that are proportional to the inverse central charge $1/c$. We then minimize its value when acting on the remaining protected contribution to crossing. The upper bound that we obtain in this way for $1/c$ then corresponds to a lower bound on the central charge.\footnote{Bounds obtained in this way for the central charge, and more generally for OPE coefficients, have been studied in the literature starting with \cite{Rattazzi:2010gj,Poland:2010wg,Caracciolo:2009bx}.}

The results of this program are shown in Fig.~\ref{Fig:SU2-cbound}. The numerically excluded region is shaded in gray. This result was obtained with $\Lambda = 30$, \ie, by taking at most $30$ derivatives in the $z$ or $\zb$ directions. Bounds for smaller $\Lambda$ are indicated with gray curves. One interesting and very general lesson to be drawn from Fig.~\ref{Fig:SU2-cbound} is that the analytic and the numerical bounds complement each other, and the most stringent constraints can only be obtained by using both methods. For example, the numerical analysis eliminates the possibility of a unitary SCFT existing at the intersection point $(c,k) = (\frac{1}{20},\frac{2}{3})$ of the two analytic bounds given in \eqref{analytic_bound}. We also find that for all values of $k$, there exists a universal lower bound on the central charge,
%%%%%%
\begin{equation}
c \geqslant 0.055~,
\end{equation}
%%%%%%
for any $\NN = 2$ SCFT with a non-abelian flavor symmetry. For comparison we may note that for a free hypermultiplet $c = 1/12 = 0.0833\ldots$. From Fig.~\ref{Fig:SU2-cbound} it seems that there may in fact be a solution to crossing symmetry with roughly this minimum value of the central charge, because the global minimum of the exclusion curve at $1/k \simeq 0.68$ seems rather stable against increasing $\Lambda$. We are however not aware of any $\NN = 2$ SCFT (with or without non-abelian flavor symmetry) with such a low central charge.

A feature of these our numerical bounds that we will be repeated both here and in the next section is that they are non-optimal, meaning that they display substantial dependence on $\Lambda$ for the values of the cutoff considered. This is in contrast with, \eg, the three-dimensional investigations in \cite{ElShowk:2012ht}. In that paper the bounds converge much faster and on the scales that we consider here they are essentially constant at $\Lambda = 22$.\footnote{In \cite{ElShowk:2012ht} the cutoff is defined differently -- $\Lambda = 22$ here corresponds to $n_{\rm max} = 11$ there.} Notice that with $\Lambda= 30$ and three flavor symmetry channels we have a functional with $392$ components, which surpasses even the $231$ components used in the precision work on the three-dimensional Ising model \cite{El-Showk:2014dwa}. Apparently this crossing symmetry problem is numerically more expensive. We cannot currently offer a good explanation as to why this is the case.

A natural way to deal with the relatively poor convergence is to extrapolate our results from finite to infinite $\Lambda$.\footnote{We do not currently have theoretical control of the dependence of the numerical bounds on $\Lambda$, but we hope the apparent smoothness of the numerical results is enough to justify such extrapolations.} In this way we can generate a rough guess of where the best possible bound may lie. Fig.~\ref{Fig:su2_cboundkeq1} shows an example of such an extrapolation. The minimum allowed value of $c$ for $k=1$ is plotted as a function of the cutoff $\Lambda$, and a possible extrapolation to infinite cutoff is sketched. The dashed line in the figure corresponds to the central charge which saturates the analytic bound at $k=1$ (corresponding to the free hypermultiplet with $c=\frac{1}{12}$). It seems plausible that in the $\Lambda \to \infty$ limit the numerical bounds will intersect the analytic bound at this point.

%%%%%%%%%%%%%%%%%%%%%%%%%%%%%%%%%%%%%%%%%%%%%%%%%%%%%%%%%%%%%%%%%%%%%%%%%%%%%%%%%%%%%%%%%%%%%%%%%%%%%%%%%%%%
\subsubsection{Dimension bounds for $\suf(2)$}
%%%%%%%%%%%%%%%%%%%%%%%%%%%%%%%%%%%%%%%%%%%%%%%%%%%%%%%%%%%%%%%%%%%%%%%%%%%%%%%%%%%%%%%%%%%%%%%%%%%%%%%%%%%%

We now focus on the allowed region in the $(c,k)$ plane and generate bounds for the dimension of the first unprotected operator appearing in the $\hat{\BB}_1 \times \hat{\BB}_1$ OPE. In the tensor product of two copies of the adjoint representation of $\suf(2)$ one finds three irreps: the singlet, triplet, and quintuplet. We will report on bounds for the dimension of the first unprotected operator of lowest spin in each of these channels.

%%%%%%%%%%%%%%%%%%%%%%%%%%%%%%%%%%%%%%%%%%%%%%%%%%%%%%%%%%%%%%%%%%%%%%%%%%%%%%%%%%%%%%%%%%%%%%%%%%%%%%%%%%%%
\subsubsection*{Singlet channel}
%%%%%%%%%%%%%%%%%%%%%%%%%%%%%%%%%%%%%%%%%%%%%%%%%%%%%%%%%%%%%%%%%%%%%%%%%%%%%%%%%%%%%%%%%%%%%%%%%%%%%%%%%%%%

\begin{figure}[t!]
  \begin{center}
    \includegraphics[width=2.5in]{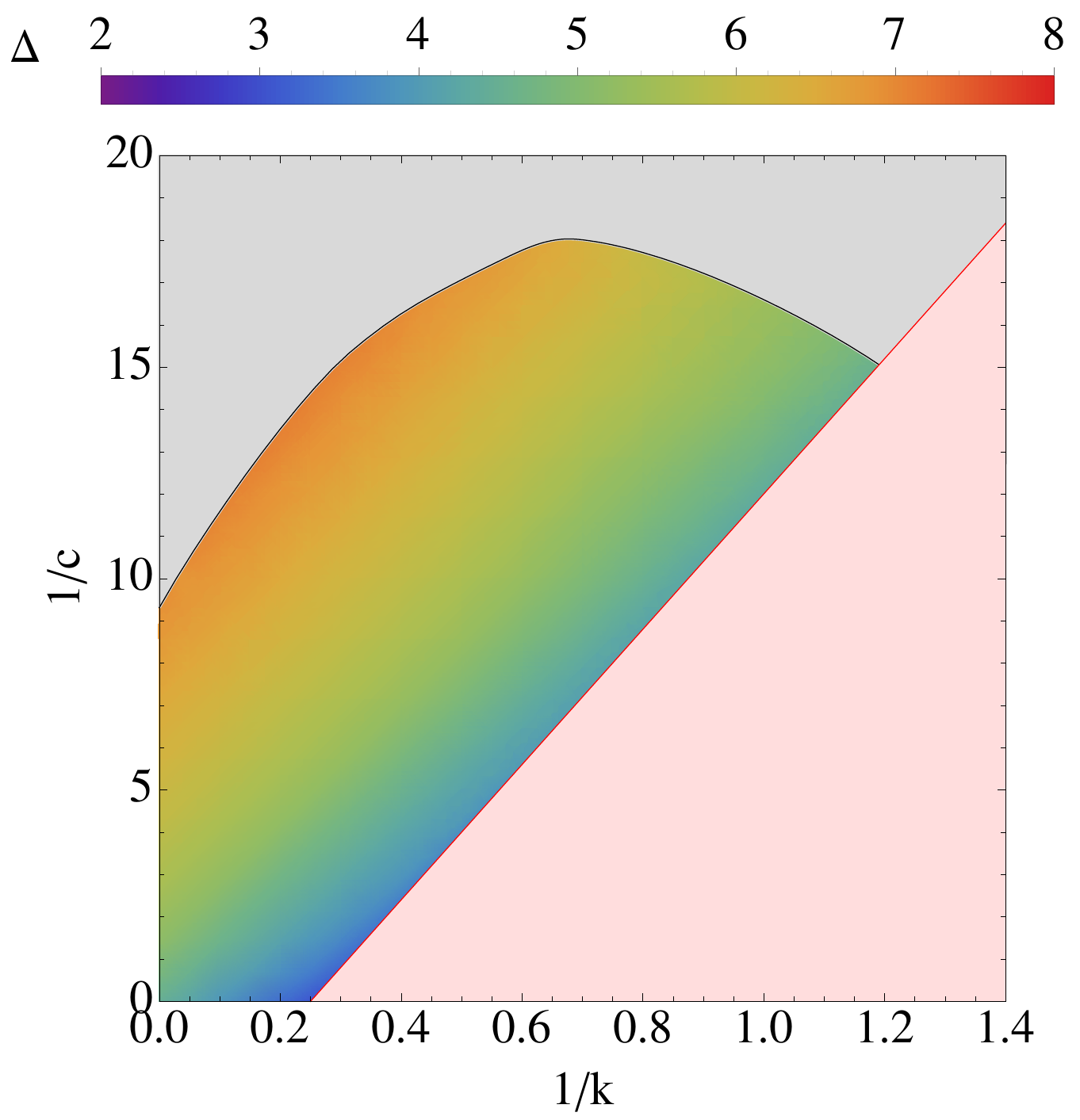}
    \includegraphics[width=3.2in]{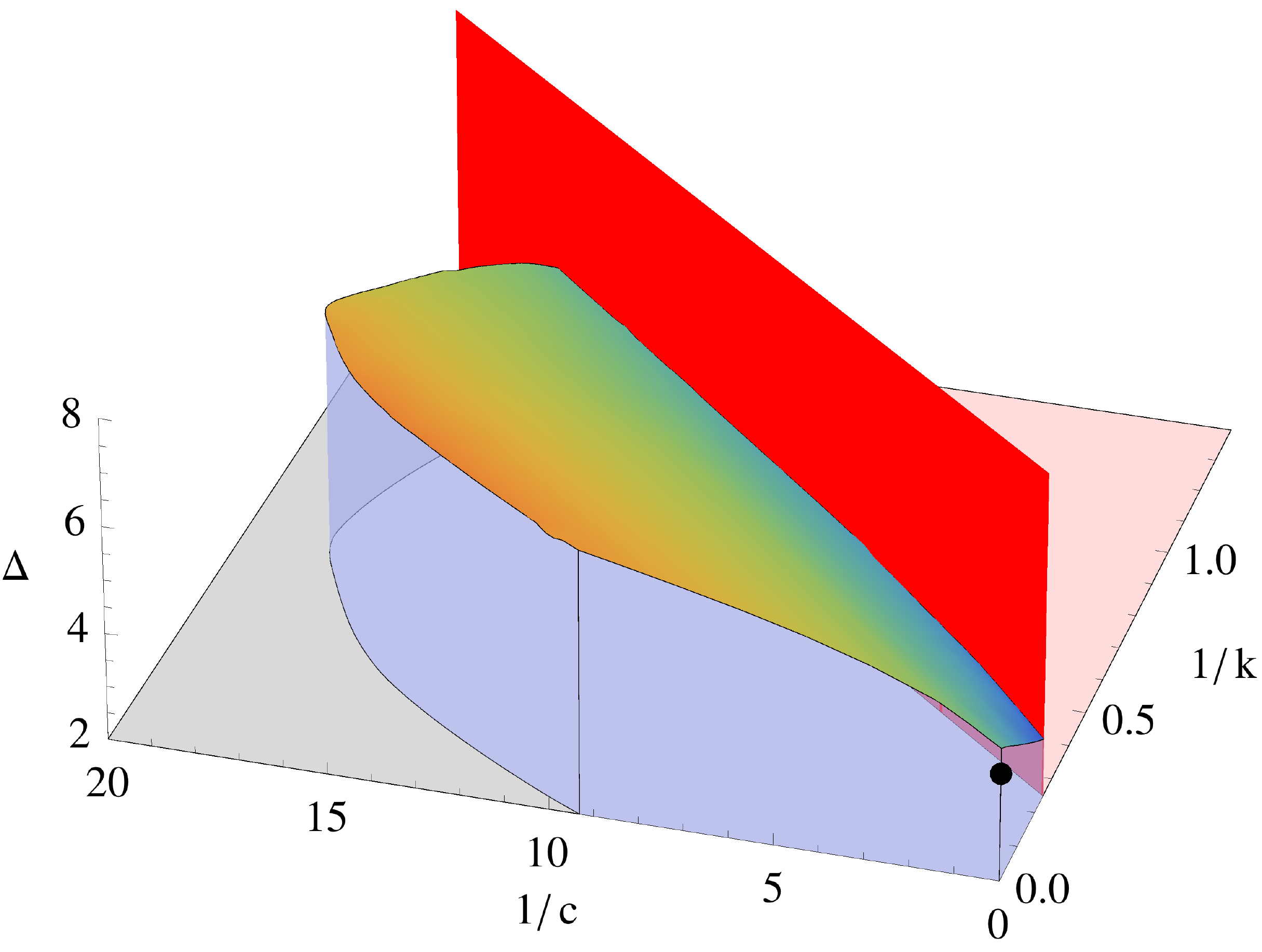}
    \caption{\label{fig:su2singletbounds}Upper bounds for the dimension of the first unprotected singlet scalar operator in theories with $\suf(2)$ flavor symmetry, as a function of $1/k$ and $1/c$. The cutoff used for this plot was $\Lambda=18$. The two- and a three-dimensional plots are generated with the same data set. The gray and light red surfaces in the figure are the excluded regions from Fig.~\ref{Fig:SU2-cbound}, and the vertical red wall is added help visualize the constraints imposed by the analytic bounds. The black dot is the generalized field theory solution to crossing.}
  \end{center}
\end{figure}

In Fig.~\ref{fig:su2singletbounds} we present the upper bound for the scaling dimension $\Delta$ of the first unprotected scalar operator in the singlet channel, for all allowed values of the central charges. The values shown are an interpolation through a total of $572$ data points, distributed on a square grid with finer resolution near the edges. The cutoff for this analysis is $\Lambda = 18$. The surface so obtained appears smooth and monotonic, with the bound getting stronger when approaching the wall that represents the analytic bound or and at large central charge.

%%%
%%%

The bounds shown in Fig.~\ref{fig:su2singletbounds} are completely universal -- any four-dimensional $\NN =2$ SCFT with at least $\suf(2)$ flavor symmetry corresponds to a point somewhere inside the allowed region. We will discuss several examples of such theories below, but as a zeroth-order check we confirm that our bounds are consistent with some elementary solutions to crossing symmetry.

At the infinite point $(1/c,1/k)=(0,0)$ the stress tensor and the flavor current decouple, their OPE coefficients being $\lambda_T \sim \frac{1}{c}$ and $\lambda_J \sim \frac{1}{k}$ respectively. A well-known solution to crossing symmetry for which these operators are absent is generalized free field theory, for which the four-point function is a sum of disconnected pieces,
%%%%%%
\begin{equation}
\langle\phi_1(x_1) \ldots \phi_4(x_4) \rangle = \langle\phi_1(x_1) \phi_2(x_2)\rangle \langle \phi_3(x_3) \phi_4(x_4) \rangle + {\rm two~permutations}~.
\end{equation}
%%%%%%
Specializing this solution to the four-point function of moment map operators, we find that the first operator in the conformal block decomposition has dimension four. As is indicated in Fig.~\ref{fig:su2singletbounds}, the generalized free field solution is consistent with the numerical upper bound which gives $\Delta \leqslant 4.47$ at this point. The numerical bound is similarly consistent with the theory of a free hypermultiplet with $(c,k) = (\frac{1}{12},1)$, since the first unprotected singlet scalar in the corresponding four-point function again has dimension four and numerically we have $\Delta \leqslant 4.38$. Finally, we can take a linear combination of the two solution with positive weights that sum to one. This results in a valid solution to crossing symmetry along the straight line in Fig.~\ref{fig:su2singletbounds} that runs from the origin to the free-field point, with a first unprotected singlet scalar operator that always has dimension four. Again, this is consistent with the numerical bound which is greater than four everywhere above this line. Much like the bound on $c$ sketched in Fig.~\ref{Fig:su2_cboundkeq1}, we expect these bounds to decrease substantially as $\Lambda$ is increased, and to converge to four along this line as $\Lambda \to \infty$. An extrapolation in $\Lambda$ for $(1/c,1/k)=(0,0)$ (not shown) bolsters this intuition. Similar extrapolation experiments suggest that the bound should end up below $4$ for all values in the $(c,k)$ plane between the analytic bound and the interpolating solution of the previous paragraph.

Although we have presented the two results in Figs.~\ref{Fig:SU2-cbound} and \ref{fig:su2singletbounds} as independent results, they are in fact related. Indeed, the bound on the first scalar operator drops sharply to the unitarity bound when we venture inside the numerically excluded region of Fig.~\ref{Fig:SU2-cbound}. Such a drop indicates that there does not exist any spectrum that is simultaneously consistent with unitarity and crossing, and delineating the region where this happens is another way to obtain the numerically excluded region in Fig.~\ref{Fig:SU2-cbound}. The $c$-minimization approach used to generate Fig.~\ref{Fig:SU2-cbound} is much more efficient, and could consequently be performed at higher values of $\Lambda$.

%%%
\begin{figure}[t!]
             \begin{center}    
              \includegraphics[width=2.5in]{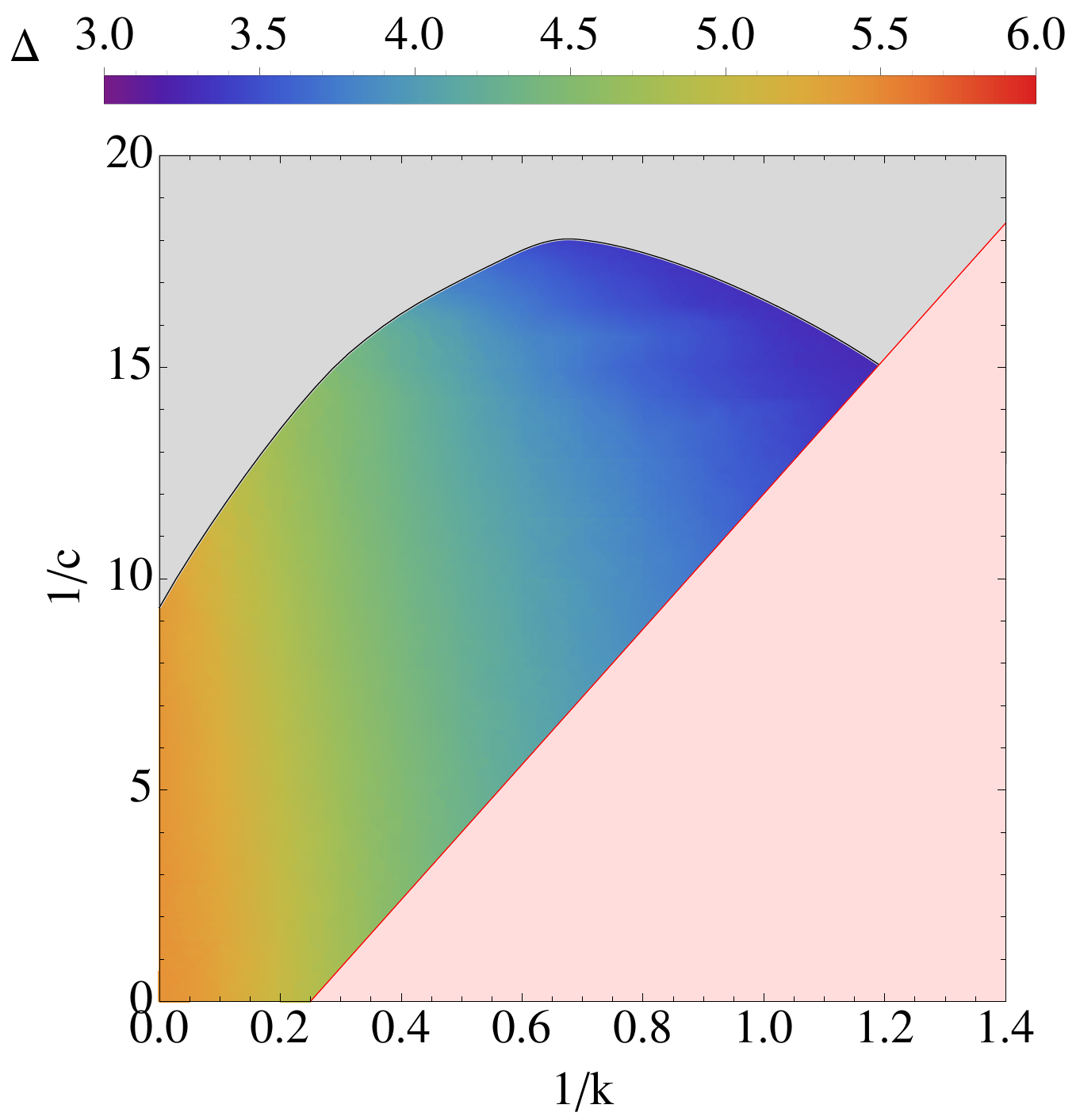}
              \includegraphics[width=3.2in]{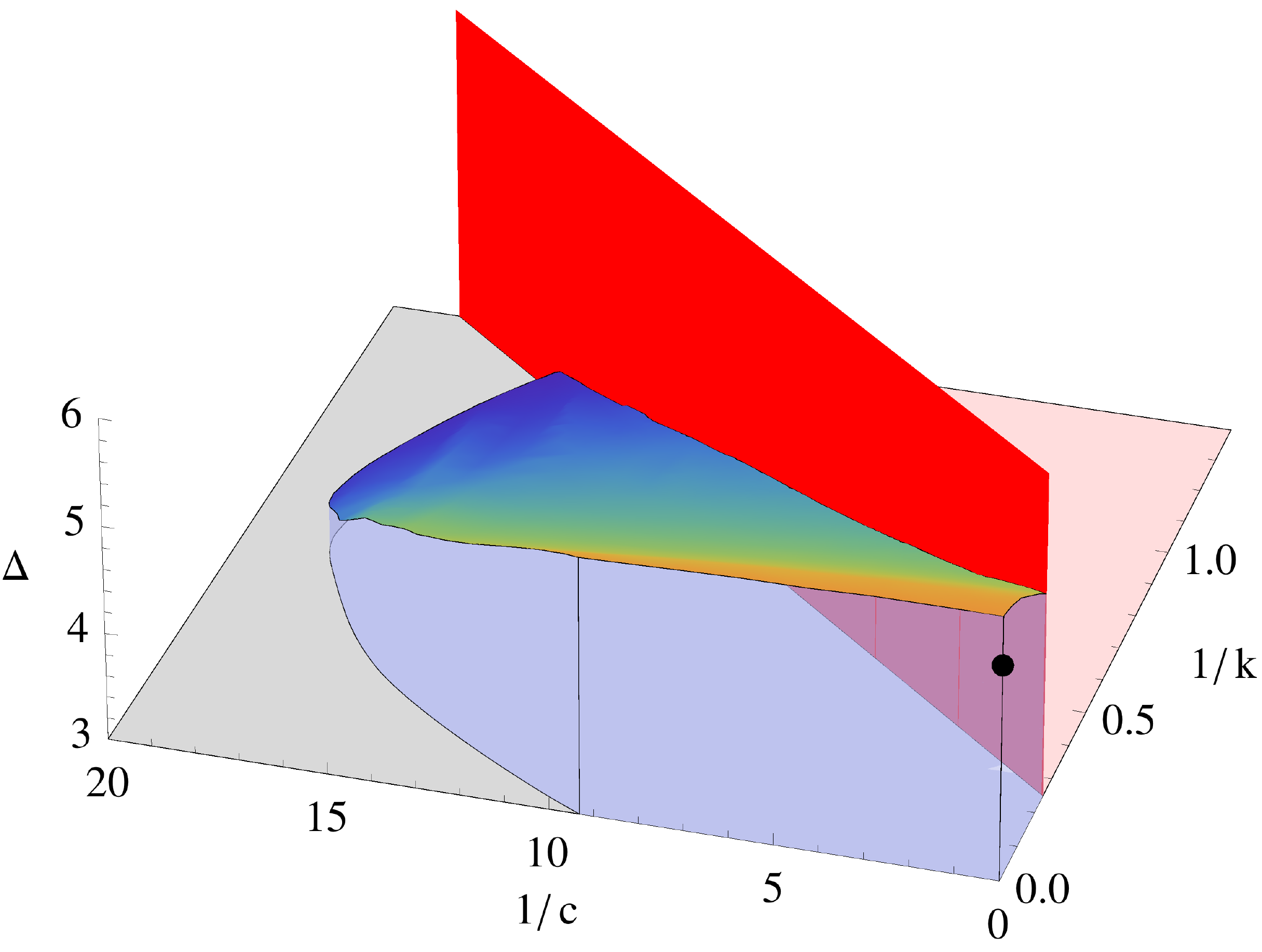}
              \caption{Upper bounds for the dimension of the first unprotected spin one multiplet in the triplet channel of a theory with $\suf(2)$ flavor symmetry, for all allowed values of $c$ and $k$, presented both as a three-dimensional plot and as a density plot. The gray and light red surfaces in the figure are the excluded regions from Fig.~\ref{Fig:SU2-cbound}. These bounds were obtained with $\Lambda=18$  and $547$ data points in the $(c,k)$ plane.}
              \label{Fig:triplet}
            \end{center}
             \begin{center}           
              \includegraphics[width=2.5in]{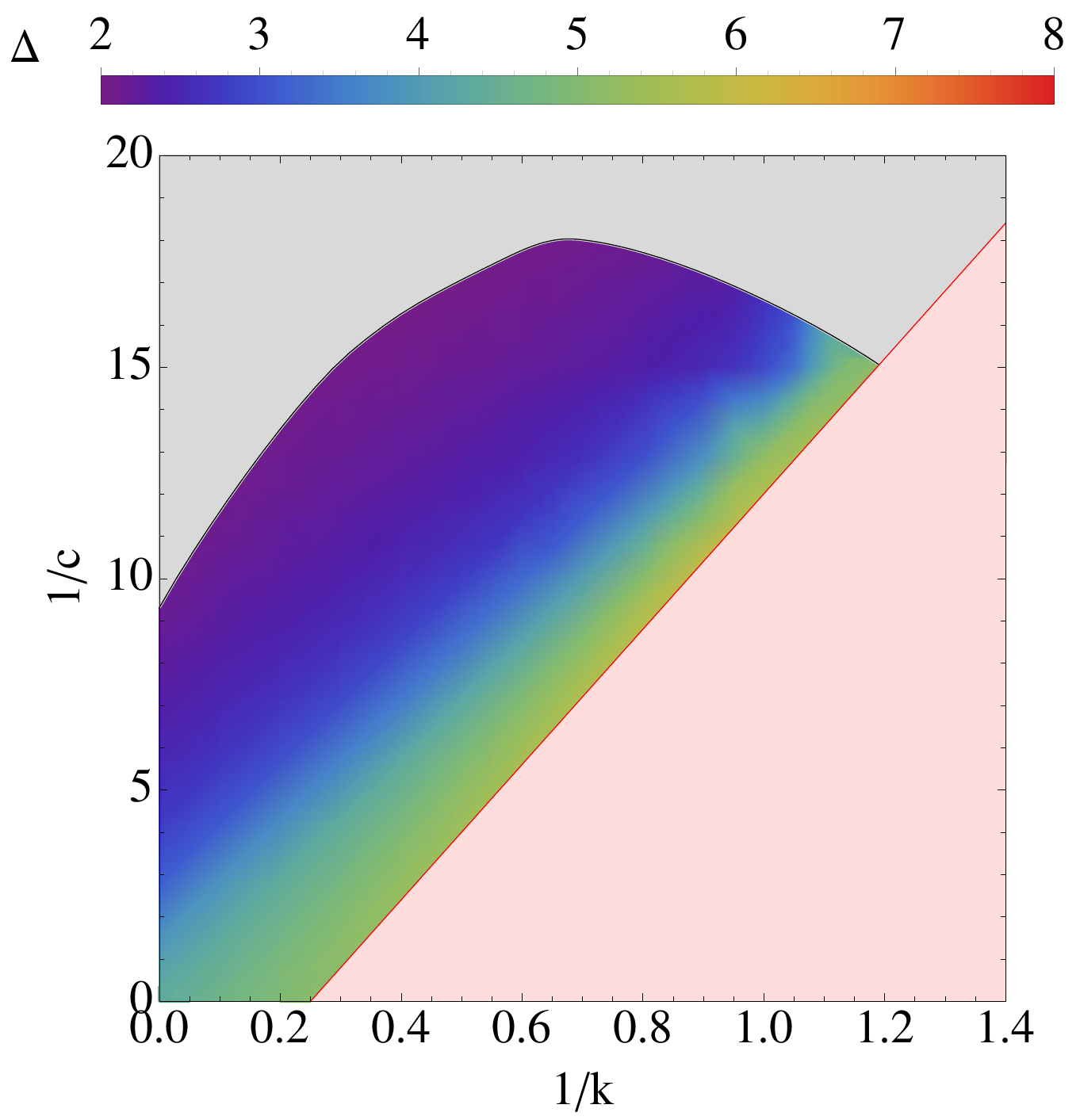}
              \includegraphics[width=3.2in]{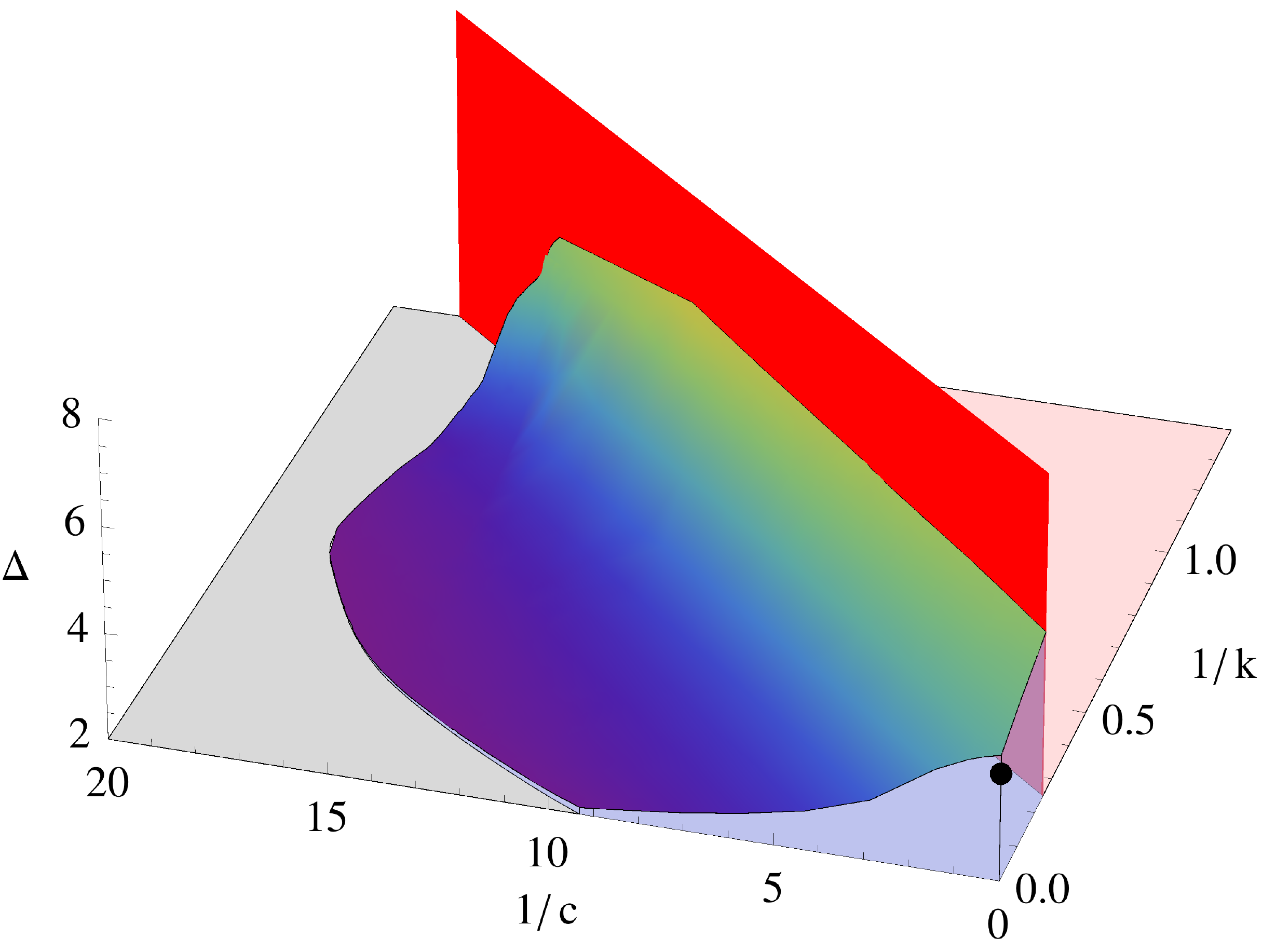}
              \caption{Upper bounds for the dimension of the first unprotected scalar in the quintuplet channel of a theory with $\suf(2)$ flavor symmetry, for all allowed values of $c$ and $k$, presented both as a three-dimensional plot and as a density plot. The gray and light red surfaces in the figure are the excluded regions from Fig.~\ref{Fig:SU2-cbound}. This plot was obtained with $\Lambda=18$ and $398$ data points in the $(c,k)$ plane.}
              \label{Fig:quintuplet}
            \end{center}
\end{figure}
%%%

%%%%%%%%%%%%%%%%%%%%%%%%%%%%%%%%%%%%%%%%%%%%%%%%%%%%%%%%%%%%%%%%%%%%%%%%%%%%%%%%%%%%%%%%%%%%%%%%%%%%%%%%%%%%
\subsubsection*{Triplet and quintuplet channels}
%%%%%%%%%%%%%%%%%%%%%%%%%%%%%%%%%%%%%%%%%%%%%%%%%%%%%%%%%%%%%%%%%%%%%%%%%%%%%%%%%%%%%%%%%%%%%%%%%%%%%%%%%%%%

We now present numerical results for the triplet and quintuplet channels. The triplet appears in the antisymmetric combination of two adjoints, so only odd spins can be exchanged. In this case we bound the dimension of the first unprotected spin one operator appearing in the $\hat{\BB}_1 \times \hat{\BB}_1$ OPE. This bound is shown in Fig.~\ref{Fig:triplet} for the allowed range of $c$ and $k$. Note that the unitarity bound for a spin one multiplet is $\Delta\geqslant3$. The numerical upper bound is again represented by a smooth surface, with weaker bounds appearing at larger values of $k$. In the limit where both $c$ and $k$ go to infinity the bound is close to $5$, which is the value for generalized free field theory.

The quintuplet channel is again symmetric, so the exchanged operators will have even spin as they did in the singlet channel. We have generated upper bounds for the dimension of the first scalar operator. These are shown in Fig.~\ref{Fig:quintuplet} as a three-dimensional plot and a density plot. The behavior of the bounds when approaching the minimum allowed values of $c$ and $k$ is different from the other two channels -- in this case the bound drops smoothly to the unitarity bound at $\Delta=2$. As either $c$ or $k$ are increased the bound gets weaker, and when they both go to infinity the bound is near $\Delta=4$, which is the correct value in generalized free field theory.

We note that the triplet and quintuplet bounds approach the unitarity bound near the minimum of the exclusion curve of Fig.~\ref{Fig:SU2-cbound} at $1/k \simeq 0.68$. This is a strong indication that the solution to crossing symmetry at that point has higher spin currents, which we would generally associate to a free theory. Because the central charge is not that of a free hypermultiplet, one may suspect that this point is not related to a physical theory.

%%%%%%%%%%%%%%%%%%%%%%%%%%%%%%%%%%%%%%%%%%%%%%%%
\subsubsection{Bounds for theories of interest}
%%%%%%%%%%%%%%%%%%%%%%%%%%%%%%%%%%%%%%%%%%%%%%%%

In the previous subsections we discussed bounds on operator dimensions for the entire $(c,k)$ plane that were obtained with a cutoff $\Lambda = 18$. We will now turn to a discussion of stronger bounds, obtained with $\Lambda = 22$, which we computed only for specific values of $c$ and $k$ that correspond to theories of interest. In this subsection we present operator dimension bounds along the curves in the $(c,k)$ plane that correspond to the $H_0$ and $H_1$ theories shown in Fig.~\ref{fig:su2landscape}. In the next subsection we will discuss the defect theory at infinite $c$ and $k = 4$ that corresponds to the dotted line in Fig.~\ref{fig:su2landscape}.

\begin{figure}[htbp!]
  \begin{center}
    \includegraphics[scale=0.4]{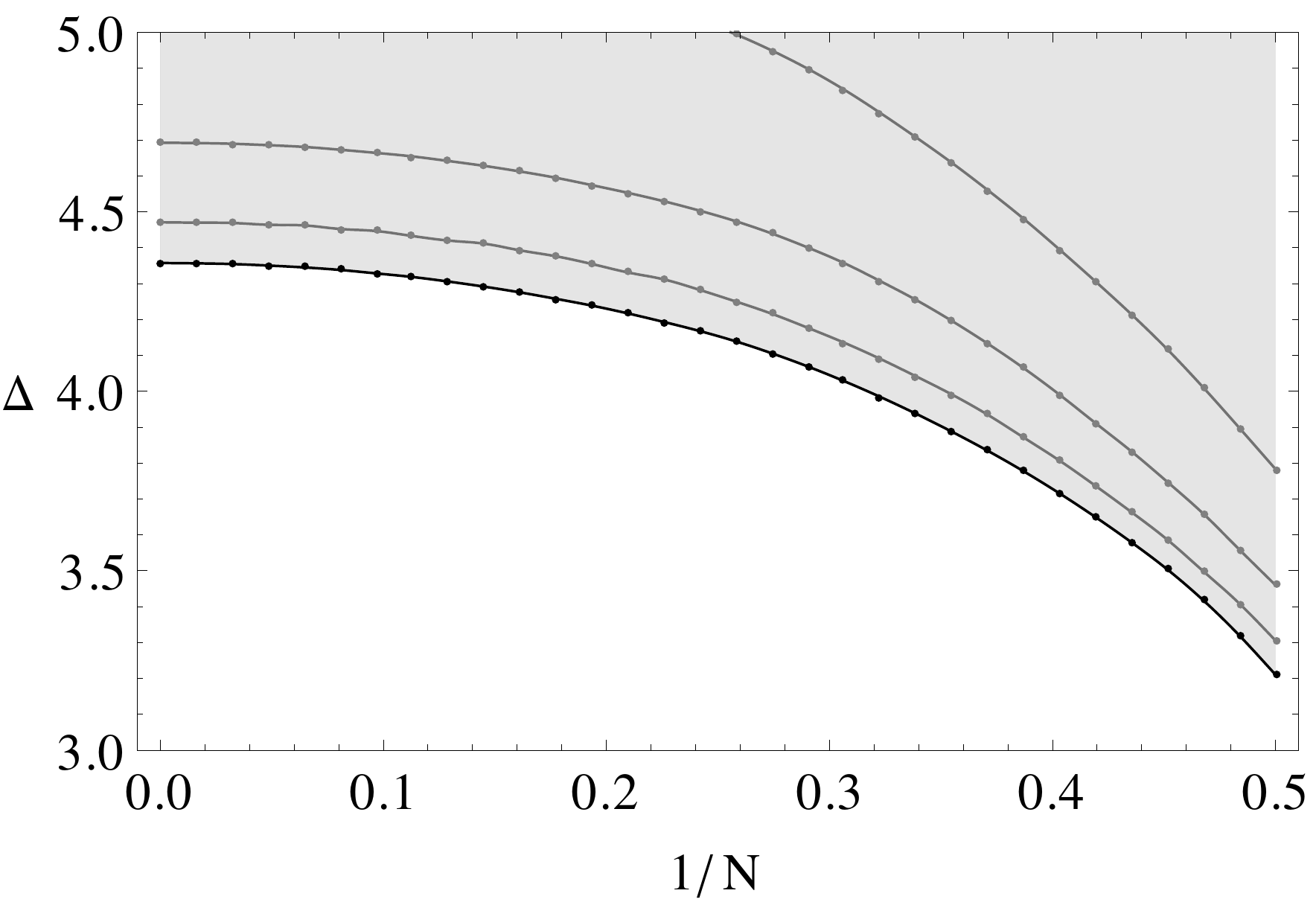}
  \end{center}
  \vspace{-.15in}
  \caption{\label{h0singletbound}Upper bounds for the first unprotected scalar in the theories of type $H_0$ as a function of the inverse rank. The bounds are extracted from the four-point function of the $\suf(2)_L$ flavor symmetry moment map and are valid only for $N \geqslant 2$. The different lines correspond to cutoff values $\Lambda = 10,14,\ldots,22$, with the strongest bound shown as the black line.}
\end{figure}

\begin{figure}[b!]
            \begin{center}
            \begin{subfigure}[b]{0.4\textwidth}
                \includegraphics[width=\textwidth]{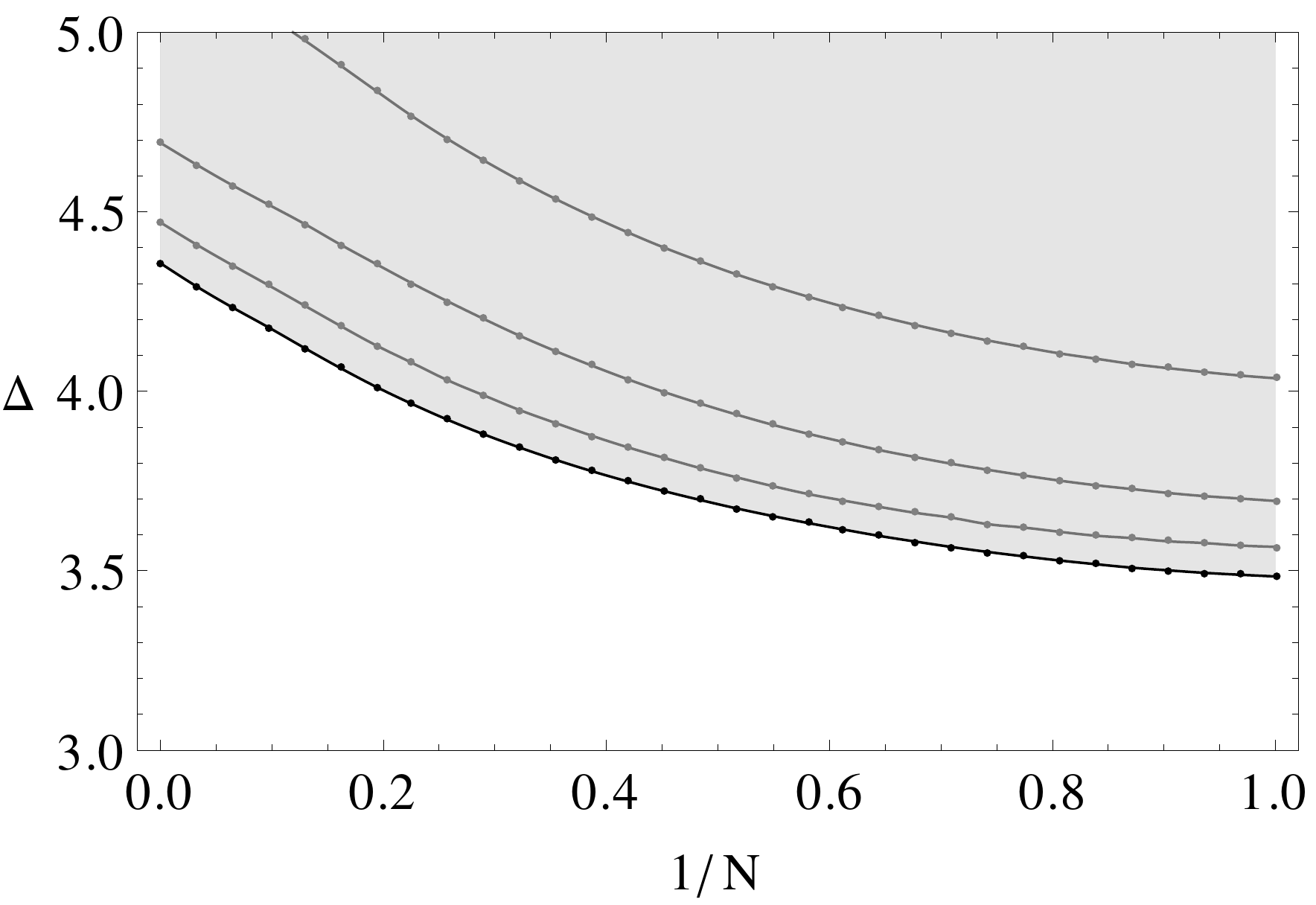}
                \label{Fig:h1_rank_NSU2}
            \end{subfigure}
            \hspace{1cm}
            \begin{subfigure}[b]{0.4\textwidth}
                \includegraphics[width=\textwidth]{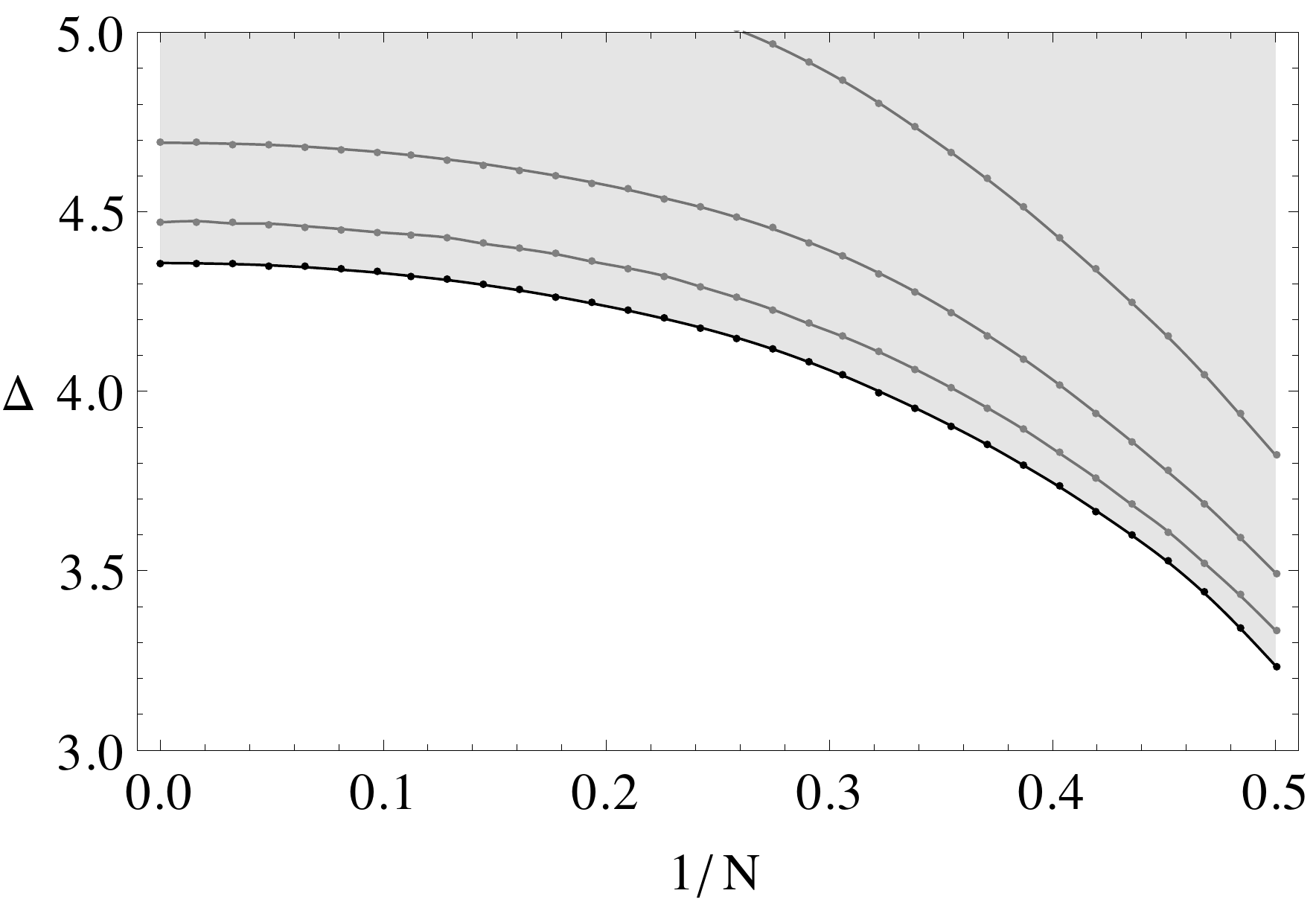}
                \label{Fig:h1_rank_NSU2L}
            \end{subfigure}
            \end{center}
            \vspace{-.25in}
            \caption{Bounds for the dimension of the first unprotected spin zero multiplet in the singlet channel for the $H_1$ theories, as a function of the inverse of the rank of the theories. The left plot comes from studying the four-point function of the ordinary $\suf(2)$ flavor symmetry moment map and is valid for all $N \geqslant 1$. The right plot comes from the four-point function of the $\suf(2)_L$ flavor symmetry moment map and valid only for $N \geqslant 2$. The different lines correspond to $\Lambda = 10,14,\ldots,22$, with the strongest bound shown as the black line.}
            \label{Fig:h1_rank_N}
\end{figure}

For the $H_0$ theories with $N\geqslant2$ the only flavor symmetry is $\suf(2)_L$. We can trace the results shown in Fig.~\ref{fig:su2singletbounds} along the $H_0$ curve in Fig.~\ref{fig:su2landscape} to recover upper bounds for the dimension of the first unprotected scalar singlet in these theories. This slice is displayed in  Fig.~\ref{h0singletbound}.
%%%
%%%
The $H_1$ theories with $N\geqslant2$ have two independent $\suf(2)$ symmetries with different flavor central charges. We derived bounds for the two different cases by following the two different curved labelled $H_1$ in Fig.~\ref{fig:su2landscape}. Both of the singlet scalar bounds so obtained are shown in Fig.~\ref{Fig:h1_rank_N}. 

%%%
%%%

In all of our plots corresponding to lines of interesting theories, we have shown the progression of the bounds as a function of the cutoff. This gives a feeling for how close to the optimal bound we have gotten -- information that is absent from the plots of the previous section where all the results came from analyses with $\Lambda=18$ was shown. In general, there seems to be some distance yet to go before the bounds will have effectively converged. In particular, in the infinite-rank limit $N \rightarrow \infty$ the stress tensor and flavor current decouple from the OPE expansion and the bounds should reach the generalized free field theory value $\Delta=4$. The difference between the $\Lambda=22$ bounds at large $N$ and the generalized free field theory value offer a simple proxy for how far we have yet to go.

Despite slow convergence, we may naively extrapolate our bounds to generate estimates for their optimal values. In particular, for the rank one $H_1$ theory there is a single $\suf(2)$ flavor factor and we might expect that the bound generated by studying the corresponding four-point function of moment maps to be saturated by this theory. Extrapolation for this value of $c$ and $k$ leads to a conjectural optimal bound in the range of $3.2-3.4$. Moving on to the rank two case, there are now two independent bounds extracted from the two $\suf(2)$ flavor symmetries. These two bounds could conceivably apply to the same operator. In other words, the same unprotected scalar singlet has no particular reason not to appear in both moment map four-point functions. However, the two bounds appear to be unrelated. The $\suf(2)_L$ bound dominates at low ranks, while the ordinary $\suf(2)$ bound dominates for higher ranks.

Similar bounds to those derived here can be obtained for the triplet and the quintuplet channels by the same methods, though we have not done so here.

%%%
\begin{figure}[t!]
  \begin{center}           
    \includegraphics[scale=0.5]{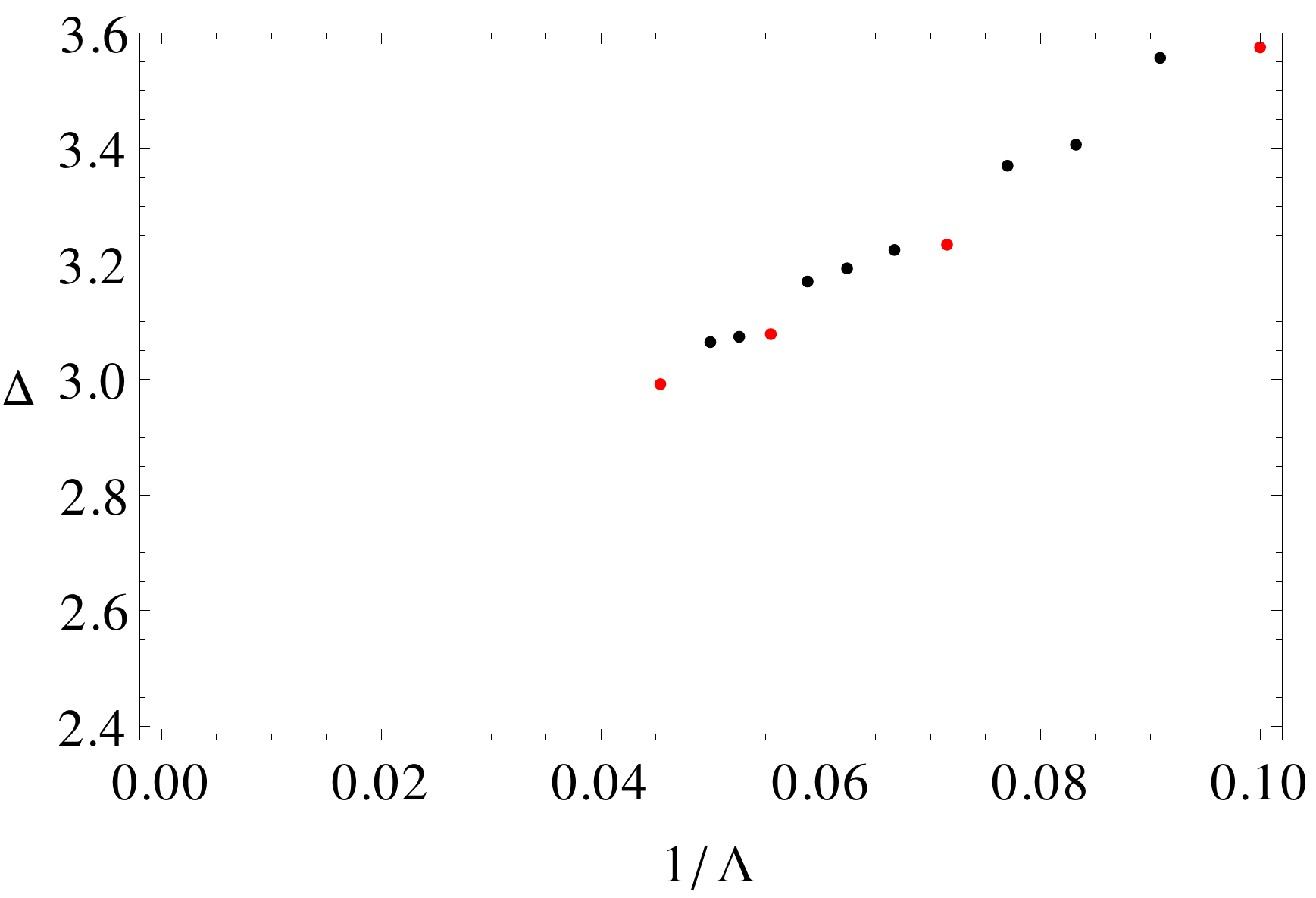}
  \end{center}
  \vspace{-.1in}
  \caption{Bound on the first unprotected scalar in the singlet channel for a theory with $k=4$ and infinite central charge, as a function of the cutoff. The red dots correspond to derivatives $10$ to $22$ in steps of four, and the black dots to the remaining values of $\Lambda$ ranging from $10$ to $20$.\label{Fig:su2critk}}
\end{figure}
%%%

%%%%%%%%%%%%%%%%%%%%%%%%%%%%%%%%%%%%%%%%%%%%%%%%%%%%%%%%%%%%%%%%%%%%%%%%%%%%%%%%%%%%%%%%%%%%%%%%%%%%%%%%%%%%
\subsubsection{Bounds for defect SCFTs}
\label{Sec:critk}
%%%%%%%%%%%%%%%%%%%%%%%%%%%%%%%%%%%%%%%%%%%%%%%%%%%%%%%%%%%%%%%%%%%%%%%%%%%%%%%%%%%%%%%%%%%%%%%%%%%%%%%%%%%%

An interesting aspect of the analytic bound \eqref{analytic_bound} is that as $c \to \infty$ the bound on $k$ stays finite and we have $k \geqslant 4 = 2 h^\vee$. The limit where $c \to \infty$ and $k$ remains finite should correspond to a theory without a stress tensor but with conserved global symmetry current. This kind of physics can be found on certain defects or interfaces in higher-dimensional theories where the global symmetry is confined to the defect but energy can leak into the bulk. There is in fact a natural set of defects that preserves $\NN = 2$ superconformal invariance in four dimensions, namely the codimension two defects in the six-dimensional $(2,0)$ SCFTs (see, \eg, \cite{Chacaltana:2012zy} and references therein). For a $(2,0)$ theory of type $\gf \in \{A_n,D_n,E_n\}$, the possible defects are labelled by an embedding $\rho: \slf(2)\to \gf$. The degrees of freedom localized on the defect carry a flavor symmetry $\mathfrak{h}$ which is the commutant of the image of $\rho$. When $\rho$ is trivial, the flavor symmetry is just $\gf$ and the corresponding flavor central charge is then given by $k = 2 h^\vee$. The bounds that we obtain at the point $k = 2 h^\vee$ with $c = \infty$ therefore constrain the spectrum of unprotected operators living on such a surface operator. Since we consider $\suf(2)$ flavor symmetry, this bound is valid for the defects of the $(2,0)$ theory of type $A_1$.

In Fig.~\ref{Fig:su2critk} we show the upper bound for scalar singlets in the defect theory as a function of the inverse cutoff. The best bound is given by $2.99$, and naive extrapolation suggests a relatively low value for the optimal bound somewhere between $2.5$ and $2.9$. It is natural to suspect that this is indeed the value of the first unprotected singlet scalar on the defect.

We notice that the bound in Fig.~\ref{Fig:su2critk} displays a step-like behavior whenever $\Lambda - 2$ is a multiple of four, corresponding to the red dots in the figure. Given our lack of theoretical control over the behavior of the bound as a function of the cutoff, we cannot currently offer any theoretical explanation for this quasi-periodicity. It however suggests an extrapolation scheme based on a restricted data set where $\Lambda$ increases in steps of four. This is what was done in generating Fig.~\ref{Fig:su2_cboundkeq1}.

%%%%%%%%%%%%%%%%%%%%%%%%%%%%%%%%%%%%%%%%%%%%%%%%%%%%%%%%%%%%%%%%%%%%%%%%%%%%%%%%%%%%%%%%%%%%%%%%%%%%%%%%%%%%
\subsection{\texorpdfstring{$\ef_6$}{e6} global symmetry}\label{subsec:e6_numerics}
%%%%%%%%%%%%%%%%%%%%%%%%%%%%%%%%%%%%%%%%%%%%%%%%%%%%%%%%%%%%%%%%%%%%%%%%%%%%%%%%%%%%%%%%%%%%%%%%%%%%%%%%%%%%

\begin{figure}[t!]
  \begin{center}
  \includegraphics[scale=0.43]{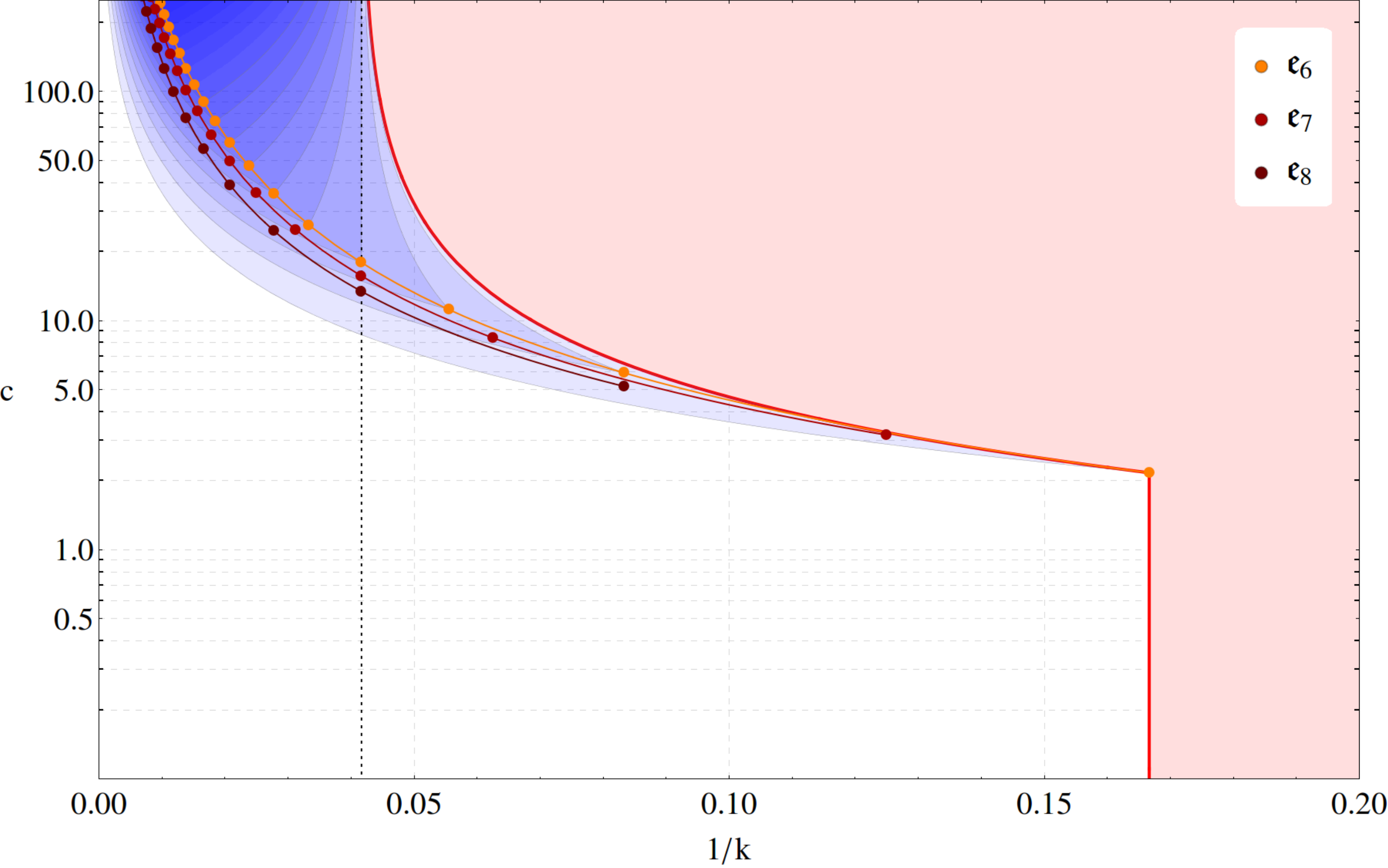}
  \caption{An overview of the of known theories with $\ef_6$ flavor symmetry, shown here as points in the plane spanned by their $c$ and $k$ central charges. The red region is excluded by analytic central charge bounds. The vertical dotted line designates the value $k=6$ with is the value of the central charge for the maximal defect SCFT in the six-dimensional $(2,0)$ theory of type $\ef$. The blue wedges with vertices at each of the $\ef_6$ theories are the region of the plane for which solutions of crossing symmetry can be realized as linear combinations of the four-point function for the theory at the vertex and those of generalized free field theory and the defect theory.\label{Fig:e6_landscape}}
  \end{center}
\end{figure}

Our second investigation focuses on theories with $\ef_6$ global symmetry. Let us again begin by making a rough sketch of the landscape of such theories as seen by the moment map four-point function. We show such a sketch in Fig.~\ref{Fig:e6_landscape}. There are analytic bounds for the central charges of theories with $\ef_6$ global symmetry arising from the chiral algebra of \cite{Beem:2013sza}. These are given by
%%%%%%
\begin{equation}\label{analytic_boundE6}
k \geqslant 6~, \qquad k \geqslant \frac{48c}{13 + 2c}~.
\end{equation}
%%%%%%
The region excluded by these bounds is shown in red in Fig.~\ref{Fig:e6_landscape}. We have also plotted several known families of theories whose flavor symmetry contains an $\ef_6$ factor, namely the theories originating from $F$-theory singularities of type $\ef_n$ for $n=6,7,8$ and for all ranks. The existence of these theories gives a collection of solutions to crossing symmetry with various values of $c$ and $k$. By taking linear combinations of these solutions, one can find solutions of crossing symmetry with $(c,k)$ values anywhere inside the blue region in Fig.~\ref{Fig:e6_landscape}. In particular, for each irreducible solution there is a wedge corresponding to linear combinations of that solution with the generalized free field theory solution and the defect solution at $k=6$ and $c\to\infty$. These wedges are shown for the $\ef_6$ theories.

For the purposes of numerical analysis, the fact that there are now five irreps in the tensor product of two copies of the adjoint representations makes the search space larger than the $\suf(2)$ case for a given value of $\Lambda$. As such, the maximum value of $\Lambda$ that we were able to reach is lower and the $\suf(2)$ case.

%%%%%%%%%%%%%%%%%%%%%%%%%%%%%%%%%%%%%%%%%%%%%%%%%%%%%%%%%%%%%%%%%%%%%%%%%%%%%%%%%%%%%%%%%%%%%%%%%%%%%%%%%%%%
\subsubsection{Constraints on \texorpdfstring{$c$}{c} and \texorpdfstring{$k$}{k}}
%%%%%%%%%%%%%%%%%%%%%%%%%%%%%%%%%%%%%%%%%%%%%%%%%%%%%%%%%%%%%%%%%%%%%%%%%%%%%%%%%%%%%%%%%%%%%%%%%%%%%%%%%%%%

\begin{figure}[t!]
             \begin{center}           
              \includegraphics[scale=0.43]{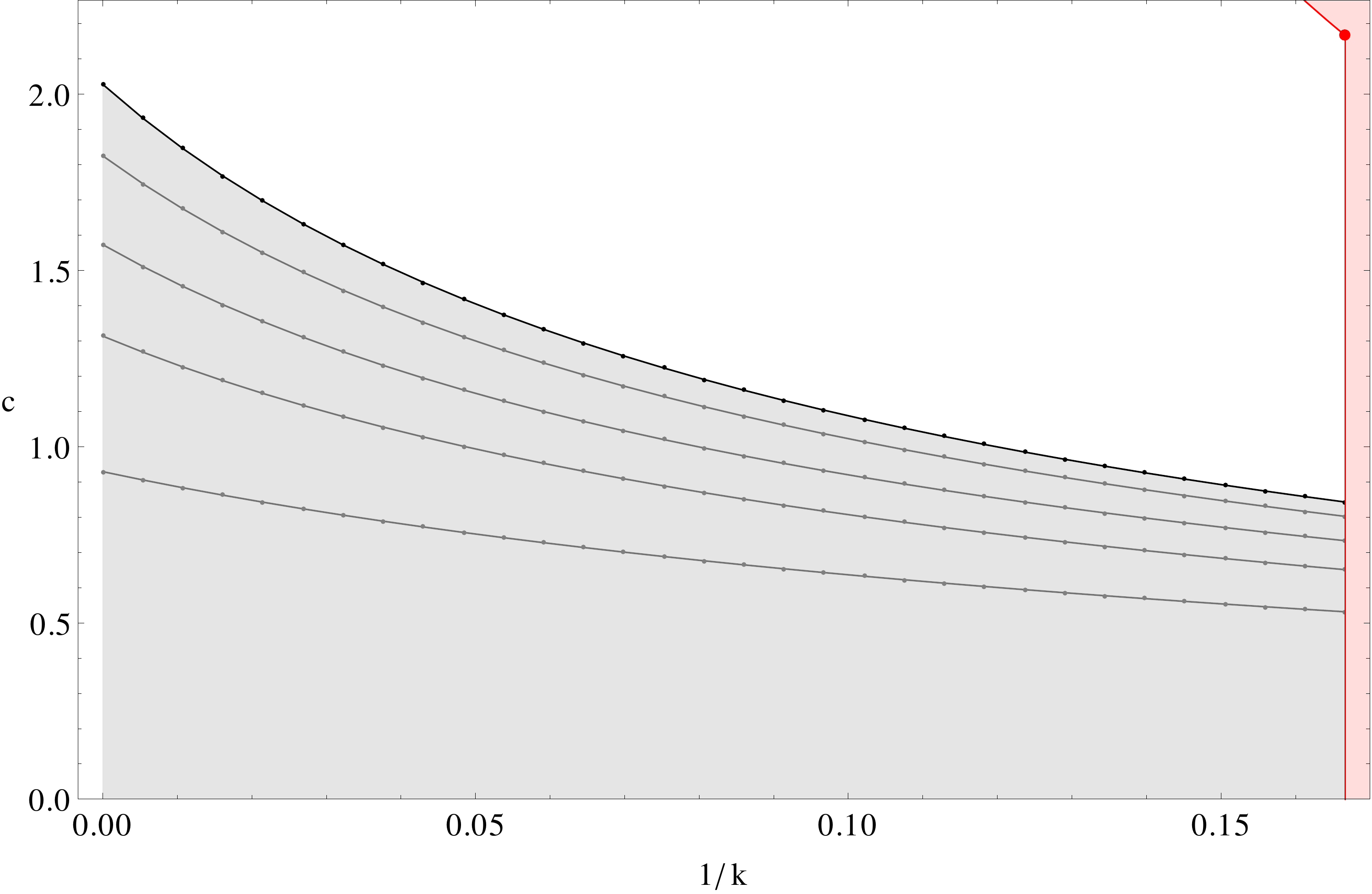}
              \caption{Bound on the central charge $c$ of a theory with $\ef_6$ flavor symmetry as a function of the flavor central charge $k$, obtained from the $\hat{\BB}_1$ four-point function. The shaded red regions on the right are the analytic bounds given in \eqref{analytic_boundE6}, and the shaded gray region at the bottom is the numerically excluded region. The gray and black lines correspond to the numerical bounds, shown for $10$ to $26$ derivatives in steps of four, with the strongest bound (black line) corresponding to $26$ derivatives. The red dot at the intersection of the two analytic bound corresponds to the rank one $\ef_6$ theory \cite{Minahan:1996fg}.}
              \label{Fig:e6_cbound}
            \end{center}
\end{figure}

We have obtained numerical lower bounds on $c$ as a function of $k$ following the same approach as in the $\suf(2)$ case. Here we considered a maximum cutoff of $\Lambda=26$. The lower bound is displayed in Fig.~\ref{Fig:e6_cbound} as a function of the (inverse) flavor current central charge. The regions shaded in red are again the ones ruled out by the analytic bounds \eqref{analytic_boundE6}. We see that independent of $k$, any $\NN = 2$ SCFT with at least $\ef_6$ flavor symmetry has
%%%%%%
\begin{equation}
c \geqslant 0.83~.
\end{equation}
%%%%%%

\begin{figure}[b!]
             \begin{center}
              \includegraphics[scale=0.45]{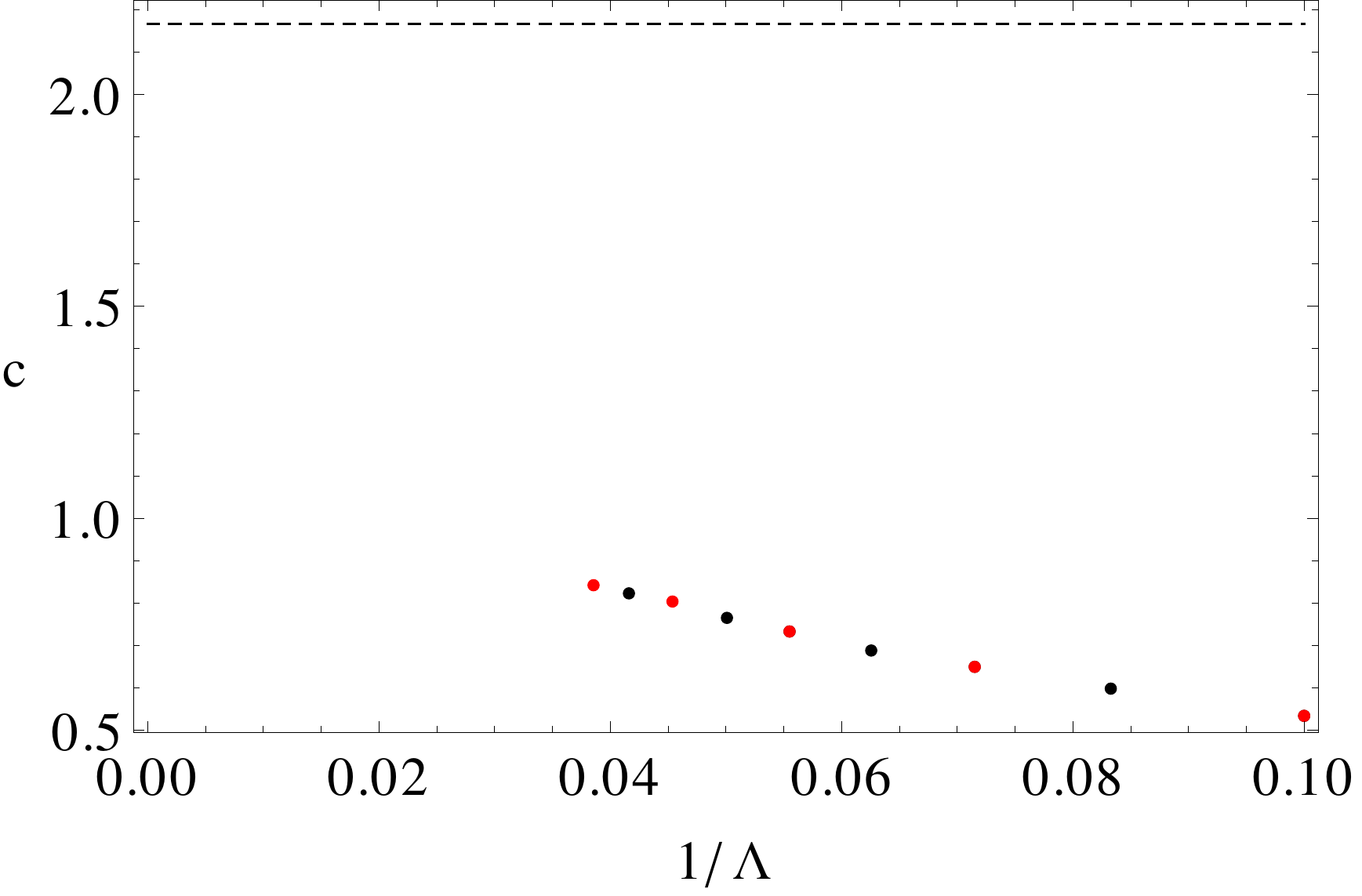}
              \vspace{-.05in}
              \caption{\label{Fig:E6_cboundkeq6}Lower bounds on $c$ for a theory with $\ef_6$ flavor symmetry and $k=6$ as a function the inverse cutoff $\Lambda$. The red dots correspond to derivatives $10,14,\ldots,26$, while the black dots show the remaining even values of $\Lambda$ ranging from $10$ to $24$. The dashed line at $c=\frac{13}{6}$ marks the central charge of the rank one $\ef_6$ theory.}
            \end{center}
\end{figure}

In contrast to the case of $\suf(2)$ global symmetry, for $\ef_6$ there is a theory living at the intersection of the two analytic bounds. This is the rank one $\ef_6$ theory of Minahan and Nemeschansky. One may wonder whether there is another theory with $k=6$ but with a lower value of $c$. In Fig.~\ref{Fig:E6_cboundkeq6} we show the lower bound on $c$ for $k=6$ derived from the moment map four-point function as a function of $\Lambda$. Though the bounds still seem to be improving, it appears unlikely that the optimal bound will reach the value $c=\frac{13}{6}$ (the value of the rank one $\ef_6$ theory). Instead, our best estimate for the optimal value of the central charge bound is somewhere between $1.1$ and $1.2$. We are not aware of a theory with such a low central charge and (at least) $\ef_6$ flavor symmetry. It would be interesting to determine whether the solution to crossing symmetry being approximated here contains higher spin currents.

%%%%%%%%%%%%%%%%%%%%%%%%%%%%%%%%%%%%%%%%%%%%%%%%%%%%%%%%%%%%%%%%%%%%%%%%%%%%%%%%%%%%%%%%%%%%%%%%%%%%%%%%%%%%
\subsubsection{Dimension bounds in the singlet channel}
%%%%%%%%%%%%%%%%%%%%%%%%%%%%%%%%%%%%%%%%%%%%%%%%%%%%%%%%%%%%%%%%%%%%%%%%%%%%%%%%%%%%%%%%%%%%%%%%%%%%%%%%%%%%

\begin{figure}[t!]
             \begin{center}           
              \includegraphics[width=2.5in]{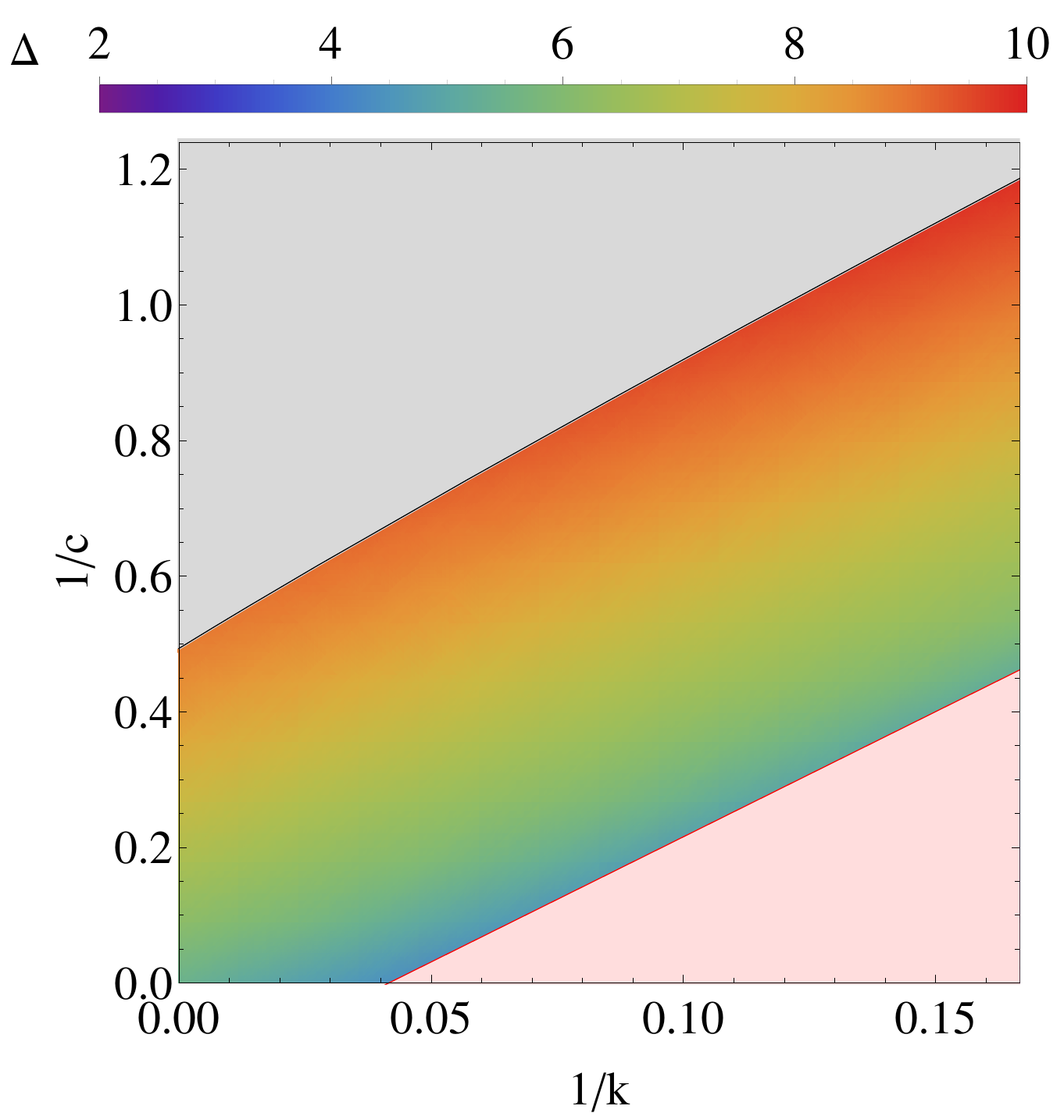}
              \includegraphics[width=3.2in]{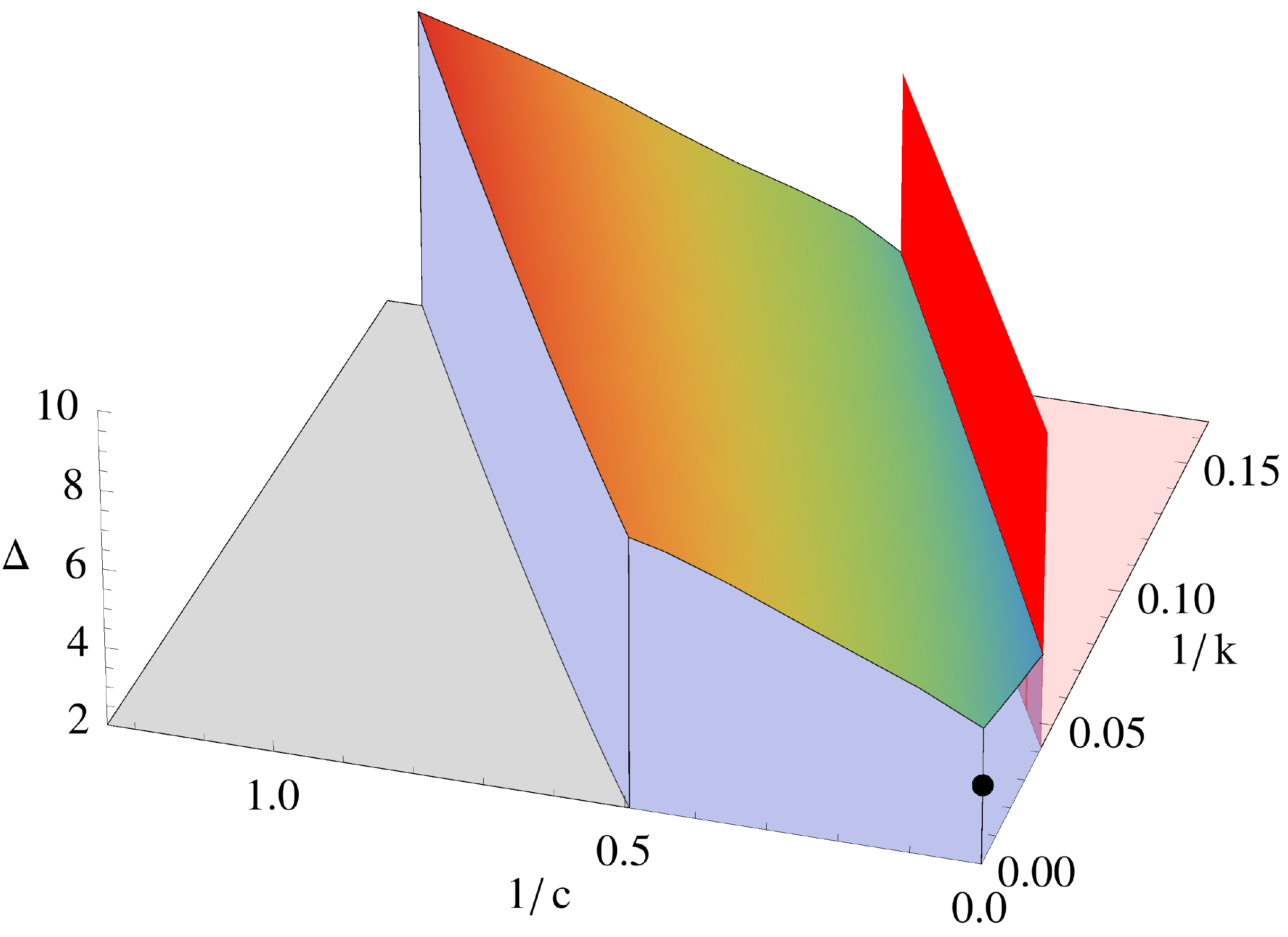}                                
              \caption{\label{e6ck}Upper bounds for the dimension of the first unprotected scalar in the singlet channel of a theory with $\ef_6$ flavor symmetry as a function of the inverse of the central charges. These bounds were generated with $\Lambda=16$. The vertical red wall corresponds to the second analytic unitarity bounds in \eqref{analytic_boundE6}, with the excluded region being the top right corner. The plot starts at $\frac{1}{k}=\frac{1}{6}$, where the first analytic unitarity bound is saturated.}
            \end{center}
\end{figure}

Bounds for the first unprotected scalar in the singlet channel as a function of the (inverse) central charges are shown in Fig.~\ref{e6ck}. The range of central charges allowed by unitarity is limited by \eqref{analytic_boundE6}. Our plot therefore starts at $k=6$, and the vertical red wall delimits the region allowed by the second bound in \eqref{analytic_boundE6}. The gray region in Fig.~\ref{e6ck} for low values of the central charge represents central charges excluded by the numerical bounds of the previous section. As both central charges go to infinity we expect the bound to go to the generalized free field theory value of $\Delta=4$, which we denoted with a black dot in the figure. This point is consistent with the numerical bounds, and naive extrapolation of the numerical results (not shown) suggests convergence towards $\Delta=4$.

%%%%%%%%%%%%%%%%%%%%%%%%%%%%%%%%%%%%%%%%%%%%%%%%%%%%%%%%%%%%%%%%%%%%%%%%%%%%%%%%%%%%%%%%%%%%%%%%%%%%%%%%%%%%
\subsubsection{Bounds for theories of interest}
\label{Sec:ef_6_F-theory}
%%%%%%%%%%%%%%%%%%%%%%%%%%%%%%%%%%%%%%%%%%%%%%%%%%%%%%%%%%%%%%%%%%%%%%%%%%%%%%%%%%%%%%%%%%%%%%%%%%%%%%%%%%%%

We now specialize to the values in the $(c,k)$ plane that correspond to the theories of $D3$ branes probing F-theory singularities with $\ef_6$ flavor symmetry. The central charges for these theories, shown in orange in Fig.~\ref{Fig:e6_landscape}, are given by \cite{Aharony:2007dj}
%%%%%%
\begin{equation}
\begin{split}
c & = \frac{3}{4} N^2 + \frac{3}{2} N -\frac{1}{12}~,\\
k & = 6 N~,
\end{split}
\end{equation}
%%%%%%
where $N$ is the rank of the theory. All theories with rank $N \geqslant 2$ have an extra $\suf(2)_L$ flavor symmetry, with $k_L = 3N^2 - 2N - 1$.

%\afterpage{ 
\begin{figure}[htpb]
        \begin{subfigure}[b]{0.5\textwidth}
                \includegraphics[width=\textwidth]{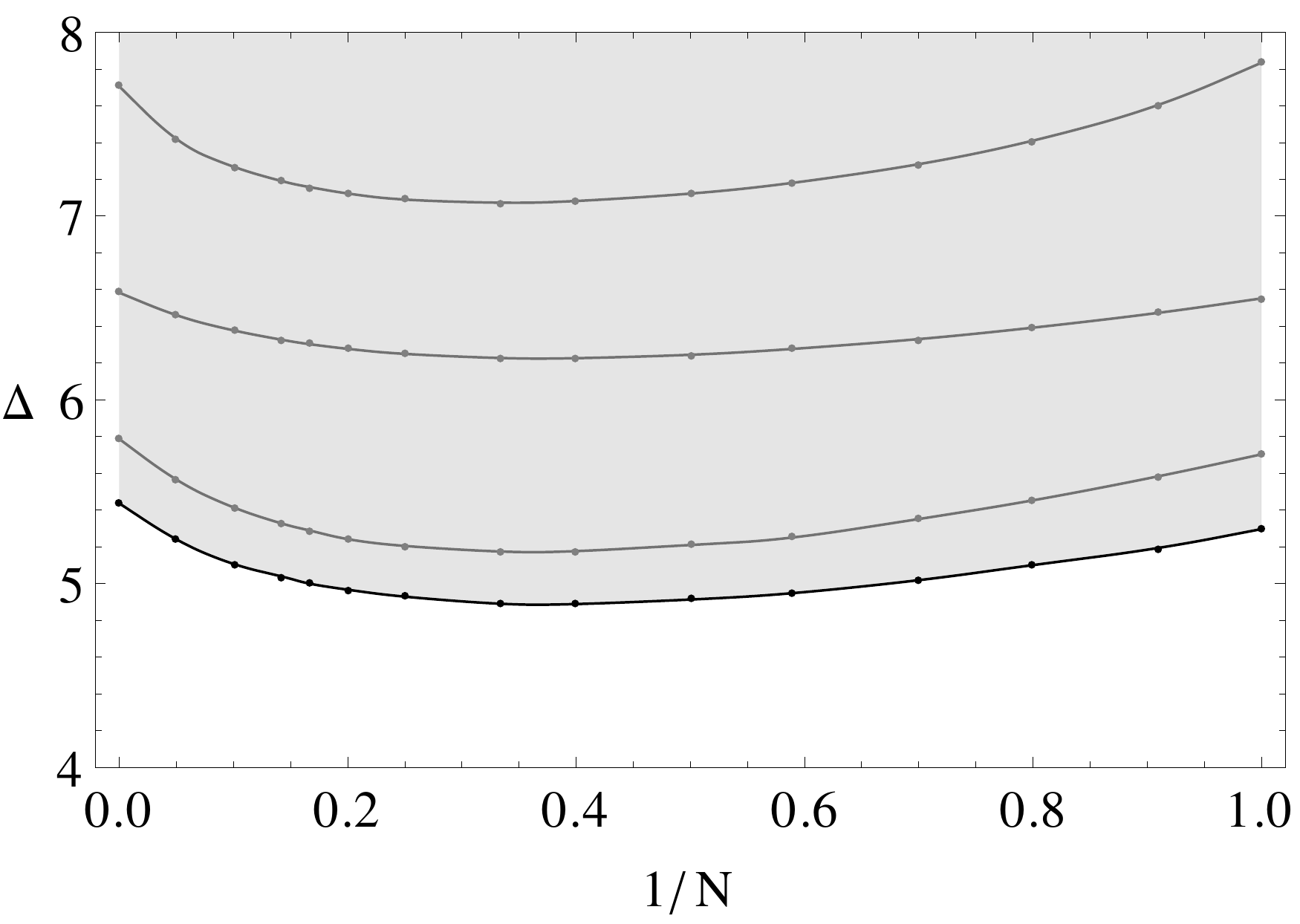}
                \caption{Singlet, scalar}
                \label{Fig:singet}
        \end{subfigure}
        ~
        \begin{subfigure}[b]{0.5\textwidth}
                \includegraphics[width=\textwidth]{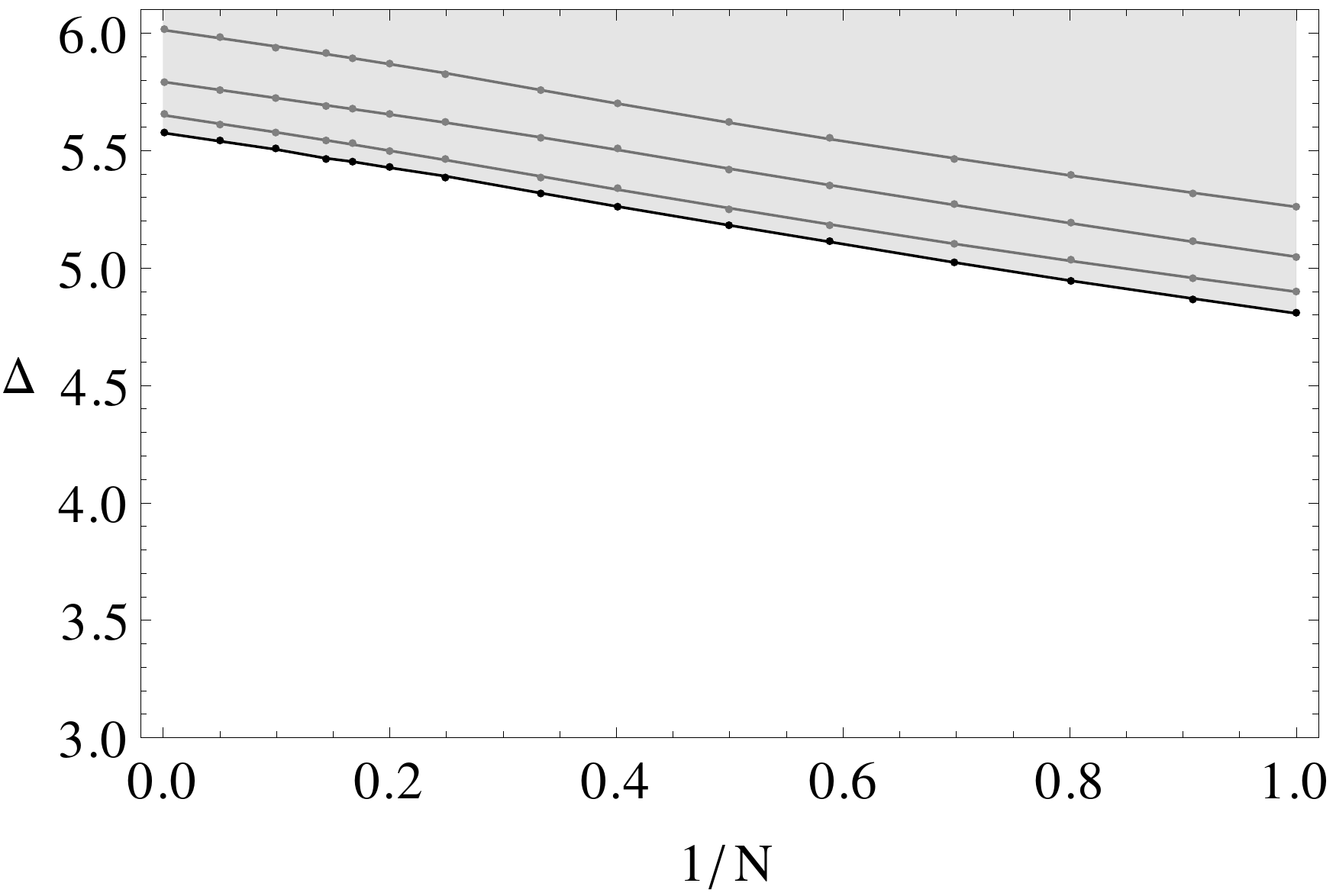}
                \caption{$\mathbf{78}$, spin $1$}
                \label{Fig:78}
        \end{subfigure}
        \\
        \begin{subfigure}[b]{0.5\textwidth}
                \includegraphics[width=\textwidth]{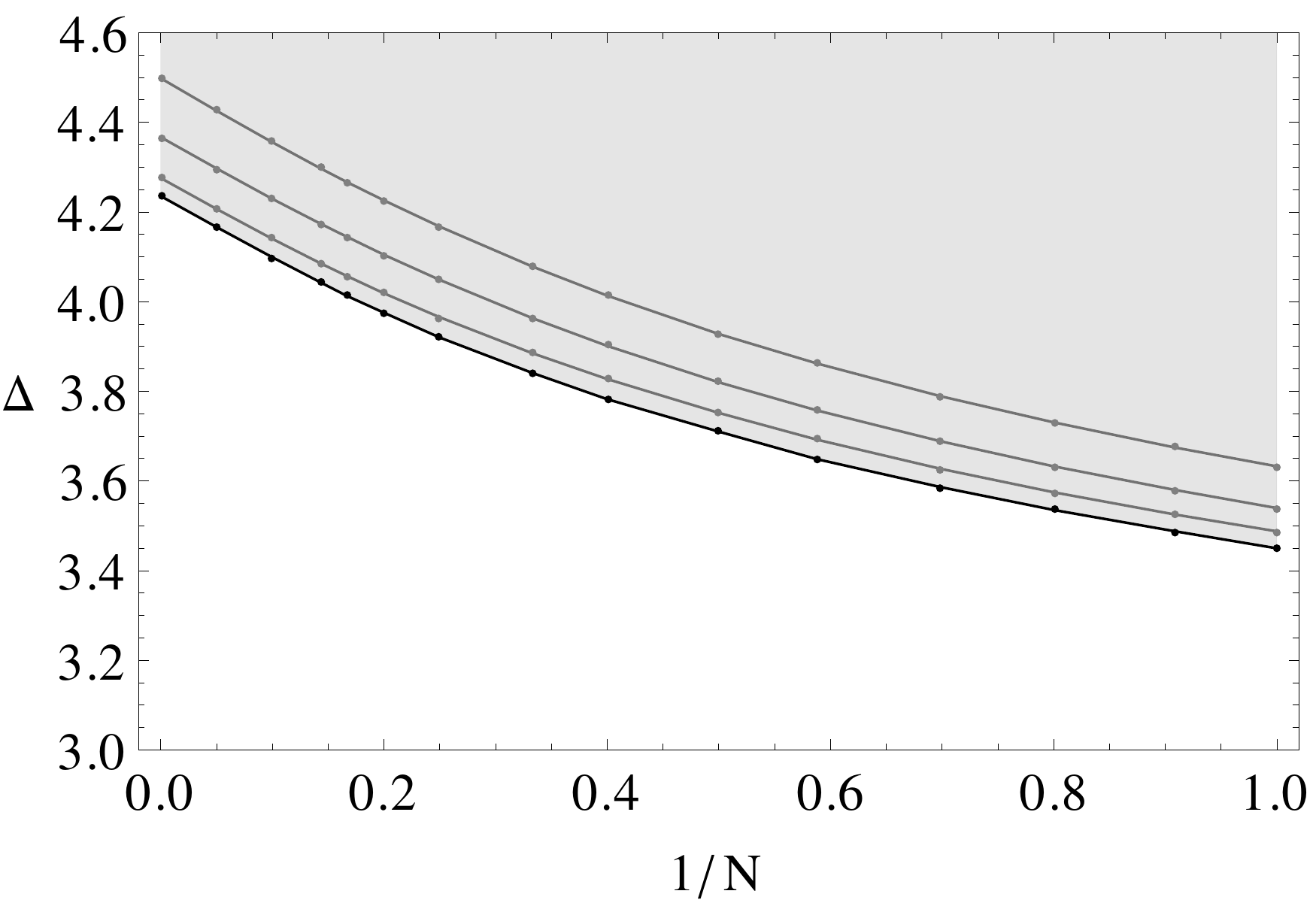}
                \caption{$\mathbf{650}$, scalar}
                \label{Fig:650}
        \end{subfigure}
        ~
        \begin{subfigure}[b]{0.5\textwidth}
                \includegraphics[width=\textwidth]{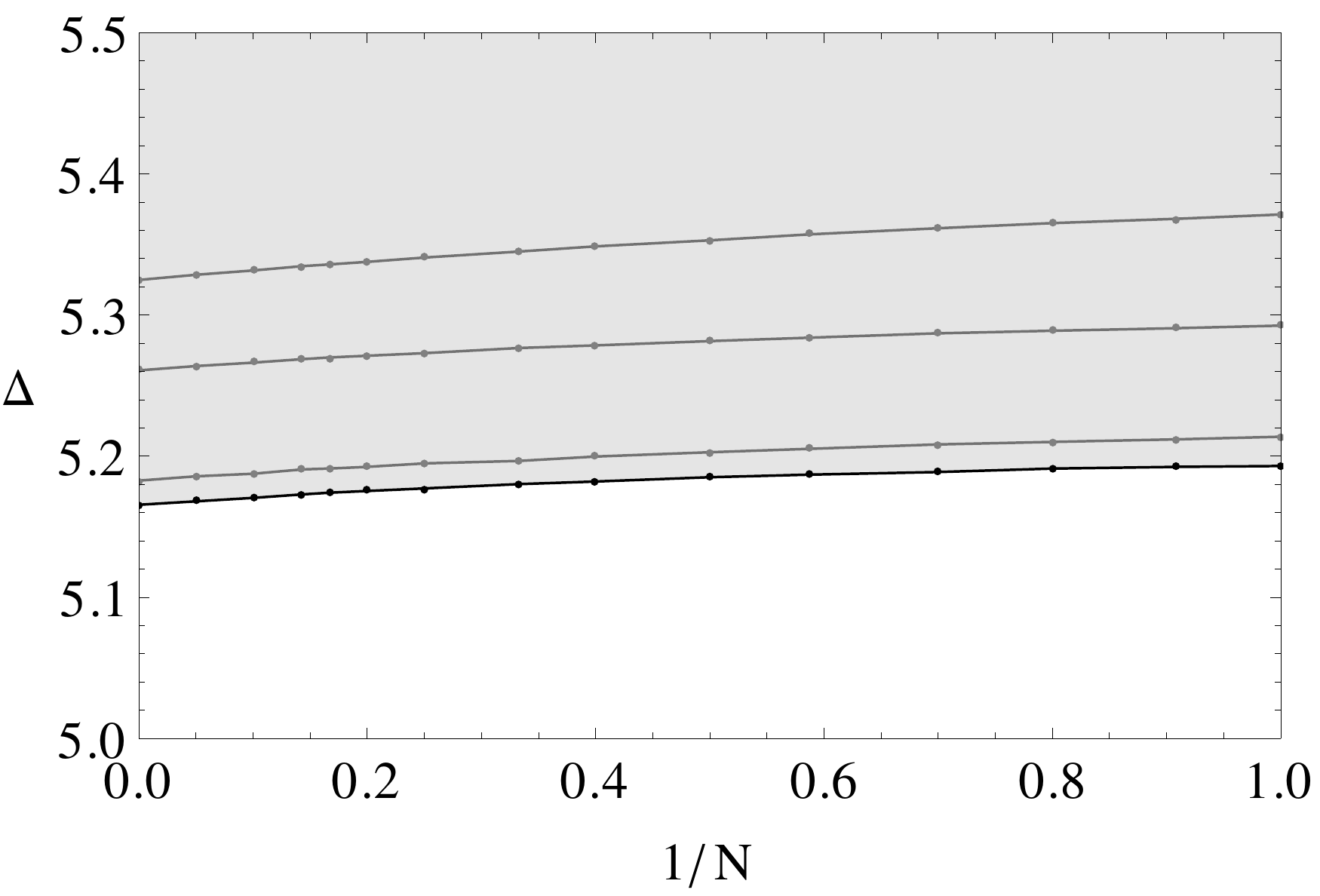}
                \caption{$\mathbf{2925}$, spin $1$}
                \label{Fig:2925}
        \end{subfigure}
        \\
        \begin{subfigure}[b]{0.5\textwidth}
                \includegraphics[width=\textwidth]{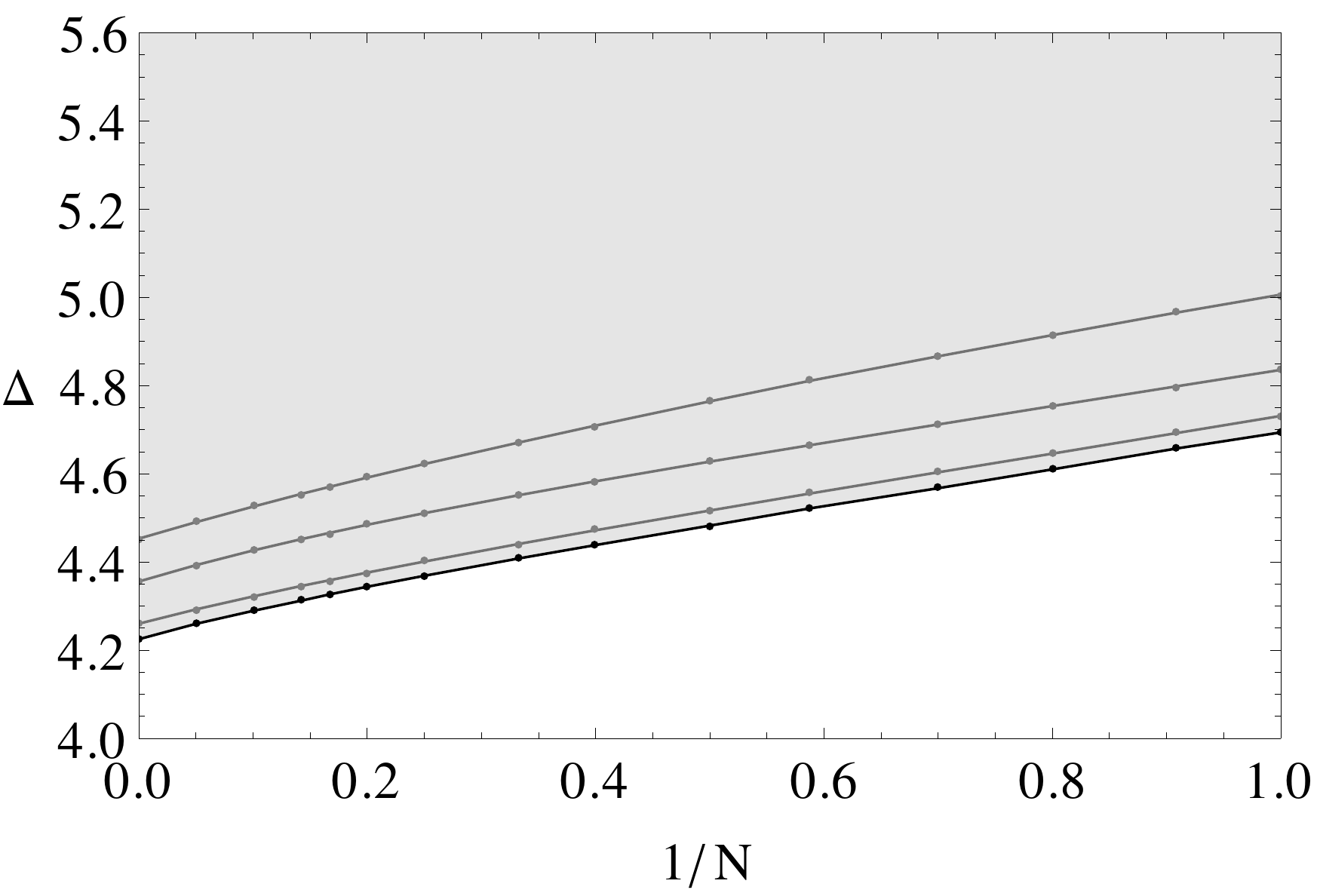}
                \caption{$\mathbf{2430}$, scalar}
                \label{Fig:2430}
        \end{subfigure}
        \caption{Bound for the dimension of the first unprotected spin zero multiplet in the singlet, $\mathbf{78}$, $\mathbf{650}$, $\mathbf{2925}$ and $\mathbf{2430}$ channels for the theories with flavor symmetry $\ef_6$ arising from F-theory singularities, as a function of the inverse of the rank of the theories. The number of derivatives is increased from $10$ to $16$ in steps of two, with the strongest bound given by the black line.}
        \label{Fig:E6Fthy}
\end{figure}
%\clearpage}

We derived upper bounds for the dimensions of the first unprotected operators of lowest spin in each of the flavor symmetry channels appearing in the tensor product of two adjoints. For the case of symmetric representations (singlet, $\mathbf{650}$, $\mathbf{2430}$) we therefore obtain a bound for spin zero operators, and for antisymmetric representations ($\mathbf{78}$, $\mathbf{2925}$) we bound spin one operators. These bounds are displayed in Fig.~\ref{Fig:E6Fthy}. They are still far from optimal, but serve to give us a feeling for the general shape of things. It would be interesting to improve our numerical search power to the point where these bounds would converge. 

\begin{figure}[h!t]
            \begin{center}               
            \includegraphics[scale=0.46]{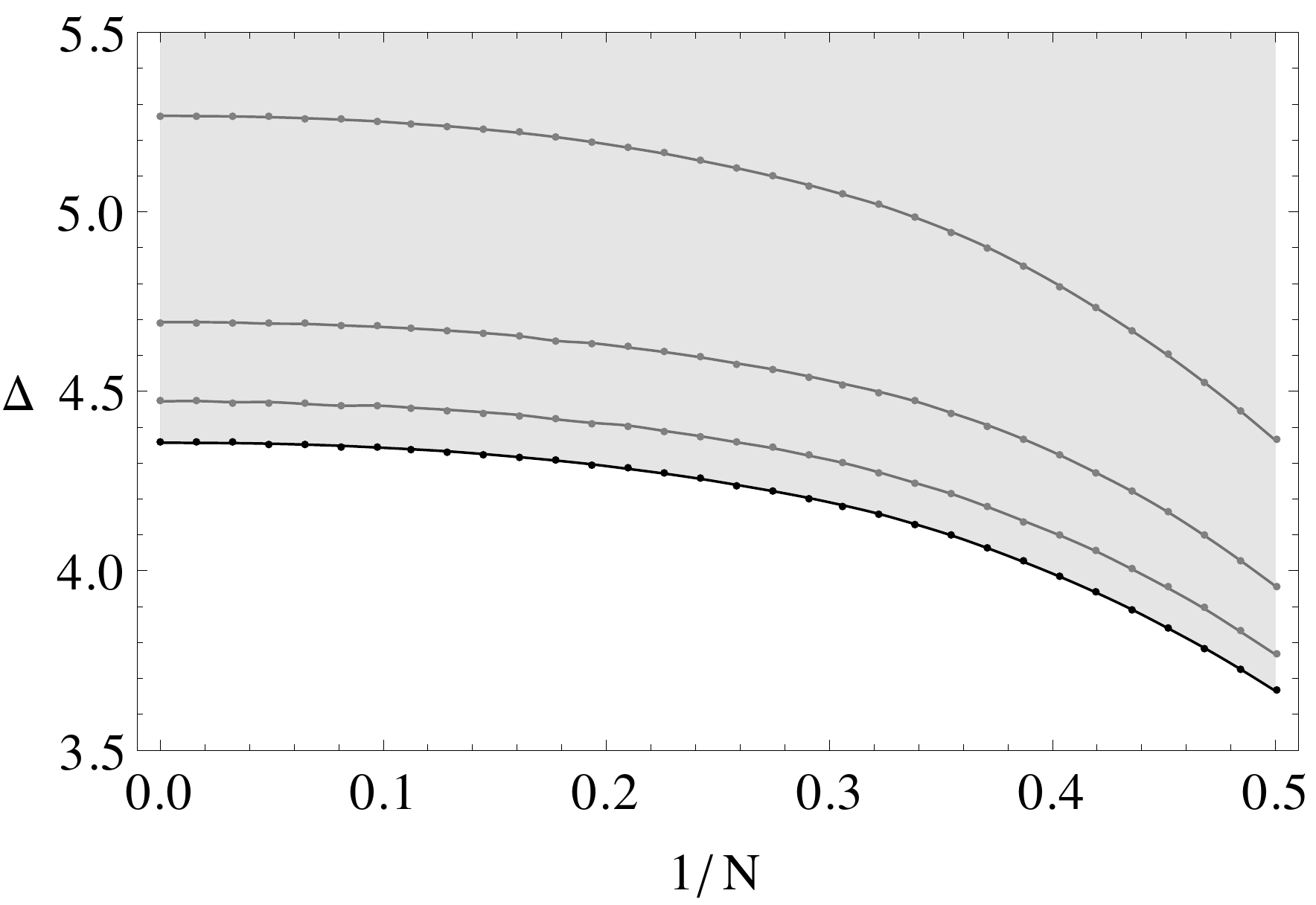}
            \end{center}
            \caption{Upper bounds for the dimension of the first unprotected scalar in the singlet channel of the $\suf(2)_L$ moment map correlator for the $\ef_6$ theories, as a function of the inverse of the rank of the theories. The cutoff is increased from $10$ to $22$ in steps of four, with the strongest bound given by the black line.}
            \label{Fig:su2_L}
\end{figure}

We should compare the singlet bounds in Fig.~\ref{Fig:singet} for rank $N \geqslant 2$ to the bounds obtained from the $\suf(2)_L$ flavor symmetry of those theories. In principle the same unprotected operators may contribute to the four-point functions of both sets of moment maps, so if the bounds recovered from both correlators are related to these rank $N$ theories then they should agree to some extent. These bounds are shown in Fig.~\ref{Fig:su2_L}. The two sets of bounds appear to have nothing in common. It is hard to say whether this is a feature of the space of solutions to crossing symmetry or a consequence of inadequate numerical power.

%%%%%%%%%%%%%%%%%%%%%%%%%%%%%%%%%%%%%%%%%%%%%%%%%%%%%%%%%%%%%%%%%%%%%%%%%%%%%%%%%%%%%%%%%%%%%%%%%%%%%%%%%%%%
\subsubsection{The rank one theory}
%%%%%%%%%%%%%%%%%%%%%%%%%%%%%%%%%%%%%%%%%%%%%%%%%%%%%%%%%%%%%%%%%%%%%%%%%%%%%%%%%%%%%%%%%%%%%%%%%%%%%%%%%%%%

We performed a higher precision analysis at the point $k = 6$ and $c = \frac{13}{6}$, which are the central charges of the $\ef_6$ Minahan-Nemeschansky theory. It is plausible that this theory is the unique theory with these central charges and $\ef_6$ flavor symmetry. What's more, because of the location of these central charges in a corner of the allowed region of the $(c,k)$ plane, there can be no pollution at this point by solutions of crossing symmetry that are linear combinations of other irreducible solutions. This gives us some room to be optimistic that the numerical bounds at this point will converge to physical values that correspond to the scaling dimensions of operators in this theory.

\begin{figure}[h!t]
             \begin{center}           
              \includegraphics[scale=0.5]{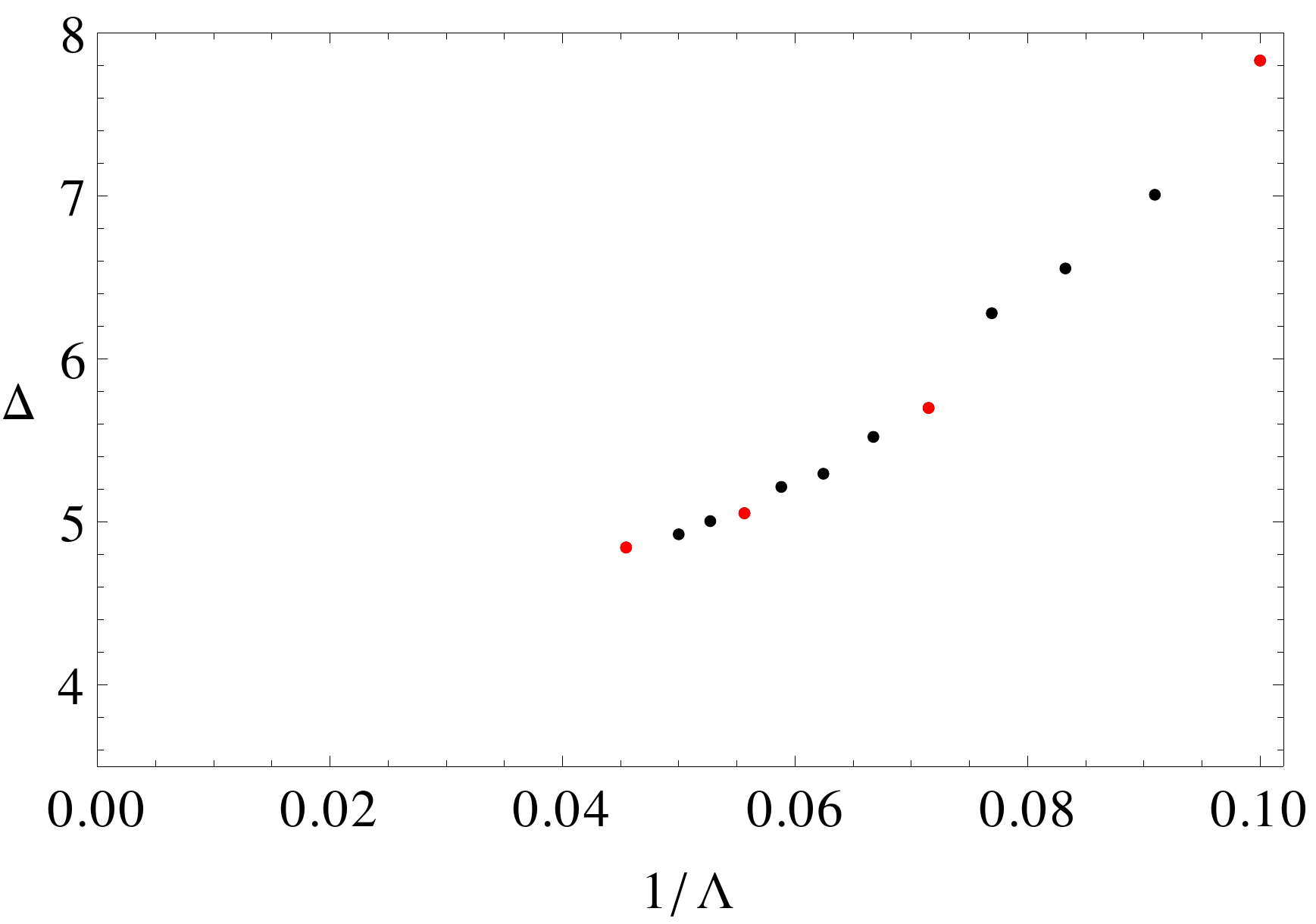}
            \end{center}
            \caption{Upper bounds for the first unprotected scalar singlet in the rank one $\ef_6$ theory as a function of the inverse cutoff. The points where $\Lambda - 2$ is a multiple of four are shown in red. For the best bound shown, the dimension of the search space in the associated semidefinite program was $366$.}
            \label{Fig:E6rank1}
\end{figure}

As a first example we may consider again the bound on the first unprotected singlet scalar. We have plotted this bound as a function of the cutoff in Fig.~\ref{Fig:E6rank1}. Naive extrapolation suggests an optimal value in the neighborhood of $\Delta \simeq 4.4$ for the first scalar singlet.

\begin{figure}[b!]
            \begin{center}           
               \includegraphics[scale=0.3]{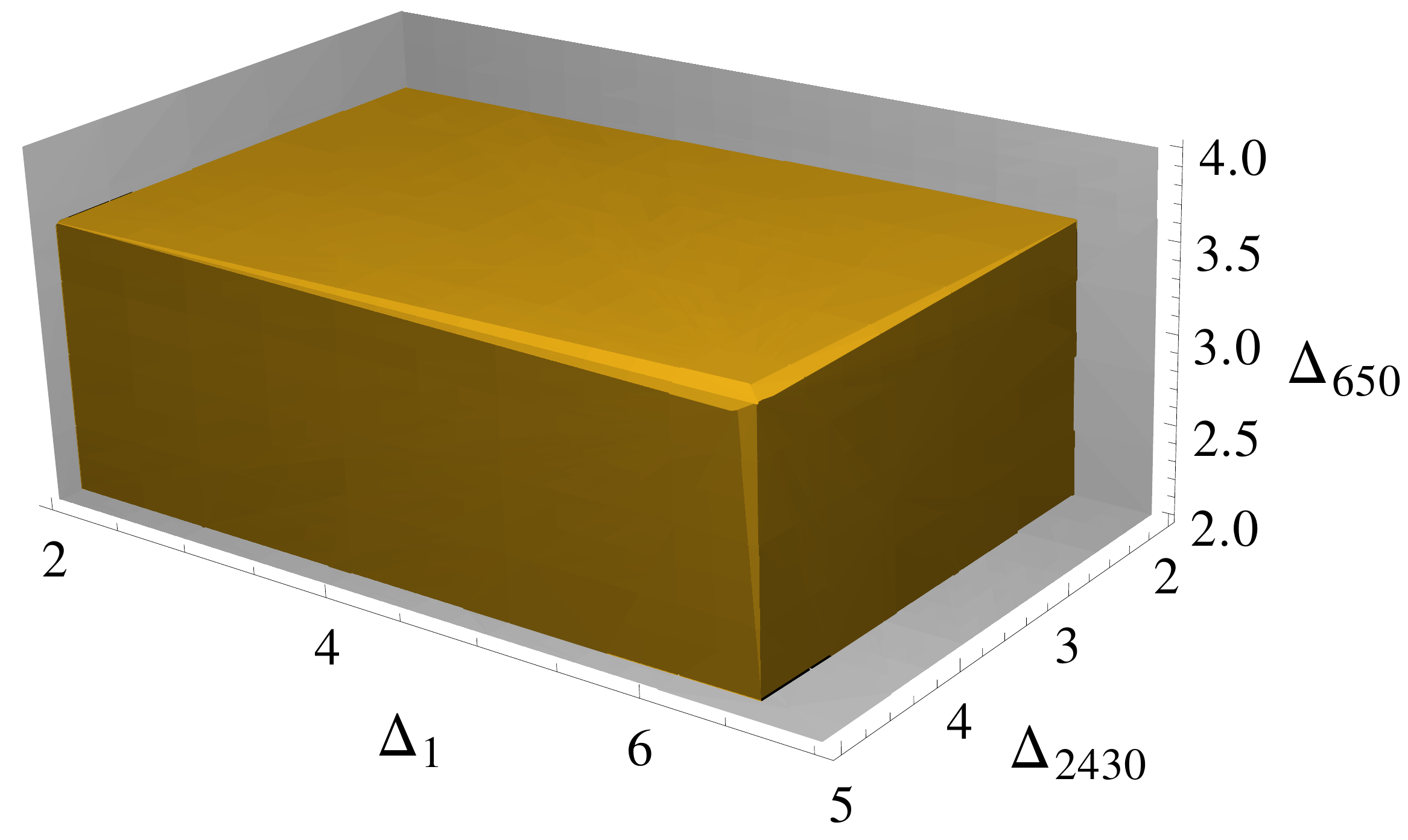}
            \end{center}
            \caption{\label{Fig:E6cube}Three-dimensional exclusion plot in the octant spanned by the scaling dimensions of the first unprotected scalar in the $\Rf={\bf 1},{\bf 650},{\bf 2430}$ representations of $\ef_6$ with $k=6$ and $c = \frac{13}{6}$. The cutoff used while generating these bounds was $\Lambda=12$.}
\end{figure}

We can also explore \emph{simultaneous bounds} for various channels by searching for functionals with $\Delta^{\star}_{\Rf_i}$ greater than the unitarity bound for several choices of flavor symmetry channel $\Rf_i$. We performed such an analysis for these central charges to derive simultaneous bounds for the first scalars in the ${\bf 1}$, ${\bf 650}$, and ${\bf 2430}$ channels. The numerics were performed with $\Lambda=12$, and the results are shown in Fig.~\ref{Fig:E6cube} in the from of an exclusion plot in the three-dimensional space spanned by the scaling dimensions $(\Delta_{\mathbf{1}},\Delta_{\mathbf{650}},\Delta_{\mathbf{2430}})$ of the first operator in those channels. The usual superconformal unitarity bounds constrain us to be in the octant where all these three dimensions are greater than two, but within this octant we have numerically carved out a further excluded region where one or more of the three operator dimensions is too high to satisfy the crossing symmetry equations.

%%%
%%%

%%%%%%%%%%%%%%%%%%%%%%%%%%%%%%%%%%%%%%%%%%%%%%%%%%%%%%%%%%%%%%%%%%%%%%%%%%%%%%%%%%%%%%%%%%%%%%%%%%%%%%%%%%%%
\subsubsection{Bounds for defect SCFTs}
%%%%%%%%%%%%%%%%%%%%%%%%%%%%%%%%%%%%%%%%%%%%%%%%%%%%%%%%%%%%%%%%%%%%%%%%%%%%%%%%%%%%%%%%%%%%%%%%%%%%%%%%%%%%

\begin{figure}[h!t]
             \begin{center}    
              \includegraphics[scale=0.5]{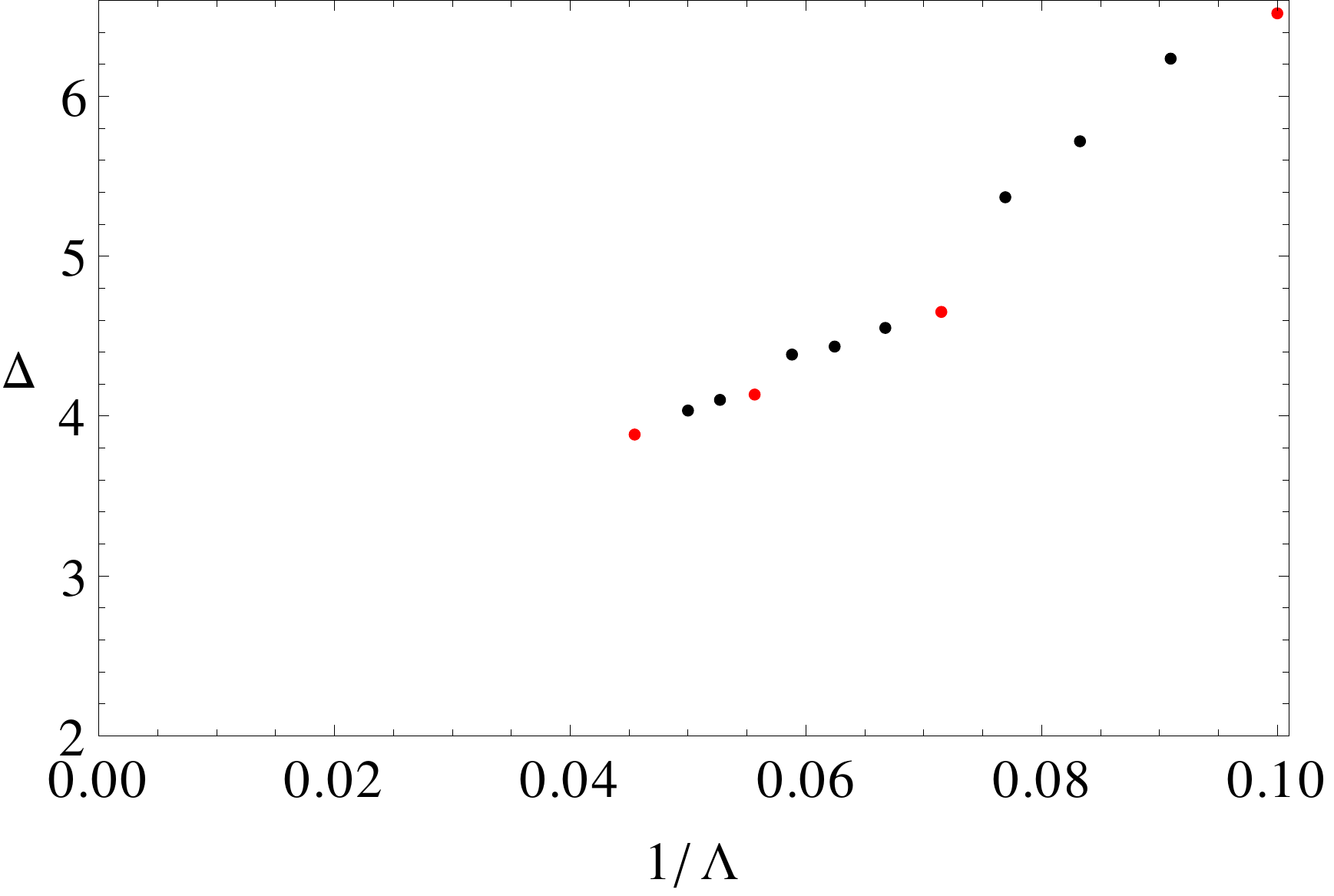}
            \end{center}
            \vspace{-.2in}
            \caption{\label{Fig:E6critk}Bound on the first unprotected scalar in the singlet channel for a theory with $k=24$ and infinite central charge as a function of the cutoff. Red dots correspond to cutoff values $\Lambda=10,14,\ldots,22$, while black dots show the remaining cutoff values ranging from $\Lambda=10$ to $\Lambda=20$.}
\end{figure}

We can again consider the limit where we send $c \to \infty$ with $k$ at the analytic bound, which gives $k = 24$ in this case. In this limit we expect to recover information about the theory living on the codimension two defect corresponding to the trivial embedding in the six-dimensional $(2,0)$ theory of type $\ef_6$. A nontrivial bound for the first unprotected scalar in the singlet channel is given in Fig.~\ref{Fig:E6critk} as a function of the cutoff. Once again we observe some quasi-periodic behavior where the bounds have a sharper jump at every fourth step in the cutoff. By naive extrapolation of the bound we arrive at a rough estimate that the optimal upper bound should be between $\Delta=3$ and $\Delta=3.2$.

%!TEX root = ../draft_maxi_Neq2.tex

\section{Results for the \texorpdfstring{$\EE_{r}$}{E(r)} four-point function}
\label{sec:eps_results}

We now turn to the numerical results obtained for the four-point function of the Coulomb branch operators $\EE_{r_0}$. Unlike in the previous section we can  vary the dimension of these operators, which we recall is given in terms of their $U(1)_r$ charge $r_0$ by $\Delta_{\mathrm{ext}} = r_0$. Unitarity requires $r_0 \geqslant 1$. We will consider four-point functions where all operators have equal dimension. A second parameter is again the $c$ central charge which appears in front of the conformal block of the stress tensor multiplet. We will therefore be able to obtain bounds as a function of $r_0$ and $c$.

%%%%%%%%%%%%%%%%%%%%%%%%%%%%%%%%%%%%%%%%%%%%%%%%%%%%%%%%%%%%%%%%%%%%%%%%
\subsection{Central charge bounds}
\label{subsec:eps_c_bounds}
%%%%%%%%%%%%%%%%%%%%%%%%%%%%%%%%%%%%%%%%%%%%%%%%%%%%%%%%%%%%%%%%%%%%%%%%

Our first bound is again a lower bound for the $c$ central charge, now as a function of the dimension $r_0$ of the Coulomb branch operators. Assuming the moduli space/chiral ring correspondence for the Coulomb branch, the Shapere-Tachikawa relation provides an analytic lower bound for $c$. More precisely this bound is obtained combining the ST sum rule \eqref{eq:ST_sum_rule} and the Hofman-Maldacena upper bound \eqref{eq:Neq2HM_bounds} on $\frac{a}{c}$. If the Coulomb branch, which is assumed to be freely generated, has dimension $N$ with generators of dimension $\{r_1,\ldots,r_N\}$, then the following bound holds,
%%%%%%
\begin{equation}
c \geqslant \frac{1}{6} \sum_{i=1}^N(2 r_i-1)~.
\end{equation}
%%%%%%
The fact that only the dimensions of \emph{generators} of the Coulomb branch chiral ring appear in this expression is important. For example, Coulomb branch operators of dimension $r_0 \geqslant 3 c + \frac{1}{2}$ are certainly allowed by this bound, they just cannot be generators. On the other hand, a theory that has any Coulomb branch at all must have $c \geqslant \frac{1}{6}$, since the dimension of a Coulomb branch generator cannot be smaller than one. Moreover, if $c=\frac{1}{6}$ then the Coulomb branch must have a single generator of dimension $r_0=1$, so will necessarily be the theory of a single free vector multiplet.

In setting up the bootstrap for this correlator, there is no straightforward way to insist that the Coulomb branch operators be generators (or that they not be generators, for that matter). Of course, any such operator with $r_0 < 2$ will necessarily a generator because it cannot be a product of two operators with dimensions above the unitarity bound. For $r_0 \geqslant 2$ if we \emph{assume} that we are dealing with a generator, then the following analytic bound will be obeyed:
%%%%%%
\begin{equation}\label{crbound}
c \geqslant \frac{1}{6} (2 r_0-1)~.
\end{equation}
%%%%%%
Notice that if we drop the generator assumption and consider four-point functions of operators that are not generators, then only the weaker bound $c \geqslant \frac{1}{6}$ applies for $r_0 \geqslant 2$. This bound is in fact saturated at any $r_0\in\Nb$ by the operators of the Coulomb branch chiral ring of the free vector multiplet.

%%%%%%%%%%%%
\begin{figure}[t!]
             \begin{center}           
              \includegraphics[scale=0.4]{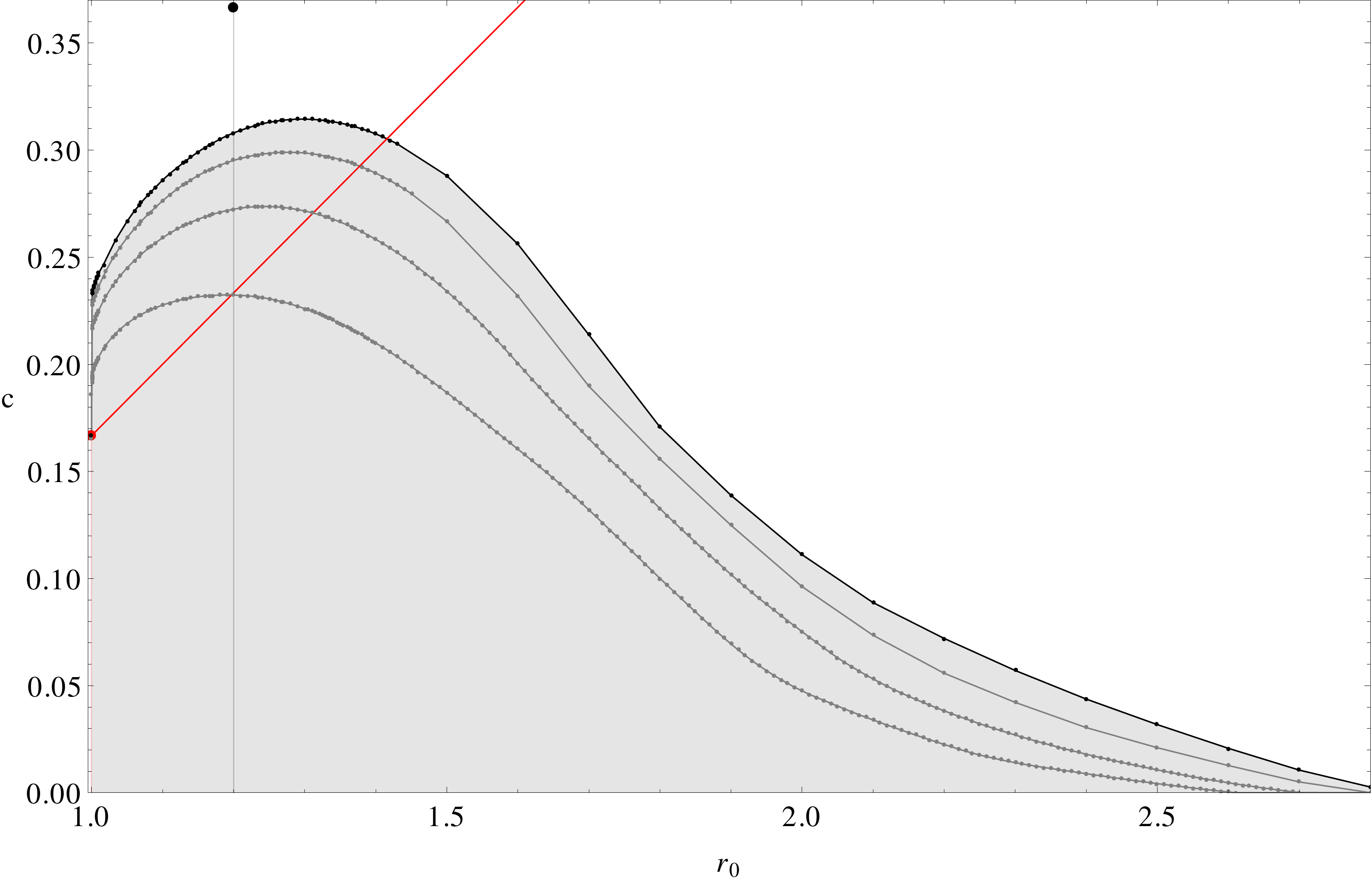}~~~~~~~~
              \caption{Lower bounds for the central charge $c$ of a theory with Coulomb branch operator $\EE_{r_0}$ as a function of its dimension $r_0$. The straight red line is the analytic bound for the case when $\EE_{r_0}$ is a Coulomb branch generator, given in \eqref{crbound}, with the excluded region lying to the right of the line. The shaded gray region is the numerically excluded region, and the gray and black lines correspond to bounds obtained with $\Lambda=10,14,18,22$, with larger $\Lambda$ giving the stronger bounds. The red dot denotes the free vector multiplet, and the black dot the rank one $H_0$ theory.}
              \label{Fig:E-cbound}
            \end{center}
\end{figure}
%%%%%%%%%%%%

In Fig.~\ref{Fig:E-cbound} we show the results of a numerical $c$-minimization procedure as a function of $r_0$. The analytic bound \eqref{crbound} is superimposed in red. For large values of $r_0$ the analytic bound always dominates over the numerical one, but for $r_0 \lesssim 1.4$ the numerical bound is dominant. Nevertheless, we would like to stress that the analytic bound is contingent upon the validity of the Coulomb branch version of the moduli space/chiral ring correspondence. If there are exceptions to this rule, then the analytic bounds will not hold, whereas the numerical bounds will still necessarily hold true.

As $r_0$ approaches one the bound drops sharply towards $c=\frac{1}{6}$, the central charge of the free vector. Though it is not clear from the figure, $c=\frac{1}{6}$ is not ruled out for $r_0=1$, where convergence with $\Lambda$ is very fast.\footnote{A similar phenomenon was observed in the context of central charge minimization in $\NN=1$ SCFTs \cite{Poland:2011ey}.} Away from $r_0=1$, convergence as a function of $\Lambda$ is slower, and the bounds presented here are still quite suboptimal. One interesting question is whether the bound at $r_0=\frac{6}{5}$ might converge to the $c\geqslant \frac{11}{30}$, with the rank one $H_0$ theory lying at the boundary. Using our methods, this would require a substantial increase in $\Lambda$. Similarly, at $r_0=2$ it seems possible that the bound may converge to the free vector value $c= \frac{1}{6}$.

%%%%%%%%%%%%%%%%%%%%%%%%%%%%%%%%%%%%%%%%%%%%%%%%%%%%%%%%%%%%%%%%%%%%%%%%
\subsection{Dimension bounds for non-chiral channel}
\label{subsec:eps_singlet_bounds}
%%%%%%%%%%%%%%%%%%%%%%%%%%%%%%%%%%%%%%%%%%%%%%%%%%%%%%%%%%%%%%%%%%%%%%%%

In the allowed region of the $(r_0,c)$ plane we can bound the dimension of the first unprotected, $R$-symmetry singlet, scalar operator appearing in the $\EE_{r_0} \bar{ \EE}_{-r_0}$ OPE. Unitarity requires that such an operator have $\Delta \geqslant 2$. When $\Delta = 2$ the long multiplet sits at the unitarity bound and decomposes into the stress tensor multiplet along with other short multiplets whose OPE coefficients vanish. In order to study local theories we should therefore add the superconformal block with $\Delta=2$ to the problem by hand and then impose a gap so that the subsequent scalar operator has dimension strictly greater than two.

This situation presents two natural options. First, we can leave the coefficient of the stress tensor block unfixed and simply require that the functional be positive when acting upon it. This approach leads to upper bounds for $\Delta_0^\star$ that are valid for any value of $c$. Alternatively, we can fix the coefficient of the stress tensor block by hand and in doing so fix the value of the central charge. We can then extract bounds on $\Delta_0^\star$ as a function of $c$.

Let us make a brief comment about the free vector theory. When $r_0=1$ we know that there exists a solution with $c = 1/6$, and in this solution there is no other scalar singlet block after the stress tensor multiplet at $\Delta=2$. Thus at this point in the $(r_0,c)$ plane our numerical procedure will never produce a nontrivial bound for the next operator, since any such bound would rule out the free field solution.\footnote{If for $r_0=1$ we do not to include the stress tensor block by hand, then the resulting bound on the first operator dimension would come be very close to two.} To avoid this singular point in our searches we have studied regions of the $(r_0,c)$ plane with $r_0 \geqslant 1.001$.

%%%%%%%%%%%%%%%%%%%%%%%%%%%%%%%%%%%%%%%%%%%%%%%%
\subsubsection*{Arbitrary central charge}
%%%%%%%%%%%%%%%%%%%%%%%%%%%%%%%%%%%%%%%%%%%%%%%%

\begin{figure}[t!]
             \begin{center}           
              \includegraphics[scale=0.43]{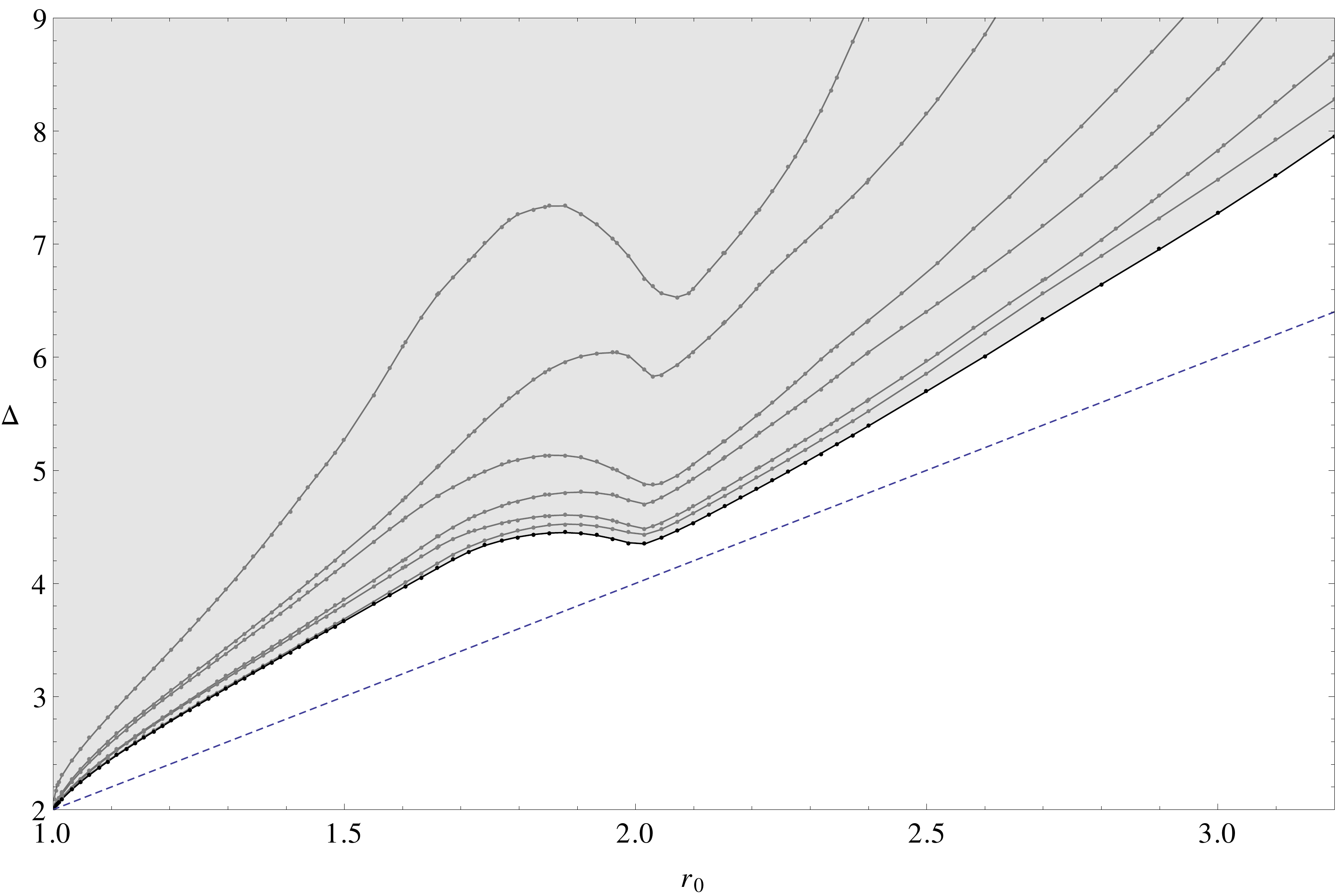}
              \caption{Bound on the first scalar in the $\EE_{r_0}\times\bar{\EE}_{-r_0}$ OPE as a function of $r_0$ for arbitrary central charge. The lines correspond to $\Lambda= 8,10,\ldots,20$, with the strongest bound being the black line. The excluded region is shaded. The dashed line corresponds to generalized free field theory solution, for which $\Delta=2r_0$.}
              \label{Fig:E-dimbound_arbc}
            \end{center}
\end{figure}

The results of the first strategy are displayed in Fig.~\ref{Fig:E-dimbound_arbc}. We find an upper bound on the dimension of the first scalar singlet as a function of $r_0 \geqslant 1.001$, with the bound at a given $r_0$ being valid for arbitrary values of $c$. Note that because there is no restriction on the value of $c$ in this approach, there may be approximate solutions to crossing symmetry that influence this plot for which the value of $c$ has been ruled out by \eqref{crbound}. Indeed, we will find below that excluded central charge values are responsible for the local maximum at $r_0$ slightly less than two. For higher values of $r_0$, it seems plausible that the bounds will converge to the generalized free field theory solution indicated in the figure with a dashed line. The results for fixed values of the central charge will shed light on the features of this bound, so we postpone further discussion of its shape to the next subsection.

%%%%%%%%%%%%
%%%%%%%%%%%%

The analogous chiral/anti-chiral OPE for $\NN=1$ SCFTs was considered in \cite{Poland:2011ey}. The exclusion curve obtained in that work for the dimension of the first unprotected scalar operator exhibited an interesting ``kink''. However, in that case a gap larger than two was being imposed, so any theory associated to the kink could not come from an $\NN = 2$ theory with a stress tensor multiplet.

%%%%%%%%%%%%%%%%%%%%%%%%%%%%%%%%%%%%%%%%%%%%%%%%
\subsubsection*{Fixed central charge}
%%%%%%%%%%%%%%%%%%%%%%%%%%%%%%%%%%%%%%%%%%%%%%%%

\begin{figure}[ht!]
             \begin{center}           
              \includegraphics[scale=0.43]{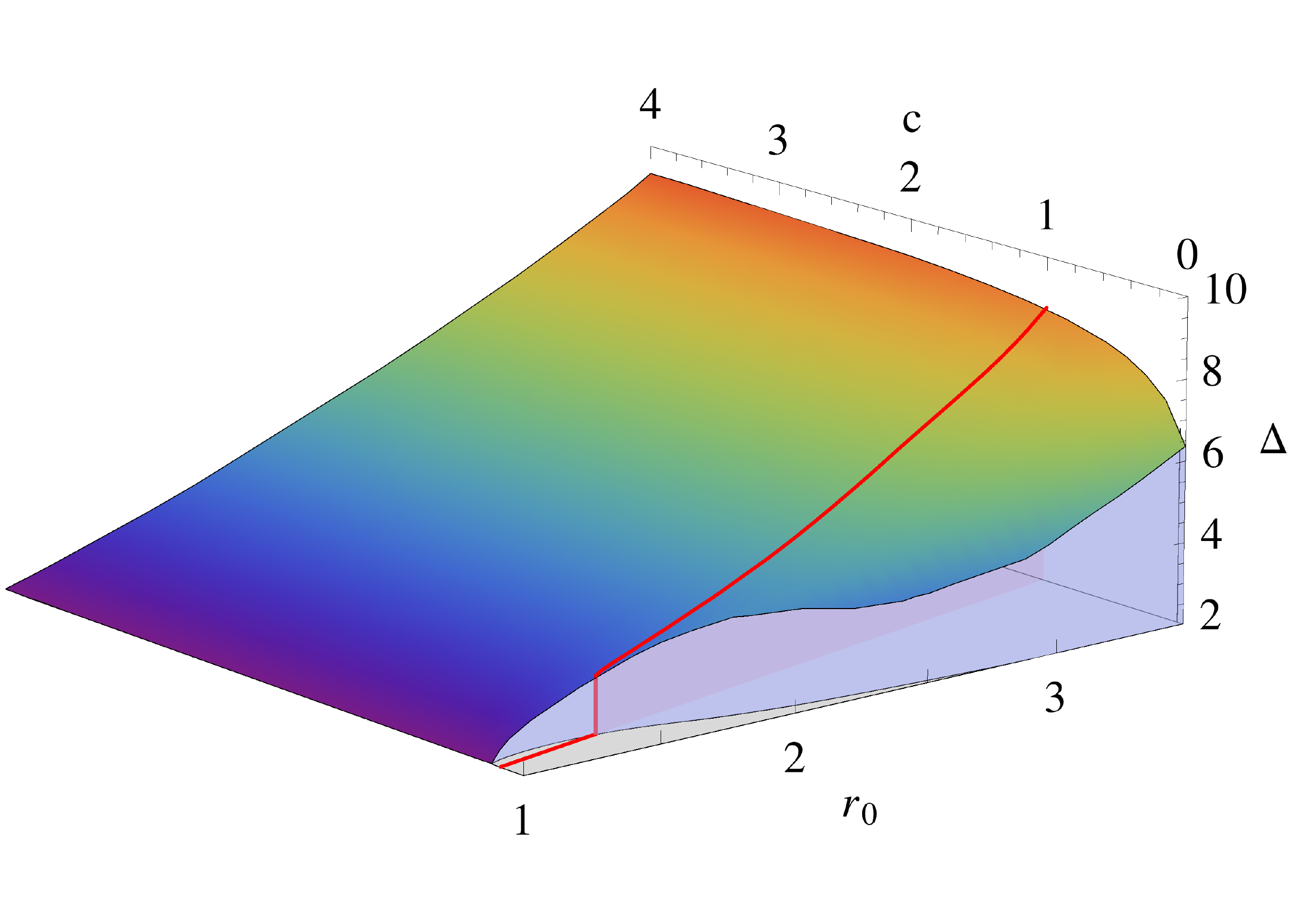}
              \caption{Bound on the first scalar in the $\EE_{r_0} \times \bar{\EE}_{-r_0}$ OPE as a function of the central charge $c$ and dimension of the external operators $r_0$. These bounds are for $\Lambda = 18$, and are obtained by imposing a gap and minimizing/maximizing the central charge value after imposing a gap in the spectrum. The gray area in the figure is a copy of the excluded region from Fig.~\ref{Fig:E-cbound}. The red line corresponds to the bound \eqref{crbound}, and the excluded region, if $\EE_{r_0}$ is to be a generator, is the one with smaller central charge.}
              \label{Fig:E-crplane}
            \end{center}
\end{figure}

We turn next to operator dimension bounds for fixed central charge. The results for fixed $\Lambda$ take the form of a function $\Delta_0^*(r_0,c)$ that is well defined for all points in the $(r,c_0)$ plane that were not excluded by the numerical bounds of Section \ref{subsec:eps_c_bounds}. This is displayed as a three-dimensional exclusion plot in Fig.~\ref{Fig:E-crplane}, which corresponds to $\Lambda=18$. The red line in Fig.~\ref{Fig:E-crplane} corresponds to the analytic bound \eqref{crbound}, but since it may not hold in all circumstances we have extracted bounds even for points in the plane that violate it.

This exclusion surface was determined in a slightly unconventional manner. Rather than fixing $c$ and $r_0$ and performing boolean searches to obtain a dimension bound, we fixed $r_0$ and imposed a gap in the scalar singlet channel and searched for upper and lower bounds on $c$ consistent with that gap.\footnote{Obtaining a lower bound for an OPE coefficient is possible as long as there is a gap between the superconformal block under consideration and the next operator, so this method can be used precisely for bounding the first scalar operator.}
In this way we were able to find bounds for the whole of the plane with only a single numerical search required for each data point. 

%%%%%%%%%%%%
%%%%%%%%%%%%

\begin{figure}[ht!]
             \begin{center}           
              \includegraphics[scale=0.43]{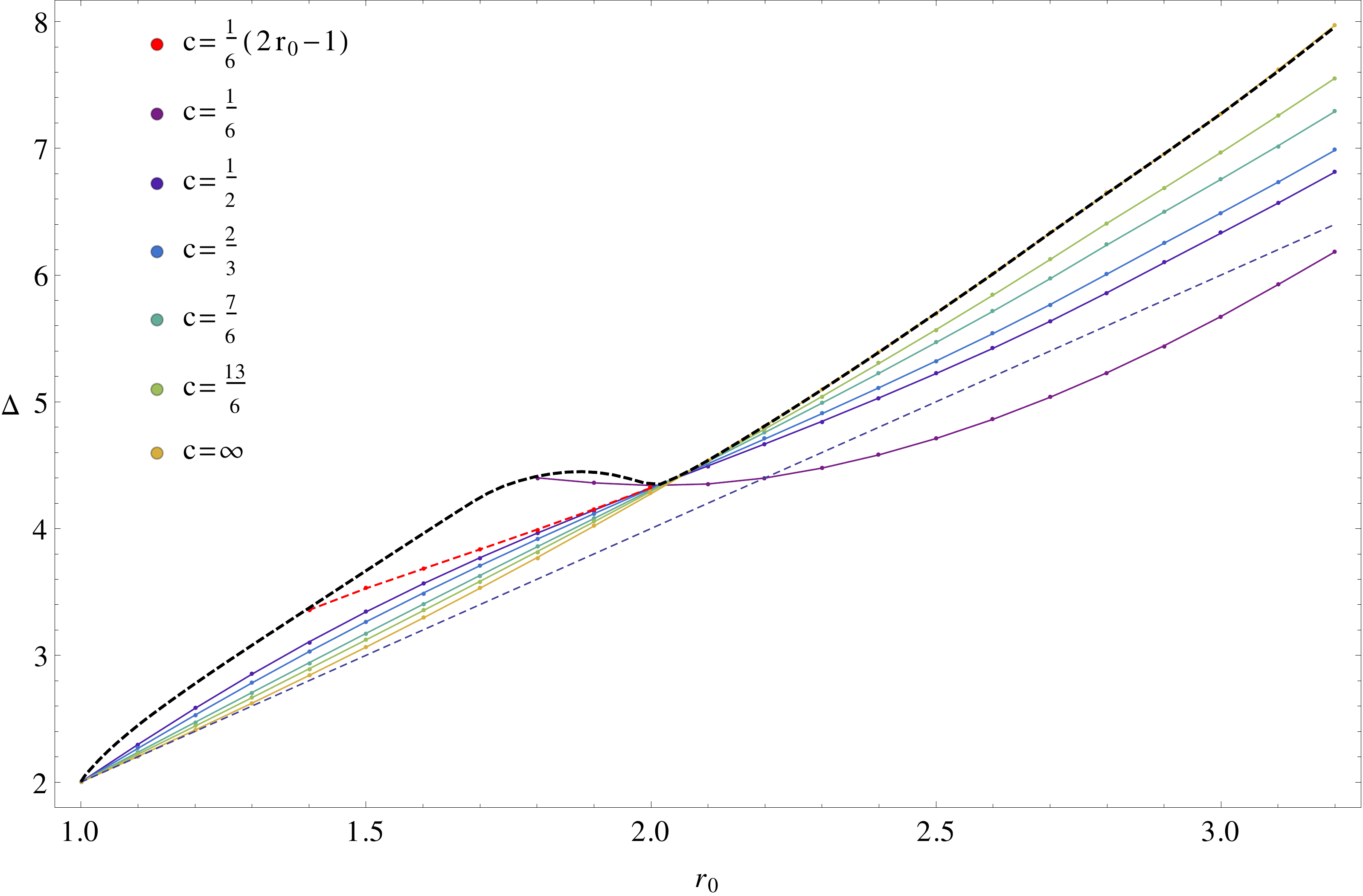}
              \caption{\label{Fig:E-crplaneslices}Bound on the first unprotected scalar in the $\EE_{r_0} \bar{\EE}_{-r_0}$ OPE as a function of $r_0$ for several different values of the central charge, obtained with $\Lambda = 20$. The dashed blue line corresponds to the generalized free field theory solution $\Delta=2r_0$, the thick dashed black line is the same as in Fig.~\ref{Fig:E-dimbound_arbc}, and the red line segment is the bound obtained for the central charge which saturates \eqref{crbound}. If the central charge of a theory is known then it must correspond to a point below the curve corresponding to that central charge. If the central charge is not known and the theory has a freely generated Coulomb branch, then equation \eqref{crbound} together with our numerics dictate that the theory must lie below both the black line and below the red line segment. If we do not know either $c$ or whether the Coulomb branch is freely generated then the theory must still lie below the black curve.}
            \end{center}
\end{figure}

By taking constant central charge slices of this surface, a feature which is not apparent in Fig.~\ref{Fig:E-crplane} comes into view. Several such slices are superimposed in Fig.~\ref{Fig:E-crplaneslices}, where the dimension bound is shown as a function of $r_0$ for various values of the central charge (including infinity). The results that are shown correspond to $\Lambda=20$. Together with these bounds there is a blue dashed straight line at $\Delta=2r_0$ corresponding to the generalized free field theory solution, and a thick dashed black line showing the $\Lambda=20$ dimension bound for arbitrary central charge. Since the latter line is the best possible bound without fixing the central charge, it envelopes all the lines for fixed $c$.

%%%%%%%%%%%%
%%%%%%%%%%%%

Although Fig.~\ref{Fig:E-crplaneslices} could have been obtained by interpolating between the data points that define the three-dimensional plane in Fig.~\ref{Fig:E-crplane}, we chose to revert to performing separate boolean searches for each point as this yields more precise results. As in the results reported in the previous sections, these boolean searches were performed by fixing the stress tensor coefficient in terms of $c$, imposing a gap in the spectrum and finding whether a functional exists as described in Section~\ref{Sec:Numerics}.

These bounds clearly have two qualitatively different regimes, depending on whether $r_0$ is greater or less than two. For $r_0 > 2$ the bound gets weaker (increases) as the central charge is increased. The weakest bound is just the $c = \infty$ line, and it coincides with the bound with unspecified central charge. In this region convergence is relatively slow, and so it is hard to guess where the bound will end up as the cutoff is lifted. Of course we cannot exclude known solutions, so for $c = \infty$ the bound will not be able to cross the generalized free field theory line. More trivially, for $c = \frac{1}{6}$ the bound will have to allow the points $r_0=2n$, $\Delta=4$ for $n\in\Nb$.

The point $r_0=2$ is particularly interesting. Here the lines for all central charges converge at a value that is close to, and seems to be approaching, $\Delta=4$. The absence of a stronger bound can be explained by the existence of a one-parameter family of four-point functions -- constructed by taking a linear combination of the free field solution and the generalized free field solution -- which can realize any $c \geqslant 1/6$ and for which the first scalar operator always has dimension four. However, recall that theories with a chiral operator with $r_0 = 2$ necessarily have a conformal manifold. If these bounds converge to $\Delta=4$, then it would follow that at any point on any conformal manifold there must be a relevant, unprotected operator with nonzero coefficient in the chiral/anti-chiral OPE. It would be interesting to check this at low order(s) in perturbation theory.

For $r_0 < 2$ the picture is reversed. The bound for infinite central charge still appears to be approaching the generalized free field theory value, but the bounds now grows stronger (decreases) as a function of the central charges. The solution to crossing symmetry along the black line corresponds to the lowest allowed value of the central charge consistent with crossing, which is precisely the bound shown in Fig.~\ref{Fig:E-cbound}. For example, the $c = 1/6$ line ends on the black curve at the same value of $r_0$ where \ref{Fig:E-cbound} begins to exclude the value $c = 1/6$, and for even smaller $r_0$ and fixed $c$ there is no unitary solution anymore.

If $\EE_{r_0}$ is a Coulomb branch generator the central charge cannot be arbitrarily small, and in particular must satisfy \eqref{crbound}. This renders part of the black curve with $r_0 < 2$ unphysical, since it corresponds to solutions with a central charge that violates \eqref{crbound}. We can correct this by assuming the central charge to be at least $\frac{1}{6}(2 r_0 - 1)$. We then obtain a correction to the black curve that is shown in Fig.~\ref{Fig:E-crplaneslices} as a red dashed line. Any unitary $\NN =2$ SCFT with a freely generated Coulomb branch must now lie below both the black curve and, because of \eqref{crbound}, also below the red line segment. This improvement removes the local maximum from Fig.~\ref{Fig:E-dimbound_arbc}.

%%%%%%%%%%%%%%%%%%%%%%%%%%%%%%%%%%%%%%%%%%%%%%%%%%%%%%%%%%%%%%%%%%%%%%%%
\subsection{\texorpdfstring{$\EE_{2r}$}{E(2r)} OPE coefficient bounds}
\label{subsec:eps_OPE_bounds}
%%%%%%%%%%%%%%%%%%%%%%%%%%%%%%%%%%%%%%%%%%%%%%%%%%%%%%%%%%%%%%%%%%%%%%%%
 
\begin{figure}[ht!]
             \begin{center}           
              \includegraphics[scale=0.35]{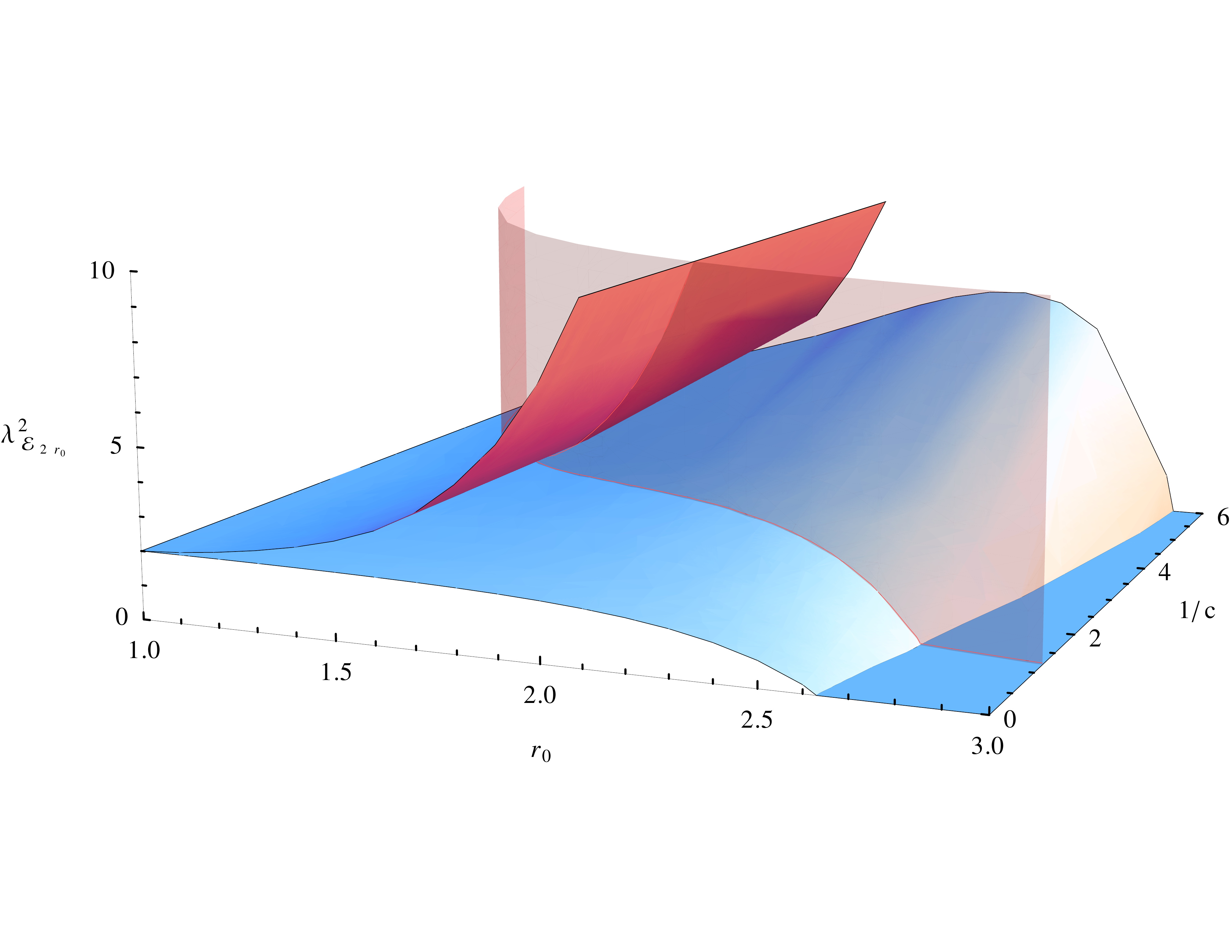}
              \caption{Upper and lower bounds on the OPE coefficient squared of $\EE_{2r_0} $ as a function of $r_0$ and $\frac{1}{c}$, corresponding to a cutoff $\Lambda = 22$. The vertical red ``wall'' corresponds to the bound \eqref{crbound}, and the excluded region if $\EE_{r_0}$ is a Coulomb branch generator is the one with smaller central charge.}
              \label{Fig:E2rbound}
            \end{center}
\end{figure}

In the chiral OPE channel it is natural to look for constraints on the (squared) OPE coefficient $\lambda^2_{\EE_{2r_0}}$ of the $\EE_{2r_0}$ multiplet. The conformal block associated to the exchange of this multiplet is given by $G_{2r_0}^{(0)}(z,\zb)$, while the next multiplet appearing in the chiral channel has $G_{2r_0+2}^{(0)}(z,\zb)$ as its conformal block. Thus the $\EE_{2r_0}$ contribution is isolated, and we can look for both upper and lower bounds on its coefficient. These bounds are displayed in Fig.~\ref{Fig:E2rbound} for $\Lambda=22$. Physical theories must lie between the two blue/red sheets. The vertical ``wall'' corresponds to $c=\frac{1}{6}(2r_0-1)$.

%%%%%%%%%%%%
%%%%%%%%%%%%

As a sanity check, we can compare these numerical bounds to some known theories. The free vector multiplet gives a solution to crossing symmetry with $r_0=1$ and $c=\frac{1}{6}$, and from the decomposition \eqref{freevectorexpansion} we can see that $\lambda^2_{\EE_{2r_0}} = 2$. This ends up being consistent with the numerical bounds, since at this point both the lower and upper bound are very close to two. Similarly, for infinite $c$ we find the generalized free field solution with an OPE coefficient that is also equal to two - again consistent with the numerical bounds.

%%%%%%%%%%%%
\begin{figure}[ht!]
             \begin{center}           
              \includegraphics[scale=0.35]{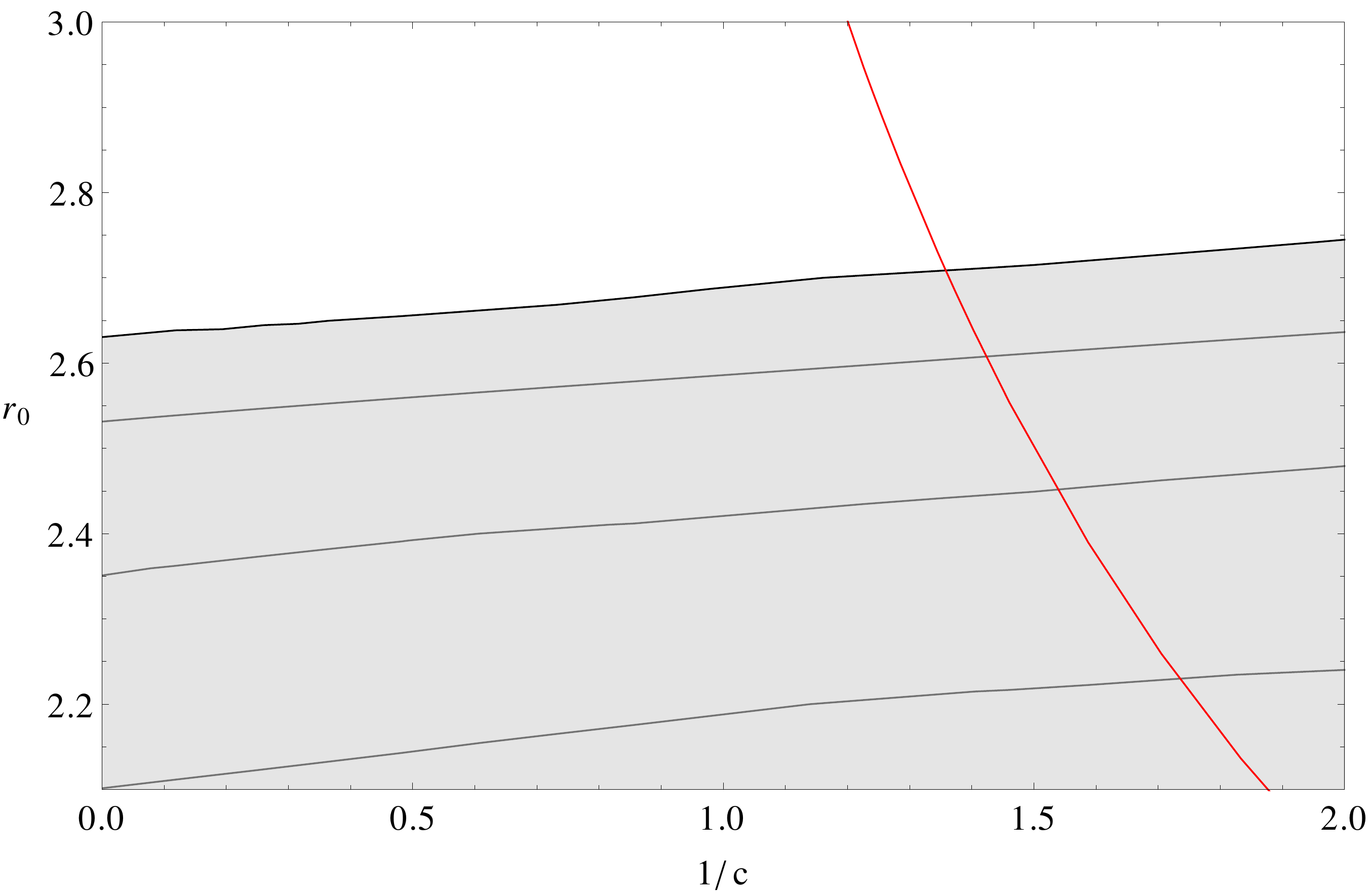}
              \caption{Region where the lower bound on the OPE coefficient squared of $\EE_{2r_0} $ is strictly positive as a function of $r_0$ and $\frac{1}{c}$, for $\Lambda$ varying from $10$ to $22$ in steps of four. The shading indicates the OPE coefficient squared is positive in that region. The red line corresponds to the unitarity bound \eqref{crbound}, and the excluded region (if $\EE_{r_0}$ is a Coulomb branch generator) is to the right of the line. Note that these results are approximate, as this plot is obtained by an interpolation procedure from results like those shown in Fig.~\ref{Fig:E2rbound}. The slight wiggles in the lines are likely due to small errors introduced by this procedure.}
              \label{Fig:E2rbound_conv}
            \end{center}
\end{figure}
%%%%%%%%%%%%

It is interesting to observe that the lower bound on this OPE coefficient is strictly positive in a large region of the $(r_0,c)$ plane. In this region, these bounds rigorously exclude the possibility of Coulomb branch chiral ring relations of the form $\EE_{r_0}\EE_{r_0} \sim 0$. The region of the plane where the lower bound is positive is displayed in Fig.~\ref{Fig:E2rbound_conv}. It is clear that the bound will improve substantially more at larger $\Lambda$.

In interpreting Figs.~\ref{Fig:E2rbound}, \ref{Fig:E2rbound_conv} there is an important subtlety. In obtaining these bounds we have fixed $c$ to a given value, which corresponds to inserting the superconformal stress tensor block with a fixed coefficient in the appropriate channel. However we have also allowed for arbitrary superconformal blocks for long multiplets in the same channel, both at and above the unitarity bound. A long block at the unitarity bound however reduces exactly to the stress tensor block and can therefore mimic the effect of the stress tensor. Since the coefficient of the stress tensor block is proportional to $1/c$, the bounds obtained for a given value of $c$ are also valid \emph{for all smaller central charges}. In other words, when increasing $c$ the bounds can never improve - instead they either worsen or stay constant. In future searches this issue could be circumvented by imposing a gap in the scalar channel.

%%%%%%%%%%%%%%%%%%%%%%%%%%%%%%%%%%%%%%%%%%%%%%%%
\subsubsection*{OPE coefficient bounds and the Zamolodchikov metric}
%%%%%%%%%%%%%%%%%%%%%%%%%%%%%%%%%%%%%%%%%%%%%%%%

The slice $r_0 = 2$ of Fig.~\ref{Fig:E2rbound} is of special interest because of its relation to the curvature of the Zamolodchikov metric on the conformal manifold \cite{Baggio:2014sna,Baggio:2014ioa}. Namely, consider an $\NN = 2$ SCFT with a moduli space $\MM$ of exactly marginal deformations. The different marginal deformations at a given point on $\MM$ are the top components of $\EE_2$ multiplets (and their complex conjugates) whose superconformal primary we will denote as $\phi_a$, $a=1,\ldots, \dim_{\Cb}(\MM)$. The Zamolodchikov metric $g_{a \bar b}$ on $\MM$ is determined by the two-point functions of these primaries,\footnote{In \cite{Baggio:2014ioa} this is the ``metric'' written as $g_{a \bar b}$, which differs from the actual metric $G_{a\bar b}$ studied in that by a factor $192$.}
%%%%%%
\begin{equation}
\langle \phi_a (x) \bar \phi_{\bar b} (0) \rangle = \frac{g_{a\bar b}}{x^4}~.
\end{equation}
%%%%%%
Unit normalizing these operators corresponds to choosing local holomorphic coordinates on $\MM$ such that $g_{a\bar b} = \delta_{a \bar b}$ at the point of interest. In these coordinates, the only non-vanishing four-point function involving the $\phi_a$ and their complex conjugates is given by
%%%%%%
\begin{equation}
\langle \phi_a (x_1) \phi_b(x_2) \bar \phi_{\bar c}(x_3)\bar \phi_{\bar d}(x_4)\rangle~.
\end{equation}
%%%%%%
The OPE of $\phi_a(x_1)$ and $\phi_b(x_2)$ is regular and correspondingly the first conformal block in the chiral channel for this four-point function is a dimension four scalar that is the superconformal primary of an $\EE_4$ multiplet. According to Eqn. (3.13) of \cite{Baggio:2014ioa}, the coefficient for this superconformal block is given by
%%%%%%
\begin{equation}
\mu_{\EE_4\, ab\bar c\bar d} = - R_{a \bar c b \bar d} + \delta_{a \bar c} \delta_{b \bar d} + \delta_{b \bar c} \delta_{a \bar d}~.
\end{equation}
%%%%%%
where $R_{a \bar c b \bar d}$ is the Riemann curvature tensor (in the aforementioned distinguished coordinates) of the Zamolodchikov metric on $\MM$.\footnote{Recall that the Zamolodchikov metric is K\"ahler and therefore $R_{a \bar c b \bar d}$ is symmetric under exchange of $a$ and $b$ (as well as exchange of $\bar c$ and $\bar d$). This is required by the braiding relation of the four-point function.}

We have obtained upper and lower bounds for the OPE coefficient in the particular four-point function with identical operators, $a = b = c = d$. In that case we have
%%%%%%
\begin{equation}
\lambda^2_{\EE_4} = \mu_{\EE_3\,aa\bar a\bar a} = 2 - R_{a \bar a a \bar a}~.
\end{equation}
%%%%%%
When $\dim_{\mathbb C}(\MM) = 1$, this expression simplifies to
%%%%%%
\begin{equation}
\lambda^2_{\EE_4} = 2 - \hf R~,
\end{equation}
%%%%%%
with $R$ the Ricci scalar of $g_{a\bar a}$. The bounds for $\lambda^2_{\EE_4}$ can therefore be interpreted as bounds for the scalar curvature of one-dimensional conformal manifolds.

%%%%%%%%%%%%
\begin{figure}[ht!]
             \begin{center}           
              \includegraphics[scale=0.4]{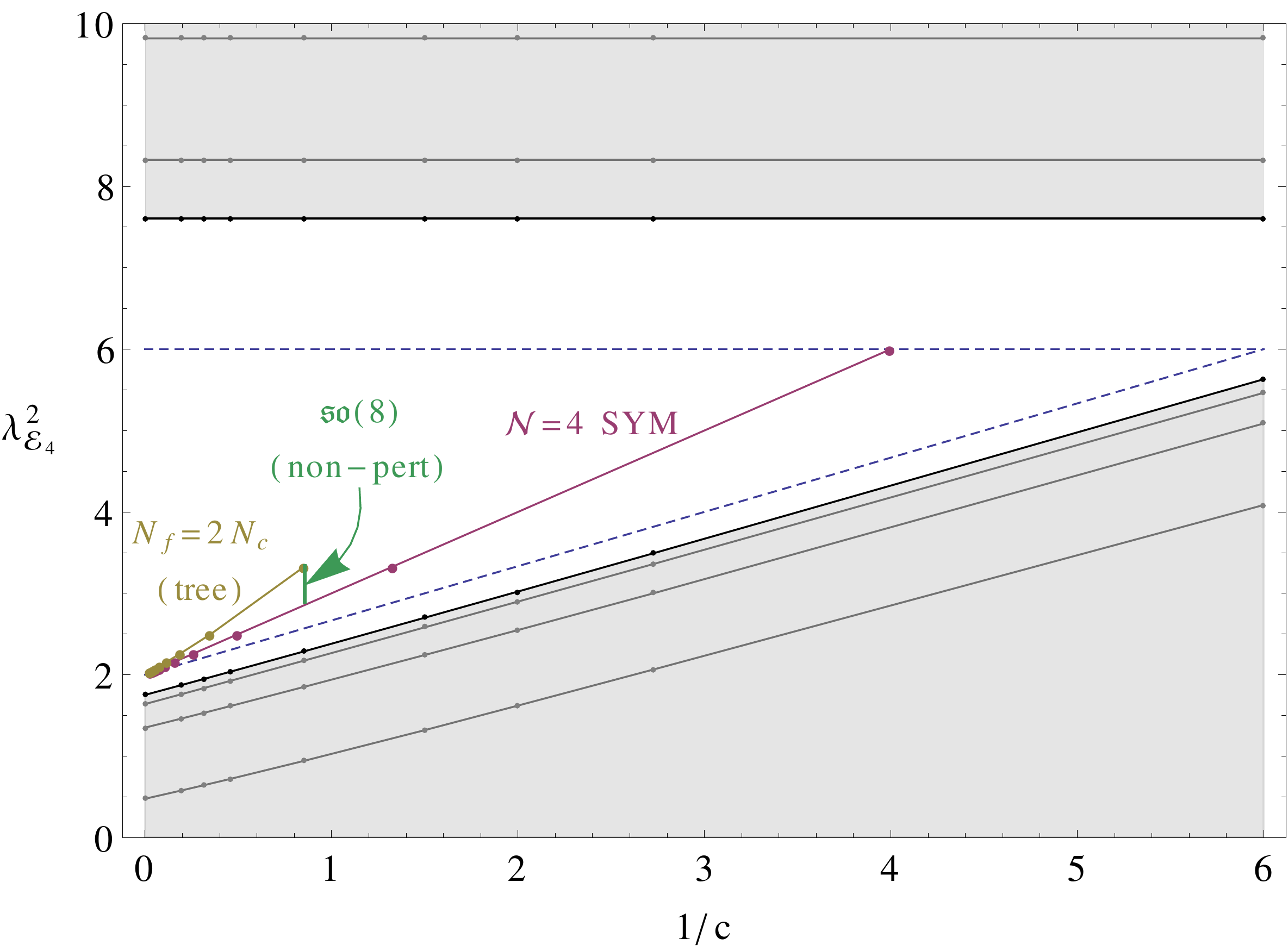}
              \caption{\label{Fig:E4bound}Upper and lower bounds on $\lambda^2_{\EE_4}$ as a function of the central charge $c$. Shaded regions are excluded by our numerics, with $\Lambda$ ranging from $10$ to $22$ in steps of four (the upper bound for $\Lambda=10$ is outside the plotted region). The dotted lines are the best possible value of the bounds as dictated by the free vector multiplet solution, and it seems likely that our bounds will converge to these values. We highlighted the known values of the coefficients for $\NN = 4$ SYM theories (which are protected), the $\NN = 2$ SCQCD theories with gauge group $SU(N_c)$ and $N_f = 2 N_c$ fundamental flavors (tree level values only), and finally the special case of $SU(2)$ SCQCD which we call the $\sof(8)$ theory. The line in the latter case shows the range of values that $\lambda^2_{\EE_4}$ takes as a function of the exactly marginal coupling, cf. the computation in Appendix~\ref{App:exactopes}. The individual dots in the colored lines correspond to gauge groups $\SU(N)$ (with $N \geqslant 2$), plus the $U(1)$ theory at $c = 1/4$ for $\NN = 4$ SYM.}
            \end{center}
\end{figure}
%%%%%%%%%%%%

The $r_0 = 2$ slice of Fig.~\ref{Fig:E2rbound} is shown in Fig.~\ref{Fig:E4bound}, with the excluded regions shaded in gray. The bounds for lower values of $\Lambda$ are also shown to indicate the cutoff-dependence. Inside the allowed region we highlighted several points and loci that correspond to known theories. The computation of $\lambda^2_{\EE_4}$ for these theories is reviewed in Appendix~\ref{App:exactopes}.

Even at infinite $\Lambda$, the the upper and lower bounds will not be able to penetrate beyond the dashed blue lines. The reason for this is as follows. In the theory of $n$ free vector multiplets one finds a chiral four-point function with $r_0=2$ for which
%%%%%%
\begin{equation}
\lambda^2_{\EE_4} = 2 + \frac{2}{3 c}~, \qquad \qquad \frac{1}{c} < 6~.
\end{equation}
%%%%%%
This is the lower dashed line in Fig.~\ref{Fig:E4bound}. The upper horizontal dashed line, on the other hand, is simply given by $\lambda^2_{\EE_4} = 6$, which is the value for the solution corresponding to a single free vector multiplet at $c = 1/6$. The numerical upper bound cannot pass this line because of the aforementioned fact that by design the bound is a non-increasing function of $1/c$.

From the dependence of the bounds on $\Lambda$ it seems natural to expect that they will eventually converge to the dashed blue lines. If this were to happen, then the purely diagonal components of the Riemann tensor would have to obey the following bound:
%%%%%%
\begin{equation}
-4 \leqslant R_{a \bar a a \bar a} \leqslant - \frac{2}{3c}~.
\end{equation}
%%%%%%
In particular, for theories with one-dimensional conformal manifolds crossing symmetry appears to dictate that their scalar curvature is always negative. To see if this is also true for higher-dimensional moduli spaces a bound for $R_{a \bar a b \bar b}$ will be necessary. We plan to investigate the corresponding four-point function in the near future.

%!TEX root = ../draft_maxi_Neq2.tex

\section{Conclusions}
\label{sec:conclusion}

The abstract operator viewpoint offers a unified language for the description of both Lagrangian and non-Lagrangian CFTs. It has also become the entry point for powerful numerical studies in the style of \cite{Rattazzi:2008pe}. In this paper we have advocated for the utility of this viewpoint in studying $\NN = 2$ superconformal field theories. We have highlighted the interplay between superconformal representation theory and interesting physics in these theories, and we identified three types of distinguished representations of particular physical interest. Our numerical investigations focused on the four-point functions of two types of multiplets, $\hat{\BB}_1$ and $\EE_r$. The result was a plethora of numerical unitarity bounds for $\NN = 2$ SCFTs involving central charges, operator dimensions, and OPE coefficients.

Our results reveal a number of interesting details about $\NN = 2$ superconformal field theories, some of which are new numerical bounds for its observables and some of which make contact with other known facts. For example, we have rigorously established that Coulomb branch chiral operators $\EE_r$ with sufficiently low values of $r$ cannot satisfy a certain type of chiral ring relations, and that theories with $\suf(2)$ flavor symmetry must have at least one flavor singlet multiplet of type $\hat{\CC}_{1,\ell = 1}$ and one flavor triplet multiplet of type $\hat{\CC}_{1,\ell = 0}$. The latter follows from our numerical exclusion of theories with $k= 2/3$ for which these multiples decouple from the $\hat \BB_1 \times \hat \BB_1$ OPE. Similarly, if our extrapolations are on track then we should be able to rule out one complex dimensional conformal manifolds like a smooth two-sphere, for which the Euler characteristic is not compatible with the sign of the curvature of the Zamolodchikov metric. In the future it would be very interesting to look for analytic arguments for some of these statements and to understand if the connection with associativity of the operator algebra can be made analytically tractable. More generally, the combination of both analytic and numerical methods appears to be the most promising way to constrain and explore the landscape of $\NN = 2$ superconformal field theories.

Throughout this work, we have observed a strong dependence on the cutoff $\Lambda$ that determines the size of the numerical problem being investigated. In other words, the bounds derived here -- though valid -- do not appear to be close to their optimal value. This is not for lack of trying: the strength of our numerical methods is completely on par with (and in some cases exceeds) the state of the art in almost all of the present literature. The strong cutoff dependence therefore appears to be an intrinsic property of bounds extracted from our specific four-point functions. In the near future we are hopeful that better numerical software tailored to the problem at hand will allow for searches with much greater reach and higher precision. Even then, however, it is not clear that the bounds presented here should be expected converge to some limiting value. For example, if the extrapolation shown in Fig.~\ref{Fig:su2_cboundkeq1} is more or less correct, then a cutoff $\Lambda$ of order $O(100)$ will be necessary to reach a value of the lower bound that is within a few percent of the asymptotic value. The corresponding search space dimension would have to be a factor ten higher than the ones used in this work. Until such methods become computationally feasible, we are stuck with the sorts of extrapolation presented in this work if we want a rough guess for the limiting value of a bound.

With additional work and the development of improved numerical methods, we see a number attractive directions for future work.

%%%%%%%%%%%%%%%%%%%%%%%%%%%%%%%%%%%%%%%%%%%%%%%
\subsubsection*{Additional correlation functions}
%%%%%%%%%%%%%%%%%%%%%%%%%%%%%%%%%%%%%%%%%%%%%%%

An obvious and interesting avenue is to analyze a more diverse collection of four-point functions. The four-point functions of $\hat \BB_1$ multiplets with a flavor symmetry algebra other than $\suf(2)$ or $\ef(6)$ is a natural choice that would involve very little groundwork on top of what we have reported here. Perhaps the most important extension will be to study the four-point functions of operators in the stress-tensor multiplet. In this case there are several natural candidates.

Recall that the superconformal primary of this multiplet is a dimension two scalar, and among its descendants we find the $R$-symmetry currents and the stress tensor. The first step towards bootstrapping any operators in this multiplet is to determine the corresponding selection rules and superconformal blocks, and this prerequisite has not yet been fulfilled. This analysis seems quite complicated for the four-point function of the stress tensor multiplet in superspace, but should be tractable for just the four-point function of the superconformal primary. An interesting case of intermediate complexity is the four-point function of $SU(2)_R$ currents, for which the chiral algebra data fixes a large number of OPE coefficients. From either of these four-point functions one may obtain bounds on the $a$ anomaly coefficient, which is a piece of data that was conspicuously absent from the four-point functions considered in this paper.

%%%%%%%%%%%%%%%%%%%%%%%%%%%%%%%%%%%%%%%%%%%%%%%
\subsubsection*{Multiple correlation functions}
%%%%%%%%%%%%%%%%%%%%%%%%%%%%%%%%%%%%%%%%%%%%%%%

The bounds reported here are valid and must be obeyed by physical theories, but they were derived as a consequence of crossing symmetry for individual correlators. In an honest CFT, crossing symmetry must hold in \emph{all} possible correlators. The simultaneous investigation of multiple correlators in a single numerical program is then a natural next step. The pioneering work of this type was \cite{Kos:2014bka}, where three-dimensional non-supersymmetric CFTs were studied. With minimal additional assumptions, the mixed correlator approach has the potential to rule out spurious solutions to single-correlator crossing symmetry that have no place in a consistent SCFT. In an optimistic scenario, this would also rule out (presumably spurious) linear combinations of solutions that may saturate the single-correlator bounds for large $\Lambda$. In the $\NN =2$ setting one should consider all mixed four-point functions containing a given subset of Coulomb or Higgs branch chiral ring generators. For the Higgs branch chiral ring, the structure of many relevant four-point functions and superconformal blocks have already been worked out in \cite{Nirschl:2004pa,Dolan:2004mu}.

%%%%%%%%%%%%%%%%%%%%%%%%%%%%%%%%%%%%%%%%%%%%%%%
\subsubsection*{Theory-specific analysis}
%%%%%%%%%%%%%%%%%%%%%%%%%%%%%%%%%%%%%%%%%%%%%%%

In this exploratory paper we have taken as general an approach as possible to the $\NN=2$ superconformal bootstrap program. In particular, we have avoided making assumptions that might not be shared by all theories. A complementary strategy is to try to specify a particular theory of interest and ``zoom in'' on that theory in the space of SCFTs. By including as much information as possible about a theory of interest, one hopes to effectively isolate the corresponding solution to crossing symmetry at a boundary of the numerically allowed region. On can then begin to solve that theory at the level of the spectrum of local operators and OPE coefficients.

The numerical results obtained here do not offer much guidance in choosing between known $\NN=2$ theories, mostly because of the absence of ``kinks'' in the bounds. Some natural candidates still present themselves upon further thought. A particularly elegant theory that we think deserves further study is $SU(2)$ SCQCD with $N_f = 4$. For this theory the exact OPE coefficients derived in \cite{Baggio:2014sna} make it possible to use the exactly marginal coupling constant $\tau$ as an \emph{input} variable, at least for the four-point function of $\EE_r$ multiplets. This opens the way towards exploring the contours of a nontrivial conformal manifold by deriving coupling constant-dependent bounds. This was not possible in the work of \cite{Beem:2013qxa} on $\NN=4$ SYM because in that case the known OPE coefficients are constant on the conformal manifold. The $SU(2)$ SCQCD also enjoys an $\sof(8)$ flavor symmetry, and it would be interesting to compare bounds for the corresponding $\hat{\BB}_1$ multiplet with those of the $\EE_r$ multiplets. More precisely, in \cite{Beem:2013qxa} it was conjectured that for certain $\NN = 4$ SYM theories the coupling-independent bounds were saturated at self-dual values of the coupling. If one can achieve reasonable convergence, it may be possible to check the equivalent conjecture for this theory.

Perhaps the most obvious candidate for targeted bootstrap analysis is the $\ef_6$ theory of Minahan and Nemeschansky \cite{Minahan:1996fg}, which lies at the intersection of two lines where analytic bounds derived from the two-dimensional chiral algebra are saturated. The current numerical analysis does not appear to be extremely constraining, but we expect the more refined strategies that we have mentioned to yield stronger results.

%!TEX root = ../draft_maxi_Neq2.tex

\acknowledgments
The authors have greatly benefited from discussions with 
P.~Argyres,
M.~Baggio,
N.~Mekareeya,
K.~Papadodimas,
S.~Rychkov,
N.~Seiberg,
D.~Simmons-Duffin,
and Y.~Tachikawa,
as well as the participants from ``Back to the Bootstrap 3'' at CERN and ``Back to the Bootstrap 4'' at the University of Porto.
C. B. gratefully acknowledges support from the Frank and Peggy Taplin Felloswhip at the IAS.
C. B. is also supported in part by the NSF through grant PHY-1314311.
M. L. is supported in part by FCT - Portugal, within the POPH/FSE programme, through grant SFRH/BD/70614/2010.
P. L. is supported by SFB 647 ``Raum-Zeit-Materie. Analytische und Geometrische Strukturen''.
L. R. gratefully acknowledges the generosity of the Simons and Guggenheim foundations and the
 wonderful hospitality of the IAS, Princeton and of the KITP, Santa Barbara during his sabbatical year.
The work of M.L. and L.R. is supported in part by NSF Grant PHY-1316617. The research leading to these results has received funding from the [European Union] Seventh Framework Programme [FP7-People-2010-IRSES] under grant agreement No 269217.

\appendix

%!TEX root = ../draft_maxi_Neq2.tex

\section{Unitary representations of the \texorpdfstring{$\NN=2$}{N=2} superconformal algebra} 
\label{App:representations}

The representation theory of the four-dimensional $\NN=2$ superconformal algebra plays a central role both in our choice of strategy and in the structure of the partial wave analysis of four-point functions. In this appendix we review the classification of unitary irreducible representations of $\suf(2,2|2)$ (\cf\ \cite{Dobrev:1985qv,Dolan:2002zh,Kinney:2005ej}).

Unitary representations of $\suf(2,2|2)$ are highest weight representations and are labelled by quantum numbers $(\Delta,j_1,j_2,r,R)$ of the highest weight state also called the \emph{superconformal primary} of the representation. A \emph{generic representation} -- also called a \emph{long representation} -- is obtained by the action of the eight Poincar\'e supercharges as well as the momentum generators and $SU(2)_R$ lowering operators on the highest weight state. \emph{Short representations} occur when a superconformal descendant state in what would otherwise be a long representation is rendered null by a conspiracy of quantum numbers. The unitarity bounds for a superconformal primary operator are given by
%%%%%%
\begin{equation}
\begin{alignedat}{3}
\Delta	&\geqslant \Delta_i~,&\qquad				&&				 j_i&\neq0~,\\
\Delta	&=	\Delta_i-&2~~\mbox{~or~}~~\Delta&\geqslant& \Delta_i~,	\qquad j_i&=0~,\\
\end{alignedat}
\end{equation}
%%%%%%
where we have defined
%%%%%%
\begin{equation}
\Delta_1\colonequals2+2j_1+2R+ r~, \qquad \Delta_2\colonequals2+2j_2+2R- r~.
\end{equation}
%%%%%%
Short representations occur when one or more of these bounds are saturated. The different ways in which this can happen correspond to different combinations of Poincar\'e supercharges that will annihilate the superconformal primary state in the representation. 

There are two types of shortening conditions, each of which has four incarnations corresponding to an $SU(2)_R$ doublet's worth of conditions for each supercharge chirality:
%%%%%%
\begin{eqnarray}\label{eq:constituent_shortening_conditions}
\BB^\II&:&\qquad \QQ^\II_{\alpha}|\psi\rangle=0~,\quad\alpha=1,2~,\\
{\bar \BB}_\II&:&\qquad \wt\QQ_{\II\dot\alpha}|\psi\rangle=0~,\quad\dot\alpha=1,2~,\\
\CC^\II&:&\qquad
\begin{cases} \epsilon^{\a\b}\QQ^\II_{\alpha}|\psi\rangle_\beta=0~,	\quad &j_1\neq0~,\\
\epsilon^{\a\b}\QQ^\II_{\alpha}\QQ^\II_{\beta}|\psi\rangle=0~,		\quad &j_1=0~,
\end{cases}\\
{\bar\CC}_\II&:&\qquad
\begin{cases} \epsilon^{\ad\bd}\wt\QQ_{\II\ad}|\psi\rangle_{\dot\beta}=0~,\quad &j_2\neq0~,\\
\epsilon^{\ad\bd}\wt\QQ_{\II\ad}\wt\QQ_{\II\bd}|\psi\rangle=0~,\quad &j_2=0~.
\end{cases}
\end{eqnarray}
%%%%%%
The different admissible combinations of shortening conditions that can be simultaneously realized by a single unitary representation are summarized in Table \ref{Tab:shortening}, where we also list the relations that must be satisfied by the quantum numbers of the superconformal primary in such a representation. We also list two common notations used to designate the different representations -- one from \cite{Dolan:2002zh} (DO) and the other from \cite{Kinney:2005ej} (KMMR).\footnote{We are adopting the the $R$-charge conventions of \cite{Dolan:2002zh}.}

%%%%%%%%%%%%%%%%%%%%%%%%%%%%%%%%%%%%%%%%%%%%%%%%
\begin{table}[t!]
\begin{centering}
\renewcommand{\arraystretch}{1.3}
\begin{tabular}{|l|l|l|l|}
\hline
Shortening & Quantum Number Relations & DO & KMMR \tabularnewline
\hline
\hline 
$\varnothing$					  & \makebox[3.8cm][l]{$\Delta\geqslant {\rm max}(\Delta_1,\Delta_2)$}												 & $\AA^\Delta_{R,r(j_1,j_2)}$ & ${\bf aa}_{\Delta,j_1,j_2,r,R}$ 	  	 \tabularnewline
\hline 
$\BB^1$ 						  & \makebox[3.8cm][l]{$\Delta=2R+r$}		 \makebox[3cm][l]{$j_1=0$}	  								 & $\BB_{R,r(0,j_2)}$ 		& ${\bf ba}_{0,j_2,r,R}$ 	  	 \tabularnewline
\hline 
$\bar\BB_2$						  & \makebox[3.8cm][l]{$\Delta=2R-r$}		 \makebox[3cm][l]{$j_2=0$}   								 & $\bar{\BB}_{R,r(j_1,0)}$ & ${\bf ab}_{j_1,0,r,R}$ 	  	 \tabularnewline
\hline 
$\BB^1\cap\BB^2$  				  & \makebox[3.8cm][l]{$\Delta=r$}  			 \makebox[3cm][l]{$R=0$}  									 & $\EE_{r(0,j_2)}$ 		& ${\bf ba}_{0,j_2,r,0}$ 	  	 \tabularnewline
\hline
$\bar\BB_1\cap\bar\BB_2$  		  & \makebox[3.8cm][l]{$\Delta=-r$}  		 \makebox[3cm][l]{$R=0$}  									 & $\bar \EE_{r(j_1,0)}$ 	& ${\bf ab}_{j_1,0,r,0}$ 	  	 \tabularnewline
\hline 
$\BB^1\cap\bar\BB_{2}$  		  & \makebox[3.8cm][l]{$\Delta=2R$}  		 \makebox[3cm][l]{$j_1=j_2=r=0$}							 & $\hat{\BB}_{R}$ 			& ${\bf bb}_{0,0,0,R}$ 		  	 \tabularnewline
\hline\hline 
$\CC^1$ 						  & \makebox[3.8cm][l]{$\Delta=2+2j_1+2R+r$}  															 & $\CC_{R,r(j_1,j_2)}$ 	& ${\bf ca}_{j_1,j_2,r,R}$ 	  	 \tabularnewline
\hline 
$\bar\CC_2$  					  & \makebox[3.8cm][l]{$\Delta=2+2 j_2+2R-r$}  															 & $\bar\CC_{R,r(j_1,j_2)}$	& ${\bf ac}_{j_1,j_2,r,R}$	  	 \tabularnewline
\hline
$\CC^1\cap\CC^2$  				  & \makebox[3.8cm][l]{$\Delta=2+2j_1+r$}  	 \makebox[3cm][l]{$R=0$}  									 & $\CC_{0,r(j_1,j_2)}$ 	& ${\bf ca}_{j_1,j_2,r,0}$ 	  	 \tabularnewline
\hline 
$\bar\CC_1\cap\bar\CC_2$		  & \makebox[3.8cm][l]{$\Delta=2+2 j_2-r$} 	 \makebox[3cm][l]{$R=0$}  	 								 & $\bar\CC_{0,r(j_1,j_2)}$	& ${\bf ac}_{j_1,j_2,r,0}$ 	  	 \tabularnewline
\hline 
$\CC^1\cap\bar\CC_2$  			  & \makebox[3.8cm][l]{$\Delta=2+2R+j_1+j_2$} \makebox[3cm][l]{$r=j_2-j_1$}  							 & $\hat{\CC}_{R(j_1,j_2)}$	& ${\bf cc}_{j_1,j_2,j_2-j_1,R}$ \tabularnewline
\hline\hline 
$\BB^1\cap\bar\CC_2$  			  & \makebox[3.8cm][l]{$\Delta=1+2R+j_2$}  	 \makebox[3cm][l]{$r=j_2+1$}   								 & $\DD_{R(0,j_2)}$ 		& ${\bf bc}_{0,j_2,j_2+1,R}$  	 \tabularnewline
\hline 
$\bar\BB_2\cap\CC^1$  			  & \makebox[3.8cm][l]{$\Delta=1+2R+j_1$}	 \makebox[3cm][l]{$-r=j_1+1$}    							 & $\bar\DD_{R(j_1,0)}$ 	& ${\bf cb}_{j_1,0,-j_1-1,R}$ 	 \tabularnewline
\hline 
$\BB^1\cap\BB^2\cap\bar\CC_2$  	  & \makebox[3.8cm][l]{$\Delta=r=1+j_2$} 	 \makebox[2.5cm][l]{$r=j_2+1$} 	\makebox[1.5cm][l]{$R=0$}	 & $\DD_{0(0,j_2)}$ 		& ${\bf bc}_{0,j_2,j_2+1,0}$  	 \tabularnewline
\hline
$\CC^1\cap\bar\BB_1\cap\bar\BB_2$ & \makebox[3.8cm][l]{$\Delta=-r=1+j_1$}  	 \makebox[2.5cm][l]{$-r=j_1+1$} \makebox[1.5cm][l]{$R=0$}	 & $\bar\DD_{0(j_1,0)}$ 	& ${\bf cb}_{j_1,0,-j_1-1,0}$ 	 \tabularnewline
\hline
\end{tabular}
\par\end{centering}
\caption{\label{Tab:shortening}Summary of unitary irreducible representations of the $\NN=2$ superconformal algebra.}
\end{table}
%%%%%%%%%%%%%%%%%%%%%%%%%%%%%%%%%%%%%%%%%%%%%%%%

In the limit where the dimension of a long representation approaches a unitarity bound, it becomes decomposable into a collection of short representations. This fact is often referred to as the existence of \emph{recombination rules} for short representations into a long representation at the unitarity bound. The generic recombination rules are as follows,
%%%%%%
\begin{eqnarray}\label{eq:recombination}
\makebox[1in][l]{$\AA_{R,r(j_1,j_2)}^{\Delta\to2R+r+2+2j_1}$}&\simeq& \CC_{R,r(j_1,j_2)}\oplus \CC_{R+\hf,r+\hf(j_1-\hf,j_2)}~,\nn\\
\makebox[1in][l]{$\AA_{R,r(j_1,j_2)}^{\Delta\to2R-r+2+2j_2}$}&\simeq& \bar\CC_{R,r(j_1,j_2)}\oplus \bar\CC_{R+\hf,r-\hf(j_1,j_2-\hf)}~,\\
\makebox[1in][l]{$\AA_{R,j_1-j_2(j_1,j_2)}^{\Delta\to2R+j_1+j_2+2}$}&\simeq& \hat\CC_{R(j_1,j_2)}\oplus \hat\CC_{R+\hf(j_1-\hf,j_2)}\oplus\hat\CC_{R+\hf(j_1,j_2-\hf)}\oplus\hat\CC_{R+1(j_1-\hf,j_2-\hf)}~.\nn
\end{eqnarray}
%%%%%%
In special cases the quantum numbers of the long multiplet at threshold are such that some Lorentz quantum numbers in \eqref{eq:recombination} would be negative and unphysical. In these cases the following exceptional recombination rules apply,
%%%%%%
\begin{eqnarray}\label{eq:special_recombination}
\makebox[.7in][l]{$\AA_{R,r(0,j_2)}^{2R+r+2}$} 			& \simeq & 		\CC_{R,r(0,j_2)} 		\oplus 		\BB_{R+1,r+\hf(0,j_2)}~,\nn\\
\makebox[.7in][l]{$\AA_{R,r(j_1,0)}^{2R-r+2}$}			& \simeq & 		\bar\CC_{R,r(j_1,0)} 	\oplus 		\bar\BB_{R+1,r-\hf(j_1,0)}~,\nn\\
\makebox[.7in][l]{$\AA_{R,-j_2(0,j_2)}^{2R+j_2+2}$}		& \simeq & 		\hat\CC_{R(0,j_2)} 		\oplus 		\DD_{R+1(0,j_2)} \oplus 		\hat\CC_{R+\hf(0,j_2-\hf)} 		\oplus 		\DD_{R+\frac{3}{2}(0,j_2-\hf)}~,\\
\makebox[.7in][l]{$\AA_{R,j_1(j_1,0)}^{2R+j_1+2}$}		& \simeq & 		\hat\CC_{R(j_1,0)} 		\oplus 		\hat\CC_{R+\hf(j_1-\hf,0)} \oplus 		\bar\DD_{R+1(j_1,0)} 	\oplus 		\bar\DD_{R+\frac{3}{2}(j_1-\hf,0)}~,\nn\\
\makebox[.7in][l]{$\AA_{R,0(0,0)}^{2R+2} $}			& \simeq & 		\hat\CC_{R(0,0)} 		\oplus 		\DD_{R+1(0,0)} 	\oplus 		\bar\DD_{R+1(0,0)} 	\oplus 		\hat\BB_{R+2}~.\nn
\end{eqnarray}
%%%%%%
The only recombinations that play a role in the analyses of this paper are the last recombinations in \eqref{eq:recombination} and \eqref{eq:special_recombination}. This is relevant for the partial wave analysis of the moment map four-point function in Section \ref{Sec:Bhat4ptfunc}.
%!TEX root = ../draft_maxi_Neq2.tex

\section{Superconformal block decompositions}
\label{App:blocks}

This appendix contains a number of technical details pertaining to the superconformal block decompositions of correlators investigated in this paper. The conventional conformal blocks of four-dimensional non-supersymmetric CFT make repeated appearances here, and for those we adopt the conventions of \cite{Dolan:2001tt}. Namely, the conformal block associated to the exchange of an $\sof(4,2)$ conformal family whose primary has dimension $\Delta$ and spin $\ell$ in the four-point function of degenerate scalars is given by $u^{\frac12(\Delta-\ell)}G_\Delta^{(\ell)}(u,v)$, where
%%%%%%
\begin{equation}\label{eq:bos_block}
\begin{split}
G_{\Delta}^{(\ell)}(u,v)&\colonequals\frac{1}{z-\zb}\left(\left(-\frac{z}{2}\right)^{\ell}z\, {}_2F_1\left(\frac{1}{2}\left(\Delta+\ell\right),\frac{1}{2}\left(\Delta+\ell\right);\Delta+\ell;z)\right) \right.\\
& \qquad\qquad\quad\left. \times\,{}_2F_1\left(\frac{1}{2}\left(\Delta-\ell-2\right),\frac{1}{2}\left(\Delta-\ell-2\right);\Delta-\ell-2;\zb)\right)-z\leftrightarrow\zb\right)~.
\end{split}
\end{equation}
%%%%%%
Here, as in the main text, we will only ever need to consider operators with $j_1 = j_2 \equalscolon j$, for which the spin $\ell$ is defined as $\ell \colonequals 2j$.

%%%%%%%%%%%%%%%%%%%%%%%%%%%%%%%%%%%%%%%%%%%%%%%%%%%%%%%%%%%%%%%%%%%%%%%%%%%%%%%%%%%%%%%%%%%%%%%%%%%%%%%%%%%%%%%%%%%%%%%%%%%%%%%%%%%%%%%%%%%%%%%%
\subsection{Superconformal blocks for the \texorpdfstring{$\hat{\BB}_1$}{B-hat-1} four-point function}
\label{App:blocksBhat}
%%%%%%%%%%%%%%%%%%%%%%%%%%%%%%%%%%%%%%%%%%%%%%%%%%%%%%%%%%%%%%%%%%%%%%%%%%%%%%%%%%%%%%%%%%%%%%%%%%%%%%%%%%%%%%%%%%%%%%%%%%%%%%%%%%%%%%%%%%%%%%%%

The superconformal blocks relevant to the partial wave decomposition of the $\hat{\BB}_1$ four-point function were derived in the beautiful work of \cite{Dolan:2001tt}. In this subsection we summarize those results. As our starting point we take the selection rule for operators appearing in the OPE of two moment map operators. These selection rules were determined in \cite{Arutyunov:2001qw} via an analysis of three-point functions in harmonic superspace.\footnote{These selection rules can also be understood as following a few simple criteria. Namely, a conformal primary can only have a non-zero three point function with two moment map operators if the superconformal primary of the same multiplet does as well. Ordinary Lorentz symmetry and $R$-symmetry selection rules then constrain the possible superconformal multiplets appearing in the OPE. A further constraint comes from the fact that any $R$-symmetry quintuplet appearing in the OPE comes from the product of two Higgs branch chiral ring operators, and so must itself be annihilated by the action of $\QQ^1_\alpha$ and $\wt\QQ_{2\dot\alpha}$.} The results can be schematically presented as follows
%%%%%%
\begin{equation}\label{eq:bhat_selection_rule}
\hat\BB_1 \times \hat\BB_1 ~~\sim~~ \mathbf{1} ~+~ \hat\BB_1 ~+~ \hat\BB_2 ~+~ \hat\CC_{0(j,j)}  ~+~ \hat\CC_{1(j,j)} ~+~ \AA^\Delta_{0,0(j,j)}~.
\end{equation}
%%%%%%
Below we outline the contribution of each of these multiplets in the superconformal partial wave expansion of a moment map four-point function. We do so in two ways. First, we describe the contribution of such a multiplet to the functions $\GG_i(z,\zb)$ and $f_i(z)$ that appear in the solution of the superconformal Ward identities described in Section \ref{Sec:Bhat4ptfunc}. This is the form of the superconformal blocks for the numerical analysis of crossing symmetry described in Section \ref{Sec:Bhatresults}. In order to make the structure of these contributions more transparent, we also list the contribution of each multiplet to the functions $a_{R,i}(u,v)$ associated with a fixed $SU(2)_R$ channel. Since these expressions are rather lengthy, we have collected them in Table \ref{Tab:moment_map_blocks}.\\

We start with the case of long multiplets. For these multiplets only the two-variable functions $\GG_i(u,v)$ are non-zero (the $f_i(z)$ is protected and only receives contributions from short and semi-short multiplets). In the long multiplets listed in \eqref{eq:bhat_selection_rule}, there is a unique conformal primary in the $\bf 5$ of $SU(2)_R$ that can appear in the OPE. This determines the contribution of a long multiplet to $a_{2,i}(u,v)$, which in turn via \eqref{aR_amplitudes} fixes the contribution of long multiplets as follows
%%%%%%
\begin{equation}\label{longcontr}
\makebox[1.2in][l]{$\AA^\Delta_{0,0(j,j)}~$ in $~\Rf_i:$}
\makebox[1.7in][l]{$\begin{cases}
\GG_i(u,v) 	&=~~6 u^{\frac{\Delta-\ell}{2}}\,G_{\Delta+2}^{(\ell)}(u,v)~,\\
f_i(z)		&=~~0~.		%\qquad \Delta \geqslant \ell+2~.
\end{cases}$}
\end{equation}
%%%%%%
The full conformal block expansion in the three $R$-symmetry channels can now be determined by inserting \eqref{longcontr} back into \eqref{aR_amplitudes} and making use of various identities for hypergeometric functions \cite{Dolan:2001tt}. The full expansion in terms of conventional conformal blocks is given in Table \ref{Tab:moment_map_blocks}.

Next we turn to the $\hat{\CC}_{0(j,j)}$ and $\hat{\BB}_1$ multiplets. These multiplets do not include \emph{any} operators that can contribute in the $R=2$ channel, from which it follows that for these multiplets $\GG_i(u,v)=0$. In the $R=1$ channel, each of these multiplets contributes exactly one conformal primary of dimension $\ell+3$ and spin $\ell+1$ (dimension $2$ and spin $0$ in the $\hat\BB_1$ case). This allows the values of the single-variable functions for these multiplets to be fixed from \eqref{aR_amplitudes}, and we find
%%%%%%
\begin{eqnarray}
\makebox[1.2in][l]{$\hat{\CC}_{0(j,j)}~$ in $~\Rf_i:$}&
\makebox[1.7in][l]{$\begin{cases}
\GG_i(u,v)&=~~0~,\\
f_i(z)&=~~ 2g_{2j+2}(z)~.
\end{cases}$}\label{C0contr}\\
\makebox[1.2in][l]{$\hat{\BB}_{1}~$ in $~\Rf_i:$}&
\makebox[1.7in][l]{$\begin{cases}
\GG_i(u,v)&=~~0~,\\
f_i(z)&=~~ 2g_{1}(z)~.
\end{cases}$}\label{B1contr}
\end{eqnarray}
%%%%%%
Again, the contributions of these multiplets to the individual $SU(2)_R$ channels is determined by \eqref{aR_amplitudes}, and the subsequent decomposition into conventional conformal blocks follows from identities for hypergeometric functions. The result is displayed in Table \ref{Tab:moment_map_blocks}. (Another operator that contributes only to $f_i(z)$ is the identity operator, which only arises in the $R=0$ channel and contributes to $f_i(z)$ as a constant.)
%%%%%%
\begin{table}[t!]
\begin{center}
\renewcommand{\arraystretch}{1.5}
\begin{tabular}{|c|l|}
\hline
~Multiplet in $\Rf_i$~		&	Contribution to $a_{R,i}(u,v)$\\
\hline
\hline
							&	$a_{0,i}(u,v) ~=~ \frac{1}{3}u G_{3}^{(1)}(u,v)$\\
$\hat\BB_1$					&	$a_{1,i}(u,v) ~=~ u G_{2}^{(0)}(u,v)$ \\
							&	$a_{2,i}(u,v) ~=~ 0$ \\
\hline
							&	$a_{0,i}(u,v) ~=~ \frac{1}{30}u^3 G_{6}^{(0)}(u,v)$ \\
$\hat\BB_2$					&	$a_{1,i}(u,v) ~=~ \frac{2}{5} u^2 G_{5}^{(1)}(u,v)$ \\
							&	$a_{2,i}(u,v) ~=~ u^2 G_{4}^{(0)}(u,v)$ \\
\hline
			 				& 	$a_{0,i}(u,v) ~=~ u G_{\ell+2}^{(\ell)}(u,v)
						+ \frac{(\ell+2)^2}{(2\ell+3)(2\ell + 5)}uG_{\ell+4}^{(\ell+2)}(u,v)$ \\
$\hat\CC_{0(j,j)}$			& 	$a_{1,i}(u,v) ~=~ u G_{\ell+3}^{(\ell+1)}(u,v)$ \\
			 				& 	$a_{2,i}(u,v) ~=~ 0$ \\
\hline
							& 	$a_{0,i}(u,v) ~=~ \frac{1}{2} u^2 G_{\ell+5}^{(\ell+1)}(u,v) + \frac{1}{8} u^3 G_{\ell+5}^{(\ell-1)}(u,v) + \frac{(\ell+3)^2}{8(2\ell+5)(2\ell+7)} u^3  G_{\ell+7}^{(\ell+1)}(u,v)$ \\
$\hat\CC_{1(j,j)}$			& 	$a_{1,i}(u,v) ~=~ \frac{3}{2} u^2  G_{\ell+4}^{(\ell)}(u,v) + \frac{3}{24} u^3  G_{\ell+6}^{(\ell)}(u,v) + \frac{3(\ell+3)^2}{2(2\ell+5)(2\ell+7)} u^2  G_{\ell+6}^{(\ell+2)}(u,v)$\\	
							& 	$a_{2,i}(u,v) ~=~ u^2 G_{\ell+5}^{(\ell+1)}(u,v)$ \\
\hline
							& 	$a_{0,i}(u,v) ~=~  u^{\frac{\Delta-\ell}{2}}\left( 6 G^{(\ell)}_{\Delta}(u,v) + \frac{3(\Delta + \ell+2)^2}{2(\Delta+\ell+1)(\Delta+\ell+3)}G^{(\ell+2)}_{\Delta +2}(u,v)  \right.$ \\
							& $\qquad\qquad\quad   \left. + \frac{3 (\Delta - \ell)^2 }{ 32 (\Delta-\ell-1)(\Delta-\ell+1)}u^2 G^{(\ell-2)}_{\Delta +2}(u,v) + \frac{1}{2} u G^{(\ell)}_{\Delta +2}(u,v)  \right.$ \\ 
							& $\qquad\qquad\quad  \left. + \frac{3(\Delta + \ell+2)^2 (\Delta - \ell)^2 }{128 (\Delta+\ell+1)(\Delta+\ell+3)(\Delta-\ell-1)(\Delta-\ell+1)} u^2 G^{(\ell)}_{\Delta +4}(u,v)\right)$\\
$\AA^{\Delta}_{0,0(j,j)}$	& 	$a_{1,i}(u,v) ~=~  3u^{\frac{\Delta-\ell}{2}}\left(2G_{\Delta+1}^{(\ell+1)}(u,v)+\frac{1}{2}G_{\Delta+1}^{(\ell-1)}(u,v) \right.$ \\
							& $\qquad\qquad\quad+ \frac{(\Delta+\ell+2)^2}{8(\Delta+\ell+1)(\Delta+\ell+3)}u G_{\Delta+3}^{(\ell+1)}(u,v) + \left.\frac{(\Delta-\ell)^2}{32(\Delta-\ell-1)(\Delta-\ell+1)}u^2 G_{\Delta+3}^{(\ell-1)}(u,v)\right)$ \\
							&	$a_{2,i}(u,v) ~=~ u^{\frac{\Delta+2-\ell}{2}}G_{\Delta+2}^{\ell}(u,v)$ \\
\hline				
\end{tabular}
\caption{Superconformal blocks for the different $\suf(2,2|2)$ representations appearing in the OPE of two moment map operators.\label{Tab:moment_map_blocks}}
\end{center}
\end{table}
%%%%%%

The superconformal blocks for the remaining two multiplets can be understood by studying the behavior of a generic long multiplet as it approaches the unitarity bound $\Delta = 2+\ell$. At the unitarity bound, the representation becomes reducible and decomposes according to the relevant rules in \eqref{eq:recombination} and \eqref{eq:special_recombination} specialized to the case $R=0$,
%%%%%%
\begin{equation}
\begin{split} 
\AA^{\Delta=2j+2}_{0,0(j,j)} &~~\simeq~~ \hat{\CC}_{0(j,j)} \oplus \hat{\CC}_{\frac{1}{2}(j-\frac{1}{2},j)} \oplus \hat{\CC}_{\frac{1}{2}(j,j-\frac{1}{2})}\oplus \hat{\CC}_{1(j-\frac{1}{2},j-\frac{1}{2})}~,\\
\AA^{\Delta=2j+2}_{0,0(0,0)} &~~\simeq~~ \hat{\CC}_{0(0,0)} \oplus \DD_{1(0,0)} \oplus \bar{\DD}_{1(0,0)}\oplus \hat{\BB}_{2}~.
\end{split}
\end{equation}
%%%%%%
In each case, only the first and last multiplet are allowed in the four-point function by the selection rules. This simplifies the task of finding superconformal blocks for $\hat{\CC}_{1(j,j)}$ and $\hat{\BB}_2$ multiplets. Namely, by subtracting six copies of the $\hat{\CC}_{0(j,j)}$ block from the long superconformal block with $\Delta=2+\ell$ one obtains the superconformal block for a $\hat{\CC}_{1(j-\frac12,j-\frac12)}$ with $j \geqslant \frac12$. Similarly, subtracting six copies of the $\hat\CC_{0(0,0)}$ block from the long superconformal block with $\Delta=2$ yields the superconformal block for the $\hat\BB_2$ representation. The result is that these multiplets contribute both to $f_i(z)$ and to $\GG_i(u,v)$ as follows,
%%%%%%
\begin{eqnarray}
\makebox[1.2in][l]{$\hat{\CC}_{1(j,j)}~$ in $~\Rf_i:$}&
\makebox[1.7in][l]{$\begin{cases}
\GG_i(u,v) 	&=\ph{-} 6 u\,G_{\ell+5}^{(\ell+1)}(u,v)~,\\
f_i(z) 		&=-12 g_{2j+3}(z)~,
\end{cases}$}\\
\makebox[1.2in][l]{$\hat{\BB}_{2}~$ in $\Rf_i:$}&
\makebox[1.7in][l]{$\begin{cases}
\GG_i(u,v) 	&=\ph{-}6 u\,G_{4}^{(0)}(u,v)~,\\
f_i(z) 		&=-12 g_{2}(z)~.
\end{cases}$}
\end{eqnarray}
%%%%%%
The decomposition in the three $SU(2)_R$ channels of all these superconformal blocks are again displayed in Table \ref{Tab:moment_map_blocks}.

Finally, there are a few extra selection rules having to do with the representation $\Rf_i$ of the flavor symmetry group in which the various multiplets can appear. For example, $\hat\BB_1$ multiplets are those containing the conserved flavor symmetry currents, so they necessarily appear only in the adjoint representation $\Rf={\rm Adj}$. In a theory with a unique stress tensor, there will be only one $\hat\CC_{0(0,0)}$ multiplet, so it will necessarily transform in the singlet representation $\Rf={\bf 1}$. In general, one may take tensor products of multiple SCFTs and violate this kind of selection rule. We will call a theory that is not decomposable as the tensor product of several theories \emph{simple}. The complete set of flavor symmetry selection rules for simple theories are displayed in Table \ref{tab:flavor_selection}.
%%%%%%
\begin{table}[t!]
\begin{center}
\renewcommand{\arraystretch}{1.5}
\begin{tabular}{|l|l|}
\hline
~Multiplet~			&	Possible $\Rf_i$ in simple theories\\
\hline
\hline
$\hat\BB_1$			&	$\Rf={\rm Adj.}$\\
\hline
$\hat\BB_2$			&	$\Rf\in{\rm Sym}^2({\rm Adj.})$\\
\hline
$\hat\CC_{0(j,j)}$	&	$\Rf={\bf 1}$ for $\ell=0$.\\
					&	None for $\ell\geqslant 1$.\\
\hline
$\hat\CC_{1(j,j)}$	&	$\Rf\in\wedge^2({\rm Adj.})$ for $\ell$ even.\\
					&	$\Rf\in{\rm Sym}^2({\rm Adj.})$ for $\ell$ odd.\\
\hline
$\AA^\Delta_{0,0(j,j)}$		&	$\Rf\in{\rm Sym}^2({\rm Adj.})$ for $\ell$ even.\\
								&	$\Rf\in\wedge^2({\rm Adj.})$ for $\ell$ odd.\\
\hline
\end{tabular}
\caption{Flavor symmetry selection rules for multiplets appearing in the $\hat\BB_1\times\hat\BB_1$ OPE in simple theories.\label{tab:flavor_selection}}
\end{center}
\end{table}

%%%%%%%%%%%%%%%%%%%%%%%%%%%%%%%%%%%%%%%%%%%%%%%%%%%%%%%%%%%%%%%%%%%%%%%%%%%%%%%%%%%%%%%%%%%%%%%%%%%%%%%%%%%%%%%%%%%%%%%%%%%%%%%%%%%%%%%%%%%%%%%%
\subsubsection*{Protected contributions to the crossing symmetry equation}
%%%%%%%%%%%%%%%%%%%%%%%%%%%%%%%%%%%%%%%%%%%%%%%%%%%%%%%%%%%%%%%%%%%%%%%%%%%%%%%%%%%%%%%%%%%%%%%%%%%%%%%%%%%%%%%%%%%%%%%%%%%%%%%%%%%%%%%%%%%%%%%%

Here we collect the contributions to the crossing symmetry equation \eqref{bootstrapeqn} coming from short multiplets and that are completely fixed following the discussion in Section~\ref{Sec:BhatWardID}.

%%%%%%%%%%%%%%%%%%%%%%%%%%%%%%%%%%%%%%%%%%%%%%%%%%%%%%%%%%%%%%%%%%%%%%%%%%%%%%%%%%%%%%%%%%%%%%%%%%%%%%%%%%%%%%%%%%%%%%%%%%%%%%%%%%%%%%%%%%%%%%%%
\subsubsection*{$\suf(2)$ global symmetry}
%%%%%%%%%%%%%%%%%%%%%%%%%%%%%%%%%%%%%%%%%%%%%%%%%%%%%%%%%%%%%%%%%%%%%%%%%%%%%%%%%%%%%%%%%%%%%%%%%%%%%%%%%%%%%%%%%%%%%%%%%%%%%%%%%%%%%%%%%%%%%%%%

For the global symmetry $\suf(2)$ the single variable functions $f_i(z)$ are shown in \eqref{f_su2}. From these single variable functions, the spectrum and OPE coefficients of short multiplets contributing to the four-point function can be determined in the manner described in Section \ref{Sec:BhatWardID}. The contributions of these short multiplets to the two-variable functions $\GG_i(z,\zb)$ are then given by infinite sums of the type displayed on the second line in \eqref{Gshort_Glong}. Performing the sums yields the following expressions,
%%%%%%
\begin{eqnarray}\label{G_short_su2}
\GG^{\rm short}_{\mathbf{1}}(z,\bar z)&=&\frac{ \log (1-\bar{z}) \left(k (6-z (z (c ((z-2) z+2)-6)+12))-8 c (z-1) z^2\right)}{c k (z-\bar{z})(z-1)^2}\nn \\
&&+\frac{\log (1-z)  \left(k (\bar{z} (\bar{z} (c ((\bar{z}-2) \bar{z}+2)-6)+12)-6)+8 c (\bar{z}-1) \bar{z}^2\right)}{c k  (z-\bar{z})(\bar{z}-1)^2}\nn \\
&&-\frac{6   \log (1-z) \log (1-\bar{z})}{c  z \bar{z} }~,\\
\GG^{\rm short}_{\mathbf{3}}(z,\bar z)&=&\frac{(z-2) z (z (k z+4)-4) \log (1-\bar{z})}{k (z-\bar{z})(z-1)^2}-\frac{(\bar{z}-2) \bar{z} (\bar{z} (k \bar{z}+4)-4) \log (1-z)}{k (z-\bar{z})(\bar{z}-1)^2}~,
\nn  \\
\GG^{\rm short}_{\mathbf{5}}(z,\bar z)&=&\frac{\bar{z}^2 (k \bar{z}^2-2 (2 + k) ( \bar{z}-1)) \log (1-z)}{k (\bar{z}-1)^2 (z-\bar{z})}  - \frac{z^2  (k z^2-2 (2 + k) (z-1)) \log (1-\bar{z})}{k (z-1)^2  (z-\bar{z})}~.\nn
\end{eqnarray}
%%%%%%
These expressions are part of the input to the ``known'' part of the amplitude denoted as $a_i(z,\zb)$ in \eqref{eq:short_long_decomp}.

%%%%%%%%%%%%%%%%%%%%%%%%%%%%%%%%%%%%%%%%%%%%%%%%%%%%%%%%%%%%%%%%%%%%%%%%%%%%%%%%%%%%%%%%%%%%%%%%%%%%%%%%%%%%%%%%%%%%%%%%%%%%%%%%%%%%%%%%%%%%%%%%
\subsubsection*{$\ef_6$ global symmetry}
%%%%%%%%%%%%%%%%%%%%%%%%%%%%%%%%%%%%%%%%%%%%%%%%%%%%%%%%%%%%%%%%%%%%%%%%%%%%%%%%%%%%%%%%%%%%%%%%%%%%%%%%%%%%%%%%%%%%%%%%%%%%%%%%%%%%%%%%%%%%%%%%

For $\ef_6$ global symmetry, the single-variable functions $f_i(z)$, obtained by acting with the appropriate projectors on \eqref{fABCD}, are given by
%%%%%%
\begin{eqnarray}\label{f_e6}
f_{\text{\bf{1}}}(z) &=&\frac{k (z (z ((z-2) z+80)-156)+78)+48 (z-1) z^2}{k (z-1)^2}~,
\nn \\
f_{\text{\bf{650}}}(z) &=&\frac{z^2 (k ((z-2) z+2)+12 (z-1))}{k (z-1)^2}~,
\nn \\
f_{\text{\bf{2430}}}(z) &=&\frac{z^2 (k ((z-2) z+2)-4 z+4)}{k (z-1)^2}~,
\\
f_{\text{\bf{78}}}(z) &=&-\frac{(z-2) z (z (k z+24)-24)}{k (z-1)^2}~,
\nn \\
f_{\text{\bf{2925}}}(z) &=&-\frac{(z-2) z^3}{(z-1)^2}~.\nn
\end{eqnarray}
%%%%%%

The functions $\GG^{\rm short}_{i}(z,\bar z)$ are again computed by fixing the OPE coefficients for all short multiplets as described in Section~\ref{Sec:BhatWardID} and performing the infinite sums like in \eqref{Gshort_Glong}. We find:
%%%%%%
\begingroup
\allowdisplaybreaks[1]
\begin{align}\label{G_short_e6}
\GG^{\rm short}_{\mathbf{1}}(z,\bar z)&=\frac{ \log (1-\bar{z}) \left(k (156-z (z (c ((z-2) z+2)-156)+312))-48 c (z-1) z^2\right)}{c k (z-\bar{z})(z-1)^2}\nn\\
&+\frac{\log (1-z)  \left(k (\bar{z} (\bar{z} (c ((\bar{z}-2) \bar{z}+2)-156)+312)-156)+48 c (\bar{z}-1) \bar{z}^2\right)}{c k  (z-\bar{z})(\bar{z}-1)^2}\nn\\
&-\frac{156 \log (1-z) \log (1-\bar{z})}{c  z \bar{z}  }~,
\nn \\
\GG^{\rm short}_{\mathbf{650}}(z,\bar z)&=\frac{ \bar{z}^2 (k \bar{z}^2+2 (k-6) (1-\bar{z})) \log (1-z)}{k  (\bar{z}-1)^2 (z-\bar{z})}- \frac{z^2 (k z^2+2 (k-6) (1-z)) \log (1-\bar{z})}{k (z-1)^2 (z-\bar{z})}~,
\\
\GG^{\rm short}_{\mathbf{2430}}(z,\bar z)&=\frac{ \bar{z}^2 (k ((\bar{z}-2) \bar{z}+2)-4 \bar{z}+4) \log (1-z)}{k  (\bar{z}-1)^2 (z-\bar{z})}-\frac{z^2  (k ((z-2) z+2)-4 z+4) \log (1-\bar{z})}{k (z-1)^2 (z-\bar{z})}~,
\nn \\
\GG^{\rm short}_{\mathbf{78}}(z,\bar z)&=\frac{(z-2) z (z (k z+24)-24) \log (1-\bar{z})}{k (z-\bar{z})(z-1)^2}-\frac{(\bar{z}-2) \bar{z} (\bar{z} (k \bar{z}+24)-24) \log (1-z)}{k (z-\bar{z})(\bar{z}-1)^2}~,
\nn \\
\GG^{\rm short}_{\mathbf{2925}}(z,\bar z)&=\frac{(z-2) z^3 \log (1-\bar{z})}{(z-\bar{z})(z-1)^2}-\frac{(\bar{z}-2) \bar{z}^3 \log (1-z)}{(z-\bar{z})(\bar{z}-1)^2}~.\nn
\end{align}
\endgroup
%%%%%%
%%%%%%%%%%%%%%%%%%%%%%%%%%%%%%%%%%%%%%%%%%%%%%%%%%%%%%%%%%%%%%%%%%%%%%%%%%%%%%%%%%%%%%%%%%%%%%%%%%%%%%%%%%%%%%%%%%%%%%%%%%%%%%%%%%%%%%%%%%%%%%%%
\subsection{Superconformal blocks for the \texorpdfstring{$\EE_r$}{Er} four-point function}
\label{subapp:eps_conformal_blocks}
%%%%%%%%%%%%%%%%%%%%%%%%%%%%%%%%%%%%%%%%%%%%%%%%%%%%%%%%%%%%%%%%%%%%%%%%%%%%%%%%%%%%%%%%%%%%%%%%%%%%%%%%%%%%%%%%%%%%%%%%%%%%%%%%%%%%%%%%%%%%%%%%

In the case of the four-point function of $\NN=2$ chiral operators described in Section \ref{sec:epsilon_correlator}, there are two qualitatively different sets of superconformal blocks corresponding to the \emph{chiral channel} and the \emph{non-chiral channel} for the double OPE (see Fig.~\ref{fig:Eps_OPE_channels}). In the first part of this appendix, we sketch the arguments that lead to the superconformal selection rules for these two OPE channels. It is explained in Section \ref{sec:epsilon_correlator} that, for the purposes of crossing symmetry, it is useful to change basis and introduce three channels $\hat{1}$, $\hat{2}$, and $\hat{3}$. In the second part of this appendix, we present the superconformal blocks for these different channels.

%%%%%%%%%%%%%%%%%%%%%%%%%%%%%%%%%%%%%%%%%%%%%%%%%%%%%%%%%%%%%%%%%%%%%%%%%%%%%%%%%%%%%%%%%%%%%%%%%%%%%%%%%%%%%%%%%%%%%%%%%%%%%%%%%%%%%%%%%%%%%%%%
\subsubsection{Selection rules in the non-chiral channel}
\label{subapp:epps_nonchiral_selection}
%%%%%%%%%%%%%%%%%%%%%%%%%%%%%%%%%%%%%%%%%%%%%%%%%%%%%%%%%%%%%%%%%%%%%%%%%%%%%%%%%%%%%%%%%%%%%%%%%%%%%%%%%%%%%%%%%%%%%%%%%%%%%%%%%%%%%%%%%%%%%%%%

The set of representations that may appear in an $\EE_{r_0}\times\bar{\EE}_{-r_0}$ OPE can be determined by means of a simple selection rule. Without loss of generality, we may focus on conformal primary operators. Then let us consider an operator $\OO(x)$ that is a conformal primary but a descendant of a superconformal primary $\OO^\prime(x)$. The selection rule that we will derive below can then be summarized as follows,
%%%%%%
\begin{equation}\label{eq:eps_selection_rule}
\big\langle\phi(x_1)\bar{\phi}(x_2)\OO(x_3)\big\rangle\neq0 ~~\implies~~ \big\langle\phi(x_1)\bar{\phi}(x_2)\OO^\prime(x_3)\big\rangle\neq0~.
\end{equation}
%%%%%%
In other words, for any operator that is a super-descendant to have a nonvanishing three-point function with an $\NN=2$ chiral primary and its conjugate, the superconformal primary for that operator must also have such a nonvanishing three-point function.

This selection rule follows from a direct application of superconformal Ward identities. The relevant Ward identities have been derived in \cite{Baggio:2012rr}, and they take the following form,
%%%%%
\begin{equation}\label{E_Wardid}
\psi^\aa (x_3) \Big\langle \phi(x_1) \bar{\phi}(x_2) \left[ \QQ^\II_\aa , \OO \right\rbrace (x_3) \Big\rangle + \partial_{\aa \aad} \psi^{\aa}(x_3) \Big\langle \phi(x_1) \bar{\phi}(x_2) \left[ \wt{\SS}^{\II,\aad} , \OO \right\rbrace(x_3) \Big\rangle =0~.
\end{equation}
%%%%%%
As in \cite{Baggio:2012rr}, the commutators appearing in the above expression should be interpreted as meaning that the relevant commutator has been computed at the origin and the resulting operator has been translated to the appropriate insertion point. An analogous identity holds with $\wt{\QQ}_{\II,\aad}$ and $\SS_{\II}^{\aa}$. Now if $\OO(x_3)$ is a superconformal primary operator itself, then the second term in \eqref{E_Wardid} vanishes, from which it follows that operators of the form $\left[\QQ^\II_\aa,\OO(x)\right.\!\!\big\}$ cannot appear in the $\phi\times\bar\phi$ OPE. If instead we take $\OO(x)= \big[\wt{\QQ}_{\JJ,\bbd}, \OO^\prime(x) \big\}$, with $\OO^\prime$ being a superconformal primary, then some algebraic manipulations lead to the following form of the Ward identity,\footnote{In this calculation we have assumed that $\OO^\prime(x_3)$ is bosonic. A similar calculation leading to the same conclusion holds in the fermionic case.}
%%%%%%
\begin{eqnarray}\label{eq:eps_ward_complicated}
\psi^\aa (x_3) \langle \phi(x_1) \bar{\phi}(x_2) &&\!\!\!\!\!\!\!\!\!\left\lbrace \QQ^\II_\aa, \left[ \wt{\QQ}_{\JJ,\bbd}, \OO^{\prime \KK_1,\ldots \KK_n}_{\aa_1 \ldots \aa_{2j} \aad_1 \ldots \aad_{2 j}} \right] \right\rbrace (x_3) \rangle = \nn\\
&-&~  \partial_{\aa \aad} \psi^{\aa}(x_3)  \left( \delta_\JJ^\II \left(j \langle \phi(x_1) \bar{\phi}(x_2)  \OO'^{\KK_1,\ldots \KK_n}_{\aa_1 \ldots \aa_{2j}  \bbd ( \aad_1 \ldots \aad_{2 j-1}} (x_3)\rangle \delta_{\aad_{2j})}^{\aad} \right. \right.  \nn \\
&+&~ \left. \left(\tfrac{\Delta-j+r-n}{2} \right)\delta^{\aad}_{\bbd} \langle \phi(x_1) \bar{\phi}(x_2)  \OO'^{\KK_1,\ldots \KK_n}_{\aa_1 \ldots \aa_{2j} \aad_1 \ldots \aad_{2 j}}(x_3) \rangle \right)\nn \\
&+&~\left. \delta^{\aad}_{\bbd} \delta_\JJ^{(\KK_1}  \langle \phi(x_1) \bar{\phi}(x_2)  \OO'^{\KK_2,\ldots \KK_n),\II}_{\aa_1 \ldots \aa_{2j} \aad_1 \ldots \aad_{2 j}} (x_3) \rangle\right)~.
\end{eqnarray}
%%%%%%
where $r$ and $\Delta$ are the $U(1)_r$ charge and dimension of $\OO^\prime$.
It follows from this identity that the three-point function including the superconformal descendant $\left\lbrace \QQ^\II_\aa, \left[ \wt{\QQ}_{\JJ,\bbd}, \OO^{\prime \KK_1,\ldots \KK_n}_{\aa_1 \ldots \aa_{2j_1} \aad_1 \ldots \aad_{2 j_2}} \right] \right\rbrace $ is fixed in terms of the three point function of the superconformal primary. Similar results can be derived for all higher descendants of $\OO^\prime(x)$ using \eqref{E_Wardid} plus the corresponding relation involving the conjugate supercharges. All told, we are left with the selection rule given above in \eqref{eq:eps_selection_rule}.

Given these selection rules, the possible superconformal representations that may appear in the $\phi\times\bar\phi$ OPE are severely restricted. Namely, only representations for which the superconformal primary has $R=r=0$ and $j\colonequals j_1=j_2$ may appear. A brief survey of the representations in Appendix \ref{App:representations} leads to the following list,
%%%%%%
\begin{equation}
\EE_{r_0(0,0)} \times \bar\EE_{-r_0(0,0)} ~~\sim~~ \mathbf{1} ~+~ \hat{\CC}_{0(j,j)} ~+~ \AA^{\Delta}_{\,0,0(j,j)}~.
\end{equation}
%%%%%%
We should note that this selection rule has only been derived here for the \emph{superconformal primaries} of the $\EE_{r_0(0,0)} $ and $\bar\EE_{-r_0(0,0)}$ multiplets.

%%%%%%%%%%%%%%%%%%%%%%%%%%%%%%%%%%%%%%%%%%%%%%%%%%%%%%%%%%%%%%%%%%%%%%%%%%%%%%%%%%%%%%%%%%%%%%%%%%%%%%%%%%%%%%%%%%%%%%%%%%%%%%%%%%%%%%%%%%%%%%%%
\subsubsection{Selection rules in the chiral channel}
\label{subapp:eps_chiral_selection}
%%%%%%%%%%%%%%%%%%%%%%%%%%%%%%%%%%%%%%%%%%%%%%%%%%%%%%%%%%%%%%%%%%%%%%%%%%%%%%%%%%%%%%%%%%%%%%%%%%%%%%%%%%%%%%%%%%%%%%%%%%%%%%%%%%%%%%%%%%%%%%%%

The selection rules for the chiral OPE can be determined by a generalization of arguments of \cite{Poland:2010wg}, where the analogous problem for $\NN=1$ SCFTs was considered. Suppose an operator $\OO(x)$ appears in the $\phi_{r_0}\times\phi_{r_0}$ OPE. Ordinary non-supersymmetric selection rules imply that $\OO$ must be an $SU(2)_R$ singlet with $r_{\OO}=2r_0$ and $j\colonequals j_1=j_2\in\Zb$. There are then additional constraints that come from the supersymmetry properties of the chiral operators that are being multiplied. Namely, we observe that for any $x$, we have
%%%%%%
\begin{equation}
[\QQ_\alpha^\II,\phi_{r_0}(x)]=0~,\qquad [\wt{\SS}^{\II,\dot\alpha},\phi_{r_0}(x)]=0~.
\end{equation}
%%%%%%
The first condition is simply a part of the definition of the $\EE_{r}$ multiplet. The latter is automatic when $x=0$ because $\phi_{r_0}$ is the superconformal primary in its representation. For $x \neq 0$, we note the following relation from the $\NN=2$ superconformal algebra,
%%%%%%
\begin{equation}
[\PP_{\alpha \dot{\alpha}},\wt{\SS}^{\II,\dot{\beta}}]=\delta_{\dot{\alpha}}^{\dot{\beta}} \QQ^\II_\alpha~.
\end{equation}
%%%%%%
It follows that when $\phi_{r_0}$ is translated away from the origin, its variation under the action of $\wt{\SS}^{\II,\dot\alpha}$ is proportional to its variation under the action of a chiral supercharge, which vanishes.

Thus we see that $\phi_{r_0}(x_1)\times\phi_{r_0}(x_2)$ itself is invariant under the action of $\QQ_\alpha^\II$ and $\wt{\SS}^{\II,\dot\alpha}$, and so must be any operator appearing in the corresponding OPE,
%%%%%%
\begin{equation}
[\QQ_\alpha^\II,\OO(x)] = 0~, \qquad [\wt{\SS}^{\II,\dot\alpha},\OO(x)]=0~.
\end{equation}
%%%%%%
The only superconformal primary operator that can appear in the chiral OPE is therefore that of an $\EE_{2r}$ multiplet, and its superconformal descendants are excluded from appearing. Any other operator that appears must be a superconformal descendant obtained by acting on a given superconformal primary with all possible supercharges $\QQ^\II_\alpha$ that do not annihilate it. Thus only one conformal family per superconformal multiplet can contribute, and the superconformal blocks in this channel will be equal to the conventional conformal blocks for that family.

Upon consulting the catalogue of $\NN=2$ superconformal multiplets reviewed in Appendix \ref{App:representations}, it is straightforward to identify the multiplets that fit the bill. (For simplicity, we temporarily assume that $r_0>1$.) To illustrate the procedure, let us consider the case of long multiplets. The above argument implies that a long multiplet may only contribute to this OPE via a descendant of the schematic form $\OO=\QQ^4 \OO^\prime$, where $\OO^\prime$ is a superconformal primary. This descendant must be an $SU(2)_R$ singlet with $r_{\OO}=r_{\OO^\prime}+2=2r_0$ and spin $\ell_{\OO}=2j=\ell_{\OO^\prime}$. The relevant long multiplet is therefore of type $\AA_{0,2r_0-2(j,j)}$. Unitarity requires that the dimension of the superconformal primary satisfies $\Delta_{\OO^\prime}\geqslant 2r_0+\ell$, so the contributing descendant will have $\Delta_{\OO} \geqslant 2r_0+\ell+2$. 

Similar reasoning leads to the complete list of short multiplets that may contribute to the OPE, with the final selection rule taking the form
%%%%%%
\begin{equation}\label{EEOPE}
\EE_{r_0 (0,0)} \times \EE_{r_0(0,0)} \sim \EE_{2r_0(0,0)} + \CC_{0,2r_0-1(j-1,j) }+ \BB_{1,2r_0-1(0,0)} + \CC_{\frac12,2r_0-\frac32 (j-\frac12,j)}+\AA_{0,2r_0-2(j,j)}~.
\end{equation}
%%%%%%
We note that again, this derivation applies only to the OPE for superconformal primaries of the $\EE_{r_0(0,0)} $ multiplets. For $r_0 = 1$ we can find additional short multiplets of types
%%%%%%
\begin{equation}
\DD_{1(0,0)}, \qquad \hat{\CC}_{\frac12(j-\frac12,j)} ,\qquad \hat{\CC}_{0(j-1,j)}~.
\end{equation}
%%%%%%
The last of these multiplets contains higher spin conserved currents, as is to be expected since the chiral operator with $r_0=1$ is a free scalar field.

%%%%%%%%%%%%%%%%%%%%%%%%%%%%%%%%%%%%%%%%%%%%%%%%%%%%%%%%%%%%%%%%%%%%%%%%%%%%%%%%%%%%%%%%%%%%%%%%%%%%%%%%%%%%%%%%%%%%%%%%%%%%%%%%%%%%%%%%%%%%%%%%
\subsubsection{Superconformal blocks in the non-chiral channel}
\label{subapp:eps_blocks_nonchiral}
%%%%%%%%%%%%%%%%%%%%%%%%%%%%%%%%%%%%%%%%%%%%%%%%%%%%%%%%%%%%%%%%%%%%%%%%%%%%%%%%%%%%%%%%%%%%%%%%%%%%%%%%%%%%%%%%%%%%%%%%%%%%%%%%%%%%%%%%%%%%%%%%

The superconformal blocks for the various representations appearing in the non-chiral channel have been determined in \cite{Fitzpatrick:2014oza}. In the language of Section \ref{sec:epsilon_correlator}, these are the superconformal blocks in the $\hat{1}$ channel. They are as follows,
%%%%%%
\begin{eqnarray}
\makebox[.57in][l]{$\GG_{\hat{1}}^{\rm Id}(z,\zb)$} 			&\colonequals& 1~,\nn\\
\makebox[.57in][l]{$\GG_{\hat{1}}^{\hat\CC,\ell}(z,\zb)$} 		&\colonequals& \frac{z \zb}{z-\zb} \left(\left(-\tfrac{z}{2}\right)^{\ell}z\, {}_2F_1\left(\ell+1,\ell+3;2\ell+4;z)\right)-z\leftrightarrow\zb\right)~,\\
\makebox[.57in][l]{$\GG_{\hat{1}}^{\Delta,\ell}(z,\zb)$} 		&\colonequals& \frac{(z \zb)^{\frac{\Delta-\ell}{2}}}{z-\zb}\left(\left(-\tfrac{z}{2}\right)^{\ell}z\, {}_2F_1\left(\tfrac{1}{2}\left(\Delta+\ell\right),\tfrac{1}{2}\left(\Delta+\ell+4\right);\Delta+\ell+2;z)\right) \right.\nn\\
&&\qquad\qquad\qquad\qquad\left. \times\;{}_2F_1\left(\tfrac{1}{2}\left(\Delta-\ell-2\right),\tfrac{1}{2}\left(\Delta-\ell+2\right);\Delta-\ell;\zb)\right)-z\leftrightarrow\zb\right)~,\nn
\end{eqnarray}
%%%%%%
Note that the superconformal block for the $\hat{\CC}_{0(j,j)}$ representation is just the specialization of the superconformal block for a long multiplet to the case $\Delta=\ell+2$. This is to be expected based on the recombination rules of Appendix \ref{App:representations}. The superconformal block for a long multiplet can be decomposed into ordinary conformal blocks, which makes manifest the collection of conformal families from this multiplet that contribute to the four-point function:
%%%%%%
\begin{equation}\label{eq:eps_superblock_decomp}
\begin{split}
&\GG_{i=\hat{1}}^{\Delta,\ell}(z,\zb)=	 																	u^{\frac{\Delta-\ell }{2}}  G_{\Delta}^{(\ell)}(u,v)
+\left(\tfrac{1}{2(\Delta-\ell)}-\tfrac{1}{4}\right) 														u^{\frac{\Delta-\ell+2}{2}} G_{\Delta+1}^{(\ell-1)}(u,v)
-\tfrac{(\Delta+\ell)}{(\Delta+\ell+2)}																		u^{\frac{\Delta-\ell }{2}}  G_{\Delta+1}^{(\ell+1)}(u,v)\\	
&+\tfrac{(\Delta+\ell)^2}{4(\Delta+\ell+1)(\Delta+\ell+3)}													u^{\frac{\Delta-\ell}{2}}   G_{\Delta+2}^{(\ell+2)}(u,v)
+\tfrac{(\Delta-\ell-2)(\Delta+\ell)}{4(\Delta-\ell)(\Delta+\ell+2)}										u^{\frac{\Delta-\ell+2}{2}} G_{\Delta+2}^{(\ell)}(u,v)\\
& +\tfrac{(\Delta-\ell-2)^2}{64 \left((\Delta-\ell)^2-1\right)}												u^{\frac{\Delta-\ell+4}{2}} G_{\Delta+2}^{(\ell-2)}(u,v)
-\tfrac{(\Delta-\ell-2)^2(\Delta+\ell)}{64(\Delta-\ell-1)(\Delta-\ell+1)(\Delta+\ell+2)}					u^{\frac{\Delta-\ell+4}{2}} G_{\Delta+3}^{(\ell-1)}(u,v) \\
&-\tfrac{(\Delta-\ell-2)(\Delta+\ell)^2}{16(\Delta-\ell)(\Delta+\ell+1)(\Delta+\ell+3)}						u^{\frac{\Delta-\ell+2}{2}} G_{\Delta+3}^{(\ell+1)}(u,v)
+\tfrac{(\Delta-\ell-2)^2(\Delta+\ell)^2}{256(\Delta-\ell-1)(\Delta-\ell+1)(\Delta+\ell+1)(\Delta+\ell+3)}	u^{\frac{\Delta-\ell+4}{2}} G_{\Delta+4}^{(\ell)}(u,v)~.
\end{split}
\end{equation}
%%%%%%
The same multiplets contributing to the non-chiral channel also contribute to the $\hat{3}$ channel via the ``braided'' version of the above superconformal blocks. The braided version is obtained by replacing each $G_{\Delta}^{(\ell)}$ by $(-1)^{\ell}\,G_{\Delta}^{(\ell)}$ in \eqref{eq:eps_superblock_decomp}.

%%%%%%%%%%%%%%%%%%%%%%%%%%%%%%%%%%%%%%%%%%%%%%%%%%%%%%%%%%%%%%%%%%%%%%%%%%%%%%%%%%%%%%%%%%%%%%%%%%%%%%%%%%%%%%%%%%%%%%%%%%%%%%%%%%%%%%%%%%%%%%%%
\subsubsection{Superconformal blocks in the chiral channel}
\label{subapp:eps_chiral_blocks}
%%%%%%%%%%%%%%%%%%%%%%%%%%%%%%%%%%%%%%%%%%%%%%%%%%%%%%%%%%%%%%%%%%%%%%%%%%%%%%%%%%%%%%%%%%%%%%%%%%%%%%%%%%%%%%%%%%%%%%%%%%%%%%%%%%%%%%%%%%%%%%%%

Because the supermultiplets appearing in the chiral channel contribute a single conformal family to the four point function, the superconformal blocks in the chiral channel (or $\hat 2$ channel in the language of Section \ref{sec:epsilon_correlator}) are just the conventional conformal blocks appropriate to those conformal families. Table~\ref{tab:eps_chiral_blocks} displays the corresponding block for each allowed supermultiplet.

%%%%%%
\begin{table}[h!t]
\begin{center}
\renewcommand{\arraystretch}{1.75}
\begin{tabular}{|l|l|l|}
\hline
~Multiplet~		&	Contribution to $\GG_{\hat{i}=\hat{2}}(u,v)$	&	Restrictions\\
\hline
\hline
$\AA_{0,2r_0-2(j,j)}$			&	$u^{\frac{\Delta-\ell}{2}}\,G_{\Delta}^{(\ell=2j)}(u,v)$ & $\Delta \geqslant 2+2r_0+\ell$ \\
\hline
$\EE_{2r_0}$					&	$u^{r_0}\,G_{2r_0}^{(0)}(u,v)$ &\\
\hline
$\CC_{0,2r_0-1(j-1,j)}$			&   $u^{r_0}\,G_{2r_0+\ell}^{(\ell)}(u,v)$ & $\ell \geqslant 2$\\
\hline
$\BB_{1,2r_0-1(0,0)}$			&	$u^{r_0+1}\,G_{2r_0+2}^ {(0)}(u,v)$ & \\
\hline
$\CC_{\frac12,2r_0-\frac32(j-\frac12,j)}$	&	$u^{r_0+1}\,G_{2r_0+\ell+2}^{(\ell)}(u,v)$ & $\ell \geqslant 2$\\
\hline\hline
$\DD_{1 (0,0)}$				&  $u^2 \,G_{\Delta=4}^ {(0)}$ & $r_0=1$\\
\hline
$\hat{\CC}_{\frac12(j-\frac12,j)}$	&  $u^{2}\,G_{\Delta=\ell+4}^{(\ell)}$ & $\ell\geqslant 2;~r_0=1$\\
\hline
$\hat{\CC}_{0(j-1,j)}$				&  $u\,G_{\Delta=\ell+2}^ {(\ell)}$ & $\ell\geqslant 2;~r_0=1$\\
\hline
\end{tabular}
\caption{Superconformal blocks for the $\EE_{r_0}$ four point function in the $\hat{2}$ channel.\label{tab:eps_chiral_blocks}}
\end{center}
\end{table}
%%%%%%

The fourth and fifth lines in Table \ref{tab:eps_chiral_blocks} correspond to short representations that lie at the unitarity bound for long multiplets. Accordingly, their superconformal blocks are simply the specializations of the long multiplet block to appropriate values of $\Delta$ and $\ell$. On the other hand, the first two classes of short representations are separated from the continuous spectrum of long multiplets by a gap. The last three representations are only present when we relax our assumption that there are no higher spin conserved currents or free fields in the theory.
%!TEX root = ../draft_maxi_Neq2.tex

\section{Semidefinite programming and polynomial inequalities}
\label{App:semidefinite}

This appendix is devoted to a review of the methods of \cite{Poland:2011ey}, whereby the search for a linear functional of the type described in Section \ref{Sec:Numerics} can be recast as a \emph{semidefinite program}. The principal observation that leads to this reformulation is that, up to a universal prefactor, any derivative of a conformal block for fixed $\ell$ can be arbitrarily well approximated by a \emph{polynomial} in the conformal dimension $\Delta$, that is
%%%%%%
\begin{equation}\label{eq:poly_approx}
\del_z^m \del_{\bar z}^n G_{\Delta}^{(\ell)}(z,\bar z)|_{z = \bar z = \frac{1}{2}} ~\approx~ \chi(\Delta,\ell) \PP^{(\ell)}_{m,n} (\Delta)~.
\end{equation}
%%%%%%
Here $\chi(\Delta,\ell)$ may be complicated, but it is positive for all physical values of $\Delta$ and $\ell$ and is independent of the choice of derivative. On the other hand, $\PP^{(\ell)}_{m,n}(\Delta)$ is a finite order polynomial in $\Delta$. For the superconformal blocks appearing in this paper, the details of this polynomial approximation are explained below in Appendix \ref{App:polyapprox}. 

With the aid of this approximation, we consider the action of a linear functional on smooth functions of $z$ and $\zb$ of the form
%%%%%%
\begin{equation}
\phi[F(z,\zb)]=\sum_{m,n=0}^{\Lambda}a_{m,n}\restr{\partial_z^m\partial_{\zb}^nF(z,\zb)}{z=\zb=\frac12}~.
\end{equation}
%%%%%%
Up to the positive prefactor described above, the action of this functional on a conformal block is now given by a finite order polynomial in the conformal dimension,
%%%%%%
\begin{equation}\label{eq:positive_poly}
\phi[G_{\Delta}^{(\ell)}(z,\zb)]=\chi(\Delta,\ell)\,\sum_{m,n=0}^{\Lambda}a_{m,n}\PP_{m,n}^{(\ell)}(\Delta)\equalscolon \goodchi(\Delta,\ell)\, \PP^{\ell}(\Delta)~.
\end{equation}
%%%%%%
The numerical problem in question (see Section \ref{Sec:Numerics}) is thus transformed into a search in the space of $a_{m,n}\in\Rb$ such that the polynomial $\PP^{\ell}(\Delta)\geqslant0$ for $\Delta\geqslant\Delta_{\ell}^{\star}$ for each $\ell$. Note that the range of values of $\Delta$ for which the polynomial must be positive is always bounded from below, either by the unitarity bound or by the chosen value $\Delta_{\ell}^\star$. 

A polynomial in $\Delta$ that is positive for all $\Delta > \Delta^{\star}$ can always be decomposed as follows,
%%%%%
\begin{equation}\label{eq:decompose_positive_poly}
\PP(\Delta)=P(\Delta) + (\Delta - \Delta^*) Q(\Delta)~,
\end{equation}
%%%%%%
where $P(\Delta)$ and $Q(\Delta)$ are polynomials that are positive for all real $\Delta$. Furthermore, in terms of the monomial vector $\vec\Delta \colonequals (1,\Delta,\Delta^2,\ldots,\Delta^N)$, such non-negative polynomials can always be written as
%%%%%
\begin{equation}\label{eq:poly_as_matrix}
P(\Delta) = \vec\Delta^t P \vec\Delta~,\qquad Q(\Delta) = \vec\Delta^t Q \vec\Delta~,
\end{equation}
%%%%%%
where $P$ and $Q$ are positive semidefinite matrices, which is notated as $P,Q \succeq 0$. We should emphasize that the matrices $P$ and $Q$ are not completely fixed in terms of $P(\Delta)$ and $Q(\Delta)$. There is a redundancy to which we will return shortly. 

The action of the functional on conformal blocks will therefore be non-negative above some dimension $\Delta_{\ell}^{\star}$ in the spin $\ell$ channel if and only if there exist two positive semidefinite matrices, $P^{(\ell)},Q^{(\ell)} \succeq 0$ such that
%%%%%
\begin{equation}\label{eq:pos_semi_def_functional}
a_{m,n} P^{(\ell)}_{m,n} (\Delta) = \vec \Delta^t P^{(\ell)} \vec \Delta + (\Delta - \Delta_{\ell}^*) \vec \Delta^t Q^{(\ell)} \vec \Delta~.
\end{equation}
%%%%%%
In words, we are demanding that the left- and right-hand sides of \eqref{eq:pos_semi_def_functional} be the same polynomial in $\Delta$, which amounts to linear relations between the coefficients of $P^{(\ell)}$ and $Q^{(\ell)}$ and the $a_{m,n}$. Such an equation must hold for each $\ell$ appearing in the crossing symmetry equation, and if there are multiple flavor symmetry channels then there will be such an equation for each channel. The problem is thus reduced to the search for a set of positive semidefinite matrices whose entries satisfy certain linear constraints. This is a prototypical instance of a semidefinite program, the basic theory of which we review next.

%%%%%%%%%%%%%%%%%%%%%%%%%%%%%%%%%%%%%%%%%%%%%%%%%%%%%%%%%%%%%%%%%%%%%%%%%%%%%%%%%%%%%%%%%%%%%%%%%%%%%%%%%%%%%%%%%%%%%%%%
\subsubsection*{Semidefinite programming}
%%%%%%%%%%%%%%%%%%%%%%%%%%%%%%%%%%%%%%%%%%%%%%%%%%%%%%%%%%%%%%%%%%%%%%%%%%%%%%%%%%%%%%%%%%%%%%%%%%%%%%%%%%%%%%%%%%%%%%%%

A \emph{semidefinite program} (SDP) is an optimization problem wherein the goal is to minimize a linear objective function over the intersection of the cone of positive semidefinite matrices with an affine space. Such a problem can be described in terms of a vector of real variables $x_i$ as follows,
%%%%%%
\begin{equation}\label{eq:semidefprimal}
\begin{aligned}
& \underset{x_i}{\text{minimize}} & & (x_i c^i)\\
& \text{such that} & & X := x_i F^i  - F^0 \succeq 0~,
\end{aligned}
\end{equation}
%%%%%%
where $c^i$ is a fixed \emph{cost vector} that defines the objective function, and $F^i$ and $F^0$ are some fixed square matrices.

This semidefinite program has a \emph{dual problem} that is defined as the search for a positive semi-definite matrix $Y$ that maximizes an appropriate objective function and satisfies certain linear constraints,
%%%%%%
\begin{equation}\label{eq:semidefdual}
\begin{aligned}
& \underset{Y}{\text{maximize}} & & \tr(F^0 \cdot Y) \\
& \text{such that} & & Y \succeq 0~,\\
&&& \tr(F^i \cdot Y) = c^i~.
\end{aligned}
\end{equation}
%%%%%%
The original problem written in \eqref{eq:semidefprimal} -- called the \emph{primal problem} -- and the dual problem of \eqref{eq:semidefdual} are not generally guaranteed to be equivalent. Indeed, given a solution $x_i$ to the primal problem and a solution $Y$ to the dual problem, a measure of the inequivalence of the solutions is the \emph{duality gap}:
%%%%%%
\begin{equation}\label{eq:duality_gap}
x_i c^i - \tr(F^0 \cdot Y) = x_i \tr(F^i \cdot Y) - \tr(F^0 \cdot Y) = \tr(X \cdot Y) \geqslant 0~,
\end{equation}
%%%%%%
where the last line holds because both matrices are positive semidefinite. 

The absence of a duality gap, and the existence of an optimal solution to the primal (dual) problem, is guaranteed if the dual (primal) problem is bounded from above (below) and has a \emph{strictly} feasible solution, \ie, there exists a matrix $Y \succ 0$ ($X \succ 0$) satisfying the relevant constraints. This is called \emph{Slater's condition}.

%%%%%%%%%%%%%%%%%%%%%%%%%%%%%%%%%%%%%%%%%%%%%%%%%%%%%%%%%%%%%%%%%%%%%%%%%%%%%%%%%%%%%%%%%%%%%%%%%%%%%%%%%%%%%%%%%%%%%%%%
\subsection{A toy model for polynomial inequalities}
%%%%%%%%%%%%%%%%%%%%%%%%%%%%%%%%%%%%%%%%%%%%%%%%%%%%%%%%%%%%%%%%%%%%%%%%%%%%%%%%%%%%%%%%%%%%%%%%%%%%%%%%%%%%%%%%%%%%%%%%

To demonstrate the application of semidefinite programming techniques to the type of crossing symmetry problem being considered in this paper, let us consider a simplified model in which the notation is less burdensome. Namely, consider the problem of studying the space of solutions to a ``crossing symmetry'' equation of the form
%%%%%%
\begin{equation}\label{eq:simplecrossing}
\sum_k \lambda_k^2 G_{\Delta_k}(z) = c(z)~,
\end{equation}
%%%%%%
where $\Delta_k$ are allowed to vary over the entire real line. We will assume that the functions $G_{\Delta}(z)$ and their derivatives can be well approximated by polynomials in $\Delta$, so we have
%%%%%%
\begin{equation}
\partial_z^i G_{\Delta}(z) \Big|_{z = 1/2} \approx \sum_{\alpha = 0}^{2N} p^i_\alpha \Delta^\alpha \equalscolon \hat P^i(\Delta)~,
\end{equation}
%%%%%%
where we have assumed that for a given range of values of $i$, each such polynomial has degree less than or equal to some fixed even number $2N$.\footnote{For the sake of comparison, we note that in the actual crossing symmetry equations encountered in this work we have an additional $\zb$ coordinate, as well as sums over spins and possibly flavor symmetry channels. Also the values of $\Delta_k$ are bounded below in a given channel by unitarity bounds. However, these complications do not conceptually change this discussion.}

%%%%%%%%%%%%%%%%%%%%%%%%%%%%%%%%%%%%%%%%%%%%%%%%%%%%%%%%%%%%%%%%%%%%%%%%%%%%%%%%%%%%%%%%%%%%%%%%%%%%%%%%%%%%%%%%%%%%%%%%
\subsubsection{The primal problem: ruling out solutions}
%%%%%%%%%%%%%%%%%%%%%%%%%%%%%%%%%%%%%%%%%%%%%%%%%%%%%%%%%%%%%%%%%%%%%%%%%%%%%%%%%%%%%%%%%%%%%%%%%%%%%%%%%%%%%%%%%%%%%%%%

To constrain the space of solutions to such a problem, we consider acting with a linear functional $\phi$ on both sides of the equality and check for contradictions. The problem can be formalized as follows,
%%%%%%
\begin{equation}
\begin{aligned}
& \underset{\phi}{\text{minimize}} & & \phi[c(z)] \\
& \text{such that} & & \phi[G_{\Delta} (z)] \geqslant 0 \qquad \forall~\Delta~.
\end{aligned}
\end{equation}
%%%%%% 
If the minimum turns out to be negative then our toy problem has no solution. Taking $\phi[f(z)] \colonequals \sum_{i=0}^{n} \restr{a_i \del^i_z f(z)}{z=1/2}$, we can reformulate the optimization problem as follows
%%%%%%
\begin{equation}\label{eq:polyopt}
\begin{aligned}
& \underset{a_i}{\text{minimize}} & & a_i c_i \\
& \text{such that} & & a_i \hat P^i(\Delta) \geqslant 0 \qquad \forall\, \Delta~.
\end{aligned}
\end{equation}
%%%%%% 
where we have defined
%%%%%%
\begin{equation}
c_i\colonequals \restr{\partial_z^i c(z)}{z=\frac12}~.
\end{equation}
%%%%%%
In terms of the vector $\vec \Delta = (1, \Delta, \Delta^2, \ldots \Delta^{N})^t$, the second line of \eqref{eq:polyopt} requires the existence of a symmetric, positive semidefinite matrix $\hat P$ such that
%%%%%%
\begin{equation}\label{eq:toy_polynomial_to_matrix}
\hat P(\Delta) = \vec \Delta^t P \vec \Delta \qquad {\text{with}} \qquad P \succeq 0~.
\end{equation}
%%%%%%
This allows us to reformulate the polynomial inequalities as a semidefinite program. 

We begin by introducing two sets of matrices in terms of which the problem is naturally reformulated. For $N > 1$, the matrix $P$ is not completely fixed by \eqref{eq:toy_polynomial_to_matrix} because there are only $2N+1$ components in $\hat P(\Delta)$ whereas $P$ has $(N+1)(N+2)/2$ independent components. This redundancy in $P$ can be parametrized by matrices $Q$ satisfying
%%%%%%
\begin{equation}
\vec \Delta^t Q \vec \Delta = 0 \qquad \qquad \forall\, \Delta~.
\end{equation}
%%%%%%
Examples of such matrices $Q$ are the $3 \times 3$ matrices with $(-1,2,-1)$ on the cross-diagonal, or the $4 \times 4$ matrix with $(1,-1,-1,1)$ on the cross-diagonal. All other matrices $Q$ take a similar form, and the first set of matrices we must introduce is a complete basis for such $Q$. We denote the elements of this basis as $Q^{\hat i}$.

The second set of matrices are in one-to-one correspondence with the polynomials $\hat P^i(\Delta)$. They take the form:
%%%%%%
\begin{equation}
P^i \colonequals 
\begin{pmatrix}
p^i_0   	& \hf p^i_1 & 0         & 0       	& \ldots\\
\hf p^i_1 	& p^i_2   	& \hf p^i_3 & 0       	& \ldots\\
0       	& \hf p^i_3 & p^i_4     & \hf p^i_5 & \ldots\\
0       	& 0         & \hf p^i_5 & p^i_6   	& \ldots\\
\vdots  	& \vdots    & \vdots    & \vdots  	& \ddots
\end{pmatrix}~.
\end{equation}
%%%%%%
By construction these matrices satisfy the condition
%%%%%%
\begin{equation}
\hat P^i(\Delta) = \vec \Delta^t P^i \vec \Delta~.
\end{equation}
%%%%%%

Armed with these matrices we can write down the most general matrix that, upon contraction from both sides with $\vec\Delta$, gives the requisite polynomial:
%%%%%%
\begin{equation}
a_i \hat{P}^i(\Delta)=\vec \Delta^t \left(a_i P^i + b_{\hat i} Q^{\hat i}\right)\vec \Delta~,
\end{equation}
%%%%%%
where the $b_{\hat i}$ are arbitrary real parameters. The optimization \eqref{eq:polyopt} can now be rephrased as
%%%%%%
\begin{equation}\label{eq:polysdoptprimal}
\begin{aligned}
& \underset{a_i, b_{\hat i}}{\text{minimize}} & & a_i c_i \\
& \text{such that} & & a_i P^i + b_{\hat i} Q^{\hat i} \succeq 0~,
\end{aligned}
\end{equation}
%%%%%% 
which we recognize to be precisely a semidefinite program of the form given in \eqref{eq:semidefprimal}, with
%%%%%%
\begin{equation}
x_i \sim (a_i,b_{\hat i})~,\qquad
F^i \sim (P^i,Q^{\hat i})~,\qquad
F^0 = 0~.
\end{equation}
%%%%%%
The constraints in \eqref{eq:polysdoptprimal} are invariant under an overall rescaling of the $(a_i,b_{\hat i})$, so the optimal value is either zero or negative infinity. To render the primal formulation bounded we can introduce an additional normalization constraint
%%%%%%
\begin{equation}\label{eq:traceconstr}
\tr(P)=a_i \tr(P^i) + b_{\hat i} \tr(Q^{\hat i}) = 1~.
\end{equation}
%%%%%%
This condition is always enforceable because a nonzero, positive semidefinite matrix has strictly positive trace. Although other normalization conditions are possible, we will see that \eqref{eq:traceconstr} is particularly natural from the perspective of the dual problem. In practice, we can simply solve  the additional constraint for, say, $a_1$ to end up with a bounded variation of \eqref{eq:polysdoptprimal}.

%%%%%%%%%%%%%%%%%%%%%%%%%%%%%%%%%%%%%%%%%%%%%%%%%%%%%%%%%%%%%%%%%%%%%%%%%%%%%%%%%%%%%%%%%%%%%%%%%%%%%%%%%%%%%%%%%%%%%%%%
\subsubsection{The dual problem: constructing solutions}
%%%%%%%%%%%%%%%%%%%%%%%%%%%%%%%%%%%%%%%%%%%%%%%%%%%%%%%%%%%%%%%%%%%%%%%%%%%%%%%%%%%%%%%%%%%%%%%%%%%%%%%%%%%%%%%%%%%%%%%%

Let us now address the dual problem to \eqref{eq:polysdoptprimal} with the additional constraint \eqref{eq:traceconstr}. After a little rewriting, the problem is as follows:
%%%%%%
\begin{equation}\label{eq:polysdoptdual}
\begin{aligned}
& \underset{\lambda, Y}{\text{maximize}} & & -\lambda & & \\
& \text{such that} & & Y + \lambda I \succeq 0\,, & &\\
&&& \tr(P^i \cdot Y) = c^i & & \forall~ i,\\
&&& \tr(Q^{\hat i} \cdot Y) = 0 & & \forall~ \hat i~.
\end{aligned}
\end{equation}
%%%%%%
This is a well-known form of a \emph{feasibility} problem, which is the search for a matrix $Y \succeq 0$ subject to linear constraints. If the optimal value of $\lambda$ comes out non-positive then such a matrix $Y$ exists (\ie, there is a \emph{feasible solution}), otherwise it does not. In standard applications the reason for introducing a variable $\lambda$ multiplying the identity matrix $I$ is to ensure that a \emph{strictly} feasible solution will always exist, because for $\lambda \gg 0$ the matrix $Y + \lambda I \succ 0$. Its appearance in \eqref{eq:polysdoptdual} is a consequence of the trace constraint \eqref{eq:traceconstr} in the primal problem.

Whereas the primal problem amounted to the search for functionals that certify the absence of solutions to crossing symmetry, dual problem is related to constructing solutions to crossing symmetry \cite{El-Showk:2014dwa}. Let us observe how this works for these semidefinite programs. We first solve the constraints $\tr(Q^{\hat i} \cdot Y) = 0$. The most general solution is given by
%%%%%%
\begin{equation}
Y = y^\alpha Y_\alpha~,\qquad \alpha=0,\ldots,2N,
\end{equation}
%%%%%%
with arbitrary coefficients $y^\alpha$ and with matrices $Y_\alpha$ defined as
%%%%%%
\begin{equation}
Y_0 =
\begin{pmatrix}
1 & ~0 & ~0 & ~0  & \cdots\\
~0 & ~0 & ~0 & ~0  & \cdots\\
~0 & ~0 & ~0 & ~0  & \cdots\\
~0 & ~0 & ~0 & ~0  & \cdots\\
~\vdots  & ~\vdots    & ~\vdots    & ~\vdots  & \ddots
\end{pmatrix}
,\quad
Y_1 =
\begin{pmatrix}
~0 & 1 & ~0 & ~0  & \cdots\\
1 & ~0 & ~0 & ~0  & \cdots\\
~0 & ~0 & ~0 & ~0  & \cdots\\
~0 & ~0 & ~0 & ~0  & \cdots\\
~\vdots  & ~\vdots    & ~\vdots    & ~\vdots  & \ddots
\end{pmatrix}
,\quad
Y_2 =
\begin{pmatrix}
~0 & ~0 & 1 & ~0  & \cdots\\
~0 & 1 & ~0 & ~0  & \cdots\\
1 & ~0 & ~0 & ~0  & \cdots\\
~0 & ~0 & ~0 & ~0  & \cdots\\
~\vdots  & ~\vdots    & ~\vdots    & ~\vdots  & \ddots
\end{pmatrix}~,\quad\cdots~.
\end{equation}
%%%%%%
Now let us choose tuples $(\lambda^2_k,\Delta_k)$ so that $y^\alpha = \sum_k \lambda^2_k (\Delta_k)^\alpha$. We then have $Y = \sum_k \lambda^2_k \vec \Delta_k \vec \Delta_k^t$ and the additional constraints of the form $\tr(P^i \cdot Y) = c^i$ become
%%%%%%
\begin{equation}
\sum_k \lambda_k^2 \hat P^i(\Delta_k) = c^i~.
\end{equation}
%%%%%%
This is precisely the crossing symmetry equation \eqref{eq:simplecrossing} after truncating to a finite number of derivatives.

Finally, let us comment on the duality gap and the interpretation of solutions to this problem. The freedom to set $\lambda$ to a large positive number ensures that the above formulation of the dual problem is strictly feasible. It is, however, not obviously bounded. From the formulation of the problem it is clear that this is related to the existence of solutions to crossing symmetry where $c(z) = 0$. More precisely, the problem is unbounded if there is a positive semidefinite matrix $Y$ that satisfies  $\tr(P^i \cdot Y) = 0$ and $\tr(Q^{\hat i} \cdot Y) = 0$ for all $i$ and $\hat i$. In the absence of such solutions the problem is bounded, Slater's condition is satisfied, and there is no duality gap, so for the optimal values we find that $-\lambda = a^i c_i$. This equation makes intuitive sense. Indeed, suppose the dual formulation does not find a solution to crossing symmetry. This happens when $- \lambda = a^i c_i < 0$ and therefore the primal formulation indeed provides a functional that proves that such a solution cannot exist. Similarly, suppose we do find a matrix $Y \succeq 0$ satisfying all the above constraints. In that case $-\lambda = a^i c_i \geqslant 0$, so no functional can be found in the primal problem.

%%%%%%%%%%%%%%%%%%%%%%%%%%%%%%%%%%%%%%%%%%%%%%%%%%%%%%%%%%%%%%%%%%%%%%%%%%%%%%%%%%%%%%%%%%%%%%%%%%%%%%%%%%%%%%%%%%%%%%%%
\subsubsection*{Extremal functionals}
%%%%%%%%%%%%%%%%%%%%%%%%%%%%%%%%%%%%%%%%%%%%%%%%%%%%%%%%%%%%%%%%%%%%%%%%%%%%%%%%%%%%%%%%%%%%%%%%%%%%%%%%%%%%%%%%%%%%%%%%

In the applications of this framework to study interesting physical theories, there are often additional parameters in the problem such as assumed gaps in the spectrum for certain spins. In such cases we are usually interested in finding the boundary in the space of such parameters between regions where crossing symmetry can and cannot be satisfied. Precisely at the boundary $- \lambda = a^i c_i = 0$. This turns out to imply that the corresponding solution to crossing symmetry is completely determined by the zeroes of the \emph{extremal functional} \cite{Poland:2010wg,ElShowk:2012hu}. This is because the absence of a duality gap implies $\tr(X \cdot Y) = 0$ which together with the above assumption on the form of $Y$ leads to
%%%%%%
\begin{equation}
a_i P^i(\Delta_k) = 0~.
\end{equation}
%%%%%%
The solution to crossing symmetry encoded in $Y$ therefore involves precisely those values of $\Delta$ for which the extremal functional vanishes. This observation leads to the following algorithm for finding the solution to crossing symmetry: one first lists the $\Delta_k$ for which the $\vec \Delta_k^t X \vec \Delta_k = 0$, and then finding the $\lambda_k^2$ reduces to solving the linear problem $y^\alpha = \sum_k \lambda_k^2 (\Delta_k)^\alpha$. Note that we require both the $X$ and the $Y$ matrix here.

%%%%%%%%%%%%%%%%%%%%%%%%%%%%%%%%%%%%%%%%%%%%%%%%%%%%%%%%%%%%%%%%%%%%%%%%%%%%%%%%%%%%%%%%%%%%%%%%%%%%%%%%%%%%%%%%%%%%%%%%
\subsection{Notes on implementation}
%%%%%%%%%%%%%%%%%%%%%%%%%%%%%%%%%%%%%%%%%%%%%%%%%%%%%%%%%%%%%%%%%%%%%%%%%%%%%%%%%%%%%%%%%%%%%%%%%%%%%%%%%%%%%%%%%%%%%%%%

In this work we have utilized the dual formulation of the semidefinite program associated to crossing symmetry. We first solved all the linear constraints analogous to those appearing in \eqref{eq:polysdoptdual}, leading to a smaller set of independent parameters that we denote $z^{\hat \alpha}$ and corresponding matrices $Z_{\hat \alpha}$. The nonzero $c_i$ lead to an inhomogeneous term that we may call $Z_{\hat 0}$. The complete semidefinite program is then as above with
%%%%%%
\begin{equation}
x^i \Rightarrow (z^{\hat \alpha}, \lambda) \,, \qquad F^i \Rightarrow (Z^{\hat \alpha},I)\,, \qquad F^0 \Rightarrow Z_{\hat 0}~,
\end{equation}
%%%%%%
and a cost vector such that only $\lambda$ is extremized. Since we were unable to rigorously show that the dual problem was bounded in all cases, we added an additional constraint $\lambda \geqslant 0$. In the primal problem this additional constraint transforms the trace equality \eqref{eq:traceconstr} into the inequality $\tr(P) \leqslant 1$. With this condition the optimal value will be zero if a solution exists and no functional is found, or strictly negative if the opposite happens.

We used SDPA and SDPA-GMP solvers \cite{Fujisawa08sdpasemidefinite,sdpasite}, which use an interior point method that simultaneously optimizes both the primal and dual problems, and that terminates when the duality gap is below a certain (small) threshold. This requires a strictly feasible solution to both the primal and the dual problem, and our formulation of the problem ensures that such strictly feasible solutions exist. Furthermore, we found that a normalization of the form given in \eqref{eq:traceconstr} improves numerical stability compared to other normalizations such as, \eg, $a^i c_i = 1$. We ascribe this difference to the fact that $a^ic_i$ naturally tends to zero in physically interesting regions, and so setting it to one as a normalization leads to large numbers elsewhere.\footnote{Our normalization is not suitable for obtaining bounds on OPE coefficients. In that case we need to normalize the functional as described in Section \ref{Sec:Numerics}.} 

%%%%%%
\begin{table}[ht!]
\begin{center}
\begin{tabular}{|l|r|}
\hline
Parameter & Value  \\
\hline
\hline 
maxIteration & $1000$ \\ 
epsilonStar & $10^{-12}$ \\ 
lambdaStar &  $10^8$ \\ 
omegaStar &  $10^6$ \\ 
lowerBound &  $-10^{30}$ \\ 
upperBound & $10^{30}$ \\ 
betaStar & 0.1 \\ 
betaBar & 0.3 \\ 
gammaStar & 0.9 \\ 
epsilonDash &  $10^{-12}$ \\ 
precision &  $200$ \\ 
\hline 
\end{tabular} 
\caption{Parameters used for the SDPA and SDPA-GMP solvers. The `precision' variable is only relevant for the SDPA-GMP solver.
\label{Tab:sdpa_param}}
\end{center}
\end{table}
%%%%%%

In order to achieve maximal numerical stability we `renormalized' many of the numbers fed into the problem. For example, the polynomials $P^i(\Delta)$ can be redefined by multiplying with an overall (positive) constant, by affine redefinitions of $\Delta$, and by choosing a different basis for the space of derivatives. Altogether these reparametrizations give us the freedom to transform the problem according to
%%%%%%
\begin{equation}
P^i (\Delta) \to M^i_j P^j(a \Delta + b)\,, \qquad \qquad c^i \to M^i_j c^j~.
\end{equation}
%%%%%%
We choose $M^i_j$, $a$, and $b$ so as to minimize the potential for numerical inaccuracies. Numerical stability can be further improved by rescaling the normalization condition $\tr(X) = 1$ to $\tr(X) = \mu$ for a positive real $\mu$. (In the dual problem $\mu$ becomes the cost vector, so this parameter is introduced through the optimization of $\mu \lambda$ instead of $\lambda$.) In order to avoid large numerical differences between the primal and the dual formulation, we choose $\mu$ large so that $X$, which is a matrix of size $O(10^3)$, can have $O(1)$ entries on its diagonal.

In previous implementations of the numerical bootstrap as a semidefinite program \cite{Poland:2011ey}, it was necessary to employ the arbitrary precision solver SDPA-GMP to avoid numerical instabilities. The setup described above, with Slater's condition satisfied and coefficients that are suitably renormalized, has allowed us to use the double precision SDPA program for low and intermediate values of $\Lambda$. Since working at machine precision is significantly faster than working at arbitrary precision, we were able to explore a much greater range of the parameter space given our computational resources. We still found it necessary to switch to SDPA-GMP for higher values of $\Lambda$, with the exact transition value somewhat dependent on the problem at hand. For example, we had to switch at $\Lambda = 16$ for the bounds on theories with $\ef_6$ flavor symmetry shown in Section \ref{Sec:Bhatresults}, but were able to obtain reliable results with double precision numerics up to $\Lambda = 22$ for some of the bounds on theories with $\suf(2)$ flavor symmetry. Typical settings for the parameters of both the SDPA and SDPA-GMP solvers can be found in Table~\ref{Tab:sdpa_param}.
%!TEX root = ../draft_maxi_Neq2.tex

\section{Polynomial approximations and conformal blocks}
\label{App:polyapprox}

The semidefinite programming approach to the numerical bootstrap depends on our ability to approximate conformal blocks of fixed spin $\ell$ and varying conformal dimension $\Delta$ by polynomials in $\Delta$ \cite{Poland:2011ey, Kos:2013tga}. This appendix includes a brief review of these approximations and some details relevant to the special cases of interest. The goal is to express the conformal blocks and their derivatives in a factorized form, with one factor being a function that can be well approximated by a polynomial in $\Delta$, and the other a non-polynomial term that is strictly positive and independent of the choice of derivative. We denote the polynomial in $\Delta$ by $\PP^{(\ell)}_{m,n} (\Delta)$ and the  non-polynomial term by $\chi(\Delta,\ell)$, so the approximation takes the following form,
%%%%%%
\begin{equation}\label{Polyappprox}
\del_z^m \del_{\bar z}^n G_{\Delta}^{(\ell)}(z,\bar z)|_{z = \bar z = \frac{1}{2}} ~\approx~ \chi(\Delta,\ell) \PP^{(\ell)}_{m,n} (\Delta)~.
\end{equation}
%%%%%%
The starting point for this approximation scheme is a recursion relation for derivatives of the hypergeometric functions appearing in conformal blocks,
%%%%%%
\begin{equation}\label{hyperg_rec_rel}
\left[ \frac{d^2}{dz^2} + \frac{1-a-b}{z-1} \frac{d}{dz}+ \frac{\beta^2 -\beta +a b z}{z^2 (z-1)}  \right] \left( z^\beta\,{}_2F_1 \left(\beta - a, \beta -b , 2 \beta , z\right) \right)=0~.
\end{equation}
%%%%%%
This recursion relation follows immediately from the fact that the ${}_2F_1$ hypergeometric function is a solution to Euler's differential equation. Using this relation, any derivative of the above hypergeometric function at fixed $z$ can be expressed as the sum of the zeroth and first order derivatives of the same hypergeometric function, each with some polynomial in $\beta$ as a prefactor. Thus the only non-polynomial feature of any derivative of the hypergeometric function can be expressed in terms of the value of the hypergeometric function itself and that of its first derivative.

To approximate conventional conformal blocks we follow exactly the same steps as in \cite{Poland:2011ey}. From \eqref{eq:bos_block} any derivative of a conformal block $\del_z^m \del_{\bar z}^n G_{\Delta}^{(\ell)}(z,\zb)$ can be rewritten, by recursive use of \eqref{hyperg_rec_rel} with $a=b=0$, in terms of the hypergeometric functions and their first derivatives. These functions encode all of the non-polynomial dependence on $\Delta$. We can then pull out factors out of the blocks such that the leftover expression can be well approximated by polynomials. To start we factor out the following term
%%%%%%
\begin{equation}
\frac{1}{\beta\bar\beta} \restr{\left(\frac{\partial}{\partial z} z^\beta {}_2 F_1(\beta,\beta,2\beta,z)\right)}{z=\frac{1}{2}}
\restr{\left(\frac{\partial}{\partial z} z^{\bar \beta} {}_2 F_1(\bar\beta,\bar\beta,2\bar\beta,z)\right)}{z=\frac{1}{2}}~.
\end{equation}
%%%%%%
Here we have $\beta=\frac{\Delta+\ell}{2}$, $\bar{\beta}=\frac{\Delta-\ell-2}{2}$. This is positive for all $\beta \geqslant -1$, and so it is positive for any conformal block appearing in a unitary theory. After factoring out this positive non-polynomial term, the remaining non-polynomial dependence is isolated in the following ratio (and a similar one for $\beta\to\bar\beta$),
%%%%%%
\begin{equation}\label{Kapprox}
K_\beta=\restr{\frac{\beta z^\beta {}_2 F_1(\beta,\beta,2\beta,z)}{\frac{\partial}{\partial z} z^\beta {}_2 F_1(\beta,\beta,2\beta,z)}}{z=\frac{1}{2}} \simeq \frac{1}{\sqrt{2}} \prod_{j=0}^{M} \frac{(\beta - r_j)}{(\beta - s_j)} \equiv\frac{N_M(\beta)}{D_M(\beta)}~.
\end{equation}
%%%%%%
The coefficient $r_j$ is the $j$--th zero of ${}_2 F_1(\beta,\beta,2\beta,z)$ and $s_j$ the $j$--th zero of $\frac{\partial}{\partial z} z^\beta {}_2 F_1(\beta,\beta,2\beta,z)$.\footnote{In practice we compute the zeros of the latter by making use of the following identity, which relates it to another hypergeometric function
%%%%%%
\begin{equation}
\frac{d^n}{dz^n}\left[ z^{\beta -a +n-1} {}_2F_1\left(\beta -a, \beta -b, 2\beta,z \right)\right] = \left(\beta -a\right)_n z^{\beta - a -1} {}_2 F_1 \left( \beta -a + n, \beta -b, 2\beta,z \right)~,
\label{hygerg_der}
\end{equation}
%%%%%%
where in this case we want to use $n=1$, and we have $a=0$.} The rational function $\frac{N_M(\beta)}{D_M(\beta)}$ is an approximation of $K_\beta$ obtained by restricting to the first $M$ zeroes of both the numerator and denominator. The approximation becomes arbitrarily good as $M$ is increased, and converges very quickly, as described in \cite{Poland:2011ey}. 

The last step is to multiply by $D(\beta)D(\bar \beta)$, which is strictly positive for the same range of $\beta$ and $\bar \beta$. In this way we have factored out all of the nonpolynomial dependence of $\del_z^m \del_{\bar z}^n G_{\Delta}^{(\ell)}(z,\zb)$, which defines $\chi(\Delta,\ell)$ in \eqref{Polyappprox}, and are left with a polynomial in $\Delta$, $\PP^{(\ell)}_{m,n} (\Delta)$, whose degree is controlled by the number of terms $M$ kept in the approximation \eqref{Kapprox}. Exactly this approximation is used for the blocks in the $\hat{2}$ channel for the $\EE_r$ correlator, and for all the blocks in the $\hat{\BB}_1$ correlator (with a shift $\Delta \to \Delta + 4$).

For superconformal blocks in the $\hat{1}$ channel given in \eqref{ESuperblock} the procedure is analogous. This time we use \eqref{hyperg_rec_rel}, where now $a=1$ and $b=-1$, to write all of the block derivatives in terms of the zeroth and second derivatives of the hypergeometric function. In this case we define $\beta=\frac{\Delta+\ell+2}{2}$ and $\bar{\beta}=\frac{\Delta-\ell}{2}$. The first step is again to factor out
%%%%%%
\begin{equation}
\restr{\left(\frac{1}{\beta (\beta-1)} \frac{\partial^2}{\partial z^2} z^\beta {}_2 F_1(\beta,\beta,2\beta,z)\right)}{z=\frac{1}{2}} \restr{\left(\frac{1}{\bar \beta (\bar \beta-1)} \frac{\partial^2}{\partial z^2} z^\beta {}_2 F_1(\bar \beta,\bar \beta,2\bar \beta,z)\right)}{z=\frac{1}{2}}~,
\end{equation}
%%%%%%
which is positive for all possible values of $\beta$ and $\bar \beta$ occurring in the relevant OPE ($\beta ,\bar\beta\geqslant 1$). The remaining nonpolynomial dependence is then encoded  by ratios of hypergeometric functions and their second derivatives. As it happens, an application of various identities for hypergeometric functions (\cf\ \cite{A&S}) allows us to express this nonpolynomial quantity in terms of the same function $K_\beta$, so we utilize the same approximation of \eqref{Kapprox} and find
%%%%%%
\begin{equation}
\frac{ {}_2F_1\left( \beta-1,\beta+1,2\beta,\frac{1}{2}\right) }{ {}_2F_1\left( \beta+1,\beta+1,2\beta,\frac{1}{2}\right) } = \frac{1+4(\beta -1)  K_\beta }{4+8(\beta -1)  K_\beta } \simeq \frac{D_M(\beta)+4(\beta -1)  N_M(\beta) }{4D_M(\beta)+8(\beta -1)  N_M(\beta) }~.
\label{K_sc}
\end{equation}
%%%%%%
Here we used \eqref{hygerg_der} to relate the second derivative of the hypergeometric function to a different hypergeometric function. A similar ratio appears for the $\bar \beta$ dependent hypergeometric functions, which we approximate in the same way.
After approximating $K_\beta$ by \eqref{Kapprox} we can again factor out another strictly positive denominator $(4D_M(\beta)+8(\beta -1)  N_M(\beta))(4D_M(\bar \beta)+8(\bar \beta -1)  N_M(\bar \beta))$ for the same range of $\beta,\bar \beta$.

The approximation for the braided superconformal block goes in the same way. (We will now ignore the $\bar \beta$ dependence since it is simply obtained by $\beta \to \bar \beta$ in the discussion below.) We start by noting that braiding the block has the following effect on the hypergeometric functions~\cite{A&S}
%%%%%%
\begin{equation}
{}_2F_1\left( \beta-1,\beta+1,2\beta,\frac{z}{z-1}\right)=(1-z)^{\beta-1} \; {}_2F_1\left( \beta-1,\beta-1,2\beta,z\right)~.
\end{equation}
%%%%%%
The next step is now to write all derivatives in terms of the zeroth and second derivatives of the hypergeometric function by means of \eqref{hyperg_rec_rel} with $a=1$, $b=1$. We can then again factor out any nonnegative and nonpolynomial terms, beginning with $z^\beta {}_2F_1(\beta -1,\beta - 1, 2\beta ,\frac{1}{2}) (\beta -1) \beta $ which is strictly positive for $\beta \geqslant 1$. The residual non-polynomial dependence is then given by
%%%%%%
\begin{equation}
\frac{ {}_2F_1\left( \beta-1,\beta+1,2\beta,\frac{1}{2}\right) }{ {}_2F_1\left( \beta-1,\beta-1,2\beta,\frac{1}{2}\right) } = 4\frac{ {}_2F_1\left( \beta-1,\beta+1,2\beta,\frac{1}{2}\right) }{ {}_2F_1\left( \beta+1,\beta+1,2\beta,\frac{1}{2}\right) } \simeq 4 \frac{D_M(\beta)+4(\beta -1)  N_M(\beta) }{4D_M(\beta)+8(\beta -1) N_M(\beta)}~,
\end{equation}
%%%%%%
where we have rewritten, through \eqref{hygerg_der}, the second derivative of the hypergeometric function as $z^{\beta-2} {}_2F_1\left( \beta-1,\beta+1,2\beta,1/2\right)$,  and used several hypergeometric identities. For the relevant range of $\beta$ the denominator in the above equation is strictly positive, and it is the final term to be factored out.

The ratio $r_j/s_j$ tends to one extremely fast, and we observed that truncating the product in \eqref{Kapprox} at $M = 4$ was already accurate enough for all $\Lambda \leqslant 22$. For $22 < \Lambda \leqslant 30$ we found that $M=5$ was sufficient. In a number of cases we repeated the numerical analysis with $M=6$ and verified that there was no change in the results.
%!TEX root = ../draft_maxi_Neq2.tex

\section{Exact OPE coefficients for the \texorpdfstring{$\NN=2$}{N=2} chiral ring}
\label{App:exactopes}

The OPE coefficients of Coulomb branch chiral ring operators in four-dimensional $\NN=2$ SCFTs satisfy four-dimensional $tt^\star$ equations \cite{Baggio:2014sna,Baggio:2014ioa}. In this appendix we limit our attention to the case of theories with a conformal manifold that has one complex dimension, \ie, theories with just a single $\EE_2$ multiplet. In such cases there is a close connection between the chiral ring OPE coefficients and the Zamolodchikov metric on the conformal manifold. After diagonalization of the fields, the OPE of the (unit normalized) chiral operators takes the form
%%%%%%
\begin{equation}
\EE_2(x)\EE_2(0) = \lambda_{\EE_4}\EE_4(0) + \ldots~,
\end{equation}
%%%%%%
and we are interested in the squared OPE coefficient $\lambda_{\EE_4}^2$. Precisely this coefficient is part of a solvable subsector of the $tt^*$ equations and it takes the form
%%%%%%
\begin{equation}
\lambda^2_{\EE_4}= 2 + \frac{\del_\tau \del_{\bar \tau} \log(g_{\tau \bar \tau})}{g_{\tau \bar \tau}} = 2 - \frac{1}{2} R[g_{\tau \bar \tau}]~,
\end{equation}
%%%%%%
where $g_{\tau \bar \tau}$ is the only nonvanishing component of the Zamolodchikov metric on the conformal manifold.\footnote{In the notations of \cite{Baggio:2014ioa}, this is the metric written as $g_{i \bar j}$. This differs from the true Zamolodchikov metric $G_{i\bar j}$ by a factor of $192$.} On the right-hand side we recognize the expression for the scalar curvature of the Zamolodchikov metric. The bounds reported in Section \ref{sec:eps_results} for $\lambda^2_{\EE_4}$ therefore provide lower and upper bounds on this curvature.

Let us consider a few examples, starting with the theory of $n$ free vector multiplets. The superconformal primary of the flavor singlet $\EE_2$ multiplet in this theory is $\varphi_a \varphi_a(x)$, with $\varphi(x)$ the scalar operator in the vector multiplet. We can compute $\lambda^2_{\EE_4}$ directly by performing Wick contractions, whereupon we find
%%%%%%
\begin{equation}
\makebox[1.85in][l]{$n$ free vector multiplets:}\makebox[1.5in][l]{$\lambda^2_{\EE_4} = 2 + \frac{4}{n} = 2 + \frac{2}{3 c}~.$}
\end{equation}
%%%%%%
In the last equality we have used the precise value of the central charge in this theory: $c = n/6$. In any $\NN=2$ superconformal gauge theory with gauge group $G$, the tree-level value for this OPE coefficient takes the same form,
%%%%%%
\begin{equation}
\makebox[1.85in][l]{tree level gauge theory:}\makebox[1.5in][l]{$\lambda^2_{\EE_4} = 2 + \frac{4}{\dim(G)} \geqslant 2 + \frac{2}{3c}~.$}
\end{equation}
%%%%%%
The inequality is a consequence of the fact that the central charge of a superconformal gauge theory is always greater than that of the vector multiplets alone. 

In $\NN=4$ supersymmetric Yang-Mills theory, the central charge is $c = \frac{1}{4} \dim(G)$. In this special case, extended supersymmetry prevents the OPE coefficient in question from being renormalized. Consequently the exact value (for all values of the complex gauge coupling) is given by the tree-level result,
%%%%%%
\begin{equation}
\makebox[1.85in][l]{$\NN=4$ super Yang-Mills:}\makebox[1.5in][l]{$\lambda^2_{\EE_4} = 2 + \frac{1}{c}~.$}
\end{equation}
%%%%%%

In many $\NN=2$ SCFTs, this OPE coefficient is made accessible by the relation between the K\"ahler metric on the conformal manifold and the $S^4$ partition function \cite{Gerchkovitz:2014gta},
%%%%%%
\begin{equation}
g_{i \bar j} = \del_i \del_{\bar k} \log(Z_{S^4})~.
\end{equation}
%%%%%%
It is frequently the case that the partition function $Z_{S^4}$ can be computed exactly using supersymmetric localization \cite{Pestun:2007rz}. As an example, consider $\NN=2$ SCQCD with $N_f = 4$ flavors (sometimes referred to in the text as \emph{the $\sof(8)$ theory}). The Nekrasov instanton partition function that features in the localization result is related to four-point Virasoro conformal blocks \cite{Alday:2009aq}. These in turn are efficiently computed using the recursion relations developed in \cite{zamolodchikov1987conformal}. Altogether, one ultimately finds the following expression for the $S^4$ partition function,
%a%%%%%
\begin{equation}
\log Z_{S^4}(q) = \log\left(\int_{-\infty}^\infty da\,a^2|16q|^{2a^2}\left|\frac{G(1+2ia)^2}{G(1+ia)^8}\right|^2 H(a,q)H(a,\bar q)\right)+f(\tau)+f(\bar\tau)~,
\end{equation}
%%%%%%
where the functions $f(\tau)$ are K\"ahler transformations that drop out in the computation of the curvature, and $G(z)$ is Barnes' $G$-function.\footnote{This function is implemented in \texttt{Mathematica} as \texttt{BarnesG[z]}.} The function $H(a,q)$ has been defined in \cite{zamolodchikov1987conformal} by means of a somewhat intricate recursion relation that we will not review here. It is a building block of the Virasoro four-point conformal block with $c=25$, all four external dimensions equal to one, and internal dimension equal to $1 + a^2$. The first few terms in its series expansion take the form
%%%%%%
\begin{equation}
H(a,q) = 1+ \frac{12 \left(a^2+2\right) q^2}{\left(4 a^2+9\right)^2}+\frac{18 \left(32 a^6+308 a^4+955 a^2+940\right) q^4}{\left(4 a^2+9\right)^2 \left(4 a^2+25\right)^2} + \cdots~.
\end{equation}
%%%%%%
One should note that the parameter $q$ is $q_{\rm IR}$ which is \emph{not} the same parameter as the parameter $q_{\rm UV}$ used in \cite{Pestun:2007rz} and in \cite{Baggio:2014sna,Baggio:2014ioa}.\footnote{An early discussion of this point can be found in \cite{Dorey:1996hu}.} The relation between the two is given in \cite{Alday:2009aq}, and also in \cite{zamolodchikov1987conformal},
%%%%%%
\begin{equation}
q_{\rm IR} = \exp(i \pi \tau_{\rm IR}) = \exp\left( - \pi \frac{K(1-q_{\rm UV})}{K(q_{\rm UV})}\right)~.
\end{equation}
%%%%%%
Here $K(m)$ is the complete elliptic integral of the first kind.\footnote{This function is implemented in \texttt{Mathematica} as \texttt{EllipticK[m]}.} The explicit form of this transformation is in fact not particularly relevant for our purposes because the scalar curvature is a diffeomorphism invariant. But it is $\tau_{\rm IR}$ that is valued in the fundamental domain for the action of $SL(2,\Zb)$ on the upper half plane. Namely, under $S$- and $T$-transformations we have 
%%%%%%
\begin{alignat}{3}
&T:&\qquad&		\tau_{\rm IR} \to - 1 /\tau_{\rm IR}~,	&\qquad		&	q_{\rm UV} \to 1 - q_{\rm UV}~,\\
&S:&\qquad&		\tau_{\rm IR} \to \tau_{\rm IR}+1~,		&\qquad		&	q_{\rm UV} \to \frac{q_{\rm UV}}{q_{\rm UV}-1}~.
\end{alignat}
%%%%%%
The transformations of $q_{\rm UV}$ describe the action of crossing symmetry on the Liouville four-point function.	

%%%%%%
\begin{figure}[t!]
    \begin{center}           
        \includegraphics[height=2.5in]{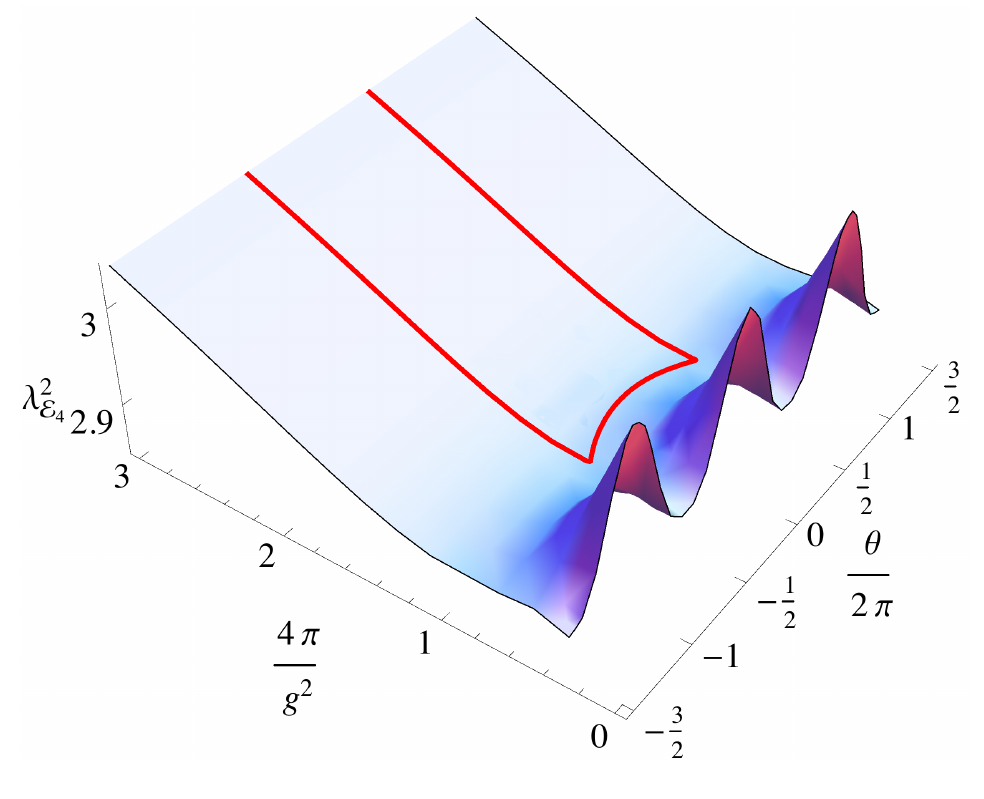}
        \caption{\label{Fig:exactlambda}The value of $\lambda^2_{\EE_4}$ for $\NN=2$ SQCD with $N_f=4$ flavors. The coupling is shown as a function of the exactly marginal complexified gauge coupling $\tau = \frac{\theta}{2\pi} + \frac{4 \pi i}{g^2}$, and the fundamental domain for the action of $SL(2,\Zb)$-duality on the coupling plane is outlined in red.}
    \end{center}
\end{figure}
%%%%%%

The value of the OPE coefficient $\lambda^2_{\EE_4}(\tau)$ can be computed numerically to arbitrary accuracy at any value of the coupling. The free-field value is given by $\lambda_{\EE_4}^2(\tau=\infty)=10/3$. The OPE coefficient decreases monotonically as a function of the gauge coupling and becomes stationary at the self-dual points. To get reasonable accuracy we need to expand $H(q)$ to order $q^8$, resulting in the following stationary values:
%%%%%%
\begin{equation}
\lambda^2_{\EE_4}(\tau = i) = 2.8983769\ldots \qquad \qquad \lambda^2_{\EE_4}(\tau = e^{i \pi /3}) = 2.8940994\ldots
\end{equation}
%%%%%%
This OPE coefficient is plotted in Fig.~\ref{Fig:exactlambda}. The stationary point at $\tau=i$ is a saddle point, while the global minimum occurs at $\tau = e^{i \pi/3}$, so the range for this OPE coefficient is given by
%%%%%%
\begin{equation}
2.8940994\ldots \leqslant \lambda_{\EE_4}^2(\tau) \leqslant \frac{10}{3}~.
\end{equation}
%%%%%%
This is the range of values that appear in Fig.~\ref{Fig:E4bound} of Section \ref{sec:eps_results}.

\bibliography{./aux/biblio}

\providecommand{\href}[2]{#2}\begingroup\raggedright\begin{thebibliography}{100}

\bibitem{Rattazzi:2008pe}
R.~Rattazzi, V.~S. Rychkov, E.~Tonni, and A.~Vichi, {\it {Bounding scalar
  operator dimensions in 4D CFT}},  {\em JHEP} {\bf 0812} (2008) 031,
  [\href{http://arxiv.org/abs/0807.0004}{{\tt arXiv:0807.0004}}].

\bibitem{Ferrara:1972cq}
S.~Ferrara, A.~Grillo, and R.~Gatto, {\it {Manifestly conformal-covariant
  expansion on the light cone}},  {\em Phys.Rev.} {\bf D5} (1972) 3102--3108.

\bibitem{Ferrara:1973vz}
S.~Ferrara, A.~Grillo, G.~Parisi, and R.~Gatto, {\it {Covariant expansion of
  the conformal four-point function}},  {\em Nucl.Phys.} {\bf B49} (1972)
  77--98.

\bibitem{Ferrara:1973yt}
S.~Ferrara, A.~Grillo, and R.~Gatto, {\it {Tensor representations of conformal
  algebra and conformally covariant operator product expansion}},  {\em Annals
  Phys.} {\bf 76} (1973) 161--188.

\bibitem{Ferrara:1974pt}
S.~Ferrara, R.~Gatto, and A.~Grillo, {\it {Positivity Restrictions on Anomalous
  Dimensions}},  {\em Phys.Rev.} {\bf D9} (1974) 3564.

\bibitem{Ferrara:1974nf}
S.~Ferrara, A.~Grillo, R.~Gatto, and G.~Parisi, {\it {Analyticity properties
  and asymptotic expansions of conformal covariant green's functions}},  {\em
  Nuovo Cim.} {\bf A19} (1974) 667--695.

\bibitem{Ferrara:1974ny}
S.~Ferrara, R.~Gatto, and A.~Grillo, {\it {Properties of Partial Wave
  Amplitudes in Conformal Invariant Field Theories}},  {\em Nuovo Cim.} {\bf
  A26} (1975) 226.

\bibitem{Polyakov:1974gs}
A.~Polyakov, {\it {Nonhamiltonian approach to conformal quantum field theory}},
   {\em Zh.Eksp.Teor.Fiz.} {\bf 66} (1974) 23--42.

\bibitem{yuji}
L.~Bhardwaj and Y.~Tachikawa, {\it {Classification of 4d N=2 gauge theories}},
  {\em JHEP} {\bf 1312} (2013) 100, [\href{http://arxiv.org/abs/1309.5160}{{\tt
  arXiv:1309.5160}}].

\bibitem{Gaiotto:2009we}
D.~Gaiotto, {\it {N=2 dualities}},  {\em JHEP} {\bf 1208} (2012) 034,
  [\href{http://arxiv.org/abs/0904.2715}{{\tt arXiv:0904.2715}}].

\bibitem{Gaiotto:2009hg}
D.~Gaiotto, G.~W. Moore, and A.~Neitzke, {\it {Wall-crossing, Hitchin Systems,
  and the WKB Approximation}},  \href{http://arxiv.org/abs/0907.3987}{{\tt
  arXiv:0907.3987}}.

\bibitem{ElShowk:2012ht}
S.~El-Showk, M.~F. Paulos, D.~Poland, S.~Rychkov, D.~Simmons-Duffin, and
  A.~Vichi, {\it {Solving the 3D Ising Model with the Conformal Bootstrap}},
  {\em Phys. Rev.} {\bf D86} (2012) 025022,
  [\href{http://arxiv.org/abs/1203.6064}{{\tt arXiv:1203.6064}}].

\bibitem{El-Showk:2014dwa}
S.~El-Showk, M.~F. Paulos, D.~Poland, S.~Rychkov, D.~Simmons-Duffin, and
  A.~Vichi, {\it {Solving the 3d Ising Model with the Conformal Bootstrap II.
  c-Minimization and Precise Critical Exponents}},  {\em J. Stat. Phys.} {\bf
  157} (2014) 869, [\href{http://arxiv.org/abs/1403.4545}{{\tt
  arXiv:1403.4545}}].

\bibitem{Kos:2014bka}
F.~Kos, D.~Poland, and D.~Simmons-Duffin, {\it {Bootstrapping Mixed Correlators
  in the 3D Ising Model}},  \href{http://arxiv.org/abs/1406.4858}{{\tt
  arXiv:1406.4858}}.

\bibitem{Hitchin:1986ea}
N.~J. Hitchin, A.~Karlhede, U.~Lindstrom, and M.~Rocek, {\it {Hyperkahler
  Metrics and Supersymmetry}},  {\em Commun.Math.Phys.} {\bf 108} (1987) 535.

\bibitem{Argyres:2007cn}
P.~C. Argyres and N.~Seiberg, {\it {S-duality in N=2 supersymmetric gauge
  theories}},  {\em JHEP} {\bf 0712} (2007) 088,
  [\href{http://arxiv.org/abs/0711.0054}{{\tt arXiv:0711.0054}}].

\bibitem{Argyres:1995jj}
P.~C. Argyres and M.~R. Douglas, {\it {New phenomena in SU(3) supersymmetric
  gauge theory}},  {\em Nucl.Phys.} {\bf B448} (1995) 93--126,
  [\href{http://arxiv.org/abs/hep-th/9505062}{{\tt hep-th/9505062}}].

\bibitem{Xie:2012hs}
D.~Xie, {\it {General Argyres-Douglas Theory}},  {\em JHEP} {\bf 1301} (2013)
  100, [\href{http://arxiv.org/abs/1204.2270}{{\tt arXiv:1204.2270}}].

\bibitem{Buican:2014hfa}
M.~Buican, S.~Giacomelli, T.~Nishinaka, and C.~Papageorgakis, {\it
  {Argyres-Douglas Theories and S-Duality}},
  \href{http://arxiv.org/abs/1411.6026}{{\tt arXiv:1411.6026}}.

\bibitem{slavalectures}
S.~Rychkov, ``{EPFL Lectures on Conformal Field Theory in $D \geqslant 3$
  Dimensions}.'' \url{https://sites.google.com/site/slavarychkov/home},
  December 2012.

\bibitem{Pappadopulo:2012jk}
D.~Pappadopulo, S.~Rychkov, J.~Espin, and R.~Rattazzi, {\it {OPE Convergence in
  Conformal Field Theory}},  {\em Phys.Rev.} {\bf D86} (2012) 105043,
  [\href{http://arxiv.org/abs/1208.6449}{{\tt arXiv:1208.6449}}].

\bibitem{Maldacena:2011jn}
J.~Maldacena and A.~Zhiboedov, {\it {Constraining Conformal Field Theories with
  A Higher Spin Symmetry}},  {\em J.Phys.} {\bf A46} (2013) 214011,
  [\href{http://arxiv.org/abs/1112.1016}{{\tt arXiv:1112.1016}}].

\bibitem{Alba:2013yda}
V.~Alba and K.~Diab, {\it {Constraining conformal field theories with a higher
  spin symmetry in d=4}},  \href{http://arxiv.org/abs/1307.8092}{{\tt
  arXiv:1307.8092}}.

\bibitem{Cardy:1988cwa}
J.~L. Cardy, {\it {Is There a c Theorem in Four-Dimensions?}},  {\em
  Phys.Lett.} {\bf B215} (1988) 749--752.

\bibitem{Komargodski:2011vj}
Z.~Komargodski and A.~Schwimmer, {\it {On Renormalization Group Flows in Four
  Dimensions}},  {\em JHEP} {\bf 1112} (2011) 099,
  [\href{http://arxiv.org/abs/1107.3987}{{\tt arXiv:1107.3987}}].

\bibitem{Dymarsky:2013wla}
A.~Dymarsky, {\it {On the four-point function of the stress-energy tensors in a
  CFT}},  \href{http://arxiv.org/abs/1311.4546}{{\tt arXiv:1311.4546}}.

\bibitem{Hofman:2008ar}
D.~M. Hofman and J.~Maldacena, {\it {Conformal collider physics: Energy and
  charge correlations}},  {\em JHEP} {\bf 0805} (2008) 012,
  [\href{http://arxiv.org/abs/0803.1467}{{\tt arXiv:0803.1467}}].

\bibitem{Kulaxizi:2010jt}
M.~Kulaxizi and A.~Parnachev, {\it {Energy Flux Positivity and Unitarity in
  CFTs}},  {\em Phys.Rev.Lett.} {\bf 106} (2011) 011601,
  [\href{http://arxiv.org/abs/1007.0553}{{\tt arXiv:1007.0553}}].

\bibitem{Zhiboedov:2013opa}
A.~Zhiboedov, {\it {On Conformal Field Theories With Extremal a/c Values}},
  {\em JHEP} {\bf 1404} (2014) 038, [\href{http://arxiv.org/abs/1304.6075}{{\tt
  arXiv:1304.6075}}].

\bibitem{Erdmenger:1996yc}
J.~Erdmenger and H.~Osborn, {\it {Conserved currents and the energy momentum
  tensor in conformally invariant theories for general dimensions}},  {\em
  Nucl.Phys.} {\bf B483} (1997) 431--474,
  [\href{http://arxiv.org/abs/hep-th/9605009}{{\tt hep-th/9605009}}].

\bibitem{Anselmi:1997rd}
D.~Anselmi, {\it {Central functions and their physical implications}},  {\em
  JHEP} {\bf 9805} (1998) 005, [\href{http://arxiv.org/abs/hep-th/9702056}{{\tt
  hep-th/9702056}}].

\bibitem{Rychkov:2009ij}
V.~S. Rychkov and A.~Vichi, {\it {Universal Constraints on Conformal Operator
  Dimensions}},  {\em Phys.Rev.} {\bf D80} (2009) 045006,
  [\href{http://arxiv.org/abs/0905.2211}{{\tt arXiv:0905.2211}}].

\bibitem{Vichi:2009zz}
A.~Vichi, {\it {Anomalous dimensions of scalar operators in CFT}},  {\em
  Nucl.Phys.Proc.Suppl.} {\bf 192-193} (2009) 197--198.

\bibitem{Caracciolo:2009bx}
F.~Caracciolo and V.~S. Rychkov, {\it {Rigorous Limits on the Interaction
  Strength in Quantum Field Theory}},  {\em Phys.Rev.} {\bf D81} (2010) 085037,
  [\href{http://arxiv.org/abs/0912.2726}{{\tt arXiv:0912.2726}}].

\bibitem{Poland:2010wg}
D.~Poland and D.~Simmons-Duffin, {\it {Bounds on 4D Conformal and
  Superconformal Field Theories}},  {\em JHEP} {\bf 1105} (2011) 017,
  [\href{http://arxiv.org/abs/1009.2087}{{\tt arXiv:1009.2087}}].

\bibitem{Rattazzi:2010gj}
R.~Rattazzi, S.~Rychkov, and A.~Vichi, {\it {Central Charge Bounds in 4D
  Conformal Field Theory}},  {\em Phys.Rev.} {\bf D83} (2011) 046011,
  [\href{http://arxiv.org/abs/1009.2725}{{\tt arXiv:1009.2725}}].

\bibitem{Rattazzi:2010yc}
R.~Rattazzi, S.~Rychkov, and A.~Vichi, {\it {Bounds in 4D Conformal Field
  Theories with Global Symmetry}},  {\em J.Phys.} {\bf A44} (2011) 035402,
  [\href{http://arxiv.org/abs/1009.5985}{{\tt arXiv:1009.5985}}].

\bibitem{Vichi:2011ux}
A.~Vichi, {\it {Improved bounds for CFT's with global symmetries}},  {\em JHEP}
  {\bf 1201} (2012) 162, [\href{http://arxiv.org/abs/1106.4037}{{\tt
  arXiv:1106.4037}}].

\bibitem{Poland:2011ey}
D.~Poland, D.~Simmons-Duffin, and A.~Vichi, {\it {Carving Out the Space of 4D
  CFTs}},  {\em JHEP} {\bf 1205} (2012) 110,
  [\href{http://arxiv.org/abs/1109.5176}{{\tt arXiv:1109.5176}}].

\bibitem{Liendo:2012hy}
P.~Liendo, L.~Rastelli, and B.~C. van Rees, {\it {The Bootstrap Program for
  Boundary CFT}},  {\em JHEP} {\bf 1307} (2013) 113,
  [\href{http://arxiv.org/abs/1210.4258}{{\tt arXiv:1210.4258}}].

\bibitem{ElShowk:2012hu}
S.~El-Showk and M.~F. Paulos, {\it {Bootstrapping Conformal Field Theories with
  the Extremal Functional Method}},  {\em Phys.Rev.Lett.} {\bf 111} (2013),
  no.~24 241601, [\href{http://arxiv.org/abs/1211.2810}{{\tt
  arXiv:1211.2810}}].

\bibitem{Beem:2013qxa}
C.~Beem, L.~Rastelli, and B.~C. van Rees, {\it {The N=4 Superconformal
  Bootstrap}},  {\em Phys.Rev.Lett.} {\bf 111} (2013) 071601,
  [\href{http://arxiv.org/abs/1304.1803}{{\tt arXiv:1304.1803}}].

\bibitem{Gliozzi:2013ysa}
F.~Gliozzi, {\it {More constraining conformal bootstrap}},  {\em
  Phys.Rev.Lett.} {\bf 111} (2013) 161602,
  [\href{http://arxiv.org/abs/1307.3111}{{\tt arXiv:1307.3111}}].

\bibitem{Kos:2013tga}
F.~Kos, D.~Poland, and D.~Simmons-Duffin, {\it {Bootstrapping the $O(N)$ vector
  models}},  {\em JHEP} {\bf 1406} (2014) 091,
  [\href{http://arxiv.org/abs/1307.6856}{{\tt arXiv:1307.6856}}].

\bibitem{El-Showk:2013nia}
S.~El-Showk, M.~Paulos, D.~Poland, S.~Rychkov, D.~Simmons-Duffin, and A.~Vichi,
  {\it {Conformal Field Theories in Fractional Dimensions}},  {\em Phys. Rev.
  Lett.} {\bf 112} (2014) 141601, [\href{http://arxiv.org/abs/1309.5089}{{\tt
  arXiv:1309.5089}}].

\bibitem{Alday:2013opa}
L.~F. Alday and A.~Bissi, {\it {The superconformal bootstrap for structure
  constants}},  {\em JHEP} {\bf 1409} (2014) 144,
  [\href{http://arxiv.org/abs/1310.3757}{{\tt arXiv:1310.3757}}].

\bibitem{Gaiotto:2013nva}
D.~Gaiotto, D.~Mazac, and M.~F. Paulos, {\it {Bootstrapping the 3d Ising twist
  defect}},  {\em JHEP} {\bf 1403} (2014) 100,
  [\href{http://arxiv.org/abs/1310.5078}{{\tt arXiv:1310.5078}}].

\bibitem{Berkooz:2014yda}
M.~Berkooz, R.~Yacoby, and A.~Zait, {\it {Bounds on $ \mathcal{N} $ = 1
  superconformal theories with global symmetries}},  {\em JHEP} {\bf 1408}
  (2014) 008, [\href{http://arxiv.org/abs/1402.6068}{{\tt arXiv:1402.6068}}].

\bibitem{Gliozzi:2014jsa}
F.~Gliozzi and A.~Rago, {\it {Critical exponents of the 3d Ising and related
  models from Conformal Bootstrap}},  {\em JHEP} {\bf 1410} (2014) 42,
  [\href{http://arxiv.org/abs/1403.6003}{{\tt arXiv:1403.6003}}].

\bibitem{Nakayama:2014lva}
Y.~Nakayama and T.~Ohtsuki, {\it {Approaching the conformal window of
  $O(n)\times O(m)$ symmetric Landau-Ginzburg models using the conformal
  bootstrap}},  {\em Phys.Rev.} {\bf D89} (2014), no.~12 126009,
  [\href{http://arxiv.org/abs/1404.0489}{{\tt arXiv:1404.0489}}].

\bibitem{Nakayama:2014yia}
Y.~Nakayama and T.~Ohtsuki, {\it {Five dimensional $O(N)$-symmetric CFTs from
  conformal bootstrap}},  {\em Phys.Lett.} {\bf B734} (2014) 193--197,
  [\href{http://arxiv.org/abs/1404.5201}{{\tt arXiv:1404.5201}}].

\bibitem{Alday:2014qfa}
L.~F. Alday and A.~Bissi, {\it {Generalized bootstrap equations for N=4 SCFT}},
   \href{http://arxiv.org/abs/1404.5864}{{\tt arXiv:1404.5864}}.

\bibitem{Chester:2014fya}
S.~M. Chester, J.~Lee, S.~S. Pufu, and R.~Yacoby, {\it {The $ \mathcal{N}=8 $
  superconformal bootstrap in three dimensions}},  {\em JHEP} {\bf 1409} (2014)
  143, [\href{http://arxiv.org/abs/1406.4814}{{\tt arXiv:1406.4814}}].

\bibitem{Caracciolo:2014cxa}
F.~Caracciolo, A.~C. Echeverri, B.~von Harling, and M.~Serone, {\it {Bounds on
  OPE Coefficients in 4D Conformal Field Theories}},  {\em JHEP} {\bf 1410}
  (2014) 20, [\href{http://arxiv.org/abs/1406.7845}{{\tt arXiv:1406.7845}}].

\bibitem{Paulos:2014vya}
M.~F. Paulos, {\it {JuliBootS: a hands-on guide to the conformal bootstrap}},
  \href{http://arxiv.org/abs/1412.4127}{{\tt arXiv:1412.4127}}.

\bibitem{Bae:2014hia}
J.-B. Bae and S.-J. Rey, {\it {Conformal Bootstrap Approach to O(N) Fixed
  Points in Five Dimensions}},  \href{http://arxiv.org/abs/1412.6549}{{\tt
  arXiv:1412.6549}}.

\bibitem{Costa:2011mg}
M.~S. Costa, J.~Penedones, D.~Poland, and S.~Rychkov, {\it {Spinning Conformal
  Correlators}},  {\em JHEP} {\bf 1111} (2011) 071,
  [\href{http://arxiv.org/abs/1107.3554}{{\tt arXiv:1107.3554}}].

\bibitem{Dolan:2011dv}
F.~Dolan and H.~Osborn, {\it {Conformal Partial Waves: Further Mathematical
  Results}},  \href{http://arxiv.org/abs/1108.6194}{{\tt arXiv:1108.6194}}.

\bibitem{Costa:2011dw}
M.~S. Costa, J.~Penedones, D.~Poland, and S.~Rychkov, {\it {Spinning Conformal
  Blocks}},  {\em JHEP} {\bf 1111} (2011) 154,
  [\href{http://arxiv.org/abs/1109.6321}{{\tt arXiv:1109.6321}}].

\bibitem{SimmonsDuffin:2012uy}
D.~Simmons-Duffin, {\it {Projectors, Shadows, and Conformal Blocks}},  {\em
  JHEP} {\bf 1404} (2014) 146, [\href{http://arxiv.org/abs/1204.3894}{{\tt
  arXiv:1204.3894}}].

\bibitem{Siegel:2012di}
W.~Siegel, {\it {Embedding versus 6D twistors}},
  \href{http://arxiv.org/abs/1204.5679}{{\tt arXiv:1204.5679}}.

\bibitem{Osborn:2012vt}
H.~Osborn, {\it {Conformal Blocks for Arbitrary Spins in Two Dimensions}},
  {\em Phys.Lett.} {\bf B718} (2012) 169--172,
  [\href{http://arxiv.org/abs/1205.1941}{{\tt arXiv:1205.1941}}].

\bibitem{Hogervorst:2013sma}
M.~Hogervorst and S.~Rychkov, {\it {Radial Coordinates for Conformal Blocks}},
  {\em Phys.Rev.} {\bf D87} (2013) 106004,
  [\href{http://arxiv.org/abs/1303.1111}{{\tt arXiv:1303.1111}}].

\bibitem{Fitzpatrick:2013sya}
A.~L. Fitzpatrick, J.~Kaplan, and D.~Poland, {\it {Conformal Blocks in the
  Large $D$ Limit}},  {\em JHEP} {\bf 1308} (2013) 107,
  [\href{http://arxiv.org/abs/1305.0004}{{\tt arXiv:1305.0004}}].

\bibitem{Hogervorst:2013kva}
M.~Hogervorst, H.~Osborn, and S.~Rychkov, {\it {Diagonal Limit for Conformal
  Blocks in $d$ Dimensions}},  {\em JHEP} {\bf 1308} (2013) 014,
  [\href{http://arxiv.org/abs/1305.1321}{{\tt arXiv:1305.1321}}].

\bibitem{Fitzpatrick:2014oza}
A.~L. Fitzpatrick, J.~Kaplan, Z.~U. Khandker, D.~Li, D.~Poland, and
  D.~Simmons-Duffin, {\it {Covariant Approaches to Superconformal Blocks}},
  {\em JHEP} {\bf 08} (2014) 129, [\href{http://arxiv.org/abs/1402.1167}{{\tt
  arXiv:1402.1167}}].

\bibitem{Khandker:2014mpa}
Z.~U. Khandker, D.~Li, D.~Poland, and D.~Simmons-Duffin, {\it {$ \mathcal{N} $
  = 1 superconformal blocks for general scalar operators}},  {\em JHEP} {\bf
  1408} (2014) 049, [\href{http://arxiv.org/abs/1404.5300}{{\tt
  arXiv:1404.5300}}].

\bibitem{Elkhidir:2014woa}
E.~Elkhidir, D.~Karateev, and M.~Serone, {\it {General Three-Point Functions in
  4D CFT}},  \href{http://arxiv.org/abs/1412.1796}{{\tt arXiv:1412.1796}}.

\bibitem{Costa:2014rya}
M.~S. Costa and T.~Hansen, {\it {Conformal correlators of mixed-symmetry
  tensors}},  \href{http://arxiv.org/abs/1411.7351}{{\tt arXiv:1411.7351}}.

\bibitem{Fitzpatrick:2012yx}
A.~L. Fitzpatrick, J.~Kaplan, D.~Poland, and D.~Simmons-Duffin, {\it {The
  Analytic Bootstrap and AdS Superhorizon Locality}},  {\em JHEP} {\bf 1312}
  (2013) 004, [\href{http://arxiv.org/abs/1212.3616}{{\tt arXiv:1212.3616}}].

\bibitem{Komargodski:2012ek}
Z.~Komargodski and A.~Zhiboedov, {\it {Convexity and Liberation at Large
  Spin}},  {\em JHEP} {\bf 1311} (2013) 140,
  [\href{http://arxiv.org/abs/1212.4103}{{\tt arXiv:1212.4103}}].

\bibitem{Alday:2013cwa}
L.~F. Alday and A.~Bissi, {\it {Higher-spin correlators}},  {\em JHEP} {\bf
  1310} (2013) 202, [\href{http://arxiv.org/abs/1305.4604}{{\tt
  arXiv:1305.4604}}].

\bibitem{Fitzpatrick:2014vua}
A.~L. Fitzpatrick, J.~Kaplan, and M.~T. Walters, {\it {Universality of
  Long-Distance AdS Physics from the CFT Bootstrap}},  {\em JHEP} {\bf 1408}
  (2014) 145, [\href{http://arxiv.org/abs/1403.6829}{{\tt arXiv:1403.6829}}].

\bibitem{Vos:2014pqa}
G.~Vos, {\it {Generalized Additivity in Unitary Conformal Field Theories}},
  \href{http://arxiv.org/abs/1411.7941}{{\tt arXiv:1411.7941}}.

\bibitem{Dobrev:1985qv}
V.~Dobrev and V.~Petkova, {\it {All Positive Energy Unitary Irreducible
  Representations of Extended Conformal Supersymmetry}},  {\em Phys.Lett.} {\bf
  B162} (1985) 127--132.

\bibitem{Dolan:2002zh}
F.~Dolan and H.~Osborn, {\it {On short and semi-short representations for
  four-dimensional superconformal symmetry}},  {\em Annals Phys.} {\bf 307}
  (2003) 41--89, [\href{http://arxiv.org/abs/hep-th/0209056}{{\tt
  hep-th/0209056}}].

\bibitem{Kinney:2005ej}
J.~Kinney, J.~M. Maldacena, S.~Minwalla, and S.~Raju, {\it {An Index for 4
  dimensional super conformal theories}},  {\em Commun.Math.Phys.} {\bf 275}
  (2007) 209--254, [\href{http://arxiv.org/abs/hep-th/0510251}{{\tt
  hep-th/0510251}}].

\bibitem{Buican:2014qla}
M.~Buican, T.~Nishinaka, and C.~Papageorgakis, {\it {Constraints on Chiral
  Operators in N=2 SCFTs}},  \href{http://arxiv.org/abs/1407.2835}{{\tt
  arXiv:1407.2835}}.

\bibitem{Kuzenko:1999pi}
S.~M. Kuzenko and S.~Theisen, {\it {Correlation functions of conserved currents
  in N=2 superconformal theory}},  {\em Class.Quant.Grav.} {\bf 17} (2000)
  665--696, [\href{http://arxiv.org/abs/hep-th/9907107}{{\tt hep-th/9907107}}].

\bibitem{Tachikawa:2013kta}
Y.~Tachikawa, {\it {N=2 supersymmetric dynamics for pedestrians}},  {\em
  Lect.Notes Phys.} {\bf 890} (2013) 2014,
  [\href{http://arxiv.org/abs/1312.2684}{{\tt arXiv:1312.2684}}].

\bibitem{Cecotti:2010fi}
S.~Cecotti, A.~Neitzke, and C.~Vafa, {\it {R-Twisting and 4d/2d
  Correspondences}},  \href{http://arxiv.org/abs/1006.3435}{{\tt
  arXiv:1006.3435}}.

\bibitem{Shapere:2008zf}
A.~D. Shapere and Y.~Tachikawa, {\it {Central charges of N=2 superconformal
  field theories in four dimensions}},  {\em JHEP} {\bf 0809} (2008) 109,
  [\href{http://arxiv.org/abs/0804.1957}{{\tt arXiv:0804.1957}}].

\bibitem{Buican:2014sfa}
M.~Buican and T.~Nishinaka, {\it {Compact Conformal Manifolds}},
  \href{http://arxiv.org/abs/1410.3006}{{\tt arXiv:1410.3006}}.

\bibitem{Beem:2013sza}
C.~Beem, M.~Lemos, P.~Liendo, W.~Peelaers, L.~Rastelli, and B.~C. van Rees,
  {\it {Infinite Chiral Symmetry in Four Dimensions}},  {\em Commun. Math.
  Phys.} {\bf 336} (2015), no.~3 1359--1433,
  [\href{http://arxiv.org/abs/1312.5344}{{\tt arXiv:1312.5344}}].

\bibitem{Beem:2014rza}
C.~Beem, W.~Peelaers, L.~Rastelli, and B.~C. van Rees, {\it {Chiral algebras of
  class S}},  \href{http://arxiv.org/abs/1408.6522}{{\tt arXiv:1408.6522}}.

\bibitem{Lemos:2014lua}
M.~Lemos and W.~Peelaers, {\it {Chiral Algebras for Trinion Theories}},
  \href{http://arxiv.org/abs/1411.3252}{{\tt arXiv:1411.3252}}.

\bibitem{Sen:1996vd}
A.~Sen, {\it {F theory and orientifolds}},  {\em Nucl.Phys.} {\bf B475} (1996)
  562--578, [\href{http://arxiv.org/abs/hep-th/9605150}{{\tt hep-th/9605150}}].

\bibitem{Banks:1996nj}
T.~Banks, M.~R. Douglas, and N.~Seiberg, {\it {Probing F theory with branes}},
  {\em Phys.Lett.} {\bf B387} (1996) 278--281,
  [\href{http://arxiv.org/abs/hep-th/9605199}{{\tt hep-th/9605199}}].

\bibitem{Dasgupta:1996ij}
K.~Dasgupta and S.~Mukhi, {\it {F theory at constant coupling}},  {\em
  Phys.Lett.} {\bf B385} (1996) 125--131,
  [\href{http://arxiv.org/abs/hep-th/9606044}{{\tt hep-th/9606044}}].

\bibitem{Minahan:1996fg}
J.~A. Minahan and D.~Nemeschansky, {\it {An N=2 superconformal fixed point with
  E(6) global symmetry}},  {\em Nucl.Phys.} {\bf B482} (1996) 142--152,
  [\href{http://arxiv.org/abs/hep-th/9608047}{{\tt hep-th/9608047}}].

\bibitem{Minahan:1996cj}
J.~A. Minahan and D.~Nemeschansky, {\it {Superconformal fixed points with E(n)
  global symmetry}},  {\em Nucl.Phys.} {\bf B489} (1997) 24--46,
  [\href{http://arxiv.org/abs/hep-th/9610076}{{\tt hep-th/9610076}}].

\bibitem{Aharony:1998xz}
O.~Aharony, A.~Fayyazuddin, and J.~M. Maldacena, {\it {The Large N limit of
  N=2, N=1 field theories from three-branes in F theory}},  {\em JHEP} {\bf
  9807} (1998) 013, [\href{http://arxiv.org/abs/hep-th/9806159}{{\tt
  hep-th/9806159}}].

\bibitem{Bonelli:2011aa}
G.~Bonelli, K.~Maruyoshi, and A.~Tanzini, {\it {Wild Quiver Gauge Theories}},
  {\em JHEP} {\bf 1202} (2012) 031, [\href{http://arxiv.org/abs/1112.1691}{{\tt
  arXiv:1112.1691}}].

\bibitem{Gaiotto:2012sf}
D.~Gaiotto and J.~Teschner, {\it {Irregular singularities in Liouville theory
  and Argyres-Douglas type gauge theories, I}},  {\em JHEP} {\bf 1212} (2012)
  050, [\href{http://arxiv.org/abs/1203.1052}{{\tt arXiv:1203.1052}}].

\bibitem{Benini:2009gi}
F.~Benini, S.~Benvenuti, and Y.~Tachikawa, {\it {Webs of five-branes and N=2
  superconformal field theories}},  {\em JHEP} {\bf 0909} (2009) 052,
  [\href{http://arxiv.org/abs/0906.0359}{{\tt arXiv:0906.0359}}].

\bibitem{Argyres:2007tq}
P.~C. Argyres and J.~R. Wittig, {\it {Infinite coupling duals of N=2 gauge
  theories and new rank 1 superconformal field theories}},  {\em JHEP} {\bf
  0801} (2008) 074, [\href{http://arxiv.org/abs/0712.2028}{{\tt
  arXiv:0712.2028}}].

\bibitem{Chacaltana:2012ch}
O.~Chacaltana, J.~Distler, and Y.~Tachikawa, {\it {Gaiotto Duality for the
  Twisted $A_{2N-1}$ Series}},  \href{http://arxiv.org/abs/1212.3952}{{\tt
  arXiv:1212.3952}}.

\bibitem{Chacaltana:2011ze}
O.~Chacaltana and J.~Distler, {\it {Tinkertoys for the $D_N$ series}},  {\em
  JHEP} {\bf 1302} (2013) 110, [\href{http://arxiv.org/abs/1106.5410}{{\tt
  arXiv:1106.5410}}].

\bibitem{Argyres:2010py}
P.~C. Argyres and J.~Wittig, {\it {Mass deformations of four-dimensional, rank
  1, N=2 superconformal field theories}},  {\em J.Phys.Conf.Ser.} {\bf 462}
  (2013), no.~1 012001, [\href{http://arxiv.org/abs/1007.5026}{{\tt
  arXiv:1007.5026}}].

\bibitem{Argyres}
P.~Argyres, M.~Lotito, Y.~L{\"u}, and M.~Martone, {\it {Geometric constraints
  on the space of N=2 SCFTs I: physical constraints on relevant deformations}},
   \href{http://arxiv.org/abs/1505.04814}{{\tt arXiv:1505.04814}}.

\bibitem{Argyres:2005pp}
P.~C. Argyres, M.~Crescimanno, A.~D. Shapere, and J.~R. Wittig, {\it
  {Classification of N=2 superconformal field theories with two-dimensional
  Coulomb branches}},  \href{http://arxiv.org/abs/hep-th/0504070}{{\tt
  hep-th/0504070}}.

\bibitem{Argyres:2005wx}
P.~C. Argyres and J.~R. Wittig, {\it {Classification of N=2 superconformal
  field theories with two-dimensional Coulomb branches. II.}},
  \href{http://arxiv.org/abs/hep-th/0510226}{{\tt hep-th/0510226}}.

\bibitem{Seiberg:1994rs}
N.~Seiberg and E.~Witten, {\it {Electric - magnetic duality, monopole
  condensation, and confinement in N=2 supersymmetric Yang-Mills theory}},
  {\em Nucl.Phys.} {\bf B426} (1994) 19--52,
  [\href{http://arxiv.org/abs/hep-th/9407087}{{\tt hep-th/9407087}}].

\bibitem{Seiberg:1994aj}
N.~Seiberg and E.~Witten, {\it {Monopoles, duality and chiral symmetry breaking
  in N=2 supersymmetric QCD}},  {\em Nucl.Phys.} {\bf B431} (1994) 484--550,
  [\href{http://arxiv.org/abs/hep-th/9408099}{{\tt hep-th/9408099}}].

\bibitem{Cheung:1997id}
Y.-K.~E. Cheung, O.~J. Ganor, and M.~Krogh, {\it {Correlators of the global
  symmetry currents of 4-D and 6-D superconformal theories}},  {\em Nucl.Phys.}
  {\bf B523} (1998) 171--192, [\href{http://arxiv.org/abs/hep-th/9710053}{{\tt
  hep-th/9710053}}].

\bibitem{Aharony:2007dj}
O.~Aharony and Y.~Tachikawa, {\it {A Holographic computation of the central
  charges of d=4, N=2 SCFTs}},  {\em JHEP} {\bf 0801} (2008) 037,
  [\href{http://arxiv.org/abs/0711.4532}{{\tt arXiv:0711.4532}}].

\bibitem{Dolan:2001tt}
F.~Dolan and H.~Osborn, {\it {Superconformal symmetry, correlation functions
  and the operator product expansion}},  {\em Nucl.Phys.} {\bf B629} (2002)
  3--73, [\href{http://arxiv.org/abs/hep-th/0112251}{{\tt hep-th/0112251}}].

\bibitem{Nirschl:2004pa}
M.~Nirschl and H.~Osborn, {\it {Superconformal Ward identities and their
  solution}},  {\em Nucl.Phys.} {\bf B711} (2005) 409--479,
  [\href{http://arxiv.org/abs/hep-th/0407060}{{\tt hep-th/0407060}}].

\bibitem{Dolan:2004mu}
F.~A. Dolan, L.~Gallot, and E.~Sokatchev, {\it {On four-point functions of
  1/2-BPS operators in general dimensions}},  {\em JHEP} {\bf 0409} (2004) 056,
  [\href{http://arxiv.org/abs/hep-th/0405180}{{\tt hep-th/0405180}}].

\bibitem{Cvitanovic:2008zz}
P.~Cvitanovic, {\em {Group theory: birdtracks, Lie's and exceptional groups}}.
\newblock Princeton Univ. Press, Princeton, NJ, 2008.

\bibitem{Beem:2014kka}
C.~Beem, L.~Rastelli, and B.~C. van Rees, {\it {W Symmetry in six dimensions}},
   \href{http://arxiv.org/abs/1404.1079}{{\tt arXiv:1404.1079}}.

\bibitem{Arutyunov:2001qw}
G.~Arutyunov, B.~Eden, and E.~Sokatchev, {\it {On nonrenormalization and OPE in
  superconformal field theories}},  {\em Nucl.Phys.} {\bf B619} (2001)
  359--372, [\href{http://arxiv.org/abs/hep-th/0105254}{{\tt hep-th/0105254}}].

\bibitem{Chacaltana:2012zy}
O.~Chacaltana, J.~Distler, and Y.~Tachikawa, {\it {Nilpotent orbits and
  codimension-two defects of 6d N=(2,0) theories}},  {\em Int.J.Mod.Phys.} {\bf
  A28} (2013) 1340006, [\href{http://arxiv.org/abs/1203.2930}{{\tt
  arXiv:1203.2930}}].

\bibitem{Baggio:2014sna}
M.~Baggio, V.~Niarchos, and K.~Papadodimas, {\it {Exact correlation functions
  in SU(2) N=2 superconformal QCD}},
  \href{http://arxiv.org/abs/1409.4217}{{\tt arXiv:1409.4217}}.

\bibitem{Baggio:2014ioa}
M.~Baggio, V.~Niarchos, and K.~Papadodimas, {\it {$tt^*$ equations,
  localization and exact chiral rings in 4d N=2 SCFTs}},
  \href{http://arxiv.org/abs/1409.4212}{{\tt arXiv:1409.4212}}.

\bibitem{Baggio:2012rr}
M.~Baggio, J.~de~Boer, and K.~Papadodimas, {\it {A non-renormalization theorem
  for chiral primary 3-point functions}},  {\em JHEP} {\bf 1207} (2012) 137,
  [\href{http://arxiv.org/abs/1203.1036}{{\tt arXiv:1203.1036}}].

\bibitem{Fujisawa08sdpasemidefinite}
K.~Fujisawa, M.~Fukuda, K.~Kobayashi, M.~Kojima, K.~Nakata, M.~Nakata, and
  M.~Yamashita, {\it Sdpa (semidefinite programming algorithm) user's manual --
  version 7.0.5},  tech. rep., 2008.

\bibitem{sdpasite}
\url{http://sdpa.sourceforge.net}.

\bibitem{A&S}
M.~Abramowitz and I.~A. Stegun, {\em Handbook of Mathematical Functions with
  Formulas, Graphs, and Mathematical Tables}.
\newblock Dover, New York, ninth dover printing, tenth gpo printing~ed., 1964.

\bibitem{Gerchkovitz:2014gta}
E.~Gerchkovitz, J.~Gomis, and Z.~Komargodski, {\it {Sphere Partition Functions
  and the Zamolodchikov Metric}},  {\em JHEP} {\bf 1411} (2014) 001,
  [\href{http://arxiv.org/abs/1405.7271}{{\tt arXiv:1405.7271}}].

\bibitem{Pestun:2007rz}
V.~Pestun, {\it {Localization of gauge theory on a four-sphere and
  supersymmetric Wilson loops}},  {\em Commun.Math.Phys.} {\bf 313} (2012)
  71--129, [\href{http://arxiv.org/abs/0712.2824}{{\tt arXiv:0712.2824}}].

\bibitem{Alday:2009aq}
L.~F. Alday, D.~Gaiotto, and Y.~Tachikawa, {\it {Liouville Correlation
  Functions from Four-dimensional Gauge Theories}},  {\em Lett.Math.Phys.} {\bf
  91} (2010) 167--197, [\href{http://arxiv.org/abs/0906.3219}{{\tt
  arXiv:0906.3219}}].

\bibitem{zamolodchikov1987conformal}
A.~B. Zamolodchikov, {\it Conformal symmetry in two-dimensional space:
  recursion representation of conformal block},  {\em Theoretical and
  Mathematical Physics} {\bf 73} (1987), no.~1 1088--1093.

\bibitem{Dorey:1996hu}
N.~Dorey, V.~V. Khoze, and M.~P. Mattis, {\it {Multi-instanton calculus in N=2
  supersymmetric gauge theory}},  {\em Phys.Rev.} {\bf D54} (1996) 2921--2943,
  [\href{http://arxiv.org/abs/hep-th/9603136}{{\tt hep-th/9603136}}].

\end{thebibliography}\endgroup
\bibliographystyle{./aux/JHEP}

\end{document}